\documentclass[12pt]{article}
\usepackage{amssymb}
\usepackage{amsmath}
\usepackage{mathtools}
\usepackage{mathrsfs}
\usepackage{axodraw}
\usepackage{cite}
\usepackage[]{hyperref}
\input xy
\xyoption{all}

\numberwithin{equation}{section}

\begin{document}

	\begin{center}
			
	\vskip 1.5 cm
{\large \bf Two-dimensional supersymmetric gauge theories \\ with exceptional gauge groups }
		\vskip 1 cm 
{Zhuo Chen, Wei Gu, Hadi Parsian and Eric Sharpe}\\
		\vskip 0.5cm
{\sl Department of Physics, MC 0435 \\
850 West Campus Drive \\
Virginia Tech\\
Blacksburg, VA 24061\\  
{\tt zhuo2012@vt.edu}, {\tt weig8@vt.edu},
{\tt varzi61@vt.edu}, {\tt ersharpe@vt.edu}}

	\end{center}

	\vskip 0.5 cm
	\begin{abstract}
We apply the recent 
proposal for mirrors of nonabelian (2,2) supersymmetric 
two-dimensional gauge theories to make predictions for
two-dimensional supersymmetric gauge theories with exceptional gauge groups 
$G_2$, $F_4$, $E_6$, $E_7$, and $E_8$.  
We compute the mirror Landau-Ginzburg models and predict 
excluded Coulomb loci and Coulomb branch relations (quantum cohomology).
We also discuss the relationship between weight lattice normalizations and
theta angle periodicities in the proposal, and explore different
conventions for the mirrors.  Finally, we discuss the behavior of pure
gauge theories with exceptional gauge groups under RG flow, and describe
evidence that any pure supersymmetric two-dimensional gauge theory
with connected and simply-connected semisimple gauge group flows in the
IR to a free theory of as many twisted chiral superfields as the rank of the
gauge group, extending previous results for $SU$, $SO$, and $Sp$ gauge theories.

	\end{abstract}

\begin{flushleft}
August 2018
\end{flushleft}

	\newpage
	
	\tableofcontents
	
	\newpage

\section{Introduction}

Mirror symmetry is a well-known duality of string theory, whose original
form has been extended in a variety of ways.  For two-dimensional abelian
gauged
linear sigma models, constructive proofs and various aspects thereof
were described in \cite{HoriVafa,Morrison:1995yh}.  In particular, the paper
\cite{HoriVafa} gave an explicit construction of a Landau-Ginzburg model
mirror to many abelian gauged linear sigma models.  However, one open problem
for many years has been to find an analogous construction for two-dimensional
supersymmetric nonabelian gauged linear sigma models.

Recently, a proposal was made in \cite{GuSharpe} for mirrors to
two-dimensional supersymmetric nonabelian gauge theories.  Specifically,
it gave a construction of Landau-Ginzburg orbifolds for supersymmetric
nonabelian gauge theories.
That work
checked the proposal against a wide variety of results for
two-dimensional theories with classical gauge groups.  To further
develop the underlying machinery, in this paper we will apply the
proposed mirror construction of \cite{GuSharpe} to two-dimensional
supersymmetric nonabelian gauge theories with the exceptional
gauge groups $G_2$, $F_4$, $E_{6,7,8}$, to make predictions for
excluded loci and Coulomb branch relations (analogues of quantum
cohomology relations).

Working through these computations will also allow us to explore some
properties of those mirror superpotentials, which take the form
\begin{eqnarray}
W & = & \sum_{a=1}^r \sigma_a \left( 
\sum_{i=1}^N \rho_i^a Y_i \: - \: \sum_{\tilde{\mu}=1}^{n-r} \alpha_{\tilde{\mu}}^a 
\ln X_{\tilde{\mu}}
\: - \: t_a \right) 
\nonumber \\
& & \: + \: \sum_{i=1}^N \exp\left(-Y_i\right) \: + \:
\sum_{\tilde{\mu}=1}^{n-r} X_{\tilde{\mu}} .
\end{eqnarray}
In the expression above, $\rho_i^a$ are components of weight vectors for
matter representations of the original gauge theory, and
$\alpha_{\tilde{\mu}}^a$ are root vectors (here taken to form a sublattice
of the weight lattice).  As described in \cite{GuSharpe},
the $\sigma$s encode theta angles in the Cartan subalgebra of the
original gauge theory, and have periodicities
reflecting the weight lattice,
or at least the sublattice generated by the matter representations.
However, the weight lattice need not be normalized in the same way
as a charge lattice.  It is always possible to find a basis for the
weight lattice (in terms of fundamental weights) so that the coefficients
in the $\sigma$ terms are all integers, reflecting $2\pi$ theta angle
periodicities and standard charge lattice conventions, but one can also
consistently work in other bases as well.  For the case of $G_2$ gauge theories,
we will use a naive basis which results in nonstandard theta angle
periodicities and charge lattices.  For $F_4$, we explain in detail how to
use instead a basis of fundamental weights, which results in standard
theta angle periodicities and charge lattice normalizations, and we will
use that convention for all of the other gauge theories discussed
in this paper (except $G_2$, which we retain as an illustrative example).

In each case, we shall also study the mirror to the pure gauge theory,
to follow up observations in \cite{Aharony:2016jki}.  In particular,
\cite{Aharony:2016jki} argued that two-dimensional pure (2,2) supersymmetric
$SU(k)$ gauge theories flow in the IR to a free theory of $k-1$ twisted
chiral multiplets, which \cite{GuSharpe} checked at the level of
topological field theory computations and extended to $SO(n)$ theories
with discrete theta angles and to $Sp(k)$ gauge theories.  In each case,
for one discrete theta angle, evidence in TFT computations was given that
the theory flowed to a pure gauge theory of as many twisted chiral multiplets
as the rank of the gauge group.  We shall check the analogous claim for
pure gauge theories with exceptional gauge groups in this paper,
at the level of topological field theory computations, and will find
evidence for the same result -- that the pure gauge theories (for
simply-connected gauge groups) flow in the IR to a theory of as many
twisted chiral multiplets as the rank of the gauge group.

Combining the results of this paper with those in \cite{GuSharpe}, a simple
conjecture emerges:  a pure two-dimensional (2,2) supersymmetric gauge theory
with connected and 
simply-connected semisimple
gauge group flows in the IR to a free theory of as many
twisted chiral superfields as the rank of the gauge group.
A check of this conjecture for Spin gauge theories can be derived
from the results for $SO$ gauge theories in \cite{GuSharpe}.
Now, $SO$ groups are not simply-connected; however, we can
apply two-dimensional decomposition \cite{Hellerman:2006zs,Sharpe:2014tca}
and the results for $SO$ theories with various discrete theta angles
to argue that a pure Spin gauge theory flows in the IR to a free 
theory of as many twisted chiral superfields as the rank.  Combined
with the results in this paper for\footnote{
$G_2$, $F_4$, and $E_8$ have no center, but $E_6$ has center
${\mathbb Z}_3$ and $E_7$ has center ${\mathbb Z}_2$, so for those
groups we must specify the simply-connected cover.
See \cite{Distler:2007av}[appendix A] for further details on centers.
} pure two-dimensional supersymmetric $G_2$, $F_4$, and $E_{6,7,8}$ gauge
theories, we have the conjecture above.

We begin in section~\ref{sect:review} by reviewing the nonabelian
mirror proposal of \cite{GuSharpe}, which will be applied in this paper
to theories with exceptional gauge groups.
In section~\ref{section2} we compute the mirror Landau-Ginzburg orbifold of 
$G_2$ gauge theories with matter in copies of the
fundamental ${\bf 7}$ dimensional representation. 
In section~\ref{section3} we compute the mirror Landau-Ginzburg orbifold of 
$F_4$ gauge theories with matter in copies of the
fundamental ${\bf 26}$ representation.
In section~\ref{section4}, we compute the mirror Landau-Ginzburg orbifold of 
$E_6$ with matter in copies of the ${\bf 27}$ representation.
In sections~\ref{section5}, \ref{section6} we perform the same
analysis for $E_7$ and $E_8$ with matter fields in copies of the
${\bf 56}$ representation of $E_7$ and ${\bf 248}$ of $E_8$.

\section{Brief review of the nonabelian mirror proposal}
\label{sect:review}

The nonabelian mirror proposal of \cite{GuSharpe} is a generalization
of the abelian duality described in \cite{HoriVafa} 
(see also \cite{Morrison:1995yh}).
It takes the following form.
For an A-twisted two-dimensional (2,2) supersymmetric gauge theory
with connected gauge group $G$, the mirror is a B-twisted Landau-Ginzburg
orbifold, defined by (twisted) chiral multiplets
\begin{itemize}
\item $Y_i$, corresponding to the $N$ matter fields of the original gauge
theory,
\item $X_{\tilde{\mu}}$, corresponding to nonzero roots $\tilde{\mu}$
of the Lie algebra ${\mathfrak g}$ of $G$, of dimension $n$,
\item $\sigma_a = \overline{D}_+ D_- V_a$, as many as the rank $r$ of $G$,
corresponding to a choice of Cartan subalgebra of ${\mathfrak g}$,
the Lie algebra of $G$,
\end{itemize}
with superpotential
\begin{eqnarray}
W & = & \sum_{a=1}^r \sigma_a \left( 
\sum_{i=1}^N \rho_i^a Y_i \: - \: \sum_{\tilde{\mu}=1}^{n-r} \alpha_{\tilde{\mu}}^a 
\ln X_{\tilde{\mu}}
\: - \: t_a \right) 
\nonumber \\
& & \: + \: \sum_{i=1}^N \exp\left(-Y_i\right) \: + \:
\sum_{\tilde{\mu}=1}^{n-r} X_{\tilde{\mu}}
\: - \: \sum_i \tilde{m}_i Y_i, 
\label{eq:proposal-w}
\end{eqnarray}
In the expression above, the $\rho_i^a$ are components of weight vectors for
the matter representations appearing in the original gauge theory, and
$\alpha^a_{\tilde{\mu}}$ are components of
nonzero roots (here viewed as defining a 
sublattice of the weight lattice).  (Also, sometimes one uses $Z = - \ln X$
for simplicity.)  The $t_a$ are constants, corresponding to Fayet-Iliopoulos
parameters of the original gauge theory, and the $\tilde{m}_i$ are twisted
masses in the original gauge theory.
One then orbifolds by the Weyl group, which acts naturally on all the fields
above, and leaves the superpotential invariant.  The expression above was
written for A-twisted gauge theories without a superpotential, but can
be generalized to mirrors of gauge theories with superpotentials by
assigning suitable R-charges and changing the fundamental fields accordingly,
as explained in \cite{GuSharpe}.

In the analysis of this theory, it was argued that some loci are dynamically
excluded -- specifically, loci where any $X_{\tilde{\mu}}$ vanishes.
These loci turn out to reproduce excluded loci on Coulomb branches of the
original gauge theories.  Furthermore, critical loci of the superpotential
above obey relations which correspond to relations in the OPE ring of the
original A-twisted gauge theory.  For gauge theories with $U(1)$ factors in
$G$, one has continuous
Fayet-Iliopoulos parameters, so one can speak of weak
coupling limits, and those OPE relations are known as quantum cohomology
relations.  In cases in which $G$ has no $U(1)$ factors, so that there
are no continuous Fayet-Iliopoulos parameters, there is no weak coupling
limit, and so referring to such relations as `quantum cohomology' relations
is somewhat misleading.  In such cases, we refer to the relations as
defining the Coulomb ring or Coulomb branch ring.

The work \cite{GuSharpe} checked the predictions of this proposal for
excluded loci and Coulomb branch and quantum cohomology relations against
known results for two-dimensional gauge theories in {\it e.g.}
\cite{Witten:1993xi,Hori:2006dk,Donagi:2007hi,Hori:2011pd}, 
and gave general arguments for why correlation functions in this
B-twisted theory should match correlation functions in corresponding A-twisted
gauge theories, such as in
{\it e.g.} \cite{Nekrasov:2014xaa,Closset:2015rna,Closset:2017vvl}.  
It also studied mirrors to pure gauge theories, to test and
refine predictions for IR behavior described in \cite{Aharony:2016jki}.
In this paper, we will apply this mirror construction to make predictions
for two-dimensional (2,2) supersymmetric gauge theories with exceptional
gauge groups.
To make all of these comparisons, the paper \cite{GuSharpe} utilized
the following operator mirror map:
\begin{eqnarray}
\exp(-Y_i) & = & - \tilde{m}_i  + \sum_{a=1}^r \sigma_a \rho_i^a, 
\label{eq:op-mirror-mass-1}
\\
X_{\tilde{\mu}}  & = & \sum_{a=1}^r \sigma_a \alpha_{\tilde{\mu}}^a ,
\label{eq:op-mirror-mass-2}
\end{eqnarray}
which we shall also use in this paper.

\section{$G_2$} \label{section2}

In this section we will consider the mirror Landau-Ginzburg orbifold 
of $G_2$ gauge theory with matter fields in copies of the
${\bf 7}$ representation,
and then we compute quantum cohomology ring.

\subsection{Mirror Landau-Ginzburg orbifold}

The mirror Landau-Ginzburg model has fields
\begin{itemize}
	\item $Y_{i \beta}$, $i \in \{1, \cdots, n\}$, $\beta \in \{0, \cdots, 6\}$,
	corresponding to the matter fields in $n$ copies of the ${\bf 7}$ of
	$G_2$,
	\item $X_m$, $\tilde{X}_m$, $m \in \{1, \cdots, 6\}$, corresponding to the
	short, respectively long roots of $G_2$,
	\item $\sigma_a$, $a \in \{1, 2 \}$.
\end{itemize}

We associate the roots and weights to fields as listed in 
table~\ref{table:g2:roots-weights} and figures~\ref{fig:g2:roots},
\ref{fig:g2:weights}.

\begin{table}[h!]
\centering
\begin{tabular}{cc|cc|cc}
		Field & Short root & Field & Long root & Field & Weight \\ \hline
		$X_1$ & $(1,0)$ & $\tilde{X}_1$ & $(-3/2, \sqrt{3}/2)$ & $Y_{i1}$ & $(1,0)$ \\
		$X_2$ & $(-1,0)$ & $\tilde{X}_2$ & $(3/2, - \sqrt{3}/2)$ & $Y_{i2}$ & $(-1,0)$ \\
		$X_3$ & $(1/2, \sqrt{3}/2)$ & $\tilde{X}_3$ & $(3/2, \sqrt{3}/2)$ &
		$Y_{i3}$ & $(1/2, \sqrt{3}/2)$ \\
		$X_4$ & $(-1/2, -\sqrt{3}/2)$ & $\tilde{X}_4$ & $(-3/2,-\sqrt{3}/2)$ &
		$Y_{i4}$ & $(-1/2, -\sqrt{3}/2)$ \\
		$X_5$ & $(-1/2,\sqrt{3}/2)$ & $\tilde{X}_5$ & $(0,\sqrt{3})$ &
		$Y_{i5}$ & $(-1/2, \sqrt{3}/2)$  \\
		$X_6$ & $(1/2, -\sqrt{3}/2)$ & $\tilde{X}_6$ & $(0,-\sqrt{3})$ &
		$Y_{i6}$ & $(1/2, -\sqrt{3}/2)$ \\
		& & & & $Y_{i0}$ & $(0,0)$
	\end{tabular}
\caption{Roots and weights for $G_2$ and associated fields. 
\label{table:g2:roots-weights} }
\end{table}

\begin{figure}[h]
\centering
\begin{picture}(200,200)
	\LongArrow(100,100)(150,100)  \Text(155,105)[b]{$X_1$}
	\LongArrow(100,100)(50,100)   \Text(45,105)[b]{$X_2$}
	\LongArrow(100,100)(100,187)  \Text(110,187)[b]{$\tilde{X}_5$}
	\LongArrow(100,100)(100,13)   \Text(110,13)[b]{$\tilde{X}_6$}
	\LongArrow(100,100)(175,143)  \Text(175,148)[b]{$\tilde{X}_3$}
	\LongArrow(100,100)(125,143)  \Text(125,148)[b]{$X_3$}
	\LongArrow(100,100)(75,143)   \Text(75,148)[b]{$X_5$}
	\LongArrow(100,100)(25,143)   \Text(25,148)[b]{$\tilde{X}_1$}
	\LongArrow(100,100)(25,57)    \Text(25,63)[b]{$\tilde{X}_4$}
	\LongArrow(100,100)(75,57)    \Text(75,63)[b]{$X_4$}
	\LongArrow(100,100)(125,57)   \Text(125,63)[b]{$X_6$}
	\LongArrow(100,100)(175,57)   \Text(175,63)[b]{$\tilde{X}_2$}
	\end{picture}
\caption{Roots of $G_2$. \label{fig:g2:roots} }
\end{figure}

\begin{figure}[h]
\centering
	\begin{picture}(200,200)
	\LongArrow(100,100)(150,100)  \Text(155,105)[b]{$Y_{i1}$}
	\LongArrow(100,100)(50,100)   \Text(45,105)[b]{$Y_{i2}$}
	\LongArrow(100,100)(125,143)  \Text(125,148)[b]{$Y_{i3}$}
	\LongArrow(100,100)(75,143)   \Text(75,148)[b]{$Y_{i5}$}
	\LongArrow(100,100)(75,57)    \Text(75,63)[b]{$Y_{i4}$}
	\LongArrow(100,100)(125,57)   \Text(125,63)[b]{$Y_{i6}$}
	\CArc(100,100)(2,0,360)   \Text(105,105)[b]{$Y_{i0}$}
	\end{picture}
\caption{Weights of ${\bf 7}$ of $G_2$. \label{fig:g2:weights} }
\end{figure}

The mirror superpotential takes the form
\begin{eqnarray}
W & = & 
\sigma_1 \Biggl( \sum_i \left( Y_{i1} - Y_{i2} + (1/2) Y_{i3} - (1/2) Y_{i4}
- (1/2) Y_{i5} + (1/2) Y_{i6} \right)
\nonumber \nonumber \\
& & \hspace*{0.5in}
\: + \:
\left( Z_1 - Z_2 + (1/2) Z_3  - (1/2) Z_4 - (1/2) Z_5 + (1/2) Z_6 \right)
\nonumber \nonumber  \\
& & \hspace*{0.5in}
\: + \:
\left(- (3/2) \tilde{Z}_1 + (3/2) \tilde{Z}_2 + (3/2) \tilde{Z}_3
- (3/2) \tilde{Z}_4 \right)
\Biggr)
\nonumber \nonumber \\
& &
\: + \: \sigma_2 \Biggl( 
(\sqrt{3}/2) \sum_i  \left( Y_{i3} - Y_{i4} +
Y_{i5} - Y_{i6} \right)
\: + \:
(\sqrt{3}/2) \left( Z_3 - Z_4 + Z_5 - Z_6 \right)
\nonumber \nonumber \\
& & \hspace*{0.5in}
\: + \:
(\sqrt{3}/2) \left( \tilde{Z}_1 - \tilde{Z}_2 +
\tilde{Z}_3 - \tilde{Z}_4 + 2 \tilde{Z}_5 - 
2\tilde{Z}_6 \right) \Biggr)
\nonumber \nonumber \\
& &  \: + \:
\sum_{i} \sum_{\alpha=0}^6 \exp\left( - Y_{i \alpha} \right)
\: + \: \sum_{m=1}^6 X_m \: + \: \sum_{m=1}^6 \tilde{X}_m - \sum_{i,\alpha} \tilde{m}_i Y_{i\alpha}, \label{eq-g2-sp}
\end{eqnarray}
where $X_m = \exp(-Z_m)$, $\tilde{X}_m = \exp(-\tilde{Z}_m)$,
with $X_m$, $\tilde{X}_m$ the fundamental fields and $\tilde{m}_i$ are the twisted masses.

The logic of the assignments above is that $X_{\rm odd}$,
$\tilde{X}_{\rm odd}$ correspond to
positive roots, $X_{\rm even}$, $\tilde{X}_{\rm even}$ correspond to
their opposites, and the weight vectors are associated to
matter fields similarly.
We follow the conventions of \cite{fh}[chapter 22]:
short roots are given by
\begin{displaymath}
( \pm 1, 0), \: \: \:
\pm (+1/2, \sqrt{3}/2), \: \: \:
\pm (-1/2, \sqrt{3}/2),
\end{displaymath}
long roots are given by
\begin{displaymath}
\pm(-3/2, \sqrt{3}/2), \: \: \:
\pm(+3/2, \sqrt{3}/2), \: \: \:
\pm(0, \sqrt{3}),
\end{displaymath}
and the weights of the ${\bf 7}$ are given by
\begin{displaymath}
\pm(1,0), \: \: \: \pm (1/2, \sqrt{3}/2), 
\pm (-1/2, \sqrt{3}/2), \: \: \: (0,0).
\end{displaymath}

Before moving on, there is an important subtlety in the expression for
the mirror superpotential above, involving the theta angle periodicities.
As described in \cite{GuSharpe},
the factors multiplied by $\sigma$s are not single-valued, reflecting
the fact that the $\sigma$ terms encode theta angles in the abelian
subgroup determined by the choice of Cartan subgroup of the original
gauge group.  The periodicities\footnote{
On a noncompact worldsheet, the theta angles generate electric fields with
periodicities determined by the matter representations -- as theta increases,
the electric field density eventually becomes strong enough to allow
pair creation of matter fields.
} of these theta angles are determined by $2\pi$ times the weight
lattice, or at least the sublattice generated by the matter representations.
However, the weight lattice need not be normalized in the same way as
a charge lattice.  For example, in our conventions for the weight
lattice of $G_2$ above, the $\sigma_1$ terms determine a theta angle
periodicity of $2 \pi/2 = \pi$ rather than $2\pi$,
and the $\sigma_2$ terms determine a theta angle periodicity of
$(\sqrt{3}/2)(2\pi) = \sqrt{3} \pi$ rather than $2\pi$.

Now, on the one hand, the normalization of the charge lattice is ultimately
a convention, and so long as one is consistent, one can work with alternative
conventions.  On the other hand, it is also often helpful to work with
standard conventions.

For the case of $G_2$, we shall use the normalization above, hence
a nonstandard charge lattice normalization.  However, it is always possible
to rotate to a conventional charge lattice normalization by writing
the weights in a basis of fundamental weights, for which any other weight
is an integer linear combination.  In terms of that mathematical basis,
the theta angle periodicities determined by $\sigma$s are all $2\pi$,
reflecting a standard charge lattice normalization.  We will discuss
this alternative basis in more detail for $F_4$, and in fact will use
that alternative basis (and standard charge normalization) to study
all the other gauge theories in this paper, after $G_2$.  We study 
$G_2$ in nonstandard conventions for illustrative purposes.

\subsection{Weyl group}

Now, let us explicitly describe the action of the Weyl group on the
fields of this theory and outline explicitly
why the superpotential is invariant in this case.
(General arguments appeared in \cite{GuSharpe}, but as the Weyl group action is
more complicated here than in the examples in that paper, a more detailed
verification seems in order.)

For any root $\alpha$, recall that
the Weyl group reflection generated by $\alpha$
acts on a weight $\mu$ as follows:
\begin{equation}
\mu \: \mapsto \: \mu \: - \: \frac{2 (\alpha \cdot \mu) }{ \alpha^2 } \alpha. \label{eq-weyl}
\end{equation}
For example, for the Weyl reflection generated by $\alpha = (1,0)$,
it is straightforward to compute that the group action on
fields corresponding to roots is given by
\begin{equation}
X_1 \leftrightarrow X_2, \: \: \:
X_3 \leftrightarrow X_5, \: \: \:
X_4 \leftrightarrow X_6,
\end{equation}
\begin{equation}
\tilde{X}_1 \leftrightarrow \tilde{X}_3, \: \: \:
\tilde{X}_2 \leftrightarrow \tilde{X}_4,
\end{equation}
and $\tilde{X}_{5,6}$ are invariant.
The action on matter fields is
\begin{equation}
Y_{i1} \leftrightarrow Y_{i2}, \: \: \:
Y_{i3} \leftrightarrow Y_{i5}, \: \: \:
Y_{i4} \leftrightarrow Y_{i6},
\end{equation}
with $Y_{i7}$ invariant.
This is just a reflection about the $y$ axis, which multiplies
the first coordinate by $-1$ but leaves the second invariant.
It is straightforward to check that the superpotential will be invariant
under this reflection so long as
\begin{equation}
\sigma_1 \leftrightarrow - \sigma_1,
\end{equation}
and $\sigma_2$ is invariant.

For another example, for Weyl reflections generated by
$\alpha = (3/2,\sqrt{3}/2)$, it is straightforward to compute that
the group action on fields corresponding to roots is given by
\begin{equation}
X_1 \leftrightarrow X_4, \: \: \:
X_2 \leftrightarrow X_3, 
\end{equation}
with $X_{5, 6}$ invariant, and
\begin{equation}
\tilde{X}_1 \leftrightarrow \tilde{X}_5, \: \: \:
\tilde{X}_2 \leftrightarrow \tilde{X}_6, \: \: \:
\tilde{X}_3 \leftrightarrow \tilde{X}_4.
\end{equation}
The action on matter fields is the same as on the mirrors to the short roots:
\begin{equation}
Y_{i1} \leftrightarrow Y_{i4}, \: \: \:
Y_{i2} \leftrightarrow Y_{i3},
\end{equation}
with $Y_{i5}$, $Y_{i6}$ invariant.
The $\sigma_a$ fields are similarly rotated:
\begin{eqnarray*}
	\sigma_1 & \mapsto & - \frac{1}{2} \sigma_1 \: - \: \frac{ \sqrt{3} }{2} 
	\sigma_2, 
	\\
	\sigma_2 & \mapsto & - \frac{ \sqrt{3} }{2} \sigma_1 \: + \:
	\frac{1}{2} \sigma_2.
\end{eqnarray*}
(Note that if we describe the action above as mapping
$\vec{\sigma} \mapsto A \vec{\sigma}$ for a $2 \times 2$ matrix $A$,
then for the choice of $A$ implicit above,
it is straightforward to check $A = A^{-1}$.)
It is straightforward to check that the superpotential is invariant
under the action above.  For example, the terms
\begin{eqnarray*}
	\lefteqn{
		\sigma_1 \left( Z_1 - Z_2 + (1/2) Z_3 - (1/2) Z_4 - (1/2) Z_5 + (1/2) Z_6
		\right)
	} \\
	& &
	\: + \:
	\sigma_2 (\sqrt{3}/2) \left( Z_3 - Z_4 + Z_5 - Z_6 \right)
	\\
	& \mapsto &
	\left( - (1/2) \sigma_1 - (\sqrt{3}/2) \sigma_2 \right) \left(
	Z_4 - Z_3 + (1/2) Z_2 - (1/2) Z_1 - (1/2) Z_5 + (1/2) Z_6 \right)
	\\
	& &
	\: + \:
	\left( - (\sqrt{3}/2) \sigma_1 + (1/2) \sigma_2\right) (\sqrt{3}/2)
	\left( Z_2 - Z_1 + Z_5 - Z_6 \right),
\end{eqnarray*}
which is easily checked to be the same as the starting point,
\begin{displaymath}
\sigma_1 \left( Z_1 - Z_2 + (1/2) Z_3 - (1/2) Z_4 - (1/2) Z_5 + (1/2) Z_6
\right) \: + \:
\sigma_2 (\sqrt{3}/2) \left( Z_3 - Z_4 + Z_5 - Z_6 \right).
\end{displaymath}
Similar statements are true of other terms, and so the superpotential
is preserved.

To be thorough, we will consider one more example of a Weyl group
action, this time a reflection defined by a short root,
specifically $\alpha = (1/2, \sqrt{3}/2)$.  It is straightforward to
compute that the group action on fields corresponding to roots is given by
\begin{equation}
X_1 \leftrightarrow X_6, \: \: \:
X_2 \leftrightarrow X_5, \: \: \:
X_3 \leftrightarrow X_4,
\end{equation}
and
\begin{equation}
\tilde{X}_3 \leftrightarrow \tilde{X}_6, \: \: \:
\tilde{X}_4 \leftrightarrow \tilde{X}_5,
\end{equation}
with $\tilde{X}_{1, 2}$ invariant.  The action on the matter fields
is the same as on the mirrors to the short roots:
\begin{equation}
Y_{i1} \leftrightarrow Y_{i6}, \: \: \:
Y_{i2} \leftrightarrow Y_{i5}, \: \: \:
Y_{i3} \leftrightarrow Y_{i4}.
\end{equation}
This is another reflection about the axis pass through $X_5$ and $X_6$.
The $\sigma_a$ fields are similarly rotated:
\begin{eqnarray*}
	\sigma_1 & \mapsto & \frac{1}{2} \sigma_1 \: - \: \frac{ \sqrt{3} }{2} \sigma_2,\\
	\sigma_2 & \mapsto & - \frac{ \sqrt{3} }{2} \sigma_1 \: - \: \frac{1}{2}
	\sigma_2.
\end{eqnarray*}
It is straightforward to check that the superpotential is invariant.

\subsection{Coulomb ring relations}

Integrating out the sigma fields in the superpotential (\ref{eq-g2-sp}), we obtain two constraints:
\begin{align*}
&\sum_i \left( 2 Y_{i1} - 2 Y_{i2} + Y_{i3} - Y_{i4}
- Y_{i5} + Y_{i6} \right) + \left( 2 Z_1 - 2 Z_2 + Z_3  - Z_4 - Z_5 + Z_6 \right) +
\\
&\hspace{89mm} +
\left(- 3 \tilde{Z}_1 + 3 \tilde{Z}_2 + 3 \tilde{Z}_3
- 3 \tilde{Z}_4 \right) =0, \\
& 
\sum_i  \left( Y_{i3} - Y_{i4} +
Y_{i5} - Y_{i6} \right)
\: + \:
\left( Z_3 - Z_4 + Z_5 - Z_6 \right) +
\\
&\hspace{78mm} + \left( \tilde{Z}_1 - \tilde{Z}_2 +
\tilde{Z}_3 - \tilde{Z}_4 + 2 \tilde{Z}_5 - 
2\tilde{Z}_6 \right) =0.
\end{align*}

With the two constraints above, we are free to eliminant two fundamental 
fields, which we will take to be $Y_{n3}$ and $Y_{n6}$:
\begin{align*}
- Y_{n3} & = \sum_{i=1}^n \left( Y_{i1} - Y_{i2} - Y_{i4} \right) + \sum_{i=1}^{n-1} Y_{i3} + \left( Z_1 - Z_2 + Z_3 - Z_4\right) \\
&\hspace{60mm} + \left( - \tilde{Z}_1 + \tilde{Z}_2 +2 \tilde{Z}_3 - 2 \tilde{Z}_4 + \tilde{Z}_5 -\tilde{Z}_6 \right), \\
- Y_{n3} & = \sum_{i=1}^n \left( Y_{i1} - Y_{i2} - Y_{i5} \right) + \sum_{i=1}^{n-1} Y_{i6} + \left( Z_1 - Z_2 - Z_5 + Z_6\right) \\ 
&\hspace{60mm}+ \left( -2\tilde{Z}_1 + 2 \tilde{Z}_2 + \tilde{Z}_3 - \tilde{Z}_4 - \tilde{Z}_5 + \tilde{Z}_6 \right).
\end{align*}
For convenience, let's define:
\begin{align}
\Pi_3 &\equiv \exp \left( - Y_{n3} \right), \nonumber \\ 
&= \prod_{i=1}^n \exp \left( Y_{i1} - Y_{i2} - Y_{i4}\right) \prod_{i=1}^{n-1} \exp \left( Y_{i3} \right) \frac{X_2 X_4}{X_1 X_3} \frac{ \tilde{X}_1 \tilde{X}_4^2 \tilde{X}_6}{\tilde{X}_2 \tilde{X}_3^2 \tilde{X}_5}, \label{eq-pi3}\\
\Pi_6 &\equiv \exp \left( - Y_{n6} \right), \nonumber \\
&= \prod_{i=1}^n \exp \left( Y_{i1} - Y_{i2} - Y_{i5}\right) \prod_{i=1}^{n-1} \exp \left( Y_{i6} \right) \frac{X_2 X_5}{X_1 X_6} \frac{\tilde{X}_1^2 \tilde{X}_4 \tilde{X}_5}{ \tilde{X}_2^2 \tilde{X}_3 \tilde{X}_6}. \label{eq-pi6}
\end{align}
Then, the superpotential~(\ref{eq-g2-sp}) reduces to
\begin{align*}
W = &\sum_{i=1}^n \Big[ \exp \left( - Y_{i0} \right) + \exp \left( - Y_{i1} \right) + \exp \left( - Y_{i2} \right) + \exp \left( - Y_{i4} \right) + \exp \left( - Y_{i5} \right) \Big] \\
& \quad + \sum_{i=1}^{n-1} \Big[ \exp \left( - Y_{i3} \right) + \exp \left( - Y_{i6} \right)\Big] + \Pi_3 + \Pi_6 + \sum_{m=1}^6 \left( X_m + \tilde{X}_m \right) \\
& \quad \quad -\sum_{i=1}^n \tilde{m}_i \left( Y_{i0} + Y_{i1} + Y_{i2} + Y_{i4} + Y_{i5}\right) - \sum_{i=1}^{n-1} \tilde{m}_i \left( Y_{i3} + Y_{i6} \right) + \tilde{m}_n (\ln \Pi_3 + \ln \Pi_6).
\end{align*}
Notice that the superpotential has poles at 
$X_1 \neq 0$, $X_3 \neq 0$, $X_6 \neq 0$, $\tilde{X}_2 \neq 0$, 
$\tilde{X}_3 \neq 0$, $\tilde{X}_5 \neq 0$ and $\tilde{X}_6 \neq 0$. 
With the mirror maps,
\begin{equation}
\exp(-Y_{i \beta}) = - \tilde{m}_i + \sum_{a=1,2} \sigma_a \rho_{i \beta}^a, \quad X_m = \sum_{a=1,2} \sigma_a \alpha_m^a, \quad \tilde{X}_m = \sum_{a=1,2} \sigma_a \tilde{\alpha}_m^a, \label{eq-mm}
\end{equation}
one can get the excluded loci:
\begin{align}
&\sigma_1 \sigma_2 (\sigma_1^2 - 3 \sigma_2^2) (3 \sigma_1^2 - \sigma_2^2) \neq 0, \label{g2-el1}\\
&\prod_i (-m_i + \sigma_1) (-m_i - \sigma_1)
\left(-m_i + \frac{1}{2} \sigma_1 + \frac{\sqrt{3}}{2} \sigma_2\right)
\left(-m_i - \frac{1}{2} \sigma_1 - \frac{\sqrt{3}}{2} \sigma_2\right)
 \cdot \nonumber\\
&\hspace{50mm} \cdot \left(-m_i - \frac{1}{2} \sigma_1 + \frac{\sqrt{3}}{2} \sigma_2\right)
\left(-m_i + \frac{1}{2} \sigma_1 - \frac{\sqrt{3}}{2} \sigma_2\right) \neq 0. \label{g2-el2}
\end{align}

The critical locus is given by
\begin{align*}
& \frac{\partial W}{\partial Y_{i0} } : \exp \left( - Y_{i0} \right) = -\tilde{m}_i, \quad \text{for} \quad i=1,\cdots,n, \\
& \frac{\partial W}{\partial Y_{i1} } : \exp \left( - Y_{i1} \right) = \Pi_3 + \Pi_6 -  \tilde{m}_i +  2\tilde{m}_n,\quad \text{for} \quad i=1,\cdots,n, \\
& \frac{\partial W}{\partial Y_{i2} } : \exp \left( - Y_{i2} \right) = -\Pi_3 - \Pi_6 -  \tilde{m}_i - 2  \tilde{m}_n,\quad \text{for} \quad i=1,\cdots,n, \\
& \frac{\partial W}{\partial Y_{i3} } : \exp \left( - Y_{i3} \right) = \Pi_3 -  \tilde{m}_i +  \tilde{m}_n,\quad \text{for} \quad i=1,\cdots,n-1, \\
& \frac{\partial W}{\partial Y_{i4} } : \exp \left( - Y_{i4} \right) = -\Pi_3 -  \tilde{m}_i -  \tilde{m}_n,\quad \text{for} \quad i=1,\cdots,n, \\
& \frac{\partial W}{\partial Y_{i5} } : \exp \left( - Y_{i5} \right) =  - \Pi_6 -  \tilde{m}_i -  \tilde{m}_n,\quad \text{for} \quad i=1,\cdots,n, \\
& \frac{\partial W}{\partial Y_{i6} } : \exp \left( - Y_{i6} \right) =  \Pi_6 -  \tilde{m}_i +  \tilde{m}_n,\quad \text{for} \quad i=1,\cdots,n-1, \\
\end{align*}
\begin{align*}
& \frac{\partial W}{\partial X_1 } : X_1 = \Pi_3 + \Pi_6 + 2  \tilde{m}_n, \\
& \frac{\partial W}{\partial X_2 } : X_2 = -\Pi_3 - \Pi_6 -  2 \tilde{m}_n, \\
& \frac{\partial W}{\partial X_3 } : X_3 = \Pi_3 + \tilde{m}_n, \\
& \frac{\partial W}{\partial X_4 } : X_4 = -\Pi_3 - \tilde{m}_n, \\
& \frac{\partial W}{\partial X_5 } : X_5 = - \Pi_6 - \tilde{m}_n, \\
& \frac{\partial W}{\partial X_6 } : X_6 = \Pi_6 + \tilde{m}_n, \\
\end{align*}
\begin{align*}
& \frac{\partial W}{\partial \tilde{X}_1 } : \tilde{X}_1 = -\Pi_3 - 2 \Pi_6 - 3 \tilde{m}_n, \\
& \frac{\partial W}{\partial \tilde{X}_2 } : \tilde{X}_2 = \Pi_3 + 2 \Pi_6 + 3 \tilde{m}_n, \\
& \frac{\partial W}{\partial \tilde{X}_3 } : \tilde{X}_3 = 2 \Pi_3 + \Pi_6 + 3 \tilde{m}_n, \\
& \frac{\partial W}{\partial \tilde{X}_4 } : \tilde{X}_4 = -2 \Pi_3 - \Pi_6 - 3 \tilde{m}_n, \\
& \frac{\partial W}{\partial \tilde{X}_5 } : \tilde{X}_5 = \Pi_3 - \Pi_6, \\
& \frac{\partial W}{\partial \tilde{X}_6 } : \tilde{X}_6 = -\Pi_3 + \Pi_6, \\
\end{align*}

Plug the above equations back to (\ref{eq-pi3}), (\ref{eq-pi6}), one obtains the Coulomb branch relations:
\begin{align}
\Pi_3  = & \prod_{i=1}^{n-1} (\Pi_3 - \tilde{m}_i + \tilde{m}_n)^{-1} \boldsymbol{\cdot} \nonumber\\
&\qquad \boldsymbol{\cdot}\prod_{i=1}^n \left( \Pi_3 + \Pi_6 -  \tilde{m}_i +  2 \tilde{m}_n \right)^{-1}  (-\Pi_3 - \Pi_6 -  \tilde{m}_i - 2  \tilde{m}_n) (-\Pi_3 -  \tilde{m}_i -  \tilde{m}_n), \\
\Pi_6  = &  \prod_{i=1}^{n-1} (\Pi_6 - \tilde{m}_i + \tilde{m}_n)^{-1} \boldsymbol{\cdot} \nonumber \\
&\qquad \boldsymbol{\cdot} \prod_{i=1}^n \left( \Pi_3 + \Pi_6 -  \tilde{m}_i +  2\tilde{m}_n \right)^{-1}  (-\Pi_3 - \Pi_6 -  \tilde{m}_i - 2  \tilde{m}_n) (-\Pi_6 -  \tilde{m}_i -  \tilde{m}_n).
\end{align}

With the mirror map (\ref{eq-mm}), on the critical locus relations, one finds
\begin{equation*}
\Pi_3 = \frac{1}{2} \sigma_1 + \frac{\sqrt{3}}{2} \sigma_2 - \tilde{m}_n, \quad \Pi_6 =  \frac{1}{2} \sigma_1 - \frac{\sqrt{3}}{2} \sigma_2 - \tilde{m}_n.
\end{equation*}
Plugging them back in, 
one obtains the Coulomb (quantum cohomology) ring 
relations for $G_2$,
\begin{align}
\prod_{i=1}^n ( -\sigma_1 - \tilde{m}_i ) 
\left( - \frac{1}{2} \sigma_1 - \frac{\sqrt{3}}{2} \sigma_2 - \tilde{m}_i 
\right)    
\: = \: \prod_{i=1}^n (\sigma_1 - \tilde{m}_i) 
\left( \frac{1}{2} \sigma_1 + \frac{\sqrt{3}}{2} \sigma_2 - \tilde{m}_i 
\right), 
\label{qcr-g2-1}\\
\prod_{i=1}^n  ( -\sigma_1 - \tilde{m}_i ) 
\left( - \frac{1}{2} \sigma_1 + \frac{\sqrt{3}}{2} \sigma_2 - \tilde{m}_i
\right) 
\: = \:
\prod_{i=1}^n (\sigma_1 - \tilde{m}_i) 
\left( \frac{1}{2} \sigma_1 - \frac{\sqrt{3}}{2} \sigma_2 - \tilde{m}_i \right).
\label{qcr-g2-2}
\end{align}
Combining the above two relations, one gets
\begin{align}
&\prod_{i=1}^n (-\sigma_1 - \tilde{m}_i)^2 
\left(-\frac{1}{2} \sigma_1 - \frac{\sqrt{3}}{2}\sigma_2 - \tilde{m}_i
\right) 
\left(-\frac{1}{2} \sigma_1 + \frac{\sqrt{3}}{2}\sigma_2 - \tilde{m}_i
\right) \nonumber \\
&\hspace{35mm} = \:
\prod_{i=1}^n (\sigma_1 - \tilde{m}_i)^2 
\left(\frac{1}{2} \sigma_1 + \frac{\sqrt{3}}{2}\sigma_2 - \tilde{m}_i \right) 
\left(\frac{1}{2} \sigma_1 - \frac{\sqrt{3}}{2}\sigma_2 - \tilde{m}_i
\right) , \label{qcr-g2-3}\\
&\prod_{i=1}^n \left(\frac{1}{2} \sigma_1 + \frac{\sqrt{3}}{2}\sigma_2 - \tilde{m}_i \right)
\left(-\frac{1}{2} \sigma_1 + \frac{\sqrt{3}}{2}\sigma_2 - \tilde{m}_i \right) 
\nonumber \\
&\hspace{35mm}   = \:
\prod_{i=1}^n\left(-\frac{1}{2} \sigma_1 - \frac{\sqrt{3}}{2}\sigma_2 - \tilde{m}_i \right) 
\left(\frac{1}{2} \sigma_1 - \frac{\sqrt{3}}{2}\sigma_2 - \tilde{m}_i
\right). \label{qcr-g2-4}
\end{align}

\subsection{Vacua}

In this section, we will count the number of vacua in cases with small
numbers $n$ of fundamental fields. 
To solve the Coulomb branch (quantum cohomology) 
relations~(\ref{qcr-g2-1}), (\ref{qcr-g2-2}) in general is not easy. 
However, since the superpotential is invariant under the Weyl group, the 
Coulomb ring relations~(\ref{qcr-g2-1}), (\ref{qcr-g2-2}) will be covariant 
under the Weyl group action, which we check explicitly. 

The Weyl group of $G_2$ is the dihedral group $D_{12}$ of degree 6 and 
order 12, which can be described as \cite{curtis-reiner}[section 7]
\begin{displaymath}
D_{12} \: = \: \left\{ a^i x^j \, | \, a^6 = 1 = x^2, x a x = a^{-1} 
\right\}.
\end{displaymath} 
(See {\it e.g.} \cite{curtis-reiner}[section 47] for a discussion of
representations of the dihedral groups.)
Among the twelve elements of the Weyl group, there are six reflections,
and below we list group elements and the field corresponding to the root
about which the reflection takes place: 
\begin{align*}
X_1 \leftrightarrow  a^3 x,
 \quad 
X_3 \leftrightarrow  a^5 x,
\quad 
X_5 \leftrightarrow  a x,
 \\
\tilde{X}_1 \leftrightarrow  a^2 x,
 \quad 
\tilde{X}_3 \leftrightarrow  a^4 x,
 \quad
\tilde{X}_5 \leftrightarrow  x.
\end{align*}
Notice that the reflections are also generated by the Weyl group 
reflection~(\ref{eq-weyl}) and we denote the reflection matrices by 
the fields correspond to the positive simple roots. 
There are also five nontrivial rotations, corresponding to
$\langle a \rangle \subset D_{12}$.

Now we can start to solve for the vacua (solutions of the Coulomb ring
relations~(\ref{qcr-g2-1}), (\ref{qcr-g2-2}))
begining with the case of small 
number of fundamental matter fields.
\begin{itemize}
	\item $n=1$, the only solution is $\sigma_1 = \sigma_2 =0$ and it is excluded by the constraints (\ref{g2-el1}), (\ref{g2-el2}),
	\item $n=2$, there are seven solutions but all of them are excluded by the constraints (\ref{g2-el1}), (\ref{g2-el2}),
	\item $n=3$, there are ninteen solutions but all of them are excluded
by the constraints (\ref{g2-el1}), (\ref{g2-el2}).
\end{itemize} 

Starting with the case $n=4$,
we begin to obtain non-trivial solutions. 
First, let us analyze the case of $n=4$ in detail. For simplicity, from now on, we will take $m_i = m_j = m, \, \forall i \neq j$ and will rescale
the $\sigma_i$ fields to $\sigma_i = \sigma_i / m$. 
There are thirty-seven solutions in total and twelve of them are true vacua
(meaning, not on the excluded locus):
\begin{align*}
i=1, \cdots, 4, \quad s_i = \bigg\{ \sigma_1 = \pm i \sqrt{5}, &\quad \sigma_2 = \pm i \sqrt{3} \bigg\},  \\
i=5, \cdots, 8, \quad s_i =\bigg\{ \sigma_1 = \pm i \sqrt{\frac{7}{2}-\frac{3 \sqrt{5}}{2}}, &\quad \sigma_2 = \pm i \sqrt{\frac{3}{2} (3+\sqrt{5})}\bigg\},
\\
i=9, \cdots, 12, \quad s_i =\bigg\{ \sigma_1 = \pm i \sqrt{\frac{7}{2}+\frac{3 \sqrt{5}}{2}}, &\quad \sigma_2 = \pm i \sqrt{\frac{3}{2} (3-\sqrt{5})} \bigg\}.
\end{align*}
Signs are assigned in each group of four solutions in the order
$\{-,-\}$, $\{-,+\}$, $\{+,-\}$, $\{+,+\}$. For example, 
\begin{align*}
X_1 = \{ \sigma_1 = - i \sqrt{5}, &\quad \sigma_2 = - i \sqrt{3} \}, \quad X_2 = \{ \sigma_1 = - i \sqrt{5}, \quad \sigma_2 = + i \sqrt{3} \}, \\
X_3 = \{ \sigma_1 = + i \sqrt{5}, &\quad \sigma_2 = - i \sqrt{3} \}, \quad X_4 = \{ \sigma_1 = + i \sqrt{5}, \quad \sigma_2 = + i \sqrt{3} \}.
\end{align*}
Under the Weyl group actions, the solutions transform as in 
table~\ref{table-1}.\footnote{Note that the Weyl group acts on $\sigma$s by the inverse of the group elements. In the table, we denote the action by the original group elements instead of the inverse of the elements. Table~\ref{table-2} adopts the same notation. }
One can see that the twelve vacua are covariant and form one Weyl orbit 
under the Weyl group action. 
\begin{table}
	\centering
	\begin{tabular}{|c||cccccccccccc|}
		\hline
		& $e$ & $X_1$ & $X_3$ & $X_5$ & $\tilde{X}_1$ &  $\tilde{X}_3$ &  $\tilde{X}_5$ & $a_5$ & $a_4$ & $a_3$ & $a_2$ & $a_1$ \\
		\hline \hline
		1 & 1 & 3 & 8 & 9 & 5 & 12 & 2 & 7 & 11 & 4 & 6 & 10 \\
		2 & 2 & 4 & 10 & 7 & 11 & 6 & 1 & 9 & 5 & 3 & 12 & 8 \\
		3 & 3 & 1 & 11 & 6 & 10 & 7 & 4 & 12 & 8 & 2 & 9 & 5 \\
		4 & 4 & 2 & 5 & 12 & 8 & 9 & 3 & 6 & 10 & 1 & 7 & 11 \\
		5 & 5 & 7 & 4 & 10 & 1 & 11 & 6 & 3 & 12 & 8 & 2 & 9 \\
		6 & 6 & 8 & 9 & 3 & 12 & 2 & 5 & 10 & 1 & 7 & 11 & 4 \\
		7 & 7 & 5 & 12 & 2 & 9 & 3 & 8 & 11 & 4 & 6 & 10 & 1 \\
		8 & 8 & 6 & 1 & 11 & 4 & 10 & 7 & 2 & 9 & 5 & 3 & 12 \\
		9 & 9 & 11 & 6 & 1 & 7 & 4 & 10 & 5 & 3 & 12 & 8 & 2 \\
		10 & 10 & 12 & 2 & 5 & 3 & 8 & 9 & 1 & 7 & 11 & 4 & 6 \\
		11 & 11 & 9 & 3 & 8 & 2 & 5 & 12 & 4 & 6 & 10 & 1 & 7 \\
		12 & 12 & 10 & 7 & 4 & 6 & 1 & 11 & 8 & 2 & 9 & 5 & 3 \\
		\hline
	\end{tabular}
	\caption{Weyl group actions on the vacua of the case $n=4$}
	\label{table-1}
\end{table}

When there are five fundamental matter multiplets, there are 
sixty-one solutions 
and twenty-four of them are non-trivial. Following the same conventions, 
those non-trivial vacua are
\begin{align*}
&i=1, \cdots, 4, \quad s_i = \left\{ \sigma_1 = \pm i \sqrt{ 5 - \frac{6}{\sqrt{5}}}, \quad \sigma_2 = \pm i \sqrt{3 - \frac{6}{\sqrt{5}}}  \right\}, \\
&i=5, \cdots, 8, \quad s_i = \left\{ \sigma_1 = \pm i \sqrt{ 5 + \frac{6}{\sqrt{5}}}, \quad \sigma_2 = \pm i \sqrt{3 + \frac{6}{\sqrt{5}}}  \right\}, \\
\end{align*}
\begin{align*}
&i=9, \cdots, 12, \quad s_i = \bigg\{ \sigma_1 = \pm i \sqrt{(1/10) (35+12 \sqrt{5}+3 (185+80 \sqrt{5})^{1/2}}, \\
&\hspace{60mm}  \sigma_2 = \pm i \sqrt{(3/10) (15+4 \sqrt{5}-(185+80 \sqrt{5})^{1/2}} \, \bigg\}, \\
&i=13, \cdots, 16, \quad s_i = \bigg\{ \sigma_1 = \pm i \sqrt{(1/10) (35+12 \sqrt{5} - 3 (185+80 \sqrt{5})^{1/2}}, \\
&\hspace{60mm}  \sigma_2 = \pm i \sqrt{(3/10) (15+4 \sqrt{5} + (185+80 \sqrt{5})^{1/2}} \, \bigg\}, \\
&i=17, \cdots, 20, \quad s_i = \bigg\{ \sigma_1 = \pm i \sqrt{(1/10) (35 - 12 \sqrt{5} + 3 (185+80 \sqrt{5})^{1/2}}, \\
&\hspace{60mm}  \sigma_2 = \pm i \sqrt{(3/10) (15 - 4 \sqrt{5}-(185+80 \sqrt{5})^{1/2}} \, \bigg\}, \\
&i=21, \cdots, 24, \quad s_i = \bigg\{ \sigma_1 = \pm i \sqrt{(1/10) (-35+12 \sqrt{5}+3 (185+80 \sqrt{5})^{1/2}}, \\
&\hspace{60mm}  \sigma_2 = \pm i \sqrt{(3/10) (15 - 4 \sqrt{5} + (185+80 \sqrt{5})^{1/2}} \, \bigg\}.
\end{align*}
The vacua form two Weyl orbits, 
each of which contains twelve elements. 
The first orbit consists of the first through fourth solutions, 
the seventeenth through twentieth solutions, 
and the twenty-first through twenty-fourth solutions. 
The rest of the solutions form the second Weyl orbit. 
We summarize the results for the Weyl group actions 
in table~\ref{table-2}.

We checked one more case, $n=6$.
In this case,
there are ninety-one solutions in total, of which forty-eight solutions are 
not on the excluded locus.
As expected, these vacua form four Weyl orbits under the 
Weyl group action and each orbit contain twelve vacua. 
The solutions in this case are much more complicated, and so we do not
list them explicitly.

\begin{table}
	\centering
	\begin{tabular}{|c||cccccccccccc|}
		\hline
		& $e$ & $X_1$ & $X_3$ & $X_5$ & $\tilde{X}_1$ &  $\tilde{X}_3$ &  $\tilde{X}_5$ & $a_5$ & $a_4$ & $a_3$ & $a_2$ & $a_1$ \\
		\hline
		1 & 1 & 3 & 22 & 17 & 23 & 20 & 2 & 21 & 19 & 4 & 24 & 18 \\
		2 & 2 & 4 & 18 & 21 & 19 & 24 & 1 & 17 & 23 & 3 & 20 & 22 \\
		3 & 3 & 1 & 19 & 24 & 18 & 21 & 4 & 20 & 22 & 2 & 17 & 23 \\ 
		4 & 4 & 2 & 23 & 20 & 22 & 17 & 3 & 24 & 18 & 1 & 21 & 19 \\
		17 & 17 & 19 & 24 & 1 & 21 & 4 & 18 & 23 & 3 & 20 & 22 & 2 \\
		18 & 18 & 20 & 2 & 23 & 3 & 22 & 17 & 1 & 21 & 19 & 4 & 24 \\
		19 & 19 & 17 & 3 & 22 & 2 & 23 & 20 & 4 & 24 & 18 & 1 & 21 \\
		20 & 20 & 18 & 21 & 4 & 24 & 1 & 19 & 22 & 2 & 17 & 23 & 3 \\
		21 & 21 & 23 & 20 & 2 & 17 & 3 & 22 & 19 & 4 & 24 & 18 & 1 \\
		22 & 22 & 24 & 1 & 19 & 4 & 18 & 21 & 2 & 17 & 23 & 3 & 20 \\
		23 & 23 & 21 & 4 & 18 & 1 & 19 & 24 & 3 & 20 & 22 & 2 & 17 \\
		24 & 24 & 22 & 17 & 3 & 20 & 2 & 23 & 18 & 1 & 21 & 19 & 4 \\
		5 & 5 & 7 & 16 & 9 & 13 & 12 & 6 & 15 & 11 & 8 & 14 & 10 \\
		6 & 6 & 8 & 10 & 15 & 11 & 14 & 5 & 9 & 13 & 7 & 12 & 16 \\
		7 & 7 & 5 & 11 & 14 & 10 & 15 & 8 & 12 & 16 & 6 & 9 & 13 \\ 
		8 & 8 & 6 & 13 & 12 & 16 & 9 & 7 & 14 & 10 & 5 & 15 & 11 \\
		9 & 9 & 11 & 14 & 5 & 15 & 8 & 10 & 13 & 7 & 12 & 16 & 6 \\
		10 & 10 & 12 & 6 & 13 & 7 & 16 & 9 & 5 & 15 & 11 & 8 & 14 \\
		11 & 11 & 9 & 7 & 16 & 6 & 13 & 12 & 8 & 14 & 10 & 5 & 15 \\
		12 & 12 & 10 & 15 & 8 & 14 & 5 & 11 & 16 & 6 & 9 & 13 & 7 \\
		13 & 13 & 15 & 8 & 10 & 5 & 11 & 14 & 7 & 12 & 16 & 6 & 9 \\ 
		14 & 14 & 16 & 9 & 7 & 12 & 6 & 13 & 10 & 5 & 15 & 11 & 8 \\
		15 & 15 & 13 & 12 & 6 & 9 & 7 & 16 & 11 & 8 & 14 & 10 & 5 \\
		16 & 16 & 14 & 5 & 11 & 8 & 10 & 15 & 6 & 9 & 13 & 7 & 12 \\
		
		\hline
	\end{tabular}
	\caption{Weyl group actions on the vacua of five fundamental matter multiplets}
	\label{table-2}
\end{table}

\subsection{Pure gauge theory}

In this section, we will check (at the level of topological field
theory computations)  that the pure $G_2$ theory flows in the IR to 
a free theory of two chiral multiplets. 
The superpotential of the pure gauge theory is 
\begin{align*}
W =& \sum_{a=1}^2 \sigma_a \left( \alpha_m^a Z_m + \tilde{\alpha}_m^a \tilde{Z}_m \right)  + \sum_m (X_m + \tilde{X}_m) , \\
=& \sigma_1 \bigg[ \Big(Z_1 - Z_2 + (1/2) Z_3  - (1/2) Z_4 - (1/2) Z_5 + (1/2) Z_6 \Big) \\
& \quad \quad + \Big(- (3/2) \tilde{Z}_1 + (3/2) \tilde{Z}_2 + (3/2) \tilde{Z}_3
- (3/2) \tilde{Z}_4 \Big) \bigg]
\\
& +  \sigma_2 (\sqrt{3}/2) \biggl( 
  Z_3 - Z_4 + Z_5 - Z_6 
+
  \tilde{Z}_1 - \tilde{Z}_2 +
\tilde{Z}_3 - \tilde{Z}_4 + 2 \tilde{Z}_5 - 
2\tilde{Z}_6  \biggr)
\\
& + \sum_{m=1}^6(X_m + \tilde{X}_m).
\end{align*}

Integrating out $X_m$ and $\tilde{X}_m$, one obtains the constraints,
\begin{align*}
& X_m = \sum_a \sigma_a \alpha_m^a, \\
& \tilde{X}_m = \sum_a \sigma \tilde{\alpha}_m^a.
\end{align*}
The point is that all the $X_m$ fields and $\tilde{X}_m$ fields 
correspond to the nonzero roots of $G_2$ which come in pairs, 
positive roots and their negatives.
As a result, pluging the constraints above back into the superpoential, 
one gets $W=0$. Therefore, the pure gauge theory indeed flows to a
free theory of two twisted chiral multiplies in the IR limit.

On the other hand, integrating out $\sigma_1$ and $\sigma_2$, one obtains the constraints,
\begin{align*}
&- \ln X_1 + \ln X_2 - (1/2)\ln X_3 + (1/2) \ln X_4 + (1/2) \ln X_5 - (1/2) \ln X_6 \\
&\quad \quad  + (3/2) \ln \tilde{X}_1 -(3/2) \ln \tilde{X}_2 - (3/2)\tilde{X}_4 + (3/2)\tilde{X}_4 = 0, \\
&- \frac{\sqrt{3}}{2} ( \ln X_3 - \ln X_4 + \ln X_5 - \ln X_6) - \frac{\sqrt{3}}{2} (\tilde{X}_1 - \tilde{X}_2 + \tilde{X}_3 - \tilde{X}_4 + 2 \tilde{X}_5- 2 \tilde{X}_6) = 0
\end{align*}
With those two constraints, one can eliminate two fields in the superpotential,
\begin{align*}
X_4 = a \frac{ X_1 X_3 \tilde{X}_2 \tilde{X}_3^2 \tilde{X}_5}{X_2 \tilde{X}_1 \tilde{X}_4^2 \tilde{X}_6}, \quad X_5 = b \frac{ X_1 X_6 \tilde{X}_2^2 \tilde{X}_3 \tilde{X}_6}{X_2 \tilde{X}_1^2 \tilde{X}_4 \tilde{X}_5}, 
\end{align*}
with $a= \pm 1$ and $b= \pm 1$.
Plugging this back into the superpotential, we get
\begin{align}
W =& X_1 + X_2 + X_3 + X_6 + \tilde{X}_ 1+ \tilde{X}_2 + \tilde{X}_3 + \tilde{X}_4 + \tilde{X}_5 + \tilde{X}_6 \nonumber \\
&\hspace{50mm} + a \frac{ X_1 X_3 \tilde{X}_2 \tilde{X}_3^2 \tilde{X}_5}{X_2 \tilde{X}_1 \tilde{X}_4^2 \tilde{X}_6} + b \frac{ X_1 X_6 \tilde{X}_2^2 \tilde{X}_3 \tilde{X}_6}{X_2 \tilde{X}_1^2 \tilde{X}_4 \tilde{X}_5}. \label{sp-p-s1s2}
\end{align}
The critical loci are given by
\begin{align*}
& a = b =1, \\
& X_1 = - X_2 = X_3 + X_6, \\
& \tilde{X}_1 = - \tilde{X}_2 = - X_3 - 2 X_6, \\
& \tilde{X}_3 = - \tilde{X}_4 = 2 X_3 + X_6, \\
& \tilde{X}_5 = - \tilde{X}_6 = X_3 - X_6.
\end{align*}
One can easily see that, on the critical locus, the above superpotential (\ref{sp-p-s1s2}) vanishes with two free fields $X_3$ and $X_6$. Therefore, the pure gauge theory again flows to free theories of two chiral multiplies in the IR.

\subsection{Comparison with A model results}

In this section, we will discuss the A-twisted gauge theory 
with gauge group $G_2$ and $n$ chiral superfields in the ${\bf 7}$,
and compare it to results from
our proposed mirror, as a check of our methods. 
In principle, this should necessarily work, for reasons discussed
in \cite{GuSharpe}[section 3]; however, we will check for the special
case of $G_2$ that indeed everything works as it should, which will also
give us the opportunity to discuss the role of theta angle periodicities
and charge lattice normalizations.

The one-loop effective twisted superpotential $\tilde{W}_{\rm eff}$ of the 
A-twisted gauge theory is given by
\cite{Nekrasov:2014xaa}[equ'ns (2.17), (2.19)],
\cite{Closset:2015rna}[equ'ns (3.16), (3.17)]
\begin{align*}
\tilde{W}_{\rm eff} =& 
- \sum_{i,\alpha} ( \sigma_1 \rho^{' 1}_{i,\alpha} + \sigma_2 \rho^{' 2}_{i,\alpha} - \tilde{m}_i) \left( \ln(  \sigma_1 \rho^{' 1}_{i,\alpha} + \sigma_2 \rho^{' 2}_{i,\alpha} - \tilde{m}_i) - 1 \right) \\
&\quad + \sum_m (\sigma_1 \alpha^{'1}_m + \sigma_2 \alpha^{'2}_m)
\left( \ln(\sigma_1 \alpha^{' 1}_m + \sigma_2 \alpha^{' 2}_m) -1 \right)
\\
& \quad
 + \sum_m  (\sigma_1 \tilde{\alpha}^{'1}_m + \sigma_2 \tilde{\alpha}^{'2}_m)
\left(
\ln(\sigma_1 \tilde{\alpha}^{'1}_m + \sigma_2 \tilde{\alpha}^{'2}_m) -1 \right).
\end{align*}
Since the logarithm branch cuts in the expressions above are supposed
to reflect (standard) theta angle periodicities of $2 \pi$, 
we have rescaled $\rho$ and
$\alpha$ to $\rho'$ and $\alpha'$.  Specifically, we have rescaled all the
charges under $\sigma_1$ by a factor of $2$ and all the charges
under $\sigma_2$ by a factor of $2 / \sqrt{3}$.

Since
the roots and weights of $G_2$ come in positive/negative pairs,
we can further simplify the effective superpotential:
\begin{align*}
\tilde{W}_{\rm eff} =&
 - \sum_{i,\alpha} ( \sigma_1 \rho^{'1}_{i,\alpha} + \sigma_2 \rho^{'2}_{i,\alpha} - \tilde{m}_i)
  \ln(  \sigma_1 \rho^{'1}_{i,\alpha} + \sigma_2 \rho^{'2}_{i,\alpha} - \tilde{m}_i)
  - 7 \sum_i \tilde{m}_i + 6 \pi i (\sigma_1 + \sigma_2).
\end{align*}
The vacua are given by
\begin{align*}
&\prod_{i,\alpha} (\sigma_1 \rho^{'1}_{i,\alpha} + \sigma_2 \rho^{'2}_{i,\alpha} - \tilde{m}_i)^{\rho_{i,\alpha}^{'1}} = 1,\\
&\prod_{i,\alpha} (\sigma_1 \rho^{'1}_{i,\alpha} + \sigma_2 \rho^{'2}_{i,\alpha} - \tilde{m}_i)^{\rho_{i,\alpha}^{'2}} = 1.
\end{align*}
Plugging in the charges, we get
\begin{align*}
&\prod_i (2 \sigma_1 - \tilde{m}_i)^2 (\sigma_1 + \sigma_2 - \tilde{m}_i) (\sigma_1 - \sigma_2 - \tilde{m}_i) \\
&\hspace{40mm} = \prod_i (- 2 \sigma_1 - \tilde{m}_i)^2 (-\sigma_1 - \sigma_2 - \tilde{m}_i) (-\sigma_1 + \sigma_2 - \tilde{m}_i), \\
&\prod_i (\sigma_1 + \sigma_2 - \tilde{m}_i) (-\sigma_1 + \sigma_2 -  \tilde{m}_i)  \\
&\hspace{40mm} = \prod_i (-\sigma_1 - \sigma_2 - \tilde{m}_i) (\sigma_1 - \sigma_2 - \tilde{m}_i).
\end{align*}
The relations above are the same as the Coulomb ring relations (\ref{qcr-g2-3}),
(\ref{qcr-g2-4}) we derived from the B model with a suitable rescaling of the $\sigma$ fields,
\begin{equation*}
\sigma_1 \to \frac{1}{2} \sigma_1, \quad \sigma_2 \to \frac{\sqrt{3}}{2} \sigma_2. 
\end{equation*}
Thus, we see that A model results match those of the B model mirror,
as expected, after correctly taking into account subtleties in theta angle
periodicities.

\subsection{Comparison to other bases for weight lattice}

So far in this section, we have used a particular basis for the weight
lattice for $G_2$.
In principle, other bases are related by field redefinitions.
To make this more explicit, in this section we will outline corresponding
results in a different basis for the weight lattice, specifically a basis
of fundamental weights.  We will describe this basis in greater detail
in the section on $F_4$, as it will be used for the rest of the exceptional
gauge groups in this paper, but for the moment, will content ourselves to
briefly outline results.

In terms of that basis of fundamental weights, it can be shown that
the roots and pertinent weights of $G_2$ are expanded as given
in table~\ref{table:g2:alt-basis}.

 \begin{table}[h!]
        \centering
        \begin{tabular}{cc|cc|cc}
                Field & Short root & Field & Long root & Field & Weight \\ \hline
                $X_1$ & $\textbf{(}1,0\textbf{)}$ & $\tilde{X}_1$ & $\textbf{(}0,1\textbf{)}$ & $Y_{i1}$ & $\textbf{(}1,0\textbf{)}$ \\
                $X_2$ & $\textbf{(}-1,1\textbf{)}$ & $\tilde{X}_2$ & $\textbf{(}3,-1\textbf{)}$ & $Y_{i2}$ & $\textbf{(}-1,1\textbf{)}$ \\
                $X_3$ & $\textbf{(}2,-1\textbf{)}$ & $\tilde{X}_3$ & $\textbf{(}-3,2\textbf{)}$ &
                $Y_{i3}$ & $\textbf{(}2,-1\textbf{)}$ \\
                $X_4$ & $\textbf{(}-1,0\textbf{)}$ & $\tilde{X}_4$ & $\textbf{(}0,-1\textbf{)}$ &
                $Y_{i4}$ & $\textbf{(}-1,0\textbf{)}$ \\
                $X_5$ & $\textbf{(}1,-1\textbf{)}$ & $\tilde{X}_5$ & $\textbf{(}-3,1\textbf{)}$ &
                $Y_{i5}$ & $\textbf{(}1,-1\textbf{)}$  \\
                $X_6$ & $\textbf{(}-2,1\textbf{)}$ & $\tilde{X}_6$ & $\textbf{(}3,-2\textbf{)}$ &
                $Y_{i6}$ & $\textbf{(}-2,1\textbf{)}$ \\
                & & & & $Y_{i0}$ & $\textbf{(}0,0\textbf{)}$
        \end{tabular}
        \caption{Roots and weights for $G_2$ and associated fields. 
                \label{table:g2:alt-basis} } 
 \end{table}

Repeating the same mirror analysis as described earlier in this
section, we derive the Coulomb branch relations
\begin{eqnarray}
\prod_{i=1}^{n} (\sigma^\prime_1-\tilde{m}_i)(2\sigma^\prime_1-\sigma^\prime_2-\tilde{m}_i)
& = &
\prod_{i=1}^{n} (-\sigma^\prime_1-\tilde{m}_i)(-2\sigma^\prime_1+\sigma^\prime_2-\tilde{m}_i) ,
\\
\prod_{i=1}^{n} (\sigma^\prime_1-\sigma^\prime_2-\tilde{m}_i)(2\sigma^\prime_1-\sigma^\prime_2-\tilde{m}_i)
& = &
\prod_{i=1}^{n} (-\sigma^\prime_1+\sigma^\prime_2-\tilde{m}_i)(-2\sigma^\prime_1+\sigma^\prime_2-\tilde{m}_i) ,
\end{eqnarray}
and excluded loci
\begin{equation}
\sigma^\prime_1 \sigma^\prime_2 (\sigma^\prime_1-\sigma^\prime_2)
(2\sigma^\prime_1-\sigma^\prime_2)(3\sigma^\prime_1-\sigma^\prime_2)
(3\sigma^\prime_1-2\sigma^\prime_2) \: \neq  \:0,
\end{equation}
\begin{eqnarray}
\lefteqn{
\prod_{i=1}^{n} (\sigma^\prime_1-\tilde{m}_i)(-\sigma^\prime_1+\sigma^\prime_2-\tilde{m}_i)(2\sigma^\prime_1-\sigma^\prime_2-\tilde{m}_i)
} \nonumber \\
& & \cdot
(-\sigma^\prime_1-\tilde{m}_i)(\sigma^\prime_1-\sigma^\prime_2-\tilde{m}_i)(-2\sigma^\prime_1+\sigma^\prime_2-\tilde{m}_i) \: \neq \: 0.
\end{eqnarray}

Comparing with earlier reuslts for the critical locus~(\ref{qcr-g2-1}),
(\ref{qcr-g2-2}) and
excluded loci~(\ref{g2-el1}), (\ref{g2-el2}), 
computed in the earlier basis, we find that the
results above are related by the following linear field redefinitions
\begin{eqnarray}
\sigma^\prime_1 & = & \frac{1}{2}\sigma_1+\frac{\sqrt{3}}{2}\sigma_2 ,
\\
\sigma^\prime_2 & = & \sqrt{3}\sigma_2,
\end{eqnarray}
or equivalently
\begin{eqnarray}
\sigma_1 & = & 2\sigma^\prime_1-\sigma^\prime_2 , 
\\
\sigma_2 & = & \frac{1}{\sqrt{3}}\sigma_2^\prime.
\end{eqnarray}

\section{$F_4$} \label{section3}

 In this section we will consider the mirror Landau-Ginzburg orbifold 
of an $F_4$ gauge theory with matter fields in the ${\bf 26}$ fundamental representation,
and then compute Coulomb branch relations.
We also consider the pure gauge theory without matter fields.

\subsection{Mirror Landau-Ginzburg orbifold}

The mirror Landau-Ginzburg model has fields
\begin{itemize}
	\item $Y_{i,\beta}$, $i \in \{1, \cdots, n\}$, $\beta \in \{1, \cdots, 26\}$,
	corresponding to the matter fields in $n$ copies of the fundamental ${\bf 26}$ dimensional representation of
	$F_4$,
	\item $X_m$, $m \in \{1, \cdots, 48\}$, corresponding to the roots of $F_4$,
	\item $\sigma_a$, $a \in \{1, 2, 3, 4 \}$.
\end{itemize}

We associate the roots, $\alpha^a_m$, to $X_m$ fields and the weights, 
$\rho^a_{i,\beta}$, of the fundamental ${\bf 26}$ representation 
to $Y_{i,\beta}$.

Now, previously for $G_2$, we worked with a basis in which the
$\theta$-angle periodicities were unusual: $\theta_1 \sim \theta_1 + \pi i$,
$\theta_2 \sim \theta_2 + 4 \pi i / \sqrt{3}$.
This essentially just corresponded to a nonstandard charge lattice
normalization.
This was convenient for relating to Lie algebras, but, is rather unusual
for physics.

Here, for $F_4$ and all the later examples we will discuss in this paper,
we would like instead to work with a basis for the roots and weights that
corresponds to an integer charge lattice, so that the $\theta$-angle
periodicities take a more nearly standard form.  In particular,
the superpotential is invariant under such basis changes, since its terms
are tensor contractions such as
\begin{displaymath}
\sum_a \sigma_a \rho^a_i Y^i.
\end{displaymath}
We can pick any basis we like, so long as we consistently change coordinates
in the tensors above.  In particular, the superpotential (for this B-twisted
Landau-Ginzburg model) does not depend explicitly upon {\it e.g.} the
Cartan matrix, so the metric on the Lie algebra is not directly relevant
in the presentation above.  Thus, we have the flexibility to pick a basis
such that the weights have integer coordinates, which yields standard
$\theta$-angle periodicities.

To be specific, we will write the weights and roots in terms of a basis
of fundamental weights.  Recall the fundamental weights are defined
as follows.
First, let $\{ \alpha_{\mu} \}$ be a basis of simple roots, normalized so
that the Cartan matrix $C_{\mu \nu}$ is given as
\begin{equation}
C_{\mu \nu} \: = \: 2 \frac{ \alpha_{\mu} \cdot \alpha_{\nu} }{
\alpha_{\nu}^2 }.
\end{equation}
The fundamental weights $\{ \omega_{\mu} \}$ are then
defined by the property that \cite{humphreys}[section 13.1]
\begin{equation}
2 \frac{ \alpha_{\mu} \cdot \omega_{\nu} }{ \alpha_{\mu}^2 } \: = \:
\delta_{\mu \nu}.
\end{equation}
Furthermore, the fundamental weights form an integer basis for the
weight lattice -- every element of the weight lattice is a linear
combination of fundamental weights with integer
coefficients \cite{humphreys}[section 13.1].
This is perfect for our purposes, as this basis yields standard
$\theta$-angle periodicities, and we will use this basis for all
computations in this and later sections.
To compute root and weight vectors as linear combinations of the
fundamental weights, as displayed in tables in this and later sections,
we used
the Mathematica package LieART \cite{lieart}.

The long roots and associated fields are listed in 
table~\ref{table:f4:longroots}.
The short roots and associated fields are listed in
table~\ref{table:f4:shortroots}.
The weights of the ${\bf 26}$ and associated fields are listed
in table~\ref{table:f4:weights}.

\begin{table}[h]
\centering
	\begin{tabular}{cc|cc}
		Field & Positive root & Field & Negative root  \\ \hline
		$X_{1}$ & $\textbf{(}1,0,0,0\textbf{)}$ & $X_{25}$ & $\textbf{(}-1,0,0,0\textbf{)}$  \\
		$X_{2}$ & $\textbf{(}-1,1,0,0\textbf{)}$ & $X_{26}$ & $\textbf{(}1,-1,0,0\textbf{)}$  \\
		$X_{3}$ & $\textbf{(}0,-1,2,0\textbf{)}$ & $X_{27}$ & $\textbf{(}0,1,-2,0\textbf{)}$  \\
		$X_{4}$ & $\textbf{(}0,1,-2,2\textbf{)}$ & $X_{28}$ & $\textbf{(}0,-1,2,-2\textbf{)}$  \\
		$X_{5}$ & $\textbf{(}1,-1,0,2\textbf{)}$ & $X_{29}$ & $\textbf{(}-1,1,0,-2\textbf{)}$  \\
		$X_{6}$ & $\textbf{(}-1,0,0,2\textbf{)}$ & $X_{30}$ & $\textbf{(}1,0,0,-2\textbf{)}$  \\
		$X_{7}$ & $\textbf{(}0,1,0,-2\textbf{)}$ & $X_{31}$ & $\textbf{(}0,-1,0,2\textbf{)}$  \\
		$X_{8}$ & $\textbf{(}1,-1,2,-2\textbf{)}$ & $X_{32}$ & $\textbf{(}-1,1,-2,2\textbf{)}$  \\
		$X_{9}$ & $\textbf{(}-1,0,2,-2\textbf{)}$ & $X_{33}$ & $\textbf{(}1,0,-2,2\textbf{)}$  \\
		$X_{10}$ & $\textbf{(}1,1,-2,0\textbf{)}$ & $X_{34}$ & $\textbf{(}-1,-1,2,0\textbf{)}$  \\
		$X_{11}$ & $\textbf{(}-1,2,-2,0\textbf{)}$ & $X_{35}$ & $\textbf{(}1,-2,2,0\textbf{)}$  \\
		$X_{12}$ & $\textbf{(}2,-1,0,0\textbf{)}$ & $X_{36}$ & $\textbf{(}-2,1,0,0\textbf{)}$  
	\end{tabular}
\caption{Long roots of $F_4$ and associated fields. 
\label{table:f4:longroots} }
\end{table}

\begin{table}[h]
\centering
	\begin{tabular}{cc|cc}
		Field & Positive root & Field & Negative root  \\ \hline
		$X_{13}$ & $\textbf{(}0,0,0,1\textbf{)}$ & $X_{37}$ & $\textbf{(}0,0,0,-1\textbf{)}$  \\
		$X_{14}$ & $\textbf{(}0,0,1,-1\textbf{)}$ & $X_{38}$ & $\textbf{(}0,0,-1,1\textbf{)}$  \\
		$X_{15}$ & $\textbf{(}0,1,-1,0\textbf{)}$ & $X_{39}$ & $\textbf{(}0,-1,1,0\textbf{)}$  \\
		$X_{16}$ & $\textbf{(}1,-1,1,0\textbf{)}$ & $X_{40}$ & $\textbf{(}-1,1,-1,0\textbf{)}$  \\
		$X_{17}$ & $\textbf{(}-1,0,1,0\textbf{)}$ & $X_{41}$ & $\textbf{(}1,0,-1,0\textbf{)}$  \\
		$X_{18}$ & $\textbf{(}1,0,-1,1\textbf{)}$ & $X_{42}$ & $\textbf{(}-1,0,1,-1\textbf{)}$  \\
		$X_{19}$ & $\textbf{(}-1,1,-1,1\textbf{)}$ & $X_{43}$ & $\textbf{(}1,-1,1,-1\textbf{)}$  \\
		$X_{20}$ & $\textbf{(}1,0,0,-1\textbf{)}$ & $X_{44}$ & $\textbf{(}-1,0,0,1\textbf{)}$  \\
		$X_{21}$ & $\textbf{(}-1,1,0,-1\textbf{)}$ & $X_{45}$ & $\textbf{(}1,-1,0,1\textbf{)}$  \\
		$X_{22}$ & $\textbf{(}0,-1,1,1\textbf{)}$ & $X_{46}$ & $\textbf{(}0,1,-1,-1\textbf{)}$  \\
		$X_{23}$ & $\textbf{(}0,-1,2,-1\textbf{)}$ & $X_{47}$ & $\textbf{(}0,1,-2,1\textbf{)}$  \\
		$X_{24}$ & $\textbf{(}0,0,-1,2\textbf{)}$ & $X_{48}$ & $\textbf{(}0,0,1,-2\textbf{)}$  
	\end{tabular}
\caption{Short roots of $F_4$ and associated fields. 
\label{table:f4:shortroots} }
\end{table}

\begin{table}[h]
\centering
	\begin{tabular}{cc|cc}
		Field & Weight & Field & Weight  \\ \hline
		$Y_{i,1}$ & $\textbf{(}0,0,0,1\textbf{)}$ & $Y_{i,13}$ & $\textbf{(}0,0,0,-1\textbf{)}$  \\
		$Y_{i,2}$ & $\textbf{(}0,0,1,-1\textbf{)}$ & $Y_{i,14}$ & $\textbf{(}0,0,-1,1\textbf{)}$  \\
		$Y_{i,3}$ & $\textbf{(}0,1,-1,0\textbf{)}$ & $Y_{i,15}$ & $\textbf{(}0,-1,1,0\textbf{)}$  \\
		$Y_{i,4}$ & $\textbf{(}1,-1,1,0\textbf{)}$ & $Y_{i,16}$ & $\textbf{(}-1,1,-1,0\textbf{)}$  \\
		$Y_{i,5}$ & $\textbf{(}-1,0,1,0\textbf{)}$ & $Y_{i,17}$ & $\textbf{(}1,0,-1,0\textbf{)}$  \\
		$Y_{i,6}$ & $\textbf{(}1,0,-1,1\textbf{)}$ & $Y_{i,18}$ & $\textbf{(}-1,0,1,-1\textbf{)}$  \\
		$Y_{i,7}$ & $\textbf{(}-1,1,-1,1\textbf{)}$ & $Y_{i,19}$ & $\textbf{(}1,-1,1,-1\textbf{)}$  \\
		$Y_{i,8}$ & $\textbf{(}1,0,0,-1\textbf{)}$ & $Y_{i,20}$ & $\textbf{(}-1,0,0,1\textbf{)}$  \\
		$Y_{i,9}$ & $\textbf{(}-1,1,0,-1\textbf{)}$ & $Y_{i,21}$ & $\textbf{(}1,-1,0,1\textbf{)}$  \\
		$Y_{i,10}$ & $\textbf{(}0,-1,1,1\textbf{)}$ & $Y_{i,22}$ & $\textbf{(}0,1,-1,-1\textbf{)}$  \\
		$Y_{i,11}$ & $\textbf{(}0,-1,2,-1\textbf{)}$ & $Y_{i,23}$ & $\textbf{(}0,1,-2,1\textbf{)}$  \\
		$Y_{i,12}$ & $\textbf{(}0,0,-1,2\textbf{)}$ & $Y_{i,24}$ & $\textbf{(}0,0,1,-2\textbf{)}$  \\
		$Y_{i,25}$ & $\textbf{(}0,0,0,0\textbf{)}$ & $Y_{i,26}$ & $\textbf{(}0,0,0,0\textbf{)}$  \\
	\end{tabular}
\caption{Weights of ${\bf 26}$ of $F_4$ and associated fields. 
\label{table:f4:weights} }
\end{table}

Now, plugging the information above into the 
mirror superpotential with twisted masses
\begin{eqnarray} \nonumber
W & = & \sum_{a=1}^{4}\sigma_a \Big( \sum_{i=1}^{n}\sum_{\beta=1}^{26} \rho^a_{i,\beta}Y_{i,\beta}+\sum_{m=1}^{48} \alpha^a_{m}Z_m  \Big)
\\
& & -\sum_{i=1}^{n}\tilde{m}_i\sum_{\beta=1}^{26}Y_{i,\beta}+\sum_{i=1}^{n}\sum_{\beta=1}^{26}\exp(-Y_{i,\beta})+\sum_{m=1}^{48}X_m ,
\end{eqnarray} 
where $X_m=\exp(-Z_m)$ and $X_m$ are the fundamental fields, we get
\begin{align}  
\hspace*{0.15in}
 W=\sum_{a=1}^{4} \sigma_a \mathcal{C}^a-\sum_{i=1}^{n}\tilde{m}_i \sum_{\beta=1}^{26} Y_{i,\beta}+\sum_{i=1}^{n} \sum_{\beta=1}^{26} \exp(-Y_{i,\beta})+\sum_{m=1}^{48}X_m   .
\end{align}
where the $\mathcal{C}^a$ are given as follows:
\begin{align}
\nonumber \mathcal{C}^1=&\sum_{i=1}^{n} \big( Y_{i,4}-Y_{i,5}+Y_{i,6}-Y_{i,7}+Y_{i,8}-Y_{i,9}+Y_{i,19}-Y_{i,20}+Y_{i,21}-Y_{i,16}+Y_{i,17}\\ \nonumber & \hspace*{4.8in}-Y_{i,18}         \big)   \\ \nonumber
&
\hspace*{.5in}
+Z_{1}-Z_{2}+Z_{5}-Z_{6}+Z_{16}-Z_{17}+Z_{8}+Z_{18}-Z_{9}-Z_{19}+Z_{20}-Z_{21}\\ \nonumber & \hspace*{0.5in} +Z_{10}-Z_{11}+2Z_{12}   
-Z_{25}+Z_{26}-Z_{29}+Z_{30}-Z_{40}+Z_{41}-Z_{32}-Z_{42}\\ \nonumber & \hspace*{2.3 in}+Z_{33}+Z_{43}-Z_{44}+Z_{45}-Z_{34}+Z_{35}-2Z_{36},
\end{align}
\begin{align} \nonumber
\mathcal{C}^2=&\sum_{i=1}^{n} \big( Y_{i,3}-Y_{i,4}+Y_{i,7}+Y_{i,9}-Y_{i,10}-Y_{i,11}-Y_{i,19}-Y_{i,21}+Y_{i,22}+Y_{i,23}-Y_{i,15} \\ \nonumber & \hspace*{4.9in} +Y_{i,16}         \big)   \\ \nonumber
&  
\hspace*{0.5in}
+Z_{2}-Z_{3}+Z_{4}+Z_{15}-Z_{5}+Z_{7}-Z_{16}-Z_{8}+Z_{19}+Z_{21}-Z_{22}+Z_{10}\\ \nonumber & \hspace*{.5in}+2Z_{11}-Z_{23}-Z_{12}   
\hspace*{0.1in}
-Z_{26}+Z_{27}-Z_{28}-Z_{39}+Z_{29}-Z_{31}+Z_{40}+Z_{32}\\ \nonumber & \hspace*{2.3in}-Z_{43}-Z_{45}  +Z_{46}-Z_{34}-2Z_{35}+Z_{47}+Z_{36}, \end{align}
\begin{align} \nonumber
\mathcal{C}^3=&  \sum_{i=1}^{n} \big( Y_{i,2}-Y_{i,3}+Y_{i,4}+Y_{i,5}-Y_{i,6}-Y_{i,7}+Y_{i,10}+2Y_{i,11}-Y_{i,12}+Y_{i,19}-Y_{i,22}\\ \nonumber &  \hspace*{1.9in}-2Y_{i,23}+Y_{i,24}-Y_{i,14}+Y_{i,15}-Y_{i,16}-Y_{i,17}+Y_{i,18}         \big)  \\ \nonumber \\ \nonumber & \hspace*{.5in}
+2Z_{3}+Z_{14}-2Z_{4}-Z_{15}+Z_{16}+Z_{17}+2Z_{8}-Z_{18}+2Z_{9}-Z_{19}+Z_{22}\\ \nonumber & \hspace*{.5in}-2Z_{10}-2Z_{11}+2Z_{23}-Z_{24}  
-2Z_{27}-Z_{38}+2Z_{28}+Z_{39}-Z_{40}-Z_{41}\\ \nonumber & \hspace*{1.2in}-2Z_{32}+Z_{42}  -2Z_{33}+Z_{43}-Z_{46}+2Z_{34}+2Z_{35}-2Z_{47}+Z_{48}  ,  
\end{align}
\begin{align} \nonumber
\mathcal{C}^4=& \sum_{i=1}^{n} \big( Y_{i,1}-Y_{i,2}+Y_{i,6}+Y_{i,7}-Y_{i,8}-Y_{i,9}+Y_{i,10}-Y_{i,11}+2Y_{i,12}-Y_{i,19}+Y_{i,20} \\ \nonumber & \hspace*{2in} +Y_{i,21}-Y_{i,22}+Y_{i,23}-2Y_{i,24}-Y_{i,13}+Y_{i,14}-Y_{i,18}       \big)  \\ \nonumber \\ \nonumber
& \hspace*{.5 in}
+Z_{13}-Z_{14}+2Z_{4}+2Z_{5}+2Z_{6}-2Z_{7}-2Z_{8}+Z_{18}-2Z_{9}+Z_{19}-Z_{20}\\ \nonumber & \hspace*{.5 in}-Z_{21}+Z_{22}-Z_{23}+2Z_{24}
-Z_{37}+Z_{38}-2Z_{28}-2Z_{29}-2Z_{30}+2Z_{31}\\ \nonumber & \hspace*{1.35 in}+2Z_{32}-Z_{42}  +2Z_{33}-Z_{43}+Z_{44}+Z_{45}-Z_{46}+Z_{47}-2Z_{48}.
\end{align}
Integrating out $\sigma_a$ fields, we get four constraints $\mathcal{C}^a=0$. So we are free to eliminate four fundamental fields. Our choice here will be $Y_{n,1}$, $Y_{n,2}$, $Y_{n,3}$ and $Y_{n,4}$.
\begin{align*}
-Y_{n,1}=&\sum_{i=1}^{n-1}Y_{i,1}+\sum_{i=1}^{n}\big( Y_{i,5}+Y_{i,7}-Y_{i,8}+Y_{i,10}+Y_{i,12}-Y_{i,19}+Y_{i,20}-Y_{i,22}-Y_{i,24}-Y_{i,13} \\ \nonumber & \hspace*{5.1in} -Y_{i,17} \big) \\
&+Z_{2}+Z_{3}+Z_{13}+Z_{4}+Z_{5}+2Z_{6}-Z_{7}+Z_{17}-Z_{8}+Z_{19}-Z_{20}+Z_{22}-Z_{10}\\ &+Z_{24}-Z_{12}-Z_{26}-Z_{27}-Z_{37}-Z_{28}-Z_{29}-2Z_{30}+Z_{31}-Z_{41}+Z_{32}-Z_{43}\\ &+Z_{44}-Z_{46}+Z_{34}-Z_{48}+Z_{36},
\end{align*}
\begin{align*}
-Y_{n,2}=&\sum_{i=1}^{n-1}Y_{i,2}+\sum_{i=1}^{n}\big( Y_{i,5}-Y_{i,6}+Y_{i,9}+Y_{i,11}-Y_{i,12}-Y_{i,21}-Y_{i,23}+Y_{i,24}-Y_{i,14}-Y_{i,17}\\ \nonumber & \hspace*{5.1in}+Y_{i,18}  \big) \\
&+Z_{2}+Z_{3}+Z_{14}-Z_{4}-Z_{5}+Z_{7}+Z_{17}+Z_{8}-Z_{18}+2Z_{9}+Z_{21}-Z_{10}+Z_{23}\\ &-Z_{24}-Z_{12}-Z_{26}-Z_{27}-Z_{38}+Z_{28}+Z_{29}-Z_{31}-Z_{41}-Z_{32}+Z_{42}-2Z_{33}\\ &-Z_{45}+Z_{34}-Z_{47}+Z_{48}+Z_{36},
\end{align*}
\begin{align*}
-Y_{n,3}=&\sum_{i=1}^{n-1}+\sum_{i=1}^{n}\big( -Y_{i,5}+Y_{i,6}+Y_{i,8}-Y_{i,10}-Y_{i,11}-Y_{i,20}+Y_{i,22}+Y_{i,23}-Y_{i,15}+Y_{i,17}\\ \nonumber & \hspace*{5.1in}-Y_{i,18} \big) \\
& +Z_{1}-Z_{3}+Z_{4}+Z_{15}-Z_{6}+Z_{7}-Z_{17}+Z_{18}-Z_{9}+Z_{20}-Z_{22}+2Z_{10}+Z_{11}\\ &-Z_{23}+Z_{12}-Z_{25}+Z_{27}-Z_{28}-Z_{39}+Z_{30}-Z_{31}+Z_{41}-Z_{42}+Z_{33}-Z_{44}\\ &+Z_{46}-2Z_{34}-Z_{35}+Z_{47}-Z_{36},
\end{align*}
\begin{align*}
-Y_{n,4}=&\sum_{i=1}^{n-1}Y_{i,4}+\sum_{i=1}^{n}\big( -Y_{i,5}+Y_{i,6}-Y_{i,7}+Y_{i,8}-Y_{i,9}+Y_{i,19}-Y_{i,20}+Y_{i,21}-Y_{i,16}+Y_{i,17}\\ \nonumber & \hspace*{5.1in} -Y_{i,18} \big) \\
& +Z_{1}-Z_{2}+Z_{5}-Z_{6}+Z_{16}-Z_{17}+Z_{8}+Z_{18}-Z_{9}-Z_{19}+Z_{20}-Z_{21}+Z_{10}\\ &-Z_{11}+2Z_{12}
-Z_{25}+Z_{26}-Z_{29}+Z_{30}-Z_{40}+Z_{41}-Z_{32}-Z_{42}+Z_{33}+Z_{43}\\ &-Z_{44}+Z_{45}-Z_{34}+Z_{35}-2Z_{36}.
\end{align*}

For convenience, we define:
\begin{align}
\nonumber \Pi_1 &\equiv  \exp(-Y_{n,1}),\\ 
\nonumber &=\prod_{i=1}^{n}\exp(Y_{i,5}+Y_{i,7}-Y_{i,8}+Y_{i,10}+Y_{i,12}-Y_{i,19}+Y_{i,20}-Y_{i,22}-Y_{i,24}-Y_{i,13}-Y_{i,17})  \\
 &\hspace*{.2in}\cdot\prod_{i=1}^{n-1}\exp(Y_{i,1}) \cdot
\frac{X_{7}X_{8}X_{20}X_{10}X_{12}X_{26}X_{27}X_{37}X_{28}X_{29}X_{30}^2X_{41}X_{43}X_{46}X_{48}}{X_{2}X_{3}X_{13}X_{4}X_{5}X_{6}^2X_{17}X_{19}X_{22}X_{24}X_{31}X_{32}X_{44}X_{34}X_{36}}, \label{Pi1}
\end{align}
\begin{align}
\nonumber \Pi_2 & \equiv \exp(-Y_{n,2}), \\ \nonumber
& = \prod_{i=1}^{n} \exp(Y_{i,5}-Y_{i,6}+Y_{i,9}+Y_{i,11}-Y_{i,12}-Y_{i,21}-Y_{i,23}+Y_{i,24}-Y_{i,14}-Y_{i,17}+Y_{i,18} ) \\ &\hspace*{.2in} \cdot \prod_{i=1}^{n-1}\exp(Y_{i,2})  \cdot
\frac{X_{4}X_{5}X_{18}X_{10}X_{24}X_{12}X_{26}X_{27}X_{38}X_{31}X_{41}X_{32}X_{33}^2X_{45}X_{47}}{X_{2}X_{3}X_{14}X_{7}X_{17}X_{8}X_{9}^2X_{21}X_{23}X_{28}X_{29}X_{42}X_{34}X_{48}X_{36}},  \label{Pi2}
\end{align}
\begin{align}
\nonumber \Pi_3 & \equiv \exp(-Y_{n,3}), \\ \nonumber
& = \prod_{i=1}^{n}\exp(-Y_{i,5}+Y_{i,6}+Y_{i,8}-Y_{i,10}-Y_{i,11}-Y_{i,20}+Y_{i,22}+Y_{i,23}-Y_{i,15}+Y_{i,17}-Y_{i,18}) \\ &\hspace*{.2in} \cdot \prod_{i=1}^{n-1}\exp(Y_{i,3}) \cdot 
\frac{X_{3}X_{6}X_{17}X_{9}X_{22}X_{23}X_{25}X_{28}X_{39}X_{31}X_{42}X_{44}X_{34}^2X_{35}X_{36}}{X_{1}X_{4}X_{15}X_{7}X_{18}X_{20}X_{10}^2X_{11}X_{12}X_{27}X_{30}X_{41}X_{33}X_{46}X_{47}},   \label{Pi3}
\end{align}
\begin{align}
\nonumber \Pi_4 & \equiv \exp(-Y_{n,4}), \\ \nonumber
& =\prod_{i=1}^{n}\exp(-Y_{i,5}+Y_{i,6}-Y_{i,7}+Y_{i,8}-Y_{i,9}+Y_{i,19}-Y_{i,20}+Y_{i,21}-Y_{i,16}+Y_{i,17}-Y_{i,18}) 
\\ &\hspace*{.2in} \cdot \prod_{i=1}^{n-1}\exp(Y_{i,4})  \cdot
\frac{X_{2}X_{6}X_{17}X_{9}X_{19}X_{21}X_{11}X_{25}X_{29}X_{40}X_{32}X_{42}X_{44}X_{34}X_{36}^2}{X_{1}X_{5}X_{16}X_{8}X_{18}X_{20}X_{10}X_{12}^2X_{26}X_{30}X_{41}X_{33}X_{43}X_{45}X_{35}}. \label{Pi4}
\end{align}

Integrating out the sigma fields and eliminating the fields, $Y_{n,1}$, $Y_{n,2}$, $Y_{n,3}$ and $Y_{n,4}$, the superpotential reduces to
\begin{align*}
W=&\sum_{i=1}^{n-1}\sum_{b=1}^{26}\big( \exp(-Y_{i,b})-\tilde{m}_iY_{i,b} \big) \\
& + \big( \Pi_1+\tilde{m}_n \ln \Pi_1 \big)+\big( \Pi_2+\tilde{m}_n \ln \Pi_2 \big)+\big( \Pi_3+\tilde{m}_n \ln \Pi_3 \big)+\big( \Pi_4+\tilde{m}_n \ln \Pi_4 \big) \\
&+\sum_{a=5}^{26}\big( \exp(-Y_{n,a})-\tilde{m}_n Y_{n,a} \big)+\sum_{m=1}^{48} X_m
\end{align*}
The superpotential is only well defined when the $X_m$ fields in the 
denominator of $\Pi_a$'s are non-zero.

The critical locus is given as follows:

For $i < n$:
\begin{eqnarray}
\frac{\partial W}{\partial Y_{i,1} } &:& \exp \left( - Y_{i,1} \right) = \Pi_1+\tilde{m}_n-\tilde{m}_i,  \\
\frac{\partial W}{\partial Y_{i,2} } &:& \exp \left( - Y_{i,2} \right) = \Pi_2+\tilde{m}_n-\tilde{m}_i, \\
\frac{\partial W}{\partial Y_{i,3} } &:& \exp \left( - Y_{i,3} \right) = \Pi_3+\tilde{m}_n-\tilde{m}_i, \\
\frac{\partial W}{\partial Y_{i,4} } &:& \exp \left( - Y_{i,4} \right) = \Pi_4+\tilde{m}_n-\tilde{m}_i, 
\end{eqnarray}
For $i \leq n$:
\begin{align}
\frac{\partial W}{\partial Y_{i,5} } &: \exp \left( - Y_{i,5} \right) = \Pi_1+\Pi_2-\Pi_3-\Pi_4-\tilde{m}_i,  \\
\frac{\partial W}{\partial Y_{i,6} } &: \exp \left( - Y_{i,6} \right) = -\Pi_2+\Pi_3+\Pi_4+\tilde{m}_n-\tilde{m}_i, \\
\frac{\partial W}{\partial Y_{i,7} } &: \exp \left( - Y_{i,7} \right) = \Pi_1-\Pi_4-\tilde{m}_i,\\
\frac{\partial W}{\partial Y_{i,8} } &: \exp \left( - Y_{i,8} \right) = -\Pi_1+\Pi_3+\Pi_4+\tilde{m}_n-\tilde{m}_i, \\
\frac{\partial W}{\partial Y_{i,9} } &: \exp \left( - Y_{i,9} \right) = \Pi_2-\Pi_4-\tilde{m}_i,  \end{align}
\begin{align}
\frac{\partial W}{\partial Y_{i,10} } &: \exp \left( - Y_{i,10} \right) = \Pi_1-\Pi_3-\tilde{m}_i, \\
\frac{\partial W}{\partial Y_{i,11} } &: \exp \left( - Y_{i,11} \right) = \Pi_2-\Pi_3-\tilde{m}_i, \\
\frac{\partial W}{\partial Y_{i,12} } &: \exp \left( - Y_{i,12} \right) = \Pi_1-\Pi_2-\tilde{m}_i, \\
\frac{\partial W}{\partial Y_{i,13} } &: \exp \left( - Y_{i,13} \right) = -\Pi_1-\tilde{m}_n-\tilde{m}_i, \\
\frac{\partial W}{\partial Y_{i,14} } &: \exp \left( - Y_{i,14} \right) = -\Pi_2-\tilde{m}_n-\tilde{m}_i,
\end{align}
\begin{align}
\frac{\partial W}{\partial Y_{i,15} } &: \exp \left( - Y_{i,15} \right) = -\Pi_3-\tilde{m}_n-\tilde{m}_i, \\
\frac{\partial W}{\partial Y_{i,16} } &: \exp \left( - Y_{i,16} \right) =-\Pi_4-\tilde{m}_n-\tilde{m}_i, \\
\frac{\partial W}{\partial Y_{i,17} } &: \exp \left( - Y_{i,17} \right) = -\Pi_1-\Pi_2+\Pi_3+\Pi_4-\tilde{m}_i, \\
\frac{\partial W}{\partial Y_{i,18} } &: \exp \left( - Y_{i,18} \right) = \Pi_2-\Pi_3-\Pi_4-\tilde{m}_n-\tilde{m}_i, \\
\frac{\partial W}{\partial Y_{i,19} } &: \exp \left( - Y_{i,19} \right) = -\Pi_1+\Pi_4-\tilde{m}_i, 
\end{align}
\begin{align}
\frac{\partial W}{\partial Y_{i,20} } &: \exp \left( - Y_{i,20} \right) = \Pi_1-\Pi_3-\Pi_4-\tilde{m}_n-\tilde{m}_i,\\
\frac{\partial W}{\partial Y_{i,21} } &: \exp \left( - Y_{i,21} \right) = -\Pi_2+\Pi_4-\tilde{m}_i, \\
\frac{\partial W}{\partial Y_{i,22} } &: \exp \left( - Y_{i,22} \right) = -\Pi_1+\Pi_3-\tilde{m}_i, \\				
\frac{\partial W}{\partial Y_{i,23} } &: \exp \left( - Y_{i,23} \right) = -\Pi_2+\Pi_3-\tilde{m}_i,\\
\frac{\partial W}{\partial Y_{i,24} } &: \exp \left( - Y_{i,24} \right) = 
-\Pi_1+\Pi_2-\tilde{m}_i, \\
\frac{\partial W}{\partial Y_{i,25} } &: \exp \left( - Y_{i,25} \right) = -\tilde{m}_i, \\
\frac{\partial W}{\partial Y_{i,26} } &: \exp \left( - Y_{i,26} \right) = -\tilde{m}_i,
\end{align}

In the same way, $\partial W/ \partial X_m $ gives:
\begin{align}
&X_{1}=\Pi_3+\Pi_4 ,  & &  X_{25}=-\Pi_3-\Pi_4, \\
&X_{2}=\Pi_1+\Pi_2-\Pi_4, & & X_{26}=-\Pi_1-\Pi_2+\Pi_4, \\
&X_{3}=\Pi_1+\Pi_2-\Pi_3, && X_{27}=-\Pi_1-\Pi_2+\Pi_3, \\
&X_{4}=\Pi_1+\Pi_3, &&  X_{28}=-\Pi_1-\Pi_3, \\
&X_{5}=\Pi_1-\Pi_2+\Pi_4, &&  X_{29}=-\Pi_1+\Pi_2-\Pi_4, 
\end{align}
\begin{align}
&X_{6}=2\Pi_1-\Pi_4,   &&  X_{30}=-2\Pi_1+\Pi_4,   \\
&X_{7}=-\Pi_1+\Pi_2+\Pi_3, &&X_{31}=\Pi_1-\Pi_2-\Pi_3, \\
&X_{8}=-\Pi_1+\Pi_2+\Pi_4, &&X_{32}=\Pi_1-\Pi_2-\Pi_4, \\
&X_{9}=2\Pi_2-\Pi_3-\Pi_4, &&X_{33}=-2\Pi_2+\Pi_3+\Pi_4,  \\
&X_{10}=-\Pi_1-\Pi_2+2\Pi_3+\Pi_4,  &&X_{34}=\Pi_1+\Pi_2-2\Pi_3-\Pi_4, 
\end{align}
\begin{align}
&X_{11}=\Pi_3-\Pi_4,  &&X_{35}=-\Pi_3+\Pi_4, \\
&X_{12}=-\Pi_1-\Pi_2+\Pi_3+2\Pi_4, &&X_{36}=\Pi_1+\Pi_2-\Pi_3-2\Pi_4, \\
&X_{13}=\Pi_1,   && X_{37}=-\Pi_1, \\
&X_{14}=\Pi_2, && X_{38}=-\Pi_2, \\
&X_{15}=\Pi_3,  &&  X_{39}=-\Pi_3,  \\
&X_{16}=\Pi_4,   &&X_{40}=-\Pi_4, 
\end{align}
\begin{align}
&X_{17}=\Pi_1+\Pi_2-\Pi_3-\Pi_4,  &&X_{41}=-\Pi_1-\Pi_2+\Pi_3+\Pi_4, \\  
&X_{18}=-\Pi_2+\Pi_3+\Pi_4, &&X_{42}=\Pi_2-\Pi_3-\Pi_4, \\
&X_{19}=\Pi_1-\Pi_4,  &&X_{43}=-\Pi_1+\Pi_4, \\
&X_{20}=-\Pi_1+\Pi_3+\Pi_4,  &&X_{44}=\Pi_1-\Pi_3-\Pi_4, \\
&X_{21}=\Pi_2-\Pi_4,   &&X_{45}=-\Pi_2+\Pi_4, \\ 
&X_{22}=\Pi_1-\Pi_3,    &&X_{46}=-\Pi_1+\Pi_3, \\
&X_{23}=\Pi_2-\Pi_3,   &&X_{47}=-\Pi_2+\Pi_3, \\
&X_{24}=\Pi_1-\Pi_2,  &&X_{48}=-\Pi_1+\Pi_2 .
\end{align}

Now, plug these constraints back into (\ref{Pi1})-(\ref{Pi4}) to get:
\begin{align*}
\Pi_1=& \prod_{i=1}^{n-1}(\Pi_1+\tilde{m}_n-\tilde{m}_i)^{-1}\prod_{i=1}^{n}(\Pi_1+\Pi_2-\Pi_3-\Pi_4-\tilde{m}_i)^{-1}(\Pi_1-\Pi_4-\tilde{m}_i)^{-1} \\
&\quad 
\cdot (-\Pi_1+\Pi_3+\Pi_4+\tilde{m}_n-\tilde{m}_i)(\Pi_1-\Pi_3-\tilde{m}_i)^{-1} (\Pi_1-\Pi_2-\tilde{m}_i)^{-1}\\
& \quad 
\cdot (-\Pi_1+\Pi_2-\tilde{m}_i)(\Pi_1-\Pi_3-\Pi_4-\tilde{m}_n-\tilde{m}_i)^{-1} (-\Pi_1+\Pi_3-\tilde{m}_i) \\ & \quad
\cdot (-\Pi_1+\Pi_2-\tilde{m}_i)(-\Pi_1-\tilde{m}_n-\tilde{m}_i)(-\Pi_1-\Pi_2+\Pi_3+\Pi_4-\tilde{m}_i) ,
\end{align*}
\begin{align*}
\Pi_2=& \prod_{i=1}^{n-1}(\Pi_2+\tilde{m}_n-\tilde{m}_i)^{-1}\prod_{i=1}^{n} (\Pi_1+\Pi_2-\Pi_3-\Pi_4-\tilde{m}_i)^{-1}(-\Pi_2+\Pi_3+\Pi_4+\tilde{m}_n-\tilde{m}_i) \\
 & \qquad
\cdot (\Pi_2-\Pi_4-\tilde{m}_i)^{-1}(\Pi_2-\Pi_3-\tilde{m}_i)^{-1}(\Pi_1-\Pi_2-\tilde{m}_i)  \\
& \qquad \cdot (-\Pi_2+\Pi_4-\tilde{m}_i)(-\Pi_2+\Pi_3-\tilde{m}_i)(-\Pi_1+\Pi_2-\tilde{m}_i)^{-1} \\& \qquad
\cdot (-\Pi_2-\tilde{m}_n-\tilde{m}_i)(-\Pi_1-\Pi_2+\Pi_3+\Pi_4-\tilde{m}_i)(\Pi_2-\Pi_3-\Pi_4-\tilde{m}_n-\tilde{m}_i)^{-1} ,
\end{align*}
\begin{align*}
\Pi_3=&
\prod_{i=1}^{n-1}(\Pi_3+\tilde{m}_n-\tilde{m}_i)^{-1}\prod_{i=1}^{n} (\Pi_1+\Pi_2-\Pi_3-\Pi_4-\tilde{m}_i)(-\Pi_2+\Pi_3+\Pi_4+\tilde{m}_n-\tilde{m}_i)^{-1} \\ & \qquad
\cdot (-\Pi_1+\Pi_3+\Pi_4+\tilde{m}_n-\tilde{m}_i)^{-1}(\Pi_1-\Pi_3-\tilde{m}_i)(\Pi_2-\Pi_3-\tilde{m}_i)\\
 & \qquad 
\cdot (\Pi_1-\Pi_3-\Pi_4-\tilde{m}_n-\tilde{m}_i)(-\Pi_1+\Pi_3-\tilde{m}_i)^{-1}(-\Pi_2+\Pi_3-\tilde{m}_i)^{-1}\\
 & \qquad 
\cdot (-\Pi_3-\tilde{m}_n-\tilde{m}_i)(-\Pi_1-\Pi_2+\Pi_3+\Pi_4-\tilde{m}_i)^{-1}(\Pi_2-\Pi_3-\Pi_4-\tilde{m}_n-\tilde{m}_i) ,
\end{align*}
\begin{align*}
\Pi_4=&
\prod_{i=1}^{n-1}(\Pi_4+\tilde{m}_n-\tilde{m}_i)^{-1}\prod_{1}^{n}(\Pi_1+\Pi_2-\Pi_3-\Pi_4-\tilde{m}_i)(-\Pi_2+\Pi_3+\Pi_4+\tilde{m}_n-\tilde{m}_i)^{-1} \\ 
& \qquad \cdot
(\Pi_1-\Pi_4-\tilde{m}_i)(-\Pi_1+\Pi_3+\Pi_4+\tilde{m}_n-\tilde{m}_i)^{-1}(\Pi_2-\Pi_4-\tilde{m}_i) \\ & \qquad
\cdot (-\Pi_1+\Pi_2-\tilde{m}_i)^{-1}(\Pi_1-\Pi_3-\Pi_4-\tilde{m}_n-\tilde{m}_i)(-\Pi_2+\Pi_4-\tilde{m}_i)^{-1} \\ & \qquad 
\cdot (-\Pi_4-\tilde{m}_n-\tilde{m}_i)(-\Pi_1-\Pi_2+\Pi_3+\Pi_4-\tilde{m}_i)^{-1}(\Pi_2-\Pi_3-\Pi_4-\tilde{m}_n-\tilde{m}_i) .
\end{align*}

The mirror maps are given by,
\begin{equation*}
\exp(-Y_{i,\beta}) \mapsto - \tilde{m}_i + \sum_{a=1}^{4} \sigma_a \rho_{i,\beta}^a, \quad X_m \mapsto \sum_{a=1}^{4} \sigma_a \alpha_m^a.
\end{equation*}
on the critical locus relations, one finds
\begin{align*}
&\Pi_1=\exp(-Y_{n,1})=\sigma_4-\tilde{m}_n ,
&&\Pi_2=\exp(-Y_{n,2})=\sigma_3-\sigma_4-\tilde{m}_n, \\
&\Pi_3=\exp(-Y_{n,3})=\sigma_2-\sigma_3-\tilde{m}_n ,
&&\Pi_4=\exp(-Y_{n,4})=\sigma_1-\sigma_2+\sigma_3-\tilde{m}_n .
\end{align*}
Plugging them back in, 
one obtains the Coulomb branch (quantum cohomology) 
ring relations for $F_4$:
\begin{align}
\nonumber	\prod_{i=1}^{n}&(-\sigma_1+\sigma_3-\tilde{m}_i)(-\sigma_1+\sigma_2-\sigma_3+\sigma_4-\tilde{m}_i)(-\sigma_2+\sigma_3+\sigma_4-\tilde{m}_i)\\
 & \cdot  \nonumber (-\sigma_3+2\sigma_4-\tilde{m}_i)(-\sigma_1+\sigma_4-\tilde{m}_i)(\sigma_4-\tilde{m}_i) \\
 \nonumber & =\prod_{i=1}^{n}(\sigma_1-\sigma_4-\tilde{m}_i)(\sigma_1-\sigma_2+\sigma_3-\sigma_4-\tilde{m}_i)(\sigma_2-\sigma_3-\sigma_4-\tilde{m}_i)\\
&  \hspace*{0.45in} \cdot (\sigma_3-2\sigma_4-\tilde{m}_i)(-\sigma_4-\tilde{m}_i)(\sigma_1-\sigma_3-\tilde{m}_i) \label{locus1} ,
\end{align}
\begin{align}
\nonumber 
\prod_{i=1}^{n}&(-\sigma_1+\sigma_3-\tilde{m}_i)(-\sigma_1+\sigma_2-\sigma_4-\tilde{m}_i)(-\sigma_2+2\sigma_3-\sigma_4-\tilde{m}_i)\\
&  \cdot \nonumber (\sigma_3-2\sigma_4-\tilde{m}_i)(-\sigma_1+\sigma_3-\sigma_4-\tilde{m}_i)(\sigma_3-\sigma_4-\tilde{m}_i)  \\
& = \nonumber \prod_{i=1}^{n}(\sigma_1-\sigma_3+\sigma_4-\tilde{m}_i)(-\sigma_3+2\sigma_4-\tilde{m}_i)(\sigma_1-\sigma_2+\sigma_4-\tilde{m}_i)\\
& \hspace*{0.45in}  \cdot (\sigma_2-2\sigma_3+\sigma_4-\tilde{m}_i)(-\sigma_3+\sigma_4-\tilde{m}_i)(\sigma_1-\sigma_3-\tilde{m}_i)    \label{locus2} ,
\end{align}
\begin{align}
\nonumber \prod_{i=1}^{n}&(\sigma_1-\sigma_3+\sigma_4-\tilde{m}_i)(\sigma_1-\sigma_4-\tilde{m}_i)(\sigma_2-\sigma_3-\sigma_4-\tilde{m}_i)\\
&  \cdot (\sigma_2-2\sigma_3+\sigma_4-\tilde{m}_i)(\sigma_1-\sigma_3-\tilde{m}_i)(\sigma_2-\sigma_3-\tilde{m}_i) \nonumber \\
& =\prod_{i=1}^{n}(-\sigma_1+\sigma_3-\tilde{m}_i)(-\sigma_2+\sigma_3+\sigma_4-\tilde{m}_i)(-\sigma_2+2\sigma_3-\sigma_4-\tilde{m}_i)\nonumber \\
& \hspace*{0.45in} \cdot (-\sigma_1+\sigma_4-\tilde{m}_i)(-\sigma_2+\sigma_3-\tilde{m}_i)(-\sigma_1+\sigma_3-\sigma_4-\tilde{m}_i) \label{locus3}  ,
\end{align}
\begin{align}
\nonumber \prod_{i=1}^{n}&(\sigma_1-\sigma_3+\sigma_4-\tilde{m}_i)(\sigma_1-\sigma_4-\tilde{m}_i)(\sigma_1-\sigma_2+\sigma_3-\sigma_4-\tilde{m}_i)\\
& \cdot \nonumber (\sigma_1-\sigma_2+\sigma_4-\tilde{m}_i)(\sigma_1-\sigma_3-\tilde{m}_i)(\sigma_1-\sigma_2+\sigma_3-\tilde{m}_i)  \\
 \nonumber & =\prod_{i=1}^{n}(-\sigma_1+\sigma_3-\tilde{m}_i)(-\sigma_1+\sigma_2-\sigma_3+\sigma_4-\tilde{m}_i)(-\sigma_1+\sigma_2-\sigma_4-\tilde{m}_i) \\
& \hspace*{0.45in} \cdot (-\sigma_1+\sigma_4-\tilde{m}_i)(-\sigma_1+\sigma_2-\sigma_3-\tilde{m}_i)(-\sigma_1+\sigma_3-\sigma_4-\tilde{m}_i)    \label{locus4} .
\end{align}

Next, let us described the excluded locus on the Coulomb branch.
As discussed previously and in \cite{GuSharpe}, part of the excluded locus
is defined by the condition $X_m \neq 0$ for all $m$.  This gives
\begin{multline}
\sigma_1(2\sigma_1-\sigma_2)(-\sigma_1+\sigma_2)(\sigma_1+\sigma_2-2\sigma_3)(-\sigma_1+2\sigma_2-2\sigma_3)(\sigma_2-\sigma_3)(-\sigma_1+\sigma_3)\\ 
\cdot (\sigma_1-\sigma_2+\sigma_3)(-\sigma_2+2\sigma_3)(\sigma_2-2\sigma_4)(-\sigma_1+2\sigma_3-2\sigma_4)(\sigma_1-\sigma_2+2\sigma_3-2\sigma_4)\\
 \cdot   (\sigma_1-\sigma_4)(-\sigma_1+\sigma_2-\sigma_4)(\sigma_3-\sigma_4)(-\sigma_2+\sigma_3-\sigma_4)\sigma_4(\sigma_1-\sigma_3+\sigma_4)(-\sigma_1+2\sigma_4) \\
 \cdot (-\sigma_1+\sigma_2-\sigma_3+\sigma_4)(-\sigma_2+\sigma_3+\sigma_4)(\sigma_1-\sigma_2+2\sigma_4)(\sigma_2-2\sigma_3+2\sigma_4)(-\sigma_3+2\sigma_4) \neq 0.  
\end{multline}

The second part of the excluded locus is determined by the condition that
$\exp(-Y) \neq 0$.  From the mirror map
\begin{displaymath}
\exp(-Y_{i,\beta})=-\tilde{m}_i+\sum_{a=1}^{4}\sigma_a \rho^a_{i,\beta},
\end{displaymath}
the excluded locus constraint becomes
\begin{displaymath}
-\tilde{m}_i+\sum_{a=1}^{4}\sigma_a \rho^a_{i,\beta} \neq 0 
\end{displaymath}
which is encoded in the expression below:
\begin{multline} 
\prod_{i=1}^{n} \left(\sigma _1-\sigma _3-\tilde{m}_i\right) \left(\sigma _2-\sigma _3-\tilde{m}_i\right) \left(-\sigma _1+\sigma _2-\sigma _3-\tilde{m}_i\right) \left(-\sigma _1+\sigma _3-\tilde{m}_i\right)
	\left(-\sigma _2+\sigma _3-\tilde{m}_i\right)  \\
 \cdot \left(\sigma _1-\sigma _2+\sigma _3-\tilde{m}_i\right) \left(\sigma _3-2 \sigma _4-\tilde{m}_i\right) \left(-\sigma _4-\tilde{m}_i\right)
	\left(\sigma _1-\sigma _4-\tilde{m}_i\right) \left(-\sigma _1+\sigma _2-\sigma _4-\tilde{m}_i\right) \\
 \cdot \left(\sigma _2-\sigma _3-\sigma _4-\tilde{m}_i\right) \left(\sigma _3-\sigma
	_4-\tilde{m}_i\right) \left(-\sigma _1+\sigma _3-\sigma _4-\tilde{m}_i\right) \left(\sigma _1-\sigma _2+\sigma _3-\sigma _4-\tilde{m}_i\right) \\
 \cdot  \left(-\sigma _2+2 \sigma _3-\sigma
	_4-\tilde{m}_i\right) \left(\sigma _4-\tilde{m}_i\right) \left(-\sigma _1+\sigma _4-\tilde{m}_i\right) \left(\sigma _1-\sigma _2+\sigma _4-\tilde{m}_i\right) \left(-\sigma _3+2 \sigma _4-\tilde{m}_i\right) \\
 \cdot  \left(-\sigma _3+\sigma _4-\tilde{m}_i\right) \left(\sigma _1-\sigma _3+\sigma _4-\tilde{m}_i\right) \left(-\sigma _1+\sigma _2-\sigma _3+\sigma _4-\tilde{m}_i\right)
	\left(-\sigma _2+\sigma _3+\sigma _4-\tilde{m}_i\right) \\
 \cdot  \left(\sigma _2-2 \sigma _3+\sigma_4
	-\tilde{m}_i\right) \neq 0.
\end{multline}

\subsection{Transformation under the Weyl group of $F_4$}

The Weyl group of $F_4$ has $1152 = 2^7 \cdot 3^2$ elements\footnote{
For the curious, information on the representation theory of the
Weyl group of $F_4$ can be found in \cite{kondo}.
} 
\cite{humphreys-refl}[table 2.2], so explicitly
listing orbits of vacua, for example, is not feasible, unlike the case of
$G_2$.  (Similarly \cite{humphreys-refl}[table 2.2], 
the order of the Weyl group of $E_6$ is 
$2^7 \cdot 3^4 \cdot 5$, the order of the Weyl group of $E_7$ is
$2^{10} \cdot 3^4 \cdot 5 \cdot 7$, and the order of the Weyl group
of $E_8$ is $2^{14} \cdot 3^5 \cdot 5^2 \cdot 7$, so we will not
be tracking orbits of vacua under the Weyl group in those cases either.)
In this section, we will instead merely check that the critical locus
equations transform into one another under Weyl reflections, a nontrivial
check of our computations.

As reviewed earlier,
the Weyl transformation acts on vectors, roots and weights:
\begin{align}
	S_{\alpha}(v^a)=v^a-2\frac{(\alpha,v)}{(\alpha,\alpha)}\alpha^a .
\end{align}
The Euclidean inner product takes the following metric matrix in this coordinate
\begin{equation}
[g_{ab}]=	\left(
	\begin{array}{cccc}
	2 & 3 & 2 & 1 \\
	3 & 6 & 4 & 2 \\
	2 & 4 & 3 & 3/2 \\
	1 & 2 & 3/2 & 1 \\
	\end{array}
	\right) . 
\end{equation}
The $\sigma_a$ transform as covectors under the same Weyl transformation.  
$F_4$ has four simple roots, which can be taken to be
\begin{align*}
	A&=\textbf{(}2,-1,0,0\textbf{)},    &&B=\textbf{(}-1,2,-2,0\textbf{)}, \\
	C&=\textbf{(}0,-1,2,-1\textbf{)},   &&D=\textbf{(}0,0,-1,2\textbf{)}.
	\end{align*}
so the Weyl group of $F_4$ has four distinguished
elements, $S_A, \cdots, S_D$, whose actions are given by
\begin{align}
\nonumber	&S_A\textit{(}v^1,v^2,v^3,v^4\textbf{)}=\textbf{(}-v^1,v^1+v^2,v^3,v^4\textbf{)}, \\
&S_A\textbf{(}\sigma_1,\sigma_2,\sigma_3,\sigma_4\textbf{)}=\textit{(}\sigma_2-\sigma_1,\sigma_2,\sigma_3,\sigma_4\textbf{)}, \\  \nonumber  \\ 
&S_B\textit{(}v^1,v^2,v^3,v^4\textbf{)}=\textbf{(}v^1+v^2,-v^2,2v^2+v^3,v^4\textbf{)}, \nonumber  \\
&S_B\textbf{(}\sigma_1,\sigma_2,\sigma_3,\sigma_4\textbf{)}=\textit{(}\sigma_1,\sigma_1-\sigma_2+2\sigma_3,\sigma_3,\sigma_4\textbf{)}, \\ \nonumber \\   
&S_C\textit{(}v^1,v^2,v^3,v^4\textbf{)}=\textbf{(}v^1,v^2+v^3,-v^3,v^3+v^4\textbf{)}, \nonumber \\
&S_C\textbf{(}\sigma_1,\sigma_2,\sigma_3,\sigma_4\textbf{)}=\textit{(}\sigma_1,\sigma_2,\sigma_2-\sigma_3+\sigma_4, \sigma_4\textbf{)}, \\  \nonumber \\  
&S_D\textit{(}v^1,v^2,v^3,v^4\textbf{)}=\textbf{(}v^1,v^2,v^3+v^4,-v^4\textbf{)}, \nonumber \\
&S_D\textbf{(}\sigma_1,\sigma_2,\sigma_3,\sigma_4\textbf{)}=\textit{(}\sigma_1,\sigma_2,\sigma_3,\sigma_3- \sigma_4\textbf{)} .
\end{align}

The superpotential is, by construction, invariant under Weyl reflections,
hence to be consistent, the critical locus equations should transform
into one another under these reflections.  We check this below.
For example, under the action of $S_D$, 
equation~(\ref{locus1})
\begin{align*}
\nonumber	\prod_{i=1}^{n}&(-\sigma_1+\sigma_3-\tilde{m}_i)(-\sigma_1+\sigma_2-\sigma_3+\sigma_4-\tilde{m}_i)(-\sigma_2+\sigma_3+\sigma_4-\tilde{m}_i)\\
& \nonumber \cdot (-\sigma_3+2\sigma_4-\tilde{m}_i)(-\sigma_1+\sigma_4-\tilde{m}_i)(\sigma_4-\tilde{m}_i) \\
 \nonumber & =\prod_{i=1}^{n}(\sigma_1-\sigma_4-\tilde{m}_i)(\sigma_1-\sigma_2+\sigma_3-\sigma_4-\tilde{m}_i)(\sigma_2-\sigma_3-\sigma_4-\tilde{m}_i)\\
& \hspace*{0.25in} \cdot (\sigma_3-2\sigma_4-\tilde{m}_i)(-\sigma_4-\tilde{m}_i)(\sigma_1-\sigma_3-\tilde{m}_i) ,
\end{align*}
transforms into
\begin{align*}
\prod_{i=1}^{n} & (-\sigma_1+\sigma_3-\tilde{m}_i)(-
\sigma_1+\sigma_2-\sigma_4-\tilde{m}_i)(-\sigma_2+2\sigma_3-\sigma_4-\tilde{m}_i) \\
& \cdot  (\sigma_3-2\sigma_4-\tilde{m}_i)(-\sigma_1+\sigma_3-\sigma_4-\tilde{m}_i)(\sigma_3-\sigma_4-\tilde{m}_i) \\
& =\prod_{i=1}^{n}  (\sigma_1-\sigma_3+\sigma_4-\tilde{m}_i)(
\sigma_1-\sigma_2+\sigma_4-\tilde{m}_i)(\sigma_2-2\sigma_3+\sigma_4-\tilde{m}_i) \\
& \hspace*{0.25in} \cdot  (-\sigma_3+2\sigma_4-\tilde{m}_i)(-\sigma_3+\sigma_4-\tilde{m}_i)(\sigma_1-\sigma_3-\tilde{m}_i),
\end{align*}
which is equation~(\ref{locus2}).

Table~\ref{table:f4:weyl:critlocus} 
schematically describes how other critical locus
equations transform under these Weyl reflections.

\begin{table}[h]
\centering
\begin{tabular}{c|c|c}
& Initial equ'n & Final equ'n \\ \hline
$A$ & (\ref{locus1}) & (\ref{locus1}) $\times$ (\ref{locus4}) \\
$A$ & (\ref{locus2}) & (\ref{locus2}) $\times$ (\ref{locus4}) \\
$A$ & (\ref{locus3}) & (\ref{locus3}) $\times$ (\ref{locus4}) \\
$A$ & (\ref{locus4}) & (\ref{locus4}) \\ \hline
$B$ & (\ref{locus1}) & (\ref{locus1}) \\
$B$ & (\ref{locus2}) & (\ref{locus2}) \\
$B$ & (\ref{locus3}) & (\ref{locus4}) \\
$B$ & (\ref{locus4}) & (\ref{locus3}) \\ \hline
$C$ & (\ref{locus1}) & (\ref{locus1}) \\
$C$ & (\ref{locus2}) & (\ref{locus3}) $\times$ (\ref{locus4}) \\
$C$ & (\ref{locus3}) & (\ref{locus2}) $\times$ (\ref{locus4}) \\
$C$ & (\ref{locus4}) & (\ref{locus4}) \\ \hline
$D$ & (\ref{locus1}) & (\ref{locus2}) \\
$D$ & (\ref{locus2}) & (\ref{locus1}) \\
$D$ & (\ref{locus3}) & (\ref{locus3}) \\
$D$ & (\ref{locus4}) & (\ref{locus4})
\end{tabular}
\caption{Transformation of critical locus equations under four Weyl 
reflections.  \label{table:f4:weyl:critlocus} }
\end{table}

The fact that the critical locus equations are closed under
Weyl reflections associated to a set of simple roots provides a nontrivial
consistency check on our results.

\subsection{Pure gauge theory}

In this section, we will consider the mirror to the pure supersymmetric
$F_4$
gauge theory. The mirror superpotential is 
\begin{align} \nonumber
W =& \sigma_1 \Bigg( Z_{1}-Z_{2}+Z_{5}-Z_{6}+Z_{16}-Z_{17}+Z_{8}+Z_{18}-Z_{9}-Z_{19}+Z_{20}-Z_{21}+Z_{10}\\ \nonumber & \hspace*{0.5in} -Z_{11}+2Z_{12}   
-Z_{25}+Z_{26}-Z_{29}+Z_{30}-Z_{40}+Z_{41}-Z_{32}-Z_{42}+Z_{33}+Z_{43} \\ \nonumber& \hspace*{.5 in} -Z_{44}+Z_{45}-Z_{34}+Z_{35}-2Z_{36}       \Bigg)
\end{align}\begin{align} \nonumber
&  +\sigma_2 \Bigg(Z_{2}-Z_{3}+Z_{4}+Z_{15}-Z_{5}+Z_{7}-Z_{16}-Z_{8}+Z_{19}+Z_{21}-Z_{22}+Z_{10}+2Z_{11}\\ \nonumber & \hspace*{.5in} -Z_{23}-Z_{12}   
\hspace*{0.1in}
-Z_{26}+Z_{27}-Z_{28}-Z_{39}+Z_{29}-Z_{31}+Z_{40}+Z_{32}-Z_{43}-Z_{45} \\ \nonumber & \hspace*{.5in} +Z_{46}-Z_{34}-2Z_{35}+Z_{47}+Z_{36}       \Bigg) \end{align}\begin{align} \nonumber
&   +\sigma_3 \Bigg(2Z_{3}+Z_{14}-2Z_{4}-Z_{15}+Z_{16}+Z_{17}+2Z_{8}-Z_{18}+2Z_{9}-Z_{19}+Z_{22}-2Z_{10} \\ \nonumber & \hspace*{.5in}  -2Z_{11}+2Z_{23}-Z_{24}  
-2Z_{27}-Z_{38}+2Z_{28}+Z_{39}-Z_{40}-Z_{41}-2Z_{32}+Z_{42} \\ \nonumber & \hspace*{.5in} -2Z_{33}+Z_{43}-Z_{46}+2Z_{34}+2Z_{35}-2Z_{47}+Z_{48}      \Bigg) \\ \nonumber
&   +\sigma_4 \Bigg(Z_{13}-Z_{14}+2Z_{4}+2Z_{5}+2Z_{6}-2Z_{7}-2Z_{8}+Z_{18}-2Z_{9}+Z_{19}-Z_{20}-Z_{21} \\ \nonumber & \hspace*{.5 in} +Z_{22}-Z_{23}+2Z_{24}
-Z_{37}+Z_{38}-2Z_{28}-2Z_{29}-2Z_{30}+2Z_{31}+2Z_{32}-Z_{42} \\  & \hspace*{.5 in} +2Z_{33}-Z_{43}+Z_{44}+Z_{45}-Z_{46}+Z_{47}-2Z_{48} \Bigg)  +\sum_{m=1}^{48}X_m   .
\end{align}

Now, let us consider the critical locus of the superpotential above.
For each root $\mu$, the fields $X_{\mu}$ and $X_{-\mu}$ appear paired with
opoosite signs coupling to each $\sigma$.  Therefore, one impliciation of
the derivatives
\begin{displaymath}
\frac{\partial W}{\partial X_{\mu} } \: = \: 0
\end{displaymath}
is that, on the critical locus, 
\begin{equation}
X_{\mu} \: = \: - X_{- \mu}.
\end{equation}
(Furthermore, on the critical locus, each $X_{\mu}$ is determined by
$\sigma$s.)
Next, each derivative 
\begin{displaymath}
\frac{\partial W}{\partial \sigma_a}
\end{displaymath}
is a product of ratios of the form
\begin{displaymath}
\frac{ X_{\mu} }{ X_{-\mu} } \: = \: -1.
\end{displaymath}
It is straightforward to check in the superpotential above that
each $\sigma_a$ is multiplied by an even number of such ratios 
({\it i.e.} the number of $Z$'s is a multiple of four).
For example, the sum of the absolute values of the coefficients of the
$Z$'s multiplying $\sigma_1$ and $\sigma_2$
is $32  =4 \cdot 8$, and the sum of the
absolute values of the coefficients of the $Z$'s multiplying
$\sigma_3$ and $\sigma_4$ is $44 = 4 \cdot 11$.
Thus, the constraint implied by the $\sigma$'s is automatically satisfied.

As a result, following the same analysis in \cite{GuSharpe},
we see in this case, that the critical locus is nonempty, and in fact
is determined by four $\sigma$s.  In other words, at the level of these
topological field theory computations, we have evidence that the pure
supersymmetric $F_4$ gauge theory in two dimensions flows in the IR
to a theory of four free twisted chiral superfields.

\section{$E_6$} \label{section4}

 In this section we will consider the mirror Landau-Ginzburg orbifold superpotential of $E_6$ gauge theory when matter fields are in ${\bf 27}$ fundamental representation of it and then we compute quantum cohomology ring of it. Also we will consider the pure theory without matter field.

\subsection{Mirror Landau-Ginzburg orbifold}

The mirror Landau-Ginzburg model has fields
\begin{itemize}
	\item $Y_{i,\beta}$, $i \in \{1, \cdots, n\}$, $\beta \in \{1, \cdots, 27\}$,
	corresponding to the matter fields in $n$ copies of the 
${\bf 27}$ representation ${\bf 27}$,
	\item $X_m$, $m \in \{1, \cdots, 72\}$, corresponding to the roots of $E_6$,
	\item $\sigma_a$, $a \in \{1, 2, 3, 4, 5, 6 \}$.
\end{itemize}

We associate the roots, $\alpha^a_m$, to $X_m$ fields and the wights, $\rho^a_{i,\beta}$, of fundamental ${\bf 27}$ representation of $E_6$ to $Y_{i,\beta}$.\\

The roots of $E_6$ and associated fields are listed
in tables~\ref{table:e6:roots1}, \ref{table:e6:roots2}.
The weights associated to the ${\bf 27}$ of $E_6$ and their associated
fields are
listed in table~\ref{table:e6:weights}.

\begin{table}[h!]
\centering
	\begin{tabular}{cc|cc}
		Field & Positive root & Field & Negative root  \\ \hline
$X_{1}$ & $\textbf{(}0,0,0,0,0,1\textbf{)}$ & $X_{37}$ & $\textbf{(}0,0,0,0,0,-1\textbf{)}$  \\
$X_{2}$ & $\textbf{(}0,0,1,0,0,-1\textbf{)}$ & $X_{38}$ & $\textbf{(}0,0,-1,0,0,1\textbf{)}$  \\
$X_{3}$ & $\textbf{(}0,1,-1,1,0,0\textbf{)}$ & $X_{39}$ & $\textbf{(}0,-1,1,-1,0,0\textbf{)}$  \\
$X_{4}$ & $\textbf{(}0,1,0,-1,1,0\textbf{)}$ & $X_{40}$ & $\textbf{(}0,-1,0,1,-1,0\textbf{)}$  \\
$X_{5}$ & $\textbf{(}1,-1,0,1,0,0\textbf{)}$ & $X_{41}$ & $\textbf{(}-1,1,0,-1,0,0\textbf{)}$  \\
$X_{6}$ & $\textbf{(}-1,0,0,1,0,0\textbf{)}$ & $X_{42}$ & $\textbf{(}1,0,0,-1,0,0\textbf{)}$  \\
$X_{7}$ & $\textbf{(}0,1,0,0,-1,0\textbf{)}$ & $X_{43}$ & $\textbf{(}0,-1,0,0,1,0\textbf{)}$  \\
$X_{8}$ & $\textbf{(}1,-1,1,-1,1,0\textbf{)}$ & $X_{44}$ & $\textbf{(}-1,1,-1,1,-1,0\textbf{)}$  \\
$X_{9}$ & $\textbf{(}-1,0,1,-1,1,0\textbf{)}$ & $X_{45}$ & $\textbf{(}1,0,-1,1,-1,0\textbf{)}$  \\
$X_{10}$ & $\textbf{(}1,-1,1,0,-1,0\textbf{)}$ & $X_{46}$ & $\textbf{(}-1,1,-1,0,1,0\textbf{)}$  \\
$X_{11}$ & $\textbf{(}1,0,-1,0,1,1\textbf{)}$ & $X_{47}$ & $\textbf{(}-1,0,1,0,-1,-1\textbf{)}$  \\
$X_{12}$ & $\textbf{(}-1,0,1,0,-1,0\textbf{)}$ & $X_{48}$ & $\textbf{(}1,0,-1,0,1,0\textbf{)}$  \\
$X_{13}$ & $\textbf{(}-1,1,-1,0,1,1\textbf{)}$ & $X_{49}$ & $\textbf{(}1,-1,1,0,-1,-1\textbf{)}$  \\
$X_{14}$ & $\textbf{(}1,0,-1,1,-1,1\textbf{)}$ & $X_{50}$ & $\textbf{(}-1,0,1,-1,1,-1\textbf{)}$  \\
$X_{15}$ & $\textbf{(}1,0,0,0,1,-1\textbf{)}$ & $X_{51}$ & $\textbf{(}-1,0,0,0,-1,1\textbf{)}$  \\
$X_{16}$ & $\textbf{(}-1,1,-1,1,-1,1\textbf{)}$ & $X_{52}$ & $\textbf{(}1,-1,1,-1,1,-1\textbf{)}$  \\
$X_{17}$ & $\textbf{(}-1,1,0,0,1,-1\textbf{)}$ & $X_{53}$ & $\textbf{(}1,-1,0,0,-1,1\textbf{)}$  \\
$X_{18}$ & $\textbf{(}0,-1,0,0,1,1\textbf{)}$ & $X_{54}$ & $\textbf{(}0,1,0,0,-1,-1\textbf{)}$  
\end{tabular}
\caption{First set of roots of $E_6$ and associated fields.
\label{table:e6:roots1} }
\end{table}

\begin{table}[h!]
\centering
        \begin{tabular}{cc|cc}
                Field & Positive root & Field & Negative root  \\ \hline
$X_{19}$ & $\textbf{(}1,0,0,-1,0,1\textbf{)}$ & $X_{55}$ & $\textbf{(}-1,0,0,1,0,-1\textbf{)}$  \\
$X_{20}$ & $\textbf{(}1,0,0,1,-1,-1\textbf{)}$ & $X_{56}$ & $\textbf{(}-1,0,0,-1,1,1\textbf{)}$  \\
$X_{21}$ & $\textbf{(}-1,1,0,-1,0,1\textbf{)}$ & $X_{57}$ & $\textbf{(}1,-1,0,1,0,-1\textbf{)}$  \\
$X_{22}$ & $\textbf{(}-1,1,0,1,-1,-1\textbf{)}$ & $X_{58}$ & $\textbf{(}1,-1,0,-1,1,1\textbf{)}$  \\
$X_{23}$ & $\textbf{(}0,-1,0,1,-1,1\textbf{)}$ & $X_{59}$ & $\textbf{(}0,1,0,-1,1,-1\textbf{)}$  \\
$X_{24}$ & $\textbf{(}0,-1,1,0,1,-1\textbf{)}$ & $X_{60}$ & $\textbf{(}0,1,-1,0,-1,1\textbf{)}$  \\
$X_{25}$ & $\textbf{(}1,0,1,-1,0,-1\textbf{)}$ & $X_{61}$ & $\textbf{(}-1,0,-1,1,0,1\textbf{)}$  \\
$X_{26}$ & $\textbf{(}-1,1,1,-1,0,-1\textbf{)}$ & $X_{62}$ & $\textbf{(}1,-1,-1,1,0,1\textbf{)}$  \\
$X_{27}$ & $\textbf{(}0,-1,1,-1,0,1\textbf{)}$ & $X_{63}$ & $\textbf{(}0,1,-1,1,0,-1\textbf{)}$  \\
$X_{28}$ & $\textbf{(}0,-1,1,1,-1,-1\textbf{)}$ & $X_{64}$ & $\textbf{(}0,1,-1,-1,1,1\textbf{)}$  \\
$X_{29}$ & $\textbf{(}0,0,-1,1,1,0\textbf{)}$ & $X_{65}$ & $\textbf{(}0,0,1,-1,-1,0\textbf{)}$  \\
$X_{30}$ & $\textbf{(}1,1,-1,0,0,0\textbf{)}$ & $X_{66}$ & $\textbf{(}-1,-1,1,0,0,0\textbf{)}$  \\
$X_{31}$ & $\textbf{(}-1,2,-1,0,0,0\textbf{)}$ & $X_{67}$ & $\textbf{(}1,-2,1,0,0,0\textbf{)}$  \\
$X_{32}$ & $\textbf{(}0,-1,2,-1,0,-1\textbf{)}$ & $X_{68}$ & $\textbf{(}0,1,-2,1,0,1\textbf{)}$  \\
$X_{33}$ & $\textbf{(}0,0,-1,0,0,2\textbf{)}$ & $X_{69}$ & $\textbf{(}0,0,1,0,0,-2\textbf{)}$  \\
$X_{34}$ & $\textbf{(}0,0,-1,2,-1,0\textbf{)}$ & $X_{70}$ & $\textbf{(}0,0,1,-2,1,0\textbf{)}$  \\
$X_{35}$ & $\textbf{(}0,0,0,-1,2,0\textbf{)}$ & $X_{71}$ & $\textbf{(}0,0,0,1,-2,0\textbf{)}$  \\
$X_{36}$ & $\textbf{(}2,-1,0,0,0,0\textbf{)}$ & $X_{72}$ & $\textbf{(}-2,1,0,0,0,0\textbf{)}$  
\end{tabular}
\caption{Second set of roots of $E_6$ and associated fields.
\label{table:e6:roots2} }
\end{table}

\begin{table}[h!]
\centering
        \begin{tabular}{cc|cc|cc}
                Field & Weight & Field & Weight & Field & Weight \\ \hline
                $Y_{i,1}$ & $\textbf{(}1 , 0 , 0 , 0 , 0 , 0\textbf{)}$ & $Y_{i,2}$ & $\textbf{(}-1 , 1 , 0 , 0 , 0 , 0\textbf{)}$ & $Y_{i,3}$ & $\textbf{(} 0 , -1 , 1 , 0 , 0 , 0 \textbf{)}$ \\
                $Y_{i,4}$ & $\textbf{(}0,0,-1,1,0,1\textbf{)}$ & $Y_{i,5}$ & $\textbf{(}0,0,0,-1,1,1\textbf{)}$ & $Y_{i,6}$ & $\textbf{(}0,0,0,1,0,-1\textbf{)}$ \\
                $Y_{i,7}$ & $\textbf{(}0,0,0,0,-1,1\textbf{)}$ & $Y_{i,8}$ & $\textbf{(}0,0,1,-1,1,-1\textbf{)}$ & $Y_{i,9}$ & $\textbf{(}0,0,1,0,-1,-1\textbf{)}$ \\
                $Y_{i,10}$ & $\textbf{(}0,1,-1,0,1,0\textbf{)}$ & $Y_{i,11}$ & $\textbf{(}0,1,-1,1,-1,0\textbf{)}$ & $Y_{i,12}$ & $\textbf{(}1,-1,0,0,1,0\textbf{)}$ \\
                $Y_{i,13}$ & $\textbf{(}-1,0,0,0,1,0\textbf{)}$ & $Y_{i,14}$ & $\textbf{(}0,1,0,-1,0,0\textbf{)}$ & $Y_{i,15}$ & $\textbf{(}1,-1,0,1,-1,0\textbf{)}$ \\
                $Y_{i,16}$ & $\textbf{(}-1,0,0,1,-1,0\textbf{)}$ & $Y_{i,17}$ & $\textbf{(}1,-1,1,-1,0,0\textbf{)}$ & $Y_{i,18}$ & $\textbf{(}-1,0,1,-1,0,0\textbf{)}$ \\
                $Y_{i,19}$ & $\textbf{(}1,0,-1,0,0,1\textbf{)}$ & $Y_{i,20}$ & $\textbf{(}-1,1,-1,0,0,1\textbf{)}$ & $Y_{i,21}$ & $\textbf{(}1,0,0,0,0,-1\textbf{)}$ \\
                $Y_{i,22}$ & $\textbf{(}-1,1,0,0,0,-1\textbf{)}$ & $Y_{i,23}$ & $\textbf{(}0,-1,0,0,0,1\textbf{)}$ & $Y_{i,24}$ & $\textbf{(}0,-1,1,0,0,-1\textbf{)}$ \\
                $Y_{i,25}$ & $\textbf{(}0,0,-1,1,0,0\textbf{)}$ & $Y_{i,26}$ & $\textbf{(}0,0,0,-1,1,0\textbf{)}$ & $Y_{i,27}$ & $\textbf{(}0,0,0,0,-1,0\textbf{)}$ \\
        \end{tabular}
\caption{Weights of ${\bf 27}$ of $E_6$ and associated fields. 
\label{table:e6:weights} }
\end{table}

The weights in the tables in this section are written as linear combinations
of the fundamental weights, computed with LieART \cite{lieart}, as discussed
earlier, so as to get conventional $\theta$-angle periodicities.

\subsection{Superpotential}

In this section, we describe the superpotential of the mirror Landau-Ginzburg
orbifold.  It is given by
\begin{equation} \nonumber
W=\sum_{a=1}^{6}\sigma_a \Big( \sum_{i=1}^{n}\sum_{\beta=1}^{27} \rho^a_{i,\beta}Y_{i,\beta}+\sum_{m=1}^{72} \alpha^a_{m}Z_m  \Big)-\sum_{i=1}^{n}\tilde{m}_i\sum_{b=1}^{27}Y_{i,\beta}+\sum_{i=1}^{n}\sum_{\beta=1}^{27}\exp(-Y_{i,\beta})+\sum_{m=1}^{72}X_m.
\end{equation} 
where $X_m=\exp(-Z_m)$ and $X_m$ are the fundamental fields. 
Using the results of the previous section we get
\begin{align}
W=\sum_{a=1}^{6} \sigma_a \mathcal{C}^a-\sum_{i=1}^{n}\tilde{m}_i \sum_{\beta=1}^{27} Y_{i,\beta}+\sum_{i=1}^{n} \sum_{\beta=1}^{27} \exp(-Y_{i,\beta})+\sum_{m=1}^{72}X_m   .
\end{align}
where $\mathcal{C}^a$ are given as follows:
\begin{align*}
   &\mathcal{C}^1=\sum_{i=1}^{n} \big(   Y_{i,1}-Y_{i,2}+Y_{i,12}-Y_{i,13}+Y_{i,15}-Y_{i,16}+Y_{i,17}-Y_{i,18}+Y_{i,19}-Y_{i,20}+Y_{i,21}-Y_{i,22} \big) \\& \hspace*{.35in} +Z_5-Z_6+Z
	_8-Z_9+Z_{10}+Z_{11}-Z_{12}-Z_{13}+Z_{14}+Z_{15}-Z_{16}-Z_{17}+Z_{19}+Z_{20}\\& \hspace*{.35in}-Z_{21}-
	Z_{22}+Z_{25}-Z_{26}+Z_{30}-Z_{31}+2
	Z_{36}-Z_{41}+Z_{42}-Z_{44}+Z_{45}-Z_{46}-Z_{47}\\& \hspace*{.35in}+Z_{48} +Z_{49}-Z_{50}-Z_{51}+Z_{52}+
	Z_{53}-Z_{55}-Z_{56}+Z_{57}+Z_{58}-Z_{61}+Z_{62}-Z_{66}\\& \hspace*{.35in}+Z_{67}-2 Z_{72} ,  \end{align*}
	\begin{align*}
	&\mathcal{C}^2= \sum_{i=1}^{n} \big(   Y_{i,2}-Y_{i,3}+Y_{i,10}+Y_{i,11}-Y_{i,12}+Y_{i,14}-Y_{i,15}-Y_{i,17}+Y_{i,20}+Y_{i,22}-Y_{i,23}-Y_{i,24} \big)\\& \hspace*{.35in} +Z_3+Z_4-Z
	_5+Z_7-Z_8-Z_{10}+Z_{13}+Z_{16}+Z_{17}-Z_{18}+Z_{21}+Z_{22}-Z_{23}-Z_{24}\\& \hspace*{.35in} +Z_{26}-Z_{
		27}-Z_{28}+Z_{30}+2
	Z_{31}-Z_{32}-Z_{36}-Z_{39}-Z_{40}+Z_{41}-Z_{43}+Z_{44}+Z_{46}\\& \hspace*{.35in}-Z_{49}-Z_{52}-Z_{53}+
	Z_{54}-Z_{57}-Z_{58}+Z_{59}+Z_{60}-Z_{62}+Z_{63}+Z_{64}-Z_{66}-2
	Z_{67}\\& \hspace*{.35in}+Z_{68}+Z_{72} ,
\end{align*}
\begin{align*}
		&\mathcal{C}^3 = \sum_{i=1}^{n} \big( Y_{i,3}-Y_{i,4}+Y_{i,8}+Y_{i,9}-Y_{i,10}-Y_{i,11}+Y_{i,17}+Y_{i,18}-Y_{i,19}-Y_{i,20}+Y_{i,24}-Y_{i,25} \big) \\& \hspace*{.35in} +Z_2-Z_3+Z_8+Z_9
		+Z_{10}-Z_{11}+Z_{12}-Z_{13}-Z_{14}-Z_{16}+Z_{24}+Z_{25}+Z_{26}+Z_{27}\\& \hspace*{.35in}+Z_{28}-Z_{29}
		-Z_{30}-Z_{31}+2
		Z_{32}-Z_{33}-Z_{34}-Z_{38}+Z_{39}-Z_{44}-Z_{45}-Z_{46}+Z_{47}\\& \hspace*{.35in}-Z_{48}+Z_{49} +Z_{50}+
		Z_{52}-Z_{60}-Z_{61}-Z_{62}-Z_{63}-Z_{64}+Z_{65}+Z_{66}+Z_{67}-2
		Z_{68} \\& \hspace*{.35in}+Z_{69}+Z_{70}  ,
\end{align*}
\begin{align*}
		&\mathcal{C}^4 = \sum_{i=1}^{n} \big( Y_{i,4}-Y_{i,5}+Y_{i,6}-Y_{i,8}+Y_{i,11}-Y_{i,14}+Y_{i,15}+Y_{i,16}-Y_{i,17}-Y_{i,18}+Y_{i,25}-Y_{i,26} \big) \\& \hspace*{.35in} +Z_3-Z_4+Z_5+Z_6
		-Z_8-Z_9+Z_{14}+Z_{16}-Z_{19}+Z_{20}-Z_{21}+Z_{22}+Z_{23}-Z_{25}\\& \hspace*{.35in}-Z_{26}-Z_{27}+Z_{28
		}+Z_{29}-Z_{32}+2
		Z_{34}-Z_{35}-Z_{39}+Z_{40}-Z_{41}-Z_{42}+Z_{44}+Z_{45}\\& \hspace*{.35in}-Z_{50}-Z_{52} +Z_{55}-Z_{56}+
		Z_{57}-Z_{58}-Z_{59}+Z_{61}+Z_{62}+Z_{63}-Z_{64}-Z_{65}+Z_{68}\\& \hspace*{.35in}-2 Z_{70}+Z_{71} ,
\end{align*}
\begin{align*}
&\mathcal{C}^5= \sum_{i=1}^{n} \big( Y_{i,5}-Y_{i,7}+Y_{i,8}-Y_{i,9}+Y_{i,10}-Y_{i,11}+Y_{i,12}+Y_{i,13}-Y_{i,15}-Y_{i,16}+Y_{i,26}-Y_{i,27} \big) \\& \hspace*{.35in} +Z_4-Z_7+Z_8+Z_9
-Z_{10}+Z_{11}-Z_{12}+Z_{13}-Z_{14}+Z_{15}-Z_{16}+Z_{17}+Z_{18}-Z_{20}\\& \hspace*{.35in} -Z_{22}  -Z_{23}
+Z_{24}-Z_{28}+Z_{29}-Z_{34}+2
Z_{35}-Z_{40}+Z_{43}-Z_{44}-Z_{45}+Z_{46}-Z_{47}\\& \hspace*{.35in} +Z_{48}-Z_{49}  +Z_{50}-Z_{51}+Z_{52}-
Z_{53}-Z_{54}+Z_{56}+Z_{58}+Z_{59}-Z_{60}+Z_{64}-Z_{65}\\& \hspace*{.35in}+Z_{70}-2 Z_{71} ,
\end{align*}
\begin{align*}
&\mathcal{C}^6 = \sum_{i=1}^{n} \big( Y_{i,4}+Y_{i,5}-Y_{i,6}+Y_{i,7}-Y_{i,8}-Y_{i,9}+Y_{i,19}+Y_{i,20}-Y_{i,21}-Y_{i,22}+Y_{i,23}-Y_{i,24} \big) \\& \hspace*{.35in}+Z_1-Z_2+Z_{11}+Z_{13}
+Z_{14}-Z_{15}+Z_{16}-Z_{17}+Z_{18}+Z_{19}-Z_{20}+Z_{21}-Z_{22}\\& \hspace*{.35in}+Z_{23}-Z_{24}-Z_{25}
-Z_{26}+Z_{27}-Z_{28}-Z_{32}+2
Z_{33}-Z_{37}+Z_{38}-Z_{47}-Z_{49}-Z_{50}\\& \hspace*{.35in}+Z_{51}-Z_{52}+Z_{53}-Z_{54}-Z_{55}+Z_{56}-
Z_{57}+Z_{58}-Z_{59}+Z_{60}+Z_{61}+Z_{62}-Z_{63}\\& \hspace*{.35in}+Z_{64}+Z_{68}-2 Z_{69} .
\end{align*}

\subsection{Coulomb ring relations}

Integrating out the $\sigma_a$ fields, 
we obtain six constraints $\mathcal{C}^a=0$. 
Exponentiating these constraints, we obtain
a series of equations from which the Coulomb ring relations will 
be derived.  For reasons of notational
sanity, we will also slightly simplify these expressions, as follows.
To make predictions for the A model, we will evaluate the ring relations
on the critical locus, where
\begin{displaymath}
\frac{ X_m }{ X_{m+63} } \: = \: -1.
\end{displaymath}
It is straightforward to see that each of the constraints ${\mathcal C}^a$
contains 22 differences of corresponding $Z$'s, so that the exponential of the
constraints contains 22 factors of the form $X_m/X_{m+63}$ -- an even
number of factors of $-1$, which will cancel out.
Therefore, since on the critical locus those factors will cancel out,
we will omit them, and solely relate the exponentiated constraints in
terms of $Y$s.
        
The exponentiated constraints are as follows.
\begin{multline} 
\prod_{i=1}^{n}\exp \big(   Y_{i,1}-Y_{i,2}+Y_{i,12}-Y_{i,13}+Y_{i,15}-Y_{i,16}+Y_{i,17}-Y_{i,18}+Y_{i,19}-Y_{i,20}+Y_{i,21}-Y_{i,22} \big) \\
=1,   \label{E6 1}
\end{multline}
\begin{multline} 
 \prod_{i=1}^{n}\exp \big( Y_{i,2}-Y_{i,3}+Y_{i,10}+Y_{i,11}-Y_{i,12}+Y_{i,14}-Y_{i,15}-Y_{i,17}+Y_{i,20}+Y_{i,22}-Y_{i,23}-Y_{i,24}  \big)
  \\
=1,  \label{E6 2}
\end{multline}
\begin{multline}
 \prod_{i=1}^{n}\exp \big( Y_{i,3}-Y_{i,4}+Y_{i,8}+Y_{i,9}-Y_{i,10}-Y_{i,11}+Y_{i,17}+Y_{i,18}-Y_{i,19}-Y_{i,20}+Y_{i,24}-Y_{i,25} \big)
\\ 
 =1 , \label{E6 3}
\end{multline}
\begin{multline} 
  \prod_{i=1}^{n}\exp \big( Y_{i,4}-Y_{i,5}+Y_{i,6}-Y_{i,8}+Y_{i,11}-Y_{i,14}+Y_{i,15}+Y_{i,16}-Y_{i,17}-Y_{i,18}+Y_{i,25}-Y_{i,26} \big)
\\
 =1 , \label{E6 4}
\end{multline}
\begin{multline} 
 \prod_{i=1}^{n}\exp  \big( Y_{i,5}-Y_{i,7}+Y_{i,8}-Y_{i,9}+Y_{i,10}-Y_{i,11}+Y_{i,12}+Y_{i,13}-Y_{i,15}-Y_{i,16}+Y_{i,26}-Y_{i,27} \big)
\\
 =1,  \label{E6 5}
\end{multline}
\begin{multline} 
 \prod_{i=1}^{n}\exp \big( Y_{i,4}+Y_{i,5}-Y_{i,6}+Y_{i,7}-Y_{i,8}-Y_{i,9}+Y_{i,19}+Y_{i,20}-Y_{i,21}-Y_{i,22}+Y_{i,23}-Y_{i,24} \big)
\\
 =1 . \label{E6 6}
\end{multline}

The mirror maps are given by
\begin{equation*}
\exp(-Y_{i,\beta}) \mapsto - \tilde{m}_i + \sum_{a=1}^{6} \sigma_a \rho_{i,\beta}^a, \quad X_m \mapsto \sum_{a=1}^{6} \sigma_a \alpha_m^a.
\end{equation*}
Applying the operator mirror maps,
the Coulomb ring relations become
\begin{multline} 
 \prod_{i=1}^{n}\left(-\sigma _1+\sigma _2-\tilde{m}_{i}\right) \left(-\sigma _1+\sigma _3-\sigma
_4-\tilde{m}_{i}\right) \left(-\sigma _1+\sigma _4-\sigma _5-\tilde{m}_{i}\right) \\
   \hspace*{.17 in} \cdot \left(-\sigma
_1+\sigma _5-\tilde{m}_{i}\right)  \left(-\sigma _1+\sigma _2-\sigma _6-\tilde{m}_{i}\right)
\left(-\sigma _1+\sigma _2-\sigma _3+\sigma _6-\tilde{m}_{i}\right)\\
  = \prod_{i=1}^{n} \left(\sigma
_1-\tilde{m}_{i}\right) \left(\sigma _1-\sigma _2+\sigma _3-\sigma _4-\tilde{m}_{i}\right) \left(\sigma
_1-\sigma _2+\sigma _4-\sigma _5-\tilde{m}_{i}\right) \\
   \hspace*{.17 in} \cdot \left(\sigma _1-\sigma _2+\sigma
_5-\tilde{m}_{i}\right) \left(\sigma _1-\sigma _6-\tilde{m}_{i}\right) \left(\sigma _1-\sigma _3+\sigma
_6-\tilde{m}_{i}\right),
\end{multline}
\begin{multline} 
 \prod_{i=1}^{n}
	\left(-\sigma _2+\sigma _3-\tilde{m}_{i}\right) \left(\sigma _1-\sigma _2+\sigma
		_3-\sigma _4-\tilde{m}_{i}\right) \left(\sigma _1-\sigma _2+\sigma _4-\sigma _5-\tilde{m}_{i}\right) \\
 \hspace*{.17 in} \cdot
		\left(\sigma _1-\sigma _2+\sigma _5-\tilde{m}_{i}\right) \left(-\sigma _2+\sigma
		_3-\sigma _6-\tilde{m}_{i}\right) \left(-\sigma _2+\sigma _6-\tilde{m}_{i}\right) \\
 = \prod_{i=1}^{n}   \left(-\sigma
		_1+\sigma _2-\tilde{m}_{i}\right) \left(\sigma _2-\sigma _4-\tilde{m}_{i}\right) \left(\sigma
		_2-\sigma _3+\sigma _4-\sigma _5-\tilde{m}_{i}\right) \\
 \hspace*{.17 in} \cdot \left(\sigma _2-\sigma _3+\sigma
		_5-\tilde{m}_{i}\right) \left(-\sigma _1+\sigma _2-\sigma _6-\tilde{m}_{i}\right) \left(-\sigma
		_1+\sigma _2-\sigma _3+\sigma _6-\tilde{m}_{i}\right),
\end{multline}
\begin{multline}
 \prod_{i=1}^{n}
	\left(-\sigma _3+\sigma _4-\tilde{m}_{i}\right) \left(\sigma _2-\sigma _3+\sigma
		_4-\sigma _5-\tilde{m}_{i}\right) \left(\sigma _2-\sigma _3+\sigma _5-\tilde{m}_{i}\right) \\
 \hspace*{.17 in} \cdot
		\left(\sigma _1-\sigma _3+\sigma _6-\tilde{m}_{i}\right) \left(-\sigma _1+\sigma
		_2-\sigma _3+\sigma _6-\tilde{m}_{i}\right) \left(-\sigma _3+\sigma _4+\sigma
		_6-\tilde{m}_{i}\right)  \\
 = 
		 \prod_{i=1}^{n} \left(-\sigma _2+\sigma _3-\tilde{m}_{i}\right) \left(-\sigma _1+\sigma
		_3-\sigma _4-\tilde{m}_{i}\right) \left(\sigma _1-\sigma _2+\sigma _3-\sigma _4-\tilde{m}_{i}\right) \\
  \hspace*{.17 in} \cdot
		\left(-\sigma _2+\sigma _3-\sigma _6-\tilde{m}_{i}\right) \left(\sigma _3-\sigma
		_5-\sigma _6-\tilde{m}_{i}\right) \left(\sigma _3-\sigma _4+\sigma _5-\sigma _6-\tilde{m}_{i}\right),
\end{multline}
\begin{multline}
 \prod_{i=1}^{n}
	\left(\sigma _2-\sigma _4-\tilde{m}_{i}\right) \left(-\sigma _1+\sigma _3-\sigma
		_4-\tilde{m}_{i}\right) \left(\sigma _1-\sigma _2+\sigma _3-\sigma _4-\tilde{m}_{i}\right)
		\\
 \hspace*{.17 in} \cdot \left(-\sigma _4+\sigma _5-\tilde{m}_{i}\right) \left(\sigma _3-\sigma _4+\sigma
		_5-\sigma _6-\tilde{m}_{i}\right) \left(-\sigma _4+\sigma _5+\sigma
		_6-\tilde{m}_{i}\right)\\
 = \prod_{i=1}^{n} \left(-\sigma _3+\sigma _4-\tilde{m}_{i}\right) \left(-\sigma _1+\sigma
		_4-\sigma _5-\tilde{m}_{i}\right) \left(\sigma _1-\sigma _2+\sigma _4-\sigma _5-\tilde{m}_{i}\right)\\ 
  \hspace*{.17 in} \cdot
		\left(\sigma _2-\sigma _3+\sigma _4-\sigma _5-\tilde{m}_{i}\right) \left(\sigma _4-\sigma
		_6-\tilde{m}_{i}\right) \left(-\sigma _3+\sigma _4+\sigma _6-\tilde{m}_{i}\right),
\end{multline}
\begin{multline} 
 \prod_{i=1}^{n}
	 \left(-\sigma _5-\tilde{m}_{i}\right) \left(-\sigma _1+\sigma _4-\sigma _5-\tilde{m}_{i}\right)
		\left(\sigma _1-\sigma _2+\sigma _4-\sigma _5-\tilde{m}_{i}\right) \\
 \hspace*{.17 in} \cdot \left(\sigma _2-\sigma_3
		+\sigma _4-\sigma _5-\tilde{m}_{i}\right) \left(\sigma _3-\sigma _5-\sigma _6-\tilde{m}_{i}\right)
		\left(-\sigma _5+\sigma _6-\tilde{m}_{i}\right)\\
 = \prod_{i=1}^{n}  \left(-\sigma _1+\sigma _5-\tilde{m}_{i}\right)
		\left(\sigma _1-\sigma _2+\sigma _5-\tilde{m}_{i}\right) \left(\sigma _2-\sigma _3+\sigma
		_5-\tilde{m}_{i}\right) \\
   \hspace*{.17 in} \cdot\left(-\sigma _4+\sigma _5-\tilde{m}_{i}\right) \left(\sigma _3-
	\sigma_4+\sigma _5-\sigma _6-\tilde{m}_{i}\right) \left(-\sigma _4+\sigma _5+\sigma _6-\tilde{m}_{i}\right),
\end{multline}
\begin{multline} 
 \prod_{i=1}^{n}
	\left(\sigma _1-\sigma _6-\tilde{m}_{i}\right) \left(-\sigma _1+\sigma _2-\sigma_6
		-\tilde{m}_{i}\right) \left(-\sigma _2+\sigma _3-\sigma _6-\tilde{m}_{i}\right)  \\
 \hspace*{.17 in} \cdot \left(\sigma_4
		-\sigma _6-\tilde{m}_{i}\right) \left(\sigma _3-\sigma _5-\sigma _6-\tilde{m}_{i}\right)
		\left(\sigma _3-\sigma _4+\sigma _5-\sigma _6-\tilde{m}_{i}\right) \\
=  \prod_{i=1}^{n} \left(-\sigma
		_2+\sigma _6-\tilde{m}_{i}\right) \left(\sigma _1-\sigma _3+\sigma _6-\tilde{m}_{i}\right)
		\left(-\sigma _1+\sigma _2-\sigma _3+\sigma _6-\tilde{m}_{i}\right) \\
  \hspace*{.17 in} \cdot \left(-\sigma_3
		+\sigma _4+\sigma _6-\tilde{m}_{i}\right) \left(-\sigma _5+\sigma _6-\tilde{m}_{i}\right)
		\left(-\sigma _4+\sigma _5+\sigma _6-\tilde{m}_{i}\right).
\end{multline}
These are the Coulomb ring relations, defining the analogue of the
quantum cohomology ring, for $E_6$. 

Part of the excluded locus 
is defined by the vanishing locus of the $X_m$, and is
given by
\begin{multline} 
\left(2 \sigma _1-\sigma _2\right) \left(\sigma _1+\sigma _2-\sigma _3\right)
\left(-\sigma _1+2 \sigma _2-\sigma _3\right) \left(\sigma _4-\sigma _1\right)
\left(\sigma _1-\sigma _2+\sigma _4\right) \left(\sigma _2-\sigma _3+\sigma
_4\right) \\
 \cdot \left(\sigma _2-\sigma _5\right) \left(-\sigma _1+\sigma _3-\sigma
_5\right) \left(\sigma _1-\sigma _2+\sigma _3-\sigma _5\right) \left(-\sigma
_3+2 \sigma _4-\sigma _5\right) \left(\sigma _2-\sigma _4+\sigma _5\right) \\
 \cdot 
\left(-\sigma _1+\sigma _3-\sigma _4+\sigma _5\right) \left(\sigma _1-\sigma
_2+\sigma _3-\sigma _4+\sigma _5\right) \left(-\sigma _3+\sigma _4+\sigma
_5\right) \left(2 \sigma _5-\sigma _4\right) \left(\sigma _3-\sigma _6\right) \\
 \cdot 
\left(\sigma _1+\sigma _3-\sigma _4-\sigma _6\right) \left(-\sigma _1+\sigma
_2+\sigma _3-\sigma _4-\sigma _6\right) \left(-\sigma _2+2 \sigma _3-\sigma
_4-\sigma _6\right) \left(\sigma _1+\sigma _4-\sigma _5-\sigma _6\right) \\
   \cdot 
\left(-\sigma _1+\sigma _2+\sigma _4-\sigma _5-\sigma _6\right) \left(-\sigma
_2+\sigma _3+\sigma _4-\sigma _5-\sigma _6\right) \left(\sigma _1+\sigma
_5-\sigma _6\right) \left(-\sigma _1+\sigma _2+\sigma _5-\sigma _6\right)  \\
  \cdot 
\left(-\sigma _2+\sigma _3+\sigma _5-\sigma _6\right) \sigma _6 \left(\sigma
_1-\sigma _4+\sigma _6\right) \left(-\sigma _1+\sigma _2-\sigma _4+\sigma
_6\right) \left(-\sigma _2+\sigma _3-\sigma _4+\sigma _6\right) \\
 \cdot  \left(-\sigma
_2+\sigma _4-\sigma _5+\sigma _6\right) \left(\sigma _1-\sigma _3+\sigma
_4-\sigma _5+\sigma _6\right) \left(-\sigma _1+\sigma _2-\sigma _3+\sigma
_4-\sigma _5+\sigma _6\right)  \\ 
  \cdot 
 \left(-\sigma _2+\sigma _5+\sigma _6\right) 
\left(\sigma _1-\sigma _3+\sigma _5+\sigma _6\right) \left(-\sigma _1+\sigma
_2-\sigma _3+\sigma _5+\sigma _6\right) \left(2 \sigma _6-\sigma _3\right)
 \: \neq \: 0       .  
\end{multline}
Similarly, 
\begin{displaymath}
\exp(-Y_{i,\beta})=-\tilde{m}_i+\sum_{a=1}^{6}\sigma_a \rho^a_{i,\beta}
\end{displaymath}  
on the critical locus, so 
\begin{displaymath}
-\tilde{m}_i+\sum_{a=1}^{6}\sigma_a \rho^a_{i,\beta} \neq 0 
\end{displaymath}
which determines the remainder of the excluded locus: 
\begin{multline}
\left(\sigma _1-\tilde{m}_i\right) \left(-\sigma _1+\sigma _2-\tilde{m}_i\right) \left(-\sigma
_2+\sigma _3-\tilde{m}_i\right) \left(\sigma _2-\sigma _4-\tilde{m}_i\right) \left(-\sigma
_1+\sigma _3-\sigma _4-\tilde{m}_i\right) \\
 \cdot  \left(\sigma _1-\sigma _2+\sigma _3-\sigma
_4-\tilde{m}_i\right) \left(-\sigma _3+\sigma _4-\tilde{m}_i\right) \left(-\sigma _5-\tilde{m}_i\right)
\left(-\sigma _1+\sigma _4-\sigma _5-\tilde{m}_i\right) \\
 \cdot  \left(\sigma _1-\sigma
_2+\sigma _4-\sigma _5-\tilde{m}_i\right) \left(\sigma _2-\sigma _3+\sigma _4-\sigma
_5-\tilde{m}_i\right) \left(-\sigma _1+\sigma _5-\tilde{m}_i\right) \left(\sigma _1-\sigma
_2+\sigma _5-\tilde{m}_i\right) \\
 \cdot  \left(\sigma _2-\sigma _3+\sigma _5-\tilde{m}_i\right)
\left(-\sigma _4+\sigma _5-\tilde{m}_i\right) \left(\sigma _1-\sigma _6-\tilde{m}_i\right)
\left(-\sigma _1+\sigma _2-\sigma _6-\tilde{m}_i\right) \\
 \cdot  \left(-\sigma _2+\sigma
_3-\sigma _6-\tilde{m}_i\right) \left(\sigma _4-\sigma _6-\tilde{m}_i\right) \left(\sigma
_3-\sigma _5-\sigma _6-\tilde{m}_i\right) \left(\sigma _3-\sigma _4+\sigma _5-\sigma
_6-\tilde{m}_i\right) \\
 \cdot  \left(-\sigma _2+\sigma _6-\tilde{m}_i\right) \left(\sigma _1-\sigma
_3+\sigma _6-\tilde{m}_i\right) \left(-\sigma _1+\sigma _2-\sigma _3+\sigma _6-\tilde{m}_i\right)
\left(-\sigma _3+\sigma _4+\sigma _6-\tilde{m}_i\right) \\
 \cdot  \left(-\sigma _5+\sigma_6
-\tilde{m}_i\right) \left(-\sigma _4+\sigma _5+\sigma _6-\tilde{m}_i\right)
\: \neq \: 0 .
\end{multline}

\subsection{Pure gauge theory}

In this part we will consider the mirror to the pure supersymmetric
$E_6$ gauge theory.
The mirror Landau-Ginzburg superpotential is 
\begin{align*}
W=&\sigma_1 \Big(Z_5-Z_6+Z
_8-Z_9+Z_{10}+Z_{11}-Z_{12}-Z_{13}+Z_{14}+Z_{15}-Z_{16}-Z_{17}+Z_{19}+Z_{20}\\
& \hspace*{.3in} -Z_{21}-
Z_{22}+Z_{25}-Z_{26}+Z_{30}-Z_{31}+2
Z_{36}-Z_{41}+Z_{42}-Z_{44}+Z_{45}-Z_{46}-Z_{47}\\
& \hspace*{.3in} +Z_{48} +Z_{49}-Z_{50}-Z_{51}+Z_{52}+
Z_{53}-Z_{55}-Z_{56}+Z_{57}+Z_{58}-Z_{61}+Z_{62}-Z_{66}
\\
& \hspace*{.3in} +Z_{67}-2 Z_{72} \Big)  \\
 +&  \sigma_2 \Big(Z_3+Z_4-Z
_5+Z_7-Z_8-Z_{10}+Z_{13}+Z_{16}+Z_{17}-Z_{18}+Z_{21}+Z_{22}-Z_{23}-Z_{24}\\
& \hspace*{.3in}
+Z_{26}
-Z_{
	27}-Z_{28}+Z_{30}+2
Z_{31}-Z_{32}-Z_{36}-Z_{39}-Z_{40}+Z_{41}-Z_{43}+Z_{44}+Z_{46}\\
& \hspace*{.3in}
-Z_{49}-Z_{52}
-Z_{53}+
Z_{54}-Z_{57}-Z_{58}+Z_{59}+Z_{60}-Z_{62}+Z_{63}+Z_{64}-Z_{66}-2
Z_{67}
\\
& \hspace*{.3in} +Z_{68}+Z_{72} \Big)  
\end{align*}
\begin{align*}
\hspace*{.3in} +&\sigma_3 \Big(Z_2-Z_3+Z_8+Z_9
+Z_{10}-Z_{11}+Z_{12}-Z_{13}-Z_{14}-Z_{16}+Z_{24}+Z_{25}+Z_{26}+Z_{27}\\& \hspace*{.3in}+Z_{28}-Z_{29}
-Z_{30}-Z_{31}+2
Z_{32}-Z_{33}-Z_{34}-Z_{38}+Z_{39}-Z_{44}-Z_{45}-Z_{46}+Z_{47}\\& \hspace*{.3in}-Z_{48}+Z_{49}  +Z_{50}+
Z_{52}-Z_{60}-Z_{61}-Z_{62}-Z_{63}-Z_{64}+Z_{65}+Z_{66}+Z_{67}-2
Z_{68}\\& \hspace*{.3in}+Z_{69}+Z_{70} \Big) 
\end{align*}
\begin{align*}
\hspace*{.3in} +&\sigma_4 \Big(Z_3-Z_4+Z_5+Z_6
-Z_8-Z_9+Z_{14}+Z_{16}-Z_{19}+Z_{20}-Z_{21}+Z_{22}+Z_{23}-Z_{25}\\& \hspace*{.3in}-Z_{26}-Z_{27}+Z_{28
}+Z_{29}-Z_{32}+2
Z_{34}-Z_{35}-Z_{39}+Z_{40}-Z_{41}-Z_{42}+Z_{44}+Z_{45}\\& \hspace*{.3in}-Z_{50}-Z_{52} +Z_{55}-Z_{56}+
Z_{57}-Z_{58}-Z_{59}+Z_{61}+Z_{62}+Z_{63}-Z_{64}-Z_{65}+Z_{68}\\& \hspace*{.3in}-2 Z_{70}+Z_{71} \Big) 
\end{align*}
\begin{align*}
\hspace*{.3in}+&\sigma_5 \Big(Z_4-Z_7+Z_8+Z_9
-Z_{10}+Z_{11}-Z_{12}+Z_{13}-Z_{14}+Z_{15}-Z_{16}+Z_{17}+Z_{18}-Z_{20}\\& \hspace*{.3in} -Z_{22} -Z_{23}
+Z_{24}-Z_{28}+Z_{29}-Z_{34}+2
Z_{35}-Z_{40}+Z_{43}-Z_{44}-Z_{45}+Z_{46}-Z_{47}\\& \hspace*{.3in} +Z_{48}-Z_{49}  +Z_{50}-Z_{51}+Z_{52}-
Z_{53}-Z_{54}+Z_{56}+Z_{58}+Z_{59}-Z_{60}+Z_{64}-Z_{65}\\& \hspace*{.3in}+Z_{70}-2 Z_{71} \Big) 
\end{align*}
\begin{align*}
\hspace*{.3in}+&\sigma_6 \Big(Z_1-Z_2+Z_{11}+Z_{13}
+Z_{14}-Z_{15}+Z_{16}-Z_{17}+Z_{18}+Z_{19}-Z_{20}+Z_{21}-Z_{22}+Z_{23}\\& \hspace*{.3in}-Z_{24}-Z_{25}
-Z_{26}+Z_{27}-Z_{28}-Z_{32}+2
Z_{33}-Z_{37}+Z_{38}-Z_{47}-Z_{49}-Z_{50}+Z_{51}\\& \hspace*{.3in}-Z_{52}+Z_{53}-Z_{54}-Z_{55}+Z_{56}-
Z_{57}+Z_{58}-Z_{59}+Z_{60}+Z_{61}+Z_{62}-Z_{63}+Z_{64}\\& \hspace*{.3in}+Z_{68}-2 Z_{69} \Big) 
\end{align*}
\begin{align}
+\sum_{m=1}^{72}X_m   ,
\end{align}
where as usual $Z = - \ln X$.

We can analyze this mirror in the same way as previous pure gauge
theory mirrors.  As discussed previously,
for each root $\mu$, the fields $X_{\mu}$ and $X_{-\mu}$ appear paired with
opoosite signs coupling to each $\sigma$.  Therefore, one impliciation of
the derivatives
\begin{displaymath}
\frac{\partial W}{\partial X_{\mu} } \: = \: 0
\end{displaymath}
is that, on the critical locus, 
\begin{equation}
X_{\mu} \: = \: - X_{- \mu}.
\end{equation}
(Furthermore, on the critical locus, each $X_{\mu}$ is determined by
$\sigma$s.) 
Next, each derivative 
\begin{displaymath}
\frac{\partial W}{\partial \sigma_a}
\end{displaymath}
is a product of ratios of the form
\begin{displaymath}
\frac{ X_{\mu} }{ X_{-\mu} } \: = \: -1.
\end{displaymath}
It is straightforward to check that, just as in the previous examples,
in the superpotential above each $\sigma$ multiplies a number of $Z$s that
is divisible by four, {\it i.e.} an even number of ratios $X_{\mu} / X_{-\mu}$.
Specifically, the sum of the absolute values of the coefficients of the
$Z$'s multiplying each $\sigma$ is $44 = 4 \cdot 11$.
Thus, the constraint implied by integrating out
the $\sigma$'s is automatically satisfied.

As a result, following the same analysis as earlier and 
\cite{GuSharpe}, the critical locus is nonempty, and is determined
by the six $\sigma$s.  Thus, at the level of these topological field
theory computations, we have evidence that the pure supersymmetric
$E_6$ gauge theory in two dimensions flows in the IR to a theory of
six free twisted chiral superfields.

 \section{$E_7$} \label{section5}    
 
 In this section we will consider the mirror Landau-Ginzburg orbifold 
to an $E_7$ gauge theory with matter fields in the ${\bf 56}$ 
fundamental representation.
As before, we will compute Coulomb branch (quantum cohomology) ring
relations and excluded loci.  We will also
study the pure $E_7$ gauge theory without matter.
 
 \subsection{Mirror Landau-Ginzburg orbifold}
 
 The mirror Landau-Ginzburg model has superfields
 \begin{itemize}
 	\item $Y_{i,\beta}$, $i \in \{1, \cdots, n\}$, $\beta \in \{1, \cdots, 56\}$,
 	corresponding to the matter fields in $n$ copies of the 
fundamental ${\bf 56}$ representation of
 	$E_7$,
 	\item $X_m$, $m \in \{1, \cdots, 126\}$, corresponding to the 
nonzero roots of $E_7$,
 	\item $\sigma_a$, $a \in \{1, \cdots, 7 \}$.
 \end{itemize}
 We associate the roots, $\alpha^a_m$, to $X_m$ fields and 
the weights, $\rho^a_{i,\beta}$ to the $Y_{i,\beta}$.

The nonzero roots of $E_7$ are listed in tables~\ref{table:e7:roots1},
\ref{table:e7:roots2}, and \ref{table:e7:roots3}.  
The weights of the ${\bf 56}$ of $E_7$ are
listed in table~\ref{table:e7:weights}.  All weights are
given as linear combinations of fundamental weights, as discussed earlier,
and computed with LieART \cite{lieart}, so as to have conventional
$\theta$-angle periodicites.

\begin{table}[h!]
\centering
	\begin{tabular}{cc|cc}
		Field & Positive root & Field & Negative root  \\ \hline
$X_{1}$ & $\textbf{(}1,0,0,0,0,0,0\textbf{)}$ & $X_{64}$ & $\textbf{(}-1,0,0,0,0,0,0\textbf{)}$  \\
$X_{2}$ & $\textbf{(}-1,1,0,0,0,0,0\textbf{)}$ & $X_{65}$ & $\textbf{(}1,-1,0,0,0,0,0\textbf{)}$  \\
$X_{3}$ & $\textbf{(}0,-1,1,0,0,0,0\textbf{)}$ & $X_{66}$ & $\textbf{(}0,1,-1,0,0,0,0\textbf{)}$  \\
$X_{4}$ & $\textbf{(}0,0,-1,1,0,0,1\textbf{)}$ & $X_{67}$ & $\textbf{(}0,0,1,-1,0,0,-1\textbf{)}$  \\
$X_{5}$ & $\textbf{(}0,0,0,-1,1,0,1\textbf{)}$ & $X_{68}$ & $\textbf{(}0,0,0,1,-1,0,-1\textbf{)}$  \\
$X_{6}$ & $\textbf{(}0,0,0,1,0,0,-1\textbf{)}$ & $X_{69}$ & $\textbf{(}0,0,0,-1,0,0,1\textbf{)}$  \\
$X_{7}$ & $\textbf{(}0,0,0,0,-1,1,1\textbf{)}$ & $X_{70}$ & $\textbf{(}0,0,0,0,1,-1,-1\textbf{)}$  \\
$X_{8}$ & $\textbf{(}0,0,1,-1,1,0,-1\textbf{)}$ & $X_{71}$ & $\textbf{(}0,0,-1,1,-1,0,1\textbf{)}$  \\
$X_{9}$ & $\textbf{(}0,0,0,0,0,-1,1\textbf{)}$ & $X_{72}$ & $\textbf{(}0,0,0,0,0,1,-1\textbf{)}$  \\
$X_{10}$ & $\textbf{(}0,0,1,0,-1,1,-1\textbf{)}$ & $X_{73}$ & $\textbf{(}0,0,-1,0,1,-1,1\textbf{)}$  \\
$X_{11}$ & $\textbf{(}0,1,-1,0,1,0,0\textbf{)}$ & $X_{74}$ & $\textbf{(}0,-1,1,0,-1,0,0\textbf{)}$  \\
$X_{12}$ & $\textbf{(}0,0,1,0,0,-1,-1\textbf{)}$ & $X_{75}$ & $\textbf{(}0,0,-1,0,0,1,1\textbf{)}$  \\
$X_{13}$ & $\textbf{(}0,1,-1,1,-1,1,0\textbf{)}$ & $X_{76}$ & $\textbf{(}0,-1,1,-1,1,-1,0\textbf{)}$  \\
$X_{14}$ & $\textbf{(}1,-1,0,0,1,0,0\textbf{)}$ & $X_{77}$ & $\textbf{(}-1,1,0,0,-1,0,0\textbf{)}$  \\
$X_{15}$ & $\textbf{(}-1,0,0,0,1,0,0\textbf{)}$ & $X_{78}$ & $\textbf{(}1,0,0,0,-1,0,0\textbf{)}$  \\
$X_{16}$ & $\textbf{(}0,1,-1,1,0,-1,0\textbf{)}$ & $X_{79}$ & $\textbf{(}0,-1,1,-1,0,1,0\textbf{)}$  \\
$X_{17}$ & $\textbf{(}0,1,0,-1,0,1,0\textbf{)}$ & $X_{80}$ & $\textbf{(}0,-1,0,1,0,-1,0\textbf{)}$  \\
$X_{18}$ & $\textbf{(}1,-1,0,1,-1,1,0\textbf{)}$ & $X_{81}$ & $\textbf{(}-1,1,0,-1,1,-1,0\textbf{)}$  \\
$X_{19}$ & $\textbf{(}-1,0,0,1,-1,1,0\textbf{)}$ & $X_{82}$ & $\textbf{(}1,0,0,-1,1,-1,0\textbf{)}$  \\
$X_{20}$ & $\textbf{(}0,1,0,-1,1,-1,0\textbf{)}$ & $X_{83}$ & $\textbf{(}0,-1,0,1,-1,1,0\textbf{)}$  \\
$X_{21}$ & $\textbf{(}1,-1,0,1,0,-1,0\textbf{)}$ & $X_{84}$ & $\textbf{(}-1,1,0,-1,0,1,0\textbf{)}$  \\
$X_{22}$ & $\textbf{(}1,-1,1,-1,0,1,0\textbf{)}$ & $X_{85}$ & $\textbf{(}-1,1,-1,1,0,-1,0\textbf{)}$  \\
$X_{23}$ & $\textbf{(}-1,0,0,1,0,-1,0\textbf{)}$ & $X_{86}$ & $\textbf{(}1,0,0,-1,0,1,0\textbf{)}$  \\
$X_{24}$ & $\textbf{(}-1,0,1,-1,0,1,0\textbf{)}$ & $X_{87}$ & $\textbf{(}1,0,-1,1,0,-1,0\textbf{)}$  \\
$X_{25}$ & $\textbf{(}0,1,0,0,-1,0,0\textbf{)}$ & $X_{88}$ & $\textbf{(}0,-1,0,0,1,0,0\textbf{)}$  \\
$X_{26}$ & $\textbf{(}1,-1,1,-1,1,-1,0\textbf{)}$ & $X_{89}$ & $\textbf{(}-1,1,-1,1,-1,1,0\textbf{)}$  
	\end{tabular}
\caption{First set of roots of $E_7$ and assocaited fields. 
\label{table:e7:roots1} }
\end{table}

\begin{table}[h!]
\centering
\begin{tabular}{cc|cc}
		Field & Positive root & Field & Negative root  \\ \hline
$X_{27}$ & $\textbf{(}1,0,-1,0,0,1,1\textbf{)}$ & $X_{90}$ & $\textbf{(}-1,0,1,0,0,-1,-1\textbf{)}$  \\
$X_{28}$ & $\textbf{(}-1,0,1,-1,1,-1,0\textbf{)}$ & $X_{91}$ & $\textbf{(}1,0,-1,1,-1,1,0\textbf{)}$  \\
$X_{29}$ & $\textbf{(}-1,1,-1,0,0,1,1\textbf{)}$ & $X_{92}$ & $\textbf{(}1,-1,1,0,0,-1,-1\textbf{)}$  \\
$X_{30}$ & $\textbf{(}1,-1,1,0,-1,0,0\textbf{)}$ & $X_{93}$ & $\textbf{(}-1,1,-1,0,1,0,0\textbf{)}$  \\
$X_{31}$ & $\textbf{(}1,0,-1,0,1,-1,1\textbf{)}$ & $X_{94}$ & $\textbf{(}-1,0,1,0,-1,1,-1\textbf{)}$  \\
$X_{32}$ & $\textbf{(}1,0,0,0,0,1,-1\textbf{)}$ & $X_{95}$ & $\textbf{(}-1,0,0,0,0,-1,1\textbf{)}$  \\
$X_{33}$ & $\textbf{(}-1,0,1,0,-1,0,0\textbf{)}$ & $X_{96}$ & $\textbf{(}1,0,-1,0,1,0,0\textbf{)}$  \\
$X_{34}$ & $\textbf{(}-1,1,-1,0,1,-1,1\textbf{)}$ & $X_{97}$ & $\textbf{(}1,-1,1,0,-1,1,-1\textbf{)}$  \\
$X_{35}$ & $\textbf{(}-1,1,0,0,0,1,-1\textbf{)}$ & $X_{98}$ & $\textbf{(}1,-1,0,0,0,-1,1\textbf{)}$  \\
$X_{36}$ & $\textbf{(}0,-1,0,0,0,1,1\textbf{)}$ & $X_{99}$ & $\textbf{(}0,1,0,0,0,-1,-1\textbf{)}$  \\
$X_{37}$ & $\textbf{(}1,0,-1,1,-1,0,1\textbf{)}$ & $X_{100}$ & $\textbf{(}-1,0,1,-1,1,0,-1\textbf{)}$  \\
$X_{38}$ & $\textbf{(}1,0,0,0,1,-1,-1\textbf{)}$ & $X_{101}$ & $\textbf{(}-1,0,0,0,-1,1,1\textbf{)}$  \\
$X_{39}$ & $\textbf{(}-1,1,-1,1,-1,0,1\textbf{)}$ & $X_{102}$ & $\textbf{(}1,-1,1,-1,1,0,-1\textbf{)}$  \\
$X_{40}$ & $\textbf{(}-1,1,0,0,1,-1,-1\textbf{)}$ & $X_{103}$ & $\textbf{(}1,-1,0,0,-1,1,1\textbf{)}$  \\
$X_{41}$ & $\textbf{(}0,-1,0,0,1,-1,1\textbf{)}$ & $X_{104}$ & $\textbf{(}0,1,0,0,-1,1,-1\textbf{)}$  \\
$X_{42}$ & $\textbf{(}0,-1,1,0,0,1,-1\textbf{)}$ & $X_{105}$ & $\textbf{(}0,1,-1,0,0,-1,1\textbf{)}$  \\
$X_{43}$ & $\textbf{(}1,0,0,-1,0,0,1\textbf{)}$ & $X_{106}$ & $\textbf{(}-1,0,0,1,0,0,-1\textbf{)}$  \\
$X_{44}$ & $\textbf{(}1,0,0,1,-1,0,-1\textbf{)}$ & $X_{107}$ & $\textbf{(}-1,0,0,-1,1,0,1\textbf{)}$  \\
$X_{45}$ & $\textbf{(}-1,1,0,-1,0,0,1\textbf{)}$ & $X_{108}$ & $\textbf{(}1,-1,0,1,0,0,-1\textbf{)}$  \\
$X_{46}$ & $\textbf{(}-1,1,0,1,-1,0,-1\textbf{)}$ & $X_{109}$ & $\textbf{(}1,-1,0,-1,1,0,1\textbf{)}$  \\
$X_{47}$ & $\textbf{(}0,-1,0,1,-1,0,1\textbf{)}$ & $X_{110}$ & $\textbf{(}0,1,0,-1,1,0,-1\textbf{)}$  \\
$X_{48}$ & $\textbf{(}0,-1,1,0,1,-1,-1\textbf{)}$ & $X_{111}$ & $\textbf{(}0,1,-1,0,-1,1,1\textbf{)}$  \\
$X_{49}$ & $\textbf{(}0,0,-1,1,0,1,0\textbf{)}$ & $X_{112}$ & $\textbf{(}0,0,1,-1,0,-1,0\textbf{)}$  \\
$X_{50}$ & $\textbf{(}1,0,1,-1,0,0,-1\textbf{)}$ & $X_{113}$ & $\textbf{(}-1,0,-1,1,0,0,1\textbf{)}$  \\
$X_{51}$ & $\textbf{(}-1,1,1,-1,0,0,-1\textbf{)}$ & $X_{114}$ & $\textbf{(}1,-1,-1,1,0,0,1\textbf{)}$  \\
$X_{52}$ & $\textbf{(}0,-1,1,-1,0,0,1\textbf{)}$ & $X_{115}$ & $\textbf{(}0,1,-1,1,0,0,-1\textbf{)}$  
  \end{tabular}
\caption{Second set of roots of $E_7$ and associated fields.  
\label{table:e7:roots2} }
\end{table}

\begin{table}[h!]
\centering
\begin{tabular}{cc|cc}
                Field & Positive root & Field & Negative root  \\ \hline
$X_{53}$ & $\textbf{(}0,-1,1,1,-1,0,-1\textbf{)}$ & $X_{116}$ & $\textbf{(}0,1,-1,-1,1,0,1\textbf{)}$  \\
$X_{54}$ & $\textbf{(}0,0,-1,1,1,-1,0\textbf{)}$ & $X_{117}$ & $\textbf{(}0,0,1,-1,-1,1,0\textbf{)}$  \\
$X_{55}$ & $\textbf{(}0,0,0,-1,1,1,0\textbf{)}$ & $X_{118}$ & $\textbf{(}0,0,0,1,-1,-1,0\textbf{)}$  \\
$X_{56}$ & $\textbf{(}1,1,-1,0,0,0,0\textbf{)}$ & $X_{119}$ & $\textbf{(}-1,-1,1,0,0,0,0\textbf{)}$  \\
$X_{57}$ & $\textbf{(}-1,2,-1,0,0,0,0\textbf{)}$ & $X_{120}$ & $\textbf{(}1,-2,1,0,0,0,0\textbf{)}$  \\
$X_{58}$ & $\textbf{(}0,-1,2,-1,0,0,-1\textbf{)}$ & $X_{121}$ & $\textbf{(}0,1,-2,1,0,0,1\textbf{)}$  \\
$X_{59}$ & $\textbf{(}0,0,-1,0,0,0,2\textbf{)}$ & $X_{122}$ & $\textbf{(}0,0,1,0,0,0,-2\textbf{)}$  \\
$X_{60}$ & $\textbf{(}0,0,-1,2,-1,0,0\textbf{)}$ & $X_{123}$ & $\textbf{(}0,0,1,-2,1,0,0\textbf{)}$  \\
$X_{61}$ & $\textbf{(}0,0,0,-1,2,-1,0\textbf{)}$ & $X_{124}$ & $\textbf{(}0,0,0,1,-2,1,0\textbf{)}$  \\
$X_{62}$ & $\textbf{(}0,0,0,0,-1,2,0\textbf{)}$ & $X_{125}$ & $\textbf{(}0,0,0,0,1,-2,0\textbf{)}$  \\
$X_{63}$ & $\textbf{(}2,-1,0,0,0,0,0\textbf{)}$ & $X_{126}$ & $\textbf{(}-2,1,0,0,0,0,0\textbf{)}$  
	\end{tabular}
\caption{Third set of roots of $E_7$ and associated fields.  
\label{table:e7:roots3} }
\end{table}

\begin{table}[h!]
\centering
\begin{tabular}{cc|cc}
		Field & Weight & Field & Weight \\ \hline
$Y_{i,1}$ & $\textbf{(}0,0,0,0,0,1,0\textbf{)}$ & $Y_{i,29}$ & $\textbf{(}0,0,0,0,0,-1,0\textbf{)}$  \\
$Y_{i,2}$ & $\textbf{(}0,0,0,0,1,-1,0\textbf{)}$ & $Y_{i,30}$ & $\textbf{(}0,0,0,0,-1,1,0\textbf{)}$  \\
$Y_{i,3}$ & $\textbf{(}0,0,0,1,-1,0,0\textbf{)}$ & $Y_{i,31}$ & $\textbf{(}0,0,0,-1,1,0,0\textbf{)}$  \\
$Y_{i,4}$ & $\textbf{(}0,0,1,-1,0,0,0\textbf{)}$ & $Y_{i,32}$ & $\textbf{(}0,0,-1,1,0,0,0\textbf{)}$  \\
$Y_{i,5}$ & $\textbf{(}0,1,-1,0,0,0,1\textbf{)}$ & $Y_{i,33}$ & $\textbf{(}0,-1,1,0,0,0,-1\textbf{)}$  \\
$Y_{i,6}$ & $\textbf{(}0,1,0,0,0,0,-1\textbf{)}$ & $Y_{i,34}$ & $\textbf{(}0,-1,0,0,0,0,1\textbf{)}$  \\
$Y_{i,7}$ & $\textbf{(}1,-1,0,0,0,0,1\textbf{)}$ & $Y_{i,35}$ & $\textbf{(}-1,1,0,0,0,0,-1\textbf{)}$  \\
$Y_{i,8}$ & $\textbf{(}-1,0,0,0,0,0,1\textbf{)}$ & $Y_{i,36}$ & $\textbf{(}1,0,0,0,0,0,-1\textbf{)}$  \\
$Y_{i,9}$ & $\textbf{(}1,-1,1,0,0,0,-1\textbf{)}$ & $Y_{i,37}$ & $\textbf{(}-1,1,-1,0,0,0,1\textbf{)}$  \\
$Y_{i,10}$ & $\textbf{(}-1,0,1,0,0,0,-1\textbf{)}$ & $Y_{i,38}$ & $\textbf{(}1,0,-1,0,0,0,1\textbf{)}$  \\
$Y_{i,11}$ & $\textbf{(}1,0,-1,1,0,0,0\textbf{)}$ & $Y_{i,39}$ & $\textbf{(}-1,0,1,-1,0,0,0\textbf{)}$  \\
$Y_{i,12}$ & $\textbf{(}-1,1,-1,1,0,0,0\textbf{)}$ & $Y_{i,40}$ & $\textbf{(}1,-1,1,-1,0,0,0\textbf{)}$  \\
$Y_{i,13}$ & $\textbf{(}1,0,0,-1,1,0,0\textbf{)}$ & $Y_{i,41}$ & $\textbf{(}-1,0,0,1,-1,0,0\textbf{)}$  \\
$Y_{i,14}$ & $\textbf{(}-1,1,0,-1,1,0,0\textbf{)}$ & $Y_{i,42}$ & $\textbf{(}1,-1,0,1,-1,0,0\textbf{)}$  \\
$Y_{i,15}$ & $\textbf{(}0,-1,0,1,0,0,0\textbf{)}$ & $Y_{i,43}$ & $\textbf{(}0,1,0,-1,0,0,0\textbf{)}$  \\
$Y_{i,16}$ & $\textbf{(}1,0,0,0,-1,1,0\textbf{)}$ & $Y_{i,44}$ & $\textbf{(}-1,0,0,0,1,-1,0\textbf{)}$  \\
$Y_{i,17}$ & $\textbf{(}-1,1,0,0,-1,1,0\textbf{)}$ & $Y_{i,45}$ & $\textbf{(}1,-1,0,0,1,-1,0\textbf{)}$  \\
$Y_{i,18}$ & $\textbf{(}0,-1,1,-1,1,0,0\textbf{)}$ & $Y_{i,46}$ & $\textbf{(}0,1,-1,1,-1,0,0\textbf{)}$  \\
$Y_{i,19}$ & $\textbf{(}1,0,0,0,0,-1,0\textbf{)}$ & $Y_{i,47}$ & $\textbf{(}-1,0,0,0,0,1,0\textbf{)}$  \\
$Y_{i,20}$ & $\textbf{(}-1,1,0,0,0,-1,0\textbf{)}$ & $Y_{i,48}$ & $\textbf{(}1,-1,0,0,0,1,0\textbf{)}$  \\
$Y_{i,21}$ & $\textbf{(}0,-1,1,0,-1,1,0\textbf{)}$ & $Y_{i,49}$ & $\textbf{(}0,1,-1,0,1,-1,0\textbf{)}$  \\
$Y_{i,22}$ & $\textbf{(}0,0,-1,0,1,0,1\textbf{)}$ & $Y_{i,50}$ & $\textbf{(}0,0,1,0,-1,0,-1\textbf{)}$  \\
$Y_{i,23}$ & $\textbf{(}0,-1,1,0,0,-1,0\textbf{)}$ & $Y_{i,51}$ & $\textbf{(}0,1,-1,0,0,1,0\textbf{)}$  \\
$Y_{i,24}$ & $\textbf{(}0,0,-1,1,-1,1,1\textbf{)}$ & $Y_{i,52}$ & $\textbf{(}0,0,1,-1,1,-1,-1\textbf{)}$  \\
$Y_{i,25}$ & $\textbf{(}0,0,0,0,1,0,-1\textbf{)}$ & $Y_{i,53}$ & $\textbf{(}0,0,0,0,-1,0,1\textbf{)}$  \\
$Y_{i,26}$ & $\textbf{(}0,0,-1,1,0,-1,1\textbf{)}$ & $Y_{i,54}$ & $\textbf{(}0,0,1,-1,0,1,-1\textbf{)}$  \\
$Y_{i,27}$ & $\textbf{(}0,0,0,-1,0,1,1\textbf{)}$ & $Y_{i,55}$ & $\textbf{(}0,0,0,1,0,-1,-1\textbf{)}$  \\
$Y_{i,28}$ & $\textbf{(}0,0,0,1,-1,1,-1\textbf{)}$ & $Y_{i,56}$ & $\textbf{(}0,0,0,-1,1,-1,1\textbf{)}$  
	\end{tabular}
\caption{Weights of ${\bf 56}$ of $E_7$ and associated fields.  
\label{table:e7:weights}}
\end{table}

\subsection{Superpotential}

Plugging into the general expression for the mirror superpotential,
we find for this case that the mirror superpotential is given by
\begin{equation} \nonumber
W=\sum_{a=1}^{7}\sigma_a \Big( \sum_{i=1}^{n}\sum_{\beta=1}^{56} \rho^a_{i,\beta}Y_{i,\beta}+\sum_{m=1}^{126} \alpha^a_{m}Z_m  \Big)-\sum_{i=1}^{n}\tilde{m}_i\sum_{\beta=1}^{56}Y_{i,\beta}+\sum_{i=1}^{n}\sum_{\beta=1}^{56}\exp(-Y_{i,\beta})+\sum_{m=1}^{126}X_m.
\end{equation} 
where $X_m=\exp(-Z_m)$ and $X_m$ are the fundamental fields, we get:
\begin{align}
W=\sum_{a=1}^{7} \sigma_a \mathcal{C}^a-\sum_{i=1}^{n}\tilde{m}_i \sum_{\beta=1}^{56} Y_{i,\beta}+\sum_{i=1}^{n} \sum_{\beta=1}^{56} \exp(-Y_{i,\beta})+\sum_{m=1}^{126}X_m   .
\end{align}
where $\mathcal{C}^a$ are given as follows:
\begin{align*} 
\mathcal{C}^1=& \sum_{i=1}^{n}  \big(   Y_{i,7}-Y_{i,8}+Y_{i,9}-Y_{i,10}+Y_{i,11}-Y_{i,12}+Y_{i,13}-Y_{i,14}+Y_{i,16}-Y_{i,17}+Y_{i,19}-Y_{i,2
	0}\\& \hspace*{.6 in} -Y_{i,35}+Y_{i,36}-Y_{i,37}+Y_{i,38}-Y_{i,39}+Y_{i,40}-Y_{i,41}+Y_{i,42}-Y_{i,44}+Y_{i,
	45}\\& \hspace*{.6 in}-Y_{i,47}+Y_{i,48} \big) \\&  \hspace*{.4in } +Z_1-Z_2+Z_{14}-Z_{15}+Z_{18}-Z_{19}+Z_{21}+Z_{22}-Z
_{23}-Z_{24}+Z_{26}+Z_{27}-Z_{28}\\&  \hspace*{.4in }-Z_{29} +Z_{30}+Z_{31}+Z_{32}-Z_{33}-
Z_{34}-Z_{35}+Z_{37}+Z_{38}-Z_{39}-Z_{40}+Z_{43}+Z_{44}\\&  \hspace*{.4in }-Z_{45}-Z_{46}
+Z_{50}-Z_{51}+Z_{56}-Z_{57}+2
Z_{63}-Z_{64}+Z_{65}-Z_{77}+Z_{78}-Z_{81}\\&  \hspace*{.4in }+Z_{82}-Z_{84} -Z_{85}+Z_{86}
+Z_{87}-Z_{89}-Z_{90}+Z_{91}+Z_{92}-Z_{93}-Z_{94}-Z_{95}+Z_{96}\\&  \hspace*{.4in }+Z_{97
}+Z_{98}-Z_{100}-Z_{101}+Z_{102}+Z_{103}-Z_{106}-Z_{107}+Z_{108}+Z_{1
	09}-Z_{113}\\&  \hspace*{.4in }+Z_{114}-Z_{119}+Z_{120}-2 Z_{126} ,
\end{align*}
\begin{align*}
	\mathcal{C}^2=& \sum_{i=1}^{n} \big(  Y_{i,5}+Y_{i,6}-Y_{i,7}-Y_{i,9}+Y_{i,12}+Y_{i,14}-Y_{i,15}+Y_{i,17}-Y_{i,18}+Y_{i,20}-Y_{i,21}-Y_{i,23} \\ & \hspace*{.6 in}
	-Y_{i,33} -Y_{i,34}+Y_{i,35}+Y_{i,37}-Y_{i,40}-Y_{i,42}+Y_{i,43}-Y_{i,45}+Y_{i,46}-Y_{i,48}
	+Y_{i,49}\\ & \hspace*{.6 in}+Y_{i,51} \big)   \\ &  \hspace*{.4in }  +Z_2-Z_3+Z_{11}+Z_{13}-Z_{14}+Z_{16}+Z_{17}-Z_{18}+Z_{2
		0}-Z_{21}-Z_{22}+Z_{25}-Z_{26}\\ &  \hspace*{.4in } +Z_{29}-Z_{30}+Z_{34}+Z_{35}-Z_{36}+Z_{
		39}+Z_{40}-Z_{41}-Z_{42}+Z_{45}+Z_{46}-Z_{47}-Z_{48}\\ &  \hspace*{.4in }+Z_{51}-Z_{52}-Z_
	{53}+Z_{56} +2
	Z_{57}-Z_{58}-Z_{63}-Z_{65}+Z_{66}-Z_{74}-Z_{76}+Z_{77}\\ &  \hspace*{.4in }-Z_{79}-Z_{80}
	+Z_{81}-Z_{83}+Z_{84}+Z_{85}-Z_{88}  +Z_{89}-Z_{92}+Z_{93}-Z_{97}-Z_{98
	}+Z_{99}\\ &  \hspace*{.4in }-Z_{102}-Z_{103}+Z_{104}+Z_{105}-Z_{108} -Z_{109}+Z_{110}+Z_{1
		11}-Z_{114}+Z_{115}+Z_{116}\\ &  \hspace*{.4in }-Z_{119}-2 Z_{120}+Z_{121}+Z_{126} ,
\end{align*}
\begin{align*}
	\mathcal{C}^3= & \sum_{i=1}^{n} \big( Y_{i,4}-Y_{i,5}+Y_{i,9}+Y_{i,10}-Y_{i,11}-Y_{i,12}+Y_{i,18}+Y_{i,21}-Y_{i,22}+Y_{i,23}-Y_{i,24}-Y_{i,2
		6}\\& \hspace*{.6in} -Y_{i,32}+Y_{i,33}-Y_{i,37}-Y_{i,38}+Y_{i,39}+Y_{i,40}-Y_{i,46}-Y_{i,49}+Y_{i,50}-Y_{i,
		51}+Y_{i,52}\\& \hspace*{.6in}+Y_{i,54}\big)  \\ & \hspace*{.4 in} +Z_3-Z_4+Z_8+Z_{10}-Z_{11}+Z_{12}-Z_{13}-Z_{16}+Z_{2
		2}+Z_{24}+Z_{26}-Z_{27}+Z_{28}\\ & \hspace*{.4 in}-Z_{29} +Z_{30}-Z_{31}+Z_{33}-Z_{34}-Z_{
		37}-Z_{39}+Z_{42}+Z_{48}-Z_{49}+Z_{50}+Z_{51}\\ & \hspace*{.4 in}+Z_{52}+Z_{53}-Z_{54} -Z_
	{56}-Z_{57} +2
	Z_{58}-Z_{59}-Z_{60}-Z_{66}+Z_{67}-Z_{71}-Z_{73}\\ & \hspace*{.4 in}+Z_{74}-Z_{75}+Z_{76}
	+Z_{79}-Z_{85} -Z_{87}-Z_{89}+Z_{90}-Z_{91}+Z_{92}-Z_{93}+Z_{94}\\ & \hspace*{.4 in}-Z_{96
	}+Z_{97}+Z_{100}+Z_{102}-Z_{105}-Z_{111}+Z_{112}-Z_{113}-Z_{114}-Z_{1
		15}-Z_{116}\\ & \hspace*{.4 in}+Z_{117}+Z_{119}+Z_{120}-2 Z_{121}+Z_{122}+Z_{123},
\end{align*}
\begin{align*}
\mathcal{C}^4=& \sum_{i=1}^{n} \big(  Y_{i,3}-Y_{i,4}+Y_{i,11}+Y_{i,12}-Y_{i,13}-Y_{i,14}+Y_{i,15}-Y_{i,18}+Y_{i,24}+Y_{i,26}-Y_{i,27}+Y
_{i,28}\\& \hspace*{.6in} -Y_{i,31}+Y_{i,32}-Y_{i,39}-Y_{i,40}+Y_{i,41}+Y_{i,42}-Y_{i,43}+Y_{i,46}-Y_{i,52}-
Y_{i,54}+Y_{i,55}\\& \hspace*{.6in}-Y_{i,56} \big) \\& \hspace*{.4in}+Z_4-Z_5+Z_6-Z_8+Z_{13}+Z_{16}-Z_{17}+Z_{18}+Z_{1
	9}-Z_{20}+Z_{21}-Z_{22}+Z_{23}\\& \hspace*{.4in}-Z_{24}-Z_{26} -Z_{28}+Z_{37}+Z_{39}-Z_{
	43}+Z_{44}-Z_{45}+Z_{46}+Z_{47}+Z_{49}-Z_{50}\\& \hspace*{.4in}-Z_{51}-Z_{52}+Z_{53}+Z_
{54}-Z_{55}-Z_{58}+2
Z_{60}-Z_{61}-Z_{67}+Z_{68}-Z_{69}+Z_{71}\\& \hspace*{.4in}-Z_{76}-Z_{79}+Z_{80}-Z_{81}
-Z_{82}+Z_{83} -Z_{84}+Z_{85}-Z_{86}+Z_{87}+Z_{89}+Z_{91}\\& \hspace*{.4in}-Z_{100}-Z_{1
	02}+Z_{106}-Z_{107}+Z_{108}-Z_{109}-Z_{110}-Z_{112}+Z_{113}+Z_{114}+Z
_{115}\\& \hspace*{.4in}-Z_{116}-Z_{117}+Z_{118}+Z_{121}-2 Z_{123}+Z_{124},
\end{align*}
\begin{align*}
\mathcal{C}^5= & \sum_{i=1}^{n} \big( Y_{i,2}-Y_{i,3}+Y_{i,13}+Y_{i,14}-Y_{i,16}-Y_{i,17}+Y_{i,18}-Y_{i,21}+Y_{i,22}-Y_{i,24}+Y_{i,25}-Y
_{i,28} \\& \hspace*{.6in} -Y_{i,30}+Y_{i,31}-Y_{i,41}-Y_{i,42}+Y_{i,44}+Y_{i,45}-Y_{i,46}+Y_{i,49}-Y_{i,50}+
Y_{i,52}-Y_{i,53}\\& \hspace*{.6in}+Y_{i,56} \big)  \\& \hspace*{.4in} +Z_5-Z_7+Z_8-Z_{10}+Z_{11}-Z_{13}+Z_{14}+Z_{15}-Z
_{18}-Z_{19}+Z_{20}-Z_{25}+Z_{26}\\& \hspace*{.4in}+Z_{28}-Z_{30}+Z_{31}-Z_{33}+Z_{34}-
Z_{37}+Z_{38}-Z_{39}+Z_{40}+Z_{41}-Z_{44}-Z_{46}-Z_{47}\\& \hspace*{.4in}+Z_{48}-Z_{53} 
+Z_{54}+Z_{55}-Z_{60}+2
Z_{61}-Z_{62}-Z_{68}+Z_{70}-Z_{71}+Z_{73}-Z_{74}\\& \hspace*{.4in}+Z_{76}-Z_{77}-Z_{78}
+Z_{81}+Z_{82}-Z_{83}+Z_{88}-Z_{89}-Z_{91}+Z_{93}-Z_{94}+Z_{96}-Z_{97
}\\& \hspace*{.4in}+Z_{100}-Z_{101}+Z_{102}-Z_{103}-Z_{104}+Z_{107}+Z_{109}+Z_{110}-Z_{
	111}+Z_{116}-Z_{117}\\& \hspace*{.4in}-Z_{118}+Z_{123}-2 Z_{124}+Z_{125} ,
\end{align*}
\begin{align*}
\mathcal{C}^6=& \sum_{i=1}^{n} \big( Y_{i,1}-Y_{i,2}+Y_{i,16}+Y_{i,17}-Y_{i,19}-Y_{i,20}+Y_{i,21}-Y_{i,23}+Y_{i,24}-Y_{i,26}+Y_{i,27}+Y
_{i,28}\\& \hspace*{.6in} -Y_{i,29}+Y_{i,30}-Y_{i,44}-Y_{i,45}+Y_{i,47}+Y_{i,48}-Y_{i,49}+Y_{i,51}-Y_{i,52}+
Y_{i,54}-Y_{i,55}\\& \hspace*{.6in}-Y_{i,56} \big)   \\& \hspace*{.4in} +Z_7-Z_9+Z_{10}-Z_{12}+Z_{13}-Z_{16}+Z_{17}+Z_{18
}+Z_{19}-Z_{20}-Z_{21}+Z_{22}-Z_{23}\\& \hspace*{.4in}+Z_{24}  -Z_{26}+Z_{27}-Z_{28}+Z_{2
	9}-Z_{31}+Z_{32}-Z_{34}+Z_{35}+Z_{36}-Z_{38}-Z_{40}-Z_{41}\\& \hspace*{.4in}+Z_{42}-Z_{
	48}  +Z_{49}-Z_{54}+Z_{55}-Z_{61}+2
Z_{62}-Z_{70}+Z_{72}-Z_{73}+Z_{75}-Z_{76}\\& \hspace*{.4in}+Z_{79}-Z_{80}-Z_{81}-Z_{82}
+Z_{83}+Z_{84}-Z_{85}+Z_{86}-Z_{87}+Z_{89}-Z_{90}+Z_{91}-Z_{92}\\& \hspace*{.4in}+Z_{94
}-Z_{95}+Z_{97}-Z_{98}-Z_{99}+Z_{101}+Z_{103}+Z_{104}-Z_{105}+Z_{111}
-Z_{112}+Z_{117}\\& \hspace*{.4in}-Z_{118}+Z_{124}-2 Z_{125} ,
\end{align*}
\begin{align*}
\mathcal{C}^7=& \sum_{i=1}^{n} \big(  Y_{i,2}-Y_{i,3}+Y_{i,13}+Y_{i,14}-Y_{i,16}-Y_{i,17}+Y_{i,18}-Y_{i,21}+Y_{i,22}-Y_{i,24}+Y_{i,25}-Y
_{i,28} \\& \hspace*{.6in} -Y_{i,30}+Y_{i,31}-Y_{i,41}-Y_{i,42}+Y_{i,44}+Y_{i,45}-Y_{i,46}+Y_{i,49}-Y_{i,50}+
Y_{i,52}-Y_{i,53}\\& \hspace*{.6in}+Y_{i,56} \big) \\& \hspace*{.4in} +Z_5-Z_7+Z_8-Z_{10}+Z_{11}-Z_{13}+Z_{14}+Z_{15}-Z
_{18}-Z_{19}+Z_{20}-Z_{25}+Z_{26}\\& \hspace*{.4in}+Z_{28} -Z_{30}+Z_{31}-Z_{33}+Z_{34}-
Z_{37}+Z_{38}-Z_{39}+Z_{40}+Z_{41}-Z_{44}-Z_{46}-Z_{47}\\& \hspace*{.4in}+Z_{48}-Z_{53} 
+Z_{54}+Z_{55}-Z_{60}+2
Z_{61}-Z_{62}-Z_{68}+Z_{70}-Z_{71}+Z_{73}-Z_{74}\\& \hspace*{.4in}+Z_{76}-Z_{77}-Z_{78}
+Z_{81}  +Z_{82}-Z_{83}+Z_{88}-Z_{89}-Z_{91}+Z_{93}-Z_{94}+Z_{96}-Z_{97
}\\& \hspace*{.4in}+Z_{100} -Z_{101}+Z_{102}-Z_{103} -Z_{104}+Z_{107}+Z_{109}+Z_{110}-Z_{
	111}+Z_{116}-Z_{117}\\& \hspace*{.4in}-Z_{118}+Z_{123}-2 Z_{124}+Z_{125}.
\end{align*}

\subsection{Coulomb ring relations}

Integrating out the $\sigma_a$ fields, we obtain seven constraints 
$\mathcal{C}^a=0$.  Exponentiating these constraints, we obtain
a series of equations from which the Coulomb ring relations will
be derived.  For reasons of notational
sanity, we will also slightly simplify these expressions, as follows.
To make predictions for the A model, we will evaluate the ring relations
on the critical locus, where
\begin{displaymath}
\frac{ X_m }{ X_{m+63} } \: = \: -1.
\end{displaymath}
It is straightforward to see that each of the constraints ${\mathcal C}^a$
contains 34 differences of corresponding $Z$'s, so that the exponential of the
constraints contains 34 factors of the form $X_m/X_{m+63}$ -- an even
number of factors of $-1$, which will cancel out.
Therefore, since on the critical locus those factors will cancel out,
we will omit them, and solely relate the exponentiated constraints in
terms of $Y$s.

The exponentiated constraints are as follows.
\begin{multline} 
\prod_{i=1}^{n}\exp \big(   Y_{i,7}-Y_{i,8}+Y_{i,9}-Y_{i,10}+Y_{i,11}-Y_{i,12}+Y_{i,13}-Y_{i,14}+Y_{i,16}-Y_{i,17}+Y_{i,19}-Y_{i,2
	0}\\
  \hspace*{.27 in} -Y_{i,35}+Y_{i,36}-Y_{i,37}+Y_{i,38}-Y_{i,39}+Y_{i,40}-Y_{i,41}+Y_{i,42}-Y_{i,44}+Y_{i,
	45}-Y_{i,47}+Y_{i,48} \big) \\
 =1  ,  \label{E7 1}
\end{multline}
\begin{multline} 
\prod_{i=1}^{n}\exp \big(  Y_{i,5}+Y_{i,6}-Y_{i,7}-Y_{i,9}+Y_{i,12}+Y_{i,14}-Y_{i,15}+Y_{i,17}-Y_{i,18}+Y_{i,20}-Y_{i,21}-Y_{i,23} \\
   \hspace*{.27 in}
-Y_{i,33} -Y_{i,34}+Y_{i,35}+Y_{i,37}-Y_{i,40}-Y_{i,42}+Y_{i,43}-Y_{i,45}+Y_{i,46}-Y_{i,48}
+Y_{i,49}+Y_{i,51} \big) \\
 =1  , \label{E7 2} 
\end{multline}
\begin{multline} 
\prod_{i=1}^{n}\exp \big( Y_{i,4}-Y_{i,5}+Y_{i,9}+Y_{i,10}-Y_{i,11}-Y_{i,12}+Y_{i,18}+Y_{i,21}-Y_{i,22}+Y_{i,23}-Y_{i,24}-Y_{i,2
	6}\\
 \hspace*{.4in} -Y_{i,32}+Y_{i,33}-Y_{i,37}-Y_{i,38}+Y_{i,39}+Y_{i,40}-Y_{i,46}-Y_{i,49}+Y_{i,50}-Y_{i,
	51}+Y_{i,52}+Y_{i,54}\big)  \\ 
 =1    ,  \label{E7 3}
\end{multline}
\begin{multline} 
\prod_{i=1}^{n}\exp \big(  Y_{i,3}-Y_{i,4}+Y_{i,11}+Y_{i,12}-Y_{i,13}-Y_{i,14}+Y_{i,15}-Y_{i,18}+Y_{i,24}+Y_{i,26}-Y_{i,27}+Y
_{i,28}\\
  \hspace*{.4in} -Y_{i,31}+Y_{i,32}-Y_{i,39}-Y_{i,40}+Y_{i,41}+Y_{i,42}-Y_{i,43}+Y_{i,46}-Y_{i,52}-
Y_{i,54}+Y_{i,55}-Y_{i,56} \big)   \\
 =1 ,  \label{E7 4}
\end{multline}
\begin{multline} 
\prod_{i=1}^{n}\exp \big( Y_{i,2}-Y_{i,3}+Y_{i,13}+Y_{i,14}-Y_{i,16}-Y_{i,17}+Y_{i,18}-Y_{i,21}+Y_{i,22}-Y_{i,24}+Y_{i,25}-Y
_{i,28} \\
 \hspace*{.4in}  -Y_{i,30}+Y_{i,31}-Y_{i,41}-Y_{i,42}+Y_{i,44}+Y_{i,45}-Y_{i,46}+Y_{i,49}-Y_{i,50}+
Y_{i,52}-Y_{i,53}+Y_{i,56} \big) \\
 =1 , \label{E7 5}
\end{multline}
\begin{multline} 
\prod_{i=1}^{n}\exp  \big( Y_{i,1}-Y_{i,2}+Y_{i,16}+Y_{i,17}-Y_{i,19}-Y_{i,20}+Y_{i,21}-Y_{i,23}+Y_{i,24}-Y_{i,26}+Y_{i,27}+Y
_{i,28}\\
 \hspace*{.4 in}  -Y_{i,29}+Y_{i,30}-Y_{i,44}-Y_{i,45}+Y_{i,47}+Y_{i,48}-Y_{i,49}+Y_{i,51}-Y_{i,52}+
Y_{i,54}-Y_{i,55}-Y_{i,56} \big) \\
 =1  , \label{E7 6}
\end{multline}
\begin{multline} 
\prod_{i=1}^{n}\exp \big(  Y_{i,2}-Y_{i,3}+Y_{i,13}+Y_{i,14}-Y_{i,16}-Y_{i,17}+Y_{i,18}-Y_{i,21}+Y_{i,22}-Y_{i,24}+Y_{i,25}-Y
_{i,28} \\
 \hspace*{.4in}  -Y_{i,30}+Y_{i,31}-Y_{i,41}-Y_{i,42}+Y_{i,44}+Y_{i,45}-Y_{i,46}+Y_{i,49}-Y_{i,50}+
Y_{i,52}-Y_{i,53}+Y_{i,56} \big)     \\
 =1 . \label{E7 7}
\end{multline}

The mirror map is given by,
\begin{equation*}
\exp(-Y_{i,\beta}) \mapsto - \tilde{m}_i + \sum_{a=1}^{7} \sigma_a \rho_{i,\beta}^a, \quad X_m \mapsto \sum_{a=1}^{7} \sigma_a \alpha_m^a.
\end{equation*}
After applying the mirror map, the constraints become
\begin{align} \nonumber
\prod_{i=1}^{n}	&\left(-\sigma _1+\sigma _3-\sigma _4-\tilde{m}_{i}\right) \left(-\sigma
		_1+\sigma _2-\sigma _3+\sigma _4-\tilde{m}_{i}\right) \left(-\sigma
		_1+\sigma _4-\sigma _5-\tilde{m}_{i}\right)  \\ \nonumber \cdot &\left(-\sigma _1+\sigma
		_2-\sigma _4+\sigma _5-\tilde{m}_{i}\right) \left(-\sigma _1+\sigma
		_2-\sigma _6-\tilde{m}_{i}\right) \left(-\sigma _1+\sigma _5-\sigma
		_6-\tilde{m}_{i}\right) \\
 \nonumber \cdot & \left(-\sigma _1+\sigma _6-\tilde{m}_{i}\right) \left(-\sigma
		_1+\sigma _2-\sigma _5+\sigma _6-\tilde{m}_{i}\right) \left(-\sigma
		_1+\sigma _2-\sigma _7-\tilde{m}_{i}\right) \\
 \nonumber \cdot & \left(-\sigma _1+\sigma
		_3-\sigma _7-\tilde{m}_{i}\right) \left(-\sigma _1+\sigma _7-\tilde{m}_{i}\right)
		\left(-\sigma _1+\sigma _2-\sigma _3+\sigma
		_7-\tilde{m}_{i}\right) \\
  \nonumber & =\prod_{i=1}^{n} \left(\sigma _1-\sigma _2+\sigma _3-\sigma
		_4-\tilde{m}_{i}\right) \left(\sigma _1-\sigma _3+\sigma _4-\tilde{m}_{i}\right)
		\left(\sigma _1-\sigma _2+\sigma _4-\sigma _5-\tilde{m}_{i}\right) \\
 \nonumber & \hspace*{0.25in}  \cdot 
		\left(\sigma _1-\sigma _4+\sigma _5-\tilde{m}_{i}\right) \left(\sigma
		_1-\sigma _6-\tilde{m}_{i}\right) \left(\sigma _1-\sigma _2+\sigma _5-\sigma
		_6-\tilde{m}_{i}\right)  \\
 \nonumber & \hspace*{0.25in}  \cdot  \left(\sigma _1-\sigma _2+\sigma _6-\tilde{m}_{i}\right)
		\left(\sigma _1-\sigma _5+\sigma _6-\tilde{m}_{i}\right) \left(\sigma
		_1-\sigma _7-\tilde{m}_{i}\right) \\
& \hspace*{0.25in} \cdot  \left(\sigma _1-\sigma _2+\sigma _3-\sigma
		_7-\tilde{m}_{i}\right) \left(\sigma _1-\sigma _2+\sigma _7-\tilde{m}_{i}\right)
		\left(\sigma _1-\sigma _3+\sigma _7-\tilde{m}_{i}\right),
\end{align}
\begin{align} \nonumber
\prod_{i=1}^{n} &	\left(\sigma _1-\sigma _2+\sigma _3-\sigma _4-\tilde{m}_{i}\right)
		\left(-\sigma _2+\sigma _4-\tilde{m}_{i}\right) \left(\sigma _1-\sigma
		_2+\sigma _4-\sigma _5-\tilde{m}_{i}\right) \\
 \nonumber \cdot & \left(-\sigma _2+\sigma
		_3-\sigma _4+\sigma _5-\tilde{m}_{i}\right) \left(-\sigma _2+\sigma
		_3-\sigma _6-\tilde{m}_{i}\right) \left(\sigma _1-\sigma _2+\sigma _5-\sigma
		_6-\tilde{m}_{i}\right) \\
 \nonumber \cdot &  \left(\sigma _1-\sigma _2+\sigma _6-\tilde{m}_{i}\right)
		\left(-\sigma _2+\sigma _3-\sigma _5+\sigma _6-\tilde{m}_{i}\right)
		\left(-\sigma _2+\sigma _3-\sigma _7-\tilde{m}_{i}\right) \\
 \nonumber \cdot & \left(\sigma
		_1-\sigma _2+\sigma _3-\sigma _7-\tilde{m}_{i}\right) \left(-\sigma
		_2+\sigma _7-\tilde{m}_{i}\right) \left(\sigma _1-\sigma _2+\sigma
		_7-\tilde{m}_{i}\right)\\
 \nonumber &  =\prod_{i=1}^{n}  \left(\sigma _2-\sigma _4-\tilde{m}_{i}\right) \left(-\sigma
		_1+\sigma _2-\sigma _3+\sigma _4-\tilde{m}_{i}\right) \left(\sigma _2-\sigma
		_3+\sigma _4-\sigma _5-\tilde{m}_{i}\right) \\
 \nonumber & \hspace*{0.25in} \cdot   \left(-\sigma _1+\sigma
		_2-\sigma _4+\sigma _5-\tilde{m}_{i}\right) \left(-\sigma _1+\sigma
		_2-\sigma _6-\tilde{m}_{i}\right) \left(\sigma _2-\sigma _3+\sigma _5-\sigma
		_6-\tilde{m}_{i}\right) \\
 \nonumber & \hspace*{0.25in} \cdot  \left(\sigma _2-\sigma _3+\sigma _6-\tilde{m}_{i}\right)
		\left(-\sigma _1+\sigma _2-\sigma _5+\sigma _6-\tilde{m}_{i}\right)
		\left(\sigma _2-\sigma _7-\tilde{m}_{i}\right) \\
 & \hspace*{0.25in} \cdot  \left(-\sigma _1+\sigma
		_2-\sigma _7-\tilde{m}_{i}\right) \left(\sigma _2-\sigma _3+\sigma
		_7-\tilde{m}_{i}\right) \left(-\sigma _1+\sigma _2-\sigma _3+\sigma
		_7-\tilde{m}_{i}\right),
\end{align}
\begin{align} \nonumber
\prod_{i=1}^{n} & \left(-\sigma _3+\sigma _4-\tilde{m}_{i}\right) \left(\sigma _1-\sigma
		_3+\sigma _4-\tilde{m}_{i}\right) \left(-\sigma _1+\sigma _2-\sigma
		_3+\sigma _4-\tilde{m}_{i}\right) \\
 \nonumber \cdot &  \left(\sigma _2-\sigma _3+\sigma _4-\sigma
		_5-\tilde{m}_{i}\right) \left(\sigma _2-\sigma _3+\sigma _5-\sigma
		_6-\tilde{m}_{i}\right) \left(\sigma _2-\sigma _3+\sigma _6-\tilde{m}_{i}\right) \\
 \nonumber \cdot & 
		\left(\sigma _1-\sigma _3+\sigma _7-\tilde{m}_{i}\right) \left(\sigma
		_2-\sigma _3+\sigma _7-\tilde{m}_{i}\right) \left(-\sigma _1+\sigma
		_2-\sigma _3+\sigma _7-\tilde{m}_{i}\right) \\
 \nonumber \cdot &  \left(-\sigma _3+\sigma
		_5+\sigma _7-\tilde{m}_{i}\right) \left(-\sigma _3+\sigma _4-\sigma
		_6+\sigma _7-\tilde{m}_{i}\right) \left(-\sigma _3+\sigma _4-\sigma
		_5+\sigma _6+\sigma _7-\tilde{m}_{i}\right)\\
&  =\nonumber \prod_{i=1}^{n}   \left(\sigma _3-\sigma
		_4-\tilde{m}_{i}\right) \left(-\sigma _1+\sigma _3-\sigma _4-\tilde{m}_{i}\right)
		\left(\sigma _1-\sigma _2+\sigma _3-\sigma _4-\tilde{m}_{i}\right) \\
 \nonumber & \hspace*{0.25in} \cdot 
		\left(-\sigma _2+\sigma _3-\sigma _4+\sigma _5-\tilde{m}_{i}\right)
		\left(-\sigma _2+\sigma _3-\sigma _6-\tilde{m}_{i}\right) \left(-\sigma
		_2+\sigma _3-\sigma _5+\sigma _6-\tilde{m}_{i}\right) \\
 \nonumber & \hspace*{0.25in} \cdot  \left(-\sigma
		_1+\sigma _3-\sigma _7-\tilde{m}_{i}\right) \left(-\sigma _2+\sigma
		_3-\sigma _7-\tilde{m}_{i}\right) \left(\sigma _1-\sigma _2+\sigma _3-\sigma
		_7-\tilde{m}_{i}\right) \\ 
\nonumber  & \hspace*{0.25in}\cdot  \left(\sigma _3-\sigma _5-\sigma _7-\tilde{m}_{i}\right)
		\left(\sigma _3-\sigma _4+\sigma _5-\sigma _6-\sigma _7-\tilde{m}_{i}\right)
	\\
& \hspace*{0.25in} \cdot	\left(\sigma _3-\sigma _4+\sigma _6-\sigma _7-\tilde{m}_{i}\right),
\end{align}
\begin{align} \nonumber
\prod_{i=1}^{n} &
	\left(\sigma _2-\sigma _4-\tilde{m}_{i}\right) \left(\sigma _3-\sigma
		_4-\tilde{m}_{i}\right) \left(-\sigma _1+\sigma _3-\sigma _4-\tilde{m}_{i}\right) \\
 \nonumber \cdot &
		\left(\sigma _1-\sigma _2+\sigma _3-\sigma _4-\tilde{m}_{i}\right)
		\left(-\sigma _4+\sigma _5-\tilde{m}_{i}\right) \left(\sigma _1-\sigma
		_4+\sigma _5-\tilde{m}_{i}\right) \\
 \nonumber \cdot & \left(-\sigma _1+\sigma _2-\sigma
		_4+\sigma _5-\tilde{m}_{i}\right) \left(-\sigma _2+\sigma _3-\sigma
		_4+\sigma _5-\tilde{m}_{i}\right) \left(\sigma _3-\sigma _4+\sigma _5-\sigma
		_6-\sigma _7-\tilde{m}_{i}\right) \\
 \nonumber \cdot & \left(\sigma _3-\sigma _4+\sigma _6-\sigma
		_7-\tilde{m}_{i}\right) \left(-\sigma _4+\sigma _5-\sigma _6+\sigma
		_7-\tilde{m}_{i}\right) \left(-\sigma _4+\sigma _6+\sigma
		_7-\tilde{m}_{i}\right)\\
 \nonumber & =\prod_{i=1}^{n}  \left(-\sigma _2+\sigma _4-\tilde{m}_{i}\right) \left(-\sigma
		_3+\sigma _4-\tilde{m}_{i}\right) \left(\sigma _1-\sigma _3+\sigma
		_4-\tilde{m}_{i}\right) \\
 \nonumber & \hspace*{0.25in} \cdot  \left(-\sigma _1+\sigma _2-\sigma _3+\sigma
		_4-\tilde{m}_{i}\right) \left(\sigma _4-\sigma _5-\tilde{m}_{i}\right) \left(-\sigma
		_1+\sigma _4-\sigma _5-\tilde{m}_{i}\right) \\
 \nonumber & \hspace*{0.25in} \cdot  \left(\sigma _1-\sigma _2+\sigma
		_4-\sigma _5-\tilde{m}_{i}\right) \left(\sigma _2-\sigma _3+\sigma _4-\sigma
		_5-\tilde{m}_{i}\right) \left(\sigma _4-\sigma _6-\sigma _7-\tilde{m}_{i}\right) \\
 \nonumber & \hspace*{0.25in}  \cdot 
		\left(\sigma _4-\sigma _5+\sigma _6-\sigma _7-\tilde{m}_{i}\right)
		\left(-\sigma _3+\sigma _4-\sigma _6+\sigma _7-\tilde{m}_{i}\right)
\\
&  \hspace*{0.25in}	\cdot	\left(-\sigma _3+\sigma _4-\sigma _5+\sigma _6+\sigma
		_7-\tilde{m}_{i}\right), 
\end{align}
\begin{align}
 \nonumber
 \prod_{i=1}^{n} &	\left(\sigma _4-\sigma _5-\tilde{m}_{i}\right) \left(-\sigma _1+\sigma
		_4-\sigma _5-\tilde{m}_{i}\right) \left(\sigma _1-\sigma _2+\sigma _4-\sigma
		_5-\tilde{m}_{i}\right) \\
 \nonumber \cdot & \left(\sigma _2-\sigma _3+\sigma _4-\sigma
		_5-\tilde{m}_{i}\right) \left(-\sigma _5+\sigma _6-\tilde{m}_{i}\right) \left(\sigma
		_1-\sigma _5+\sigma _6-\tilde{m}_{i}\right) \\
 \nonumber \cdot & \left(-\sigma _1+\sigma
		_2-\sigma _5+\sigma _6-\tilde{m}_{i}\right) \left(-\sigma _2+\sigma
		_3-\sigma _5+\sigma _6-\tilde{m}_{i}\right) \left(\sigma _3-\sigma _5-\sigma
		_7-\tilde{m}_{i}\right) \\
 \nonumber \cdot & \left(\sigma _4-\sigma _5+\sigma _6-\sigma
		_7-\tilde{m}_{i}\right) \left(-\sigma _5+\sigma _7-\tilde{m}_{i}\right) \left(-\sigma
		_3+\sigma _4-\sigma _5+\sigma _6+\sigma
		_7-\tilde{m}_{i}\right) \\
 \nonumber &  = \prod_{i=1}^{n}  \left(-\sigma _4+\sigma _5-\tilde{m}_{i}\right) \left(\sigma
		_1-\sigma _4+\sigma _5-\tilde{m}_{i}\right) \left(-\sigma _1+\sigma
		_2-\sigma _4+\sigma _5-\tilde{m}_{i}\right) \\
 \nonumber & \hspace*{0.25in} \cdot  \left(-\sigma _2+\sigma
		_3-\sigma _4+\sigma _5-\tilde{m}_{i}\right) \left(\sigma _5-\sigma
		_6-\tilde{m}_{i}\right) \left(-\sigma _1+\sigma _5-\sigma _6-\tilde{m}_{i}\right)  \\
 \nonumber & \hspace*{0.25in} \cdot 
		\left(\sigma _1-\sigma _2+\sigma _5-\sigma _6-\tilde{m}_{i}\right)
		\left(\sigma _2-\sigma _3+\sigma _5-\sigma _6-\tilde{m}_{i}\right)
		\left(\sigma _5-\sigma _7-\tilde{m}_{i}\right) \\
 \nonumber & \hspace*{0.25in} \cdot  \left(\sigma _3-\sigma
		_4+\sigma _5-\sigma _6-\sigma _7-\tilde{m}_{i}\right) \left(-\sigma
		_3+\sigma _5+\sigma _7-\tilde{m}_{i}\right)
\\
& \hspace*{0.25in} \cdot \left(-\sigma _4+\sigma
		_5-\sigma _6+\sigma _7-\tilde{m}_{i}\right), 
\end{align}
\begin{align}  \nonumber
\prod_{i=1}^{n} &
	\left(-\sigma _6-\tilde{m}_{i}\right) \left(\sigma _1-\sigma _6-\tilde{m}_{i}\right)
		\left(-\sigma _1+\sigma _2-\sigma _6-\tilde{m}_{i}\right) \\
 \nonumber \cdot & \left(-\sigma
		_2+\sigma _3-\sigma _6-\tilde{m}_{i}\right) \left(\sigma _5-\sigma
		_6-\tilde{m}_{i}\right) \left(-\sigma _1+\sigma _5-\sigma _6-\tilde{m}_{i}\right) \\
 \nonumber \cdot &
		\left(\sigma _1-\sigma _2+\sigma _5-\sigma _6-\tilde{m}_{i}\right)
		\left(\sigma _2-\sigma _3+\sigma _5-\sigma _6-\tilde{m}_{i}\right)
		\left(\sigma _4-\sigma _6-\sigma _7-\tilde{m}_{i}\right) \\
 \nonumber \cdot & \left(\sigma
		_3-\sigma _4+\sigma _5-\sigma _6-\sigma _7-\tilde{m}_{i}\right)
		\left(-\sigma _3+\sigma _4-\sigma _6+\sigma _7-\tilde{m}_{i}\right)
		\left(-\sigma _4+\sigma _5-\sigma _6+\sigma
		_7-\tilde{m}_{i}\right) \\
 \nonumber & = \prod_{i=1}^{n}  \left(\sigma _6-\tilde{m}_{i}\right) \left(-\sigma _1+\sigma
		_6-\tilde{m}_{i}\right) \left(\sigma _1-\sigma _2+\sigma _6-\tilde{m}_{i}\right)  \\
 \nonumber & \hspace*{0.25in} \cdot 
		\left(\sigma _2-\sigma _3+\sigma _6-\tilde{m}_{i}\right) \left(-\sigma
		_5+\sigma _6-\tilde{m}_{i}\right) \left(\sigma _1-\sigma _5+\sigma
		_6-\tilde{m}_{i}\right) \\
 \nonumber & \hspace*{0.25in} \cdot  \left(-\sigma _1+\sigma _2-\sigma _5+\sigma
		_6-\tilde{m}_{i}\right) \left(-\sigma _2+\sigma _3-\sigma _5+\sigma
		_6-\tilde{m}_{i}\right) \left(\sigma _3-\sigma _4+\sigma _6-\sigma
		_7-\tilde{m}_{i}\right) \\
 \nonumber & \hspace*{0.25in} \cdot  \left(\sigma _4-\sigma _5+\sigma _6-\sigma
		_7-\tilde{m}_{i}\right) \left(-\sigma _4+\sigma _6+\sigma _7-\tilde{m}_{i}\right)
\\
& \hspace*{0.25in} \cdot \left(-\sigma _3+\sigma _4-\sigma _5+\sigma _6+\sigma
		_7-\tilde{m}_{i}\right), 
\end{align}
\begin{align} 
\nonumber
\prod_{i=1}^{n} &
	\left(\sigma _1-\sigma _7-\tilde{m}_{i}\right) \left(\sigma _2-\sigma
		_7-\tilde{m}_{i}\right) \left(-\sigma _1+\sigma _2-\sigma _7-\tilde{m}_{i}\right) \\
 \nonumber \cdot &
		\left(-\sigma _1+\sigma _3-\sigma _7-\tilde{m}_{i}\right) \left(-\sigma
		_2+\sigma _3-\sigma _7-\tilde{m}_{i}\right) \left(\sigma _1-\sigma _2+\sigma
		_3-\sigma _7-\tilde{m}_{i}\right) \\
 \nonumber \cdot & \left(\sigma _3-\sigma _5-\sigma
		_7-\tilde{m}_{i}\right) \left(\sigma _5-\sigma _7-\tilde{m}_{i}\right) \left(\sigma
		_4-\sigma _6-\sigma _7-\tilde{m}_{i}\right) \\
 \nonumber \cdot & \left(\sigma _3-\sigma _4+\sigma
		_5-\sigma _6-\sigma _7-\tilde{m}_{i}\right) \left(\sigma _3-\sigma _4+\sigma
		_6-\sigma _7-\tilde{m}_{i}\right) \left(\sigma _4-\sigma _5+\sigma _6-\sigma
		_7-\tilde{m}_{i}\right) \\ 
\nonumber & = \prod_{i=1}^{n}  \left(-\sigma _1+\sigma _7-\tilde{m}_{i}\right) \left(-\sigma
		_2+\sigma _7-\tilde{m}_{i}\right) \left(\sigma _1-\sigma _2+\sigma
		_7-\tilde{m}_{i}\right) \\
 \nonumber & \hspace*{0.25in} \cdot  \left(\sigma _1-\sigma _3+\sigma _7-\tilde{m}_{i}\right)
		\left(\sigma _2-\sigma _3+\sigma _7-\tilde{m}_{i}\right) \left(-\sigma
		_1+\sigma _2-\sigma _3+\sigma _7-\tilde{m}_{i}\right) \\
 \nonumber & \hspace*{0.25in} \cdot  \left(-\sigma
		_5+\sigma _7-\tilde{m}_{i}\right) \left(-\sigma _3+\sigma _5+\sigma
		_7-\tilde{m}_{i}\right) \left(-\sigma _3+\sigma _4-\sigma _6+\sigma
		_7-\tilde{m}_{i}\right) \\
 \nonumber & \hspace*{0.25in} \cdot  \left(-\sigma _4+\sigma _5-\sigma _6+\sigma
		_7-\tilde{m}_{i}\right) \left(-\sigma _4+\sigma _6+\sigma _7-\tilde{m}_{i}\right)
\\
& \hspace*{0.25in} \cdot\left(-\sigma _3+\sigma _4-\sigma _5+\sigma _6+\sigma
		_7-\tilde{m}_{i}\right).
\end{align}
These equations define the relations in the Coulomb ring for this gauge
theory.

Part of the excluded locus is defined by the condition that the
$X_m \neq 0$.  This part of the excluded locus is encoded by
\begin{align} \nonumber
	&\sigma _1 \left(2 \sigma _1-\sigma _2\right) \left(\sigma _2-\sigma
	_1\right) \left(\sigma _1+\sigma _2-\sigma _3\right)
	\left(-\sigma _1+2 \sigma _2-\sigma _3\right) \left(\sigma
	_3-\sigma _2\right) \left(\sigma _2-\sigma _5\right)
	\left(-\sigma _1+\sigma _3-\sigma _5\right) \\ \nonumber \cdot & \left(\sigma
	_1-\sigma _2+\sigma _3-\sigma _5\right) \left(-\sigma _3+2
	\sigma _4-\sigma _5\right) \left(\sigma _5-\sigma _1\right)
	\left(\sigma _1-\sigma _2+\sigma _5\right) \left(\sigma
	_2-\sigma _3+\sigma _5\right) \\ \nonumber \cdot & \left(-\sigma _1+\sigma _4-\sigma
	_6\right)  \left(\sigma _1-\sigma _2+\sigma _4-\sigma _6\right)
	\left(\sigma _2-\sigma _3+\sigma _4-\sigma _6\right)
	\left(\sigma _2-\sigma _4+\sigma _5-\sigma _6\right)  \\ \nonumber \cdot &
	\left(-\sigma _1+\sigma _3-\sigma _4+\sigma _5-\sigma _6\right)
	\left(\sigma _1-\sigma _2+\sigma _3-\sigma _4+\sigma _5-\sigma
	_6\right) \left(-\sigma _3+\sigma _4+\sigma _5-\sigma _6\right) \\ \nonumber \cdot &
	\left(-\sigma _4+2 \sigma _5-\sigma _6\right) \left(\sigma
	_2-\sigma _4+\sigma _6\right)  \left(-\sigma _1+\sigma _3-\sigma
	_4+\sigma _6\right) \left(\sigma _1-\sigma _2+\sigma _3-\sigma
	_4+\sigma _6\right)\\ \nonumber \cdot & \left(-\sigma _3+\sigma _4+\sigma _6\right)
	\left(-\sigma _1+\sigma _4-\sigma _5+\sigma _6\right) 
	\left(\sigma _1-\sigma _2+\sigma _4-\sigma _5+\sigma _6\right)
	\left(\sigma _2-\sigma _3+\sigma _4-\sigma _5+\sigma _6\right) \\ \nonumber \cdot &
	\left(-\sigma _4+\sigma _5+\sigma _6\right) \left(2 \sigma
	_6-\sigma _5\right)  \left(\sigma _1+\sigma _3-\sigma _4-\sigma
	_7\right) \left(-\sigma _1+\sigma _2+\sigma _3-\sigma _4-\sigma
	_7\right) \\ \nonumber \cdot &  \left(-\sigma _2+2 \sigma _3-\sigma _4-\sigma _7\right)
	\left(\sigma _4-\sigma _7\right)\left(\sigma _1+\sigma
	_4-\sigma _5-\sigma _7\right) \left(-\sigma _1+\sigma _2+\sigma
	_4-\sigma _5-\sigma _7\right)\\ \nonumber \cdot &  \left(-\sigma _2+\sigma _3+\sigma
	_4-\sigma _5-\sigma _7\right)\left(\sigma _3-\sigma _4+\sigma
	_5-\sigma _7\right) \left(\sigma _3-\sigma _6-\sigma _7\right)
	\left(\sigma _1+\sigma _5-\sigma _6-\sigma _7\right) \\ \nonumber \cdot &
	\left(-\sigma _1+\sigma _2+\sigma _5-\sigma _6-\sigma _7\right)
	\left(-\sigma _2+\sigma _3+\sigma _5-\sigma _6-\sigma _7\right)
	\left(\sigma _1+\sigma _6-\sigma _7\right) \left(-\sigma
	_1+\sigma _2+\sigma _6-\sigma _7\right) \\ \nonumber \cdot & \left(-\sigma _2+\sigma
	_3+\sigma _6-\sigma _7\right) \left(\sigma _3-\sigma _5+\sigma
	_6-\sigma _7\right) \left(\sigma _1-\sigma _4+\sigma _7\right)
	\left(-\sigma _1+\sigma _2-\sigma _4+\sigma _7\right) \\ \nonumber \cdot &
	\left(-\sigma _2+\sigma _3-\sigma _4+\sigma _7\right)
	\left(-\sigma _3+\sigma _4+\sigma _7\right) \left(-\sigma
	_2+\sigma _4-\sigma _5+\sigma _7\right) \left(\sigma _1-\sigma
	_3+\sigma _4-\sigma _5+\sigma _7\right) \\ \nonumber \cdot & \left(-\sigma _1+\sigma
	_2-\sigma _3+\sigma _4-\sigma _5+\sigma _7\right) \left(-\sigma
	_4+\sigma _5+\sigma _7\right) \left(\sigma _7-\sigma _6\right)
	\left(-\sigma _2+\sigma _5-\sigma _6+\sigma _7\right) \\ \nonumber \cdot &
	\left(\sigma _1-\sigma _3+\sigma _5-\sigma _6+\sigma _7\right)
	\left(-\sigma _1+\sigma _2-\sigma _3+\sigma _5-\sigma _6+\sigma
	_7\right) \left(-\sigma _2+\sigma _6+\sigma _7\right) \\ \nonumber \cdot &
	\left(\sigma _1-\sigma _3+\sigma _6+\sigma _7\right)
	\left(-\sigma _1+\sigma _2-\sigma _3+\sigma _6+\sigma _7\right)
	\left(-\sigma _5+\sigma _6+\sigma _7\right) \left(2 \sigma
	_7-\sigma _3\right) \: \neq \: 0 .
\end{align}
The other part of the excluded locus is determined by the fact that
$\exp(-Y) \neq 0$.  Since on the critical locus,
\begin{displaymath}
\exp(-Y_{i,\beta})=-\tilde{m}_i+\sum_{a=1}^{7}\sigma_a \rho^a_{i,\beta},
\end{displaymath}
so 
\begin{displaymath}
-\tilde{m}_i+\sum_{a=1}^{7}\sigma_a \rho^a_{i,\beta} \neq 0,
\end{displaymath} 
which is given more explicitly as  
\begin{align*}
\prod_{i=1}^{n} &\left(-\tilde{m}_i+\sigma _2-\sigma _4\right) \left(-\tilde{m}_i+\sigma _3-\sigma
_4\right) \left(-\tilde{m}_i-\sigma _1+\sigma _3-\sigma _4\right)
\left(-\tilde{m}_i+\sigma _1-\sigma _2+\sigma _3-\sigma _4\right) \\ \cdot &
\left(-\tilde{m}_i-\sigma _2+\sigma _4\right) \left(-\tilde{m}_i-\sigma _3+\sigma
_4\right) \left(-\tilde{m}_i+\sigma _1-\sigma _3+\sigma _4\right)
\left(-\tilde{m}_i-\sigma _1+\sigma _2-\sigma _3+\sigma _4\right)  \\ \cdot & 
 \left(-\tilde{m}_i-\sigma _1+\sigma
_4-\sigma _5\right) \left(-\tilde{m}_i+\sigma _1-\sigma _2+\sigma
_4-\sigma _5\right) \left(-\tilde{m}_i+\sigma _2-\sigma _3+\sigma
_4-\sigma _5\right) \\ \cdot & \left(-\tilde{m}_i+\sigma _4-\sigma _5\right) \left(-\tilde{m}_i-\sigma _4+\sigma _5\right)
\left(-\tilde{m}_i+\sigma _1-\sigma _4+\sigma _5\right) \left(-\tilde{m}_i-\sigma
_1+\sigma _2-\sigma _4+\sigma _5\right)  \\ \cdot & \left(-\tilde{m}_i-\sigma
_2+\sigma _3-\sigma _4+\sigma _5\right) \left(-\tilde{m}_i-\sigma _6\right)
\left(-\tilde{m}_i+\sigma _1-\sigma _6\right) \left(-\tilde{m}_i-\sigma _1+\sigma
_2-\sigma _6\right)  \\ \cdot & \left(-\tilde{m}_i-\sigma _2+\sigma _3-\sigma _6\right)
 \left(-\tilde{m}_i-\sigma _1+\sigma
_5-\sigma _6\right) \left(-\tilde{m}_i+\sigma _1-\sigma _2+\sigma
_5-\sigma _6\right) \\ \cdot &  \left(-\tilde{m}_i+\sigma _5-\sigma _6\right) \left(-\tilde{m}_i+\sigma _2-\sigma _3+\sigma
_5-\sigma _6\right) \left(\sigma _6-\tilde{m}_i\right) \left(-\tilde{m}_i-\sigma
_1+\sigma _6\right) \\ \cdot &  \left(-\tilde{m}_i+\sigma _1-\sigma _2+\sigma _6\right)
\left(-\tilde{m}_i+\sigma _2-\sigma _3+\sigma _6\right) \left(-\tilde{m}_i-\sigma
_5+\sigma _6\right) \left(-\tilde{m}_i+\sigma _1-\sigma _5+\sigma _6\right) \\ \cdot & 
\left(-\tilde{m}_i-\sigma _1+\sigma _2-\sigma _5+\sigma _6\right)
\left(-\tilde{m}_i-\sigma _2+\sigma _3-\sigma _5+\sigma _6\right)
\left(-\tilde{m}_i+\sigma _1-\sigma _7\right)  \left(-\tilde{m}_i+\sigma _2-\sigma
_7\right) \\ \cdot &   \left(-\tilde{m}_i-\sigma _1+\sigma _2-\sigma _7\right)
\left(-\tilde{m}_i-\sigma _1+\sigma _3-\sigma _7\right) \left(-\tilde{m}_i-\sigma
_2+\sigma _3-\sigma _7\right) \\ \cdot &  \left(-\tilde{m}_i+\sigma _1-\sigma
_2+\sigma _3-\sigma _7\right)  \left(-\tilde{m}_i+\sigma _3-\sigma
_5-\sigma _7\right) \left(-\tilde{m}_i+\sigma _5-\sigma _7\right)
\left(-\tilde{m}_i+\sigma _4-\sigma _6-\sigma _7\right) \\ \cdot &  \left(-\tilde{m}_i+\sigma
_3-\sigma _4+\sigma _5-\sigma _6-\sigma _7\right)
\left(-\tilde{m}_i+\sigma _3-\sigma _4+\sigma _6-\sigma _7\right)
\left(-\tilde{m}_i+\sigma _4-\sigma _5+\sigma _6-\sigma _7\right) \\ \cdot & 
\left(-\tilde{m}_i-\sigma _1+\sigma _7\right) \left(-\tilde{m}_i-\sigma _2+\sigma
_7\right) \left(-\tilde{m}_i+\sigma _1-\sigma _2+\sigma _7\right)
\left(-\tilde{m}_i+\sigma _1-\sigma _3+\sigma _7\right) \\ \cdot &  \left(-\tilde{m}_i+\sigma
_2-\sigma _3+\sigma _7\right) \left(-\tilde{m}_i-\sigma _1+\sigma
_2-\sigma _3+\sigma _7\right) \left(-\tilde{m}_i-\sigma _5+\sigma _7\right)
\left(-\tilde{m}_i-\sigma _3+\sigma _5+\sigma _7\right) \\ \cdot &  \left(-\tilde{m}_i-\sigma
_3+\sigma _4-\sigma _6+\sigma _7\right) \left(-\tilde{m}_i-\sigma
_4+\sigma _5-\sigma _6+\sigma _7\right) \\ \cdot &  \left(-\tilde{m}_i-\sigma
_4+\sigma _6+\sigma _7\right) \left(-\tilde{m}_i-\sigma _3+\sigma
_4-\sigma _5+\sigma _6+\sigma _7\right) \: \neq \: 0 .
\end{align*}

\subsection{Pure gauge theory}

In this part we will consider the mirror to the pure $E_7$ gauge theory.
The mirror superpotential is 
\begin{align*}
W=& \sigma_1 \Big(Z_1-Z_2+Z_{14}-Z_{15}+Z_{18}-Z_{19}+Z_{21}+Z_{22}-Z
_{23}-Z_{24}+Z_{26}+Z_{27}-Z_{28}-Z_{29}\\&  \hspace*{.27in } +Z_{30}+Z_{31}+Z_{32}-Z_{33}-
Z_{34}-Z_{35}+Z_{37}+Z_{38}-Z_{39}-Z_{40}+Z_{43}+Z_{44}-Z_{45}\\&  \hspace*{.27in }-Z_{46}
+Z_{50}-Z_{51}+Z_{56}-Z_{57}+2
Z_{63}-Z_{64}+Z_{65}-Z_{77}+Z_{78}-Z_{81}+Z_{82}-Z_{84}\\&  \hspace*{.27in } -Z_{85}+Z_{86}
+Z_{87}-Z_{89}-Z_{90}+Z_{91}+Z_{92}-Z_{93}-Z_{94}-Z_{95}+Z_{96}+Z_{97
}+Z_{98}\\&  \hspace*{.27in }-Z_{100}-Z_{101}+Z_{102}+Z_{103}-Z_{106}-Z_{107}+Z_{108}+Z_{1
	09}-Z_{113}+Z_{114}-Z_{119}\\&  \hspace*{.27in }+Z_{120}-2 Z_{126}  \Big)
\end{align*}
\begin{align*}
& +\sigma_2 \Big(Z_2-Z_3+Z_{11}+Z_{13}-Z_{14}+Z_{16}+Z_{17}-Z_{18}+Z_{2
	0}-Z_{21}-Z_{22}+Z_{25}-Z_{26}+Z_{29}\\ &  \hspace*{.4in }  -Z_{30}+Z_{34}+Z_{35}-Z_{36}+Z_{
	39}+Z_{40}-Z_{41}-Z_{42}+Z_{45}+Z_{46}-Z_{47}-Z_{48}+Z_{51}\\ &  \hspace*{.4in }-Z_{52}-Z_
{53}+Z_{56} +2
Z_{57}-Z_{58}-Z_{63}-Z_{65}+Z_{66}-Z_{74}-Z_{76}+Z_{77}-Z_{79}-Z_{80} \\ &  \hspace*{.4in }
+Z_{81}-Z_{83}+Z_{84}+Z_{85}-Z_{88}  +Z_{89}-Z_{92}+Z_{93}-Z_{97}-Z_{98
}+Z_{99}-Z_{102}-Z_{103}\\ &  \hspace*{.4in }+Z_{104}+Z_{105}-Z_{108} -Z_{109}+Z_{110}+Z_{1
	11}-Z_{114}+Z_{115}+Z_{116}-Z_{119}-2 Z_{120}\\ &  \hspace*{.4in }+Z_{121}+Z_{126}   \Big)
\end{align*}
\begin{align*}
& +\sigma_3 \Big(Z_3-Z_4+Z_8+Z_{10}-Z_{11}+Z_{12}-Z_{13}-Z_{16}+Z_{2
	2}+Z_{24}+Z_{26}-Z_{27}+Z_{28}-Z_{29}\\ & \hspace*{.4 in} +Z_{30}-Z_{31}+Z_{33}-Z_{34}-Z_{
	37}-Z_{39}+Z_{42}+Z_{48}-Z_{49}+Z_{50}+Z_{51}+Z_{52}+Z_{53}\\ & \hspace*{.4 in} -Z_{54} -Z_
{56}-Z_{57} +2
Z_{58}-Z_{59}-Z_{60}-Z_{66}+Z_{67}-Z_{71}-Z_{73}+Z_{74}-Z_{75}+Z_{76}\\ & \hspace*{.4 in}
+Z_{79}-Z_{85} -Z_{87}-Z_{89}+Z_{90}-Z_{91}+Z_{92}-Z_{93}+Z_{94}-Z_{96
}+Z_{97}+Z_{100}+Z_{102}\\ & \hspace*{.4 in}-Z_{105}-Z_{111}+Z_{112}-Z_{113}-Z_{114}-Z_{1
	15}-Z_{116}+Z_{117}+Z_{119}+Z_{120}-2 Z_{121}\\ & \hspace*{.4 in}+Z_{122}+Z_{123} \Big)
\end{align*}
\begin{align*}
& +\sigma_4 \Big(Z_4-Z_5+Z_6-Z_8+Z_{13}+Z_{16}-Z_{17}+Z_{18}+Z_{1
	9}-Z_{20}+Z_{21}-Z_{22}+Z_{23}-Z_{24}\\& \hspace*{.4in}-Z_{26} -Z_{28}+Z_{37}+Z_{39}-Z_{
	43}+Z_{44}-Z_{45}+Z_{46}+Z_{47}+Z_{49}-Z_{50}-Z_{51}-Z_{52}\\& \hspace*{.4in}+Z_{53}+Z_
{54}-Z_{55}-Z_{58}+2
Z_{60}-Z_{61}-Z_{67}+Z_{68}-Z_{69}+Z_{71}-Z_{76}-Z_{79}+Z_{80}\\& \hspace*{.4in}-Z_{81}
-Z_{82}+Z_{83}-Z_{84}+Z_{85}-Z_{86}+Z_{87}+Z_{89}+Z_{91}-Z_{100}-Z_{1
	02}+Z_{106}-Z_{107}\\& \hspace*{.4in}+Z_{108}-Z_{109}-Z_{110}-Z_{112}+Z_{113}+Z_{114}+Z
_{115}-Z_{116}-Z_{117}+Z_{118}+Z_{121}\\& \hspace*{.4in}-2 Z_{123}+Z_{124} \Big)
\end{align*}
\begin{align*}
& +\sigma_5 \Big(Z_5-Z_7+Z_8-Z_{10}+Z_{11}-Z_{13}+Z_{14}+Z_{15}-Z
_{18}-Z_{19}+Z_{20}-Z_{25}+Z_{26}+Z_{28}\\& \hspace*{.4in}-Z_{30}+Z_{31}-Z_{33}+Z_{34}-
Z_{37}+Z_{38}-Z_{39}+Z_{40}+Z_{41}-Z_{44}-Z_{46}-Z_{47}+Z_{48}\\& \hspace*{.4in}-Z_{53} 
+Z_{54}+Z_{55}-Z_{60}+2
Z_{61}-Z_{62}-Z_{68}+Z_{70}-Z_{71}+Z_{73}-Z_{74}+Z_{76}-Z_{77}\\& \hspace*{.4in}-Z_{78}
+Z_{81}+Z_{82}-Z_{83}+Z_{88}-Z_{89}-Z_{91}+Z_{93}-Z_{94}+Z_{96}-Z_{97
}+Z_{100}-Z_{101}\\& \hspace*{.4in}+Z_{102}-Z_{103}-Z_{104}+Z_{107}+Z_{109}+Z_{110}-Z_{
	111}+Z_{116}-Z_{117}-Z_{118}+Z_{123}\\& \hspace*{.4in}-2 Z_{124}+Z_{125}  \Big)
\end{align*}
\begin{align*}
& +\sigma_6 \Big(Z_7-Z_9+Z_{10}-Z_{12}+Z_{13}-Z_{16}+Z_{17}+Z_{18
}+Z_{19}-Z_{20}-Z_{21}+Z_{22}-Z_{23}+Z_{24} \\& \hspace*{.4in} -Z_{26}+Z_{27}-Z_{28}+Z_{2
	9}-Z_{31}+Z_{32}-Z_{34}+Z_{35}+Z_{36}-Z_{38}-Z_{40}-Z_{41}+Z_{42}\\& \hspace*{.4in}-Z_{
	48}  +Z_{49}-Z_{54}+Z_{55}-Z_{61}+2
Z_{62}-Z_{70}+Z_{72}-Z_{73}+Z_{75}-Z_{76}+Z_{79}-Z_{80}\\& \hspace*{.4in}-Z_{81}-Z_{82}
+Z_{83}+Z_{84}-Z_{85}+Z_{86}-Z_{87}+Z_{89}-Z_{90}+Z_{91}-Z_{92}+Z_{94
}-Z_{95}\\& \hspace*{.4in}+Z_{97}-Z_{98}-Z_{99}+Z_{101}+Z_{103}+Z_{104}-Z_{105}+Z_{111}
-Z_{112}+Z_{117}-Z_{118}+Z_{124}\\& \hspace*{.4in}-2 Z_{125} \Big)
\end{align*}
\begin{align*}
& +\sigma_7 \Big(Z_5-Z_7+Z_8-Z_{10}+Z_{11}-Z_{13}+Z_{14}+Z_{15}-Z
_{18}-Z_{19}+Z_{20}-Z_{25}+Z_{26}+Z_{28}\\& \hspace*{.4in} -Z_{30}+Z_{31}-Z_{33}+Z_{34}-
Z_{37}+Z_{38}-Z_{39}+Z_{40}+Z_{41}-Z_{44}-Z_{46}-Z_{47}+Z_{48} \\& \hspace*{.4in}-Z_{53}
+Z_{54}+Z_{55}-Z_{60}+2
Z_{61}-Z_{62}-Z_{68}+Z_{70}-Z_{71}+Z_{73}-Z_{74}+Z_{76}-Z_{77}\\& \hspace*{.4in}-Z_{78}
+Z_{81} +Z_{82}-Z_{83}+Z_{88}-Z_{89}-Z_{91}+Z_{93}-Z_{94}+Z_{96}-Z_{97
}+Z_{100}-Z_{101}\\& \hspace*{.4in}+Z_{102}-Z_{103} -Z_{104}+Z_{107}+Z_{109}+Z_{110}-Z_{
	111}+Z_{116}-Z_{117}-Z_{118}+Z_{123}\\& \hspace*{.4in}-2 Z_{124}+Z_{125} \Big)
\end{align*}
\begin{align}
+\sum_{m=1}^{126}X_m   .
\end{align}

Now, we can proceed as in previous sections.
For the reasons discussed there, since each $\sigma$ is multiplied by
both $Z_{\mu}$ and $Z_{-\mu}$ with opposite signs, the critical
locus equations
\begin{displaymath}
\frac{\partial W}{\partial X_{\mu} } \: = \: 0
\end{displaymath}
imply that on the critical locus,
\begin{equation}
X_{\mu} \: = \: - X_{- \mu}.
\end{equation}
(Furthermore, on the critical locus, each $X_{\mu}$ is determined by
$\sigma$s.) 
In addition, each derivative
\begin{displaymath}
\frac{\partial W}{\partial \sigma_a}
\end{displaymath}
is a product of ratios of the form
\begin{displaymath}
\frac{ X_{\mu} }{ X_{-\mu} } \: = \: -1.
\end{displaymath}
It is straightforward to check 
in the superpotential above that
each $\sigma_a$ is multiplied by an even number of such ratios
({\it i.e.} the number of $Z$'s is a multiple of four).
Specifically, the sum of the absolute values of the $Z$'s multiplying
each $\sigma$ is $68=4 \cdot 17$.
Thus, the constraint implied by the $\sigma$'s is automatically satisfied.

As a result, following the same analysis in \cite{GuSharpe},
we see in this case, that the critical locus is nonempty, and in fact
is determined by the seven $\sigma$s.  In other words, at the level of these
topological field theory computations, we have evidence that the pure
supersymmetric $E_7$ gauge theory in two dimensions flows in the IR
to a theory of seven free twisted chiral superfields.

\section{$E_8$} \label{section6}

In this section, we will discuss the mirror theory to 
a two-dimensional $_8$ gauge theory. 
The group $E_8$ and its algebra are the largest and most complicated 
exceptional groups, we shall only list results.

\subsection{Mirror Landau-Ginzburg orbifold}

We will consider an $E_8$ gauge theory with $n$ matter fields in the
${\bf 248}$, the lowest-dimensional representation, which also happens to
be the adjoint representation.  The mirror Landau-Ginzburg model has
fields
\begin{itemize}
	\item $Y_{i\alpha}$, $i \in \{1, \cdots, n \}$, $\alpha \in \{1, \cdots, 248 \}$
	\item $X_m$, $m \in \{1,2, \cdots, 120 \}$, correponding to  
positive roots, and $X_{120+m}$, associated with the negative roots of
those associated to $X_m$,
	\item $\sigma_a$, $a \in \{ 1, 2, \cdots, 8 \}$.
\end{itemize}
As before, we work with an integer-lattice-basis for the roots and
weights, corresponding to standard theta angle periodicities.
We associate the roots and weights to fields as listed in the 
tables~\ref{table:e8:roots1}, \ref{table:e8:roots2}, and
\ref{table:e8:roots3}. 

\begin{table}[h!]
\centering
\begin{tabular}{cc|cc}
Field & Positive root/weight & Field & Positive root/weight  \\ \hline
$X_1$, $Y_{i,1}$,  &
$(0,0,0,0,0,0,1,0)$  &
$X_2$, $Y_{i,2}$  &
$(0,0,0,0,0,1,-1,0)$   \\
$X_3$, $Y_{i,3}$  &
$(0,0,0,0,1,-1,0,0)$  &
$X_4$, $Y_{i,4}$  &
$(0,0,0,1,-1,0,0,0)$  \\
$X_5$, $Y_{i,5} $ &
$(0,0,1,-1,0,0,0,0)$  &
$X_6$, $Y_{i,6}$  &
$(0,1,-1,0,0,0,0,1)$  \\
$X_7$, $Y_{i,7}$  &
$(0,1,0,0,0,0,0,-1)$  &
$X_8$, $Y_{i,8}$  &
$(1,-1,0,0,0,0,0,1)$  \\
$X_9$, $Y_{i,9}$  &
$(-1,0,0,0,0,0,0,1)$  &
$X_{10}$, $Y_{i,10}$  &
$(1,-1,1,0,0,0,0,-1)$  \\
$X_{11}$, $Y_{i,11}$  &
$(-1,0,1,0,0,0,0,-1)$  &
$X_{12}$, $Y_{i,12}$  &
$(1,0,-1,1,0,0,0,0)$  \\
$X_{13}$, $Y_{i,13}$  &
$(-1,1,-1,1,0,0,0,0)$  &
$X_{14}$, $Y_{i,14}$  &
$(1,0,0,-1,1,0,0,0)$  \\
$X_{15}$, $Y_{i,15}$  &
$(-1,1,0,-1,1,0,0,0)$  &
$X_{16}$, $Y_{i,16}$  &
$(0,-1,0,1,0,0,0,0)$  \\
$X_{17}$, $Y_{i,17}$  &
$(1,0,0,0,-1,1,0,0)$  &
$X_{18}$, $Y_{i,18}$  &
$(-1,1,0,0,-1,1,0,0)$, \\
$X_{19}$, $Y_{i,19}$  &
$(0,-1,1,-1,1,0,0,0)$  &
$X_{20}$, $Y_{i,20}$  &
$(1,0,0,0,0,-1,1,0)$  \\
$X_{21}$, $Y_{i,21}$  &
$(-1,1,0,0,0,-1,1,0)$  &
$X_{22}$, $Y_{i,22}$  &
$(0,-1,1,0,-1,1,0,0)$  \\
$X_{23}$, $Y_{i,23}$  &
$(0,0,-1,0,1,0,0,1)$  &
$X_{24}$, $Y_{i,24}$  &
$(1,0,0,0,0,0,-1,0)$  \\
$X_{25}$, $Y_{i,25}$  &
$(-1,1,0,0,0,0,-1,0)$  &
$X_{26}$, $Y_{i,26}$  &
$(0,-1,1,0,0,-1,1,0)$  \\
$X_{27}$, $Y_{i,27}$  &
$(0,0,-1,1,-1,1,0,1)$  &
$X_{28}$, $Y_{i,28}$  &
$(0,0,0,0,1,0,0,-1)$  \\
$X_{29}$, $Y_{i,29}$  &
$(0,-1,1,0,0,0,-1,0)$  &
$X_{30}$, $Y_{i,30}$  &
$(0,0,-1,1,0,-1,1,1)$  \\
$X_{31}$, $Y_{i,31}$  &
$(0,0,0,-1,0,1,0,1)$  &
$X_{32}$, $Y_{i,32}$  &
$(0,0,0,1,-1,1,0,-1)$  \\
$X_{33}$, $Y_{i,33}$  &
$(0,0,-1,1,0,0,-1,1)$  &
$X_{34}$, $Y_{i,34}$  &
$(0,0,0,-1,1,-1,1,1)$  \\
$X_{35}$, $Y_{i,35}$  &
$(0,0,0,1,0,-1,1,-1)$  &
$X_{36}$, $Y_{i,36}$  &
$(0,0,1,-1,0,1,0,-1)$  \\
$X_{37}$, $Y_{i,37}$  &
$(0,0,0,-1,1,0,-1,1)$  &
$X_{38}$, $Y_{i,38}$  &
$(0,0,0,0,-1,0,1,1)$  \\
$X_{39}$, $Y_{i,39}$  &
$(0,0,0,1,0,0,-1,-1)$  &
$X_{40}$, $Y_{i,40}$  &
$(0,0,1,-1,1,-1,1,-1)$  \\
$X_{41}$, $Y_{i,41}$  &
$(0,1,-1,0,0,1,0,0)$  &
$X_{42}$, $Y_{i,42}$  &
$(0,0,0,0,-1,1,-1,1)$  \\
$X_{43}$, $Y_{i,43}$  &
$(0,0,1,-1,1,0,-1,-1)$  &
$X_{44}$, $Y_{i,44}$  &
$(0,0,1,0,-1,0,1,-1)$  \\
$X_{45}$, $Y_{i,45}$  &
$(0,1,-1,0,1,-1,1,0)$  &
$X_{46}$, $Y_{i,46}$  &
$(1,-1,0,0,0,1,0,0)$  \\
$X_{47}$, $Y_{i,47}$  &
$(-1,0,0,0,0,1,0,0)$  &
$X_{48}$, $Y_{i,48}$  &
$(0,0,0,0,0,-1,0,1)$  
\end{tabular}
\caption{First set of roots of $E_8$ and associated fields. 
\label{table:e8:roots1} }
\end{table}

\begin{table}[h!]
\centering
\begin{tabular}{cc|cc}
Field & Positive root/weight & Field & Positive root/weight  \\ \hline
$X_{49}$, $Y_{i,49}$  &
$(0,0,1,0,-1,1,-1,-1)$  &
$X_{50}$, $Y_{i,50}$  &
$(0,1,-1,0,1,0,-1,0)$  \\
$X_{51}$, $Y_{i,51}$  &
$(0,1,-1,1,-1,0,1,0)$  &
$X_{52}$, $Y_{i,52}$  &
$(1,-1,0,0,1,-1,1,0)$  \\
$X_{53}$, $Y_{i,53}$  &
$(-1,0,0,0,1,-1,1,0)$  &
$X_{54}$, $Y_{i,54}$  &
$(0,0,1,0,0,-1,0,-1)$  \\
$X_{55}$, $Y_{i,55}$  &
$(0,1,-1,1,-1,1,-1,0)$  &
$X_{56}$, $Y_{i,56}$  &
$(0,1,0,-1,0,0,1,0)$  \\
$X_{57}$, $Y_{i,57}$  &
$(1,-1,0,0,1,0,-1,0)$  &
$X_{58}$, $Y_{i,58}$  &
$(1,-1,0,1,-1,0,1,0)$  \\
$X_{59}$, $Y_{i,59}$  &
$(-1,0,0,0,1,0,-1,0)$  &
$X_{60}$, $Y_{i,60}$  &
$(-1,0,0,1,-1,0,1,0)$  \\
$X_{61}$, $Y_{i,61}$  &
$(0,1,-1,1,0,-1,0,0)$  &
$X_{62}$, $Y_{i,62}$  &
$(0,1,0,-1,0,1,-1,0)$  \\
$X_{63}$, $Y_{i,63}$  &
$(1,-1,0,1,-1,1,-1,0)$  &
$X_{64}$, $Y_{i,64}$  &
$(1,-1,1,-1,0,0,1,0)$  \\
$X_{65}$, $Y_{i,65}$  &
$(-1,0,0,1,-1,1,-1,0)$  &
$X_{66}$, $Y_{i,66}$  &
$(-1,0,1,-1,0,0,1,0)$  \\
$X_{67}$, $Y_{i,67}$  &
$(0,1,0,-1,1,-1,0,0)$  &
$X_{68}$, $Y_{i,68}$  &
$(1,-1,0,1,0,-1,0,0)$  \\
$X_{69}$, $Y_{i,69}$  &
$(1,-1,1,-1,0,1,-1,0)$  &
$X_{70}$, $Y_{i,70}$  &
$(1,0,-1,0,0,0,1,1)$  \\
$X_{71}$, $Y_{i,71}$  &
$(-1,0,0,1,0,-1,0,0)$  &
$X_{72}$, $Y_{i,72}$  &
$(-1,0,1,-1,0,1,-1,0)$  \\
$X_{73}$, $Y_{i,73}$  &
$(-1,1,-1,0,0,0,1,1)$  &
$X_{74}$, $Y_{i,74}$  &
$(0,1,0,0,-1,0,0,0)$  \\
$X_{75}$, $Y_{i,75}$  &
$(1,-1,1,-1,1,-1,0,0)$  &
$X_{76}$, $Y_{i,76}$  &
$(1,0,-1,0,0,1,-1,1)$  \\
$X_{77}$, $Y_{i,77}$  &
$(1,0,0,0,0,0,1,-1)$  &
$X_{78}$, $Y_{i,78}$  &
$(-1,0,1,-1,1,-1,0,0)$  \\
$X_{79}$, $Y_{i,79}$  &
$(-1,1,-1,0,0,1,-1,1)$  &
$X_{80}$, $Y_{i,80}$  &
$(-1,1,0,0,0,0,1,-1)$  \\
$X_{81}$, $Y_{i,81}$  &
$(0,-1,0,0,0,0,1,1)$  &
$X_{82}$, $Y_{i,82}$  &
$(1,-1,1,0,-1,0,0,0)$  \\
$X_{83}$, $Y_{i,83}$  &
$(1,0,-1,0,1,-1,0,1)$  &
$X_{84}$, $Y_{i,84}$  &
$(1,0,0,0,0,1,-1,-1)$  \\
$X_{85}$, $Y_{i,85}$  &
$(-1,0,1,0,-1,0,0,0)$  &
$X_{86}$, $Y_{i,86}$  &
$(-1,1,-1,0,1,-1,0,1)$  \\
$X_{87}$, $Y_{i,87}$  &
$(-1,1,0,0,0,1,-1,-1)$  &
$X_{88}$, $Y_{i,88}$  &
$(0,-1,0,0,0,1,-1,1)$  \\
$X_{89}$, $Y_{i,89}$  &
$(0,-1,1,0,0,0,1,-1)$  &
$X_{90}$, $Y_{i,90}$  &
$(1,0,-1,1,-1,0,0,1)$  \\
$X_{91}$, $Y_{i,91}$  &
$(1,0,0,0,1,-1,0,-1)$  &
$X_{92}$, $Y_{i,92}$  &
$(-1,1,-1,1,-1,0,0,1)$  \\
$X_{93}$, $Y_{i,93}$  &
$(-1,1,0,0,1,-1,0,-1)$  &
$X_{94}$, $Y_{i,94}$  &
$(0,-1,0,0,1,-1,0,1)$  
\end{tabular}
\caption{Second set of roots of $E_8$ and associated fields.
\label{table:e8:roots2} }
\end{table}

\begin{table}[h!]
\centering
\begin{tabular}{cc|cc}
Field & Positive root/weight & Field & Positive root/weight  \\ \hline
$X_{95}$, $Y_{i,95}$  &
$(0,-1,1,0,0,1,-1,-1)$  &
$X_{96}$, $Y_{i,96}$  &
$(0,0,-1,1,0,0,1,0)$  \\
$X_{97}$, $Y_{i,97}$  &
$(1,0,0,-1,0,0,0,1)$  &
$X_{98}$, $Y_{i,98}$  &
$(1,0,0,1,-1,0,0,-1)$  \\
$X_{99}$, $Y_{i,99}$  &
$(-1,1,0,-1,0,0,0,1)$  &
$X_{100}$, $Y_{i,100}$  &
$(-1,1,0,1,-1,0,0,-1)$  \\
$X_{101}$, $Y_{i,101}$  &
$(0,-1,0,1,-1,0,0,1)$  &
$X_{102}$, $Y_{i,102}$  &
$(0,-1,1,0,1,-1,0,-1)$  \\
$X_{103}$, $Y_{i,103}$  &
$(0,0,-1,1,0,1,-1,0)$  &
$X_{104}$, $Y_{i,104}$  &
$(0,0,0,-1,1,0,1,0)$  \\
$X_{105}$, $Y_{i,105}$  &
$(1,0,1,-1,0,0,0,-1)$  &
$X_{106}$, $Y_{i,106}$  &
$(-1,1,1,-1,0,0,0,-1)$  \\
$X_{107}$, $Y_{i,107}$  &
$(0,-1,1,-1,0,0,0,1)$  &
$X_{108}$, $Y_{i,108}$  &
$(0,-1,1,1,-1,0,0,-1)$  \\
$X_{109}$, $Y_{i,109}$  &
$(0,0,-1,1,1,-1,0,0)$  &
$X_{110}$, $Y_{i,110}$  &
$(0,0,0,-1,1,1,-1,0)$  \\
$X_{111}$, $Y_{i,111}$  &
$(0,0,0,0,-1,1,1,0)$  &
$X_{112}$, $Y_{i,112}$  &
$(1,1,-1,0,0,0,0,0)$  \\
$X_{113}$, $Y_{i,113}$  &
$(-1,2,-1,0,0,0,0,0)$  &
$X_{114}$, $Y_{i,114}$  &
$(0,-1,2,-1,0,0,0,-1)$  \\
$X_{115}$, $Y_{i,115}$  &
$(0,0,-1,0,0,0,0,2)$  &
$X_{116}$, $Y_{i,116}$  &
$(0,0,-1,2,-1,0,0,0)$  \\
$X_{117}$, $Y_{i,117}$  &
$(0,0,0,-1,2,-1,0,0)$  &
$X_{118}$, $Y_{i,118}$  &
$(0,0,0,0,-1,2,-1,0)$  \\
$X_{119}$, $Y_{i,119}$  &
$(0,0,0,0,0,-1,2,0)$  &
$X_{120}$, $Y_{i,120}$  &
$(2,-1,0,0,0,0,0,0)$  \\
$Y_{i,241}$,$Y_{i,242}$ &
$(0,0,0,0,0,0,0,0)$ &
$Y_{i,243}$,$Y_{i,244}$ &
$(0,0,0,0,0,0,0,0)$  \\
$Y_{i,245}$,$Y_{i,246}$ &
$(0,0,0,0,0,0,0,0)$ &
$Y_{i,247}$,$Y_{i,248}$ &
$(0,0,0,0,0,0,0,0)$  
\end{tabular}
\caption{Third set of roots of $E_8$ and associated fields. 
\label{table:e8:roots3} }
\end{table}

For the rest of the fields, the roots and weights are given by 
\begin{align*}
X_{a+120} = - X_{a}, \quad a = 1, \cdots, 120, \\
Y_{i,a+120} = - Y_{i, 120}, \quad a = 1, \cdots, 120.
\end{align*}

\subsection{Superpotential}

In this section, we give the superpotential for the Landau-Ginzburg
orbifold mirror to the theory above.
\begin{align}
W =&\sum_{a=1}^{8} \sigma_a \mathcal{C}^a  - \sum_{i=1}^n \tilde{m}_i \sum_{\alpha =1}^{248} Y_{i,\alpha} + \sum_{i=1}^n \sum_{\alpha =1}^{248} \exp( -Y_{i,\alpha}) + \sum_{m=1}^{240} X_m ,
\label{eq:e8:mirror-sup}
\end{align}
where the $\mathcal{C}^a$ are:
\begin{align*}
\mathcal{C}^1= & \sum_{i=1}^n \big( Y_{i,8}-Y_{i,9}+Y_{i,10}-Y_{i,11}+Y_{i,12}-Y_{i,13}+Y_{i,14}-Y_{i,15}+Y_{i,17}-Y_{i,18}+Y_{i,20}-Y_{i,21} \\
&\quad +Y_{i,24}-Y_{i,25}+Y_{i,46}-Y_{i,47}+Y_{i,52}-Y_{i,53}+Y_{i,57}+Y_{i,58}-Y_{i,59}-Y_{i,60}+Y_{i,63}+Y_{i,64} \\
&\quad -Y_{i,65}-Y_{i,66}+Y_{i,68}+Y_{i,69}+Y_{i,70}-Y_{i,71}-Y_{i,72}-Y_{i,73}+Y_{i,75}+Y_{i,76}+Y_{i,77}-Y_{i,78} \\
&\quad -Y_{i,79}-Y_{i,80}+Y_{i,82}+Y_{i,83}+Y_{i,84}-Y_{i,85}-Y_{i,86}-Y_{i,87}+Y_{i,90}+Y_{i,91}-Y_{i,92}-Y_{i,93} \\
&\quad +Y_{i,97}+Y_{i,98}-Y_{i,99}-Y_{i,100}+Y_{i,105}-Y_{i,106}+Y_{i,112}-Y_{i,113}+2Y_{i,120}-Y_{i,128}+Y_{i,129} \\
&\quad -Y_{i,130}+Y_{i,131}-Y_{i,132}+Y_{i,133}-Y_{i,134}+Y_{i,135}-Y_{i,137}+Y_{i,138}-Y_{i,140}+Y_{i,141}\\
&\quad-Y_{i,144} +Y_{i,145}-Y_{i,166}+Y_{i,167}-Y_{i,172}+Y_{i,173}-Y_{i,177}-Y_{i,178}+Y_{179}+Y_{i,180}-Y_{i,183}\\
& \quad-Y_{i,184}  +Y_{i,185}+Y_{i,186}-Y_{i,188}-Y_{i,189}-Y_{i,190}+Y_{i,191}+Y_{i,192}+Y_{i,193}-Y_{i,195}\\
&\quad-Y_{i,196}-Y_{i,197}  +Y_{i,198}+Y_{i,199}+Y_{i,200}-Y_{i,202}-Y_{i,203}-Y_{i,204}+Y_{i,205}+Y_{i,206}\\
&\quad+Y_{i,207}-Y_{i,210}-Y_{i,211}  +Y_{i,212}+Y_{i,213}-Y_{i,217}-Y_{i,218}+Y_{i,219}+Y_{i,220}-Y_{225}+Y_{i,226}\\
&\quad-Y_{i,232}+Y_{i,233}-2Y_{i,240} \big)
\end{align*}
\begin{align*} &\quad  
 +Z_8-Z_9+Z_{10}-Z_{11}+Z_{12}-Z_{13}+Z_{14}-Z_{15}+Z_{17}-Z_{18}+Z_{20}-Z_{21}+Z_{24}-Z_{25} \\
&\quad +Z_{46} -Z_{47}+Z_{52}-Z_{53}+Z_{57}+Z_{58}-Z_{59}-Z_{60}+Z_{63}+Z_{64}-Z_{65}-Z_{66}+Z_{68} \\
&\quad +Z_{69}+Z_{70}-Z_{71}-Z_{72}-Z_{73}+Z_{75}+Z_{76}+Z_{77}-Z_{78}-Z_{79}-Z_{80}+Z_{82}+Z_{83} \\
&\quad +Z_{84}-Z_{85}-Z_{86}-Z_{87}+Z_{90}+Z_{91}-Z_{92}-Z_{93}+Z_{97}+Z_{98}-Z_{99}-Z_{100}+Z_{105} \\
&\quad -Z_{106}+Z_{112}-Z_{113}+2Z_{120}-Z_{128}+Z_{129}-Z_{130}+Z_{131}-Z_{132}+Z_{133}-Z_{134}+Z_{135}\\
&\quad -Z_{137}+Z_{138}-Z_{140}+Z_{141}-Z_{144}+Z_{145}-Z_{166}+Z_{167}-Z_{172}+Z_{173}-Z_{177}-Z_{178} \\
&\quad +Z_{179}+Z_{180}-Z_{183}-Z_{184}+Z_{185}+Z_{186}-Z_{188}-Z_{189}-Z_{190}+Z_{191}+Z_{192}+Z_{193} \\
&\quad -Z_{195}-Z_{196}-Z_{197}+Z_{198}+Z_{199}+Z_{200}-Z_{202}-Z_{203}-Z_{204}+Z_{205}+Z_{206}+Z_{207} \\
&\quad -Z_{210}-Z_{211}+Z_{212}+Z_{213}-Z_{217}-Z_{218}+Z_{219}+Z_{220}-Z_{225}+Z_{226}-Z_{232}+Z_{233} \\
&\quad -2 Z_{240} ,
\end{align*}
\begin{align*}
\mathcal{C}^2 = & \sum_{i=1}^n \big( Y_{i,6}+Y_{i,7}-Y_{i,8}-Y_{i,10}+Y_{i,13}+Y_{i,15}-Y_{i,16}+Y_{i,18}-Y_{i,19}+Y_{i,21}-Y_{i,22}+Y_{i,25} \\
&\quad -Y_{i,26}-Y_{i,29}+Y_{i,41}+Y_{i,45}-Y_{i,46}+Y_{i,50}+Y_{i,51}-Y_{i,52}+Y_{i,55}+Y_{i,56}-Y_{i,57}-Y_{i,58} \\
&\quad+Y_{i,61}+Y_{i,62}-Y_{i,63}-Y_{i,64}+Y_{i,67}-Y_{i,68}-Y_{i,69}+Y_{i,73}+Y_{i,74}-Y_{i,75}+Y_{i,79}+Y_{i,80} \\
&\quad -Y_{i,81}-Y_{i,82}+Y_{i,86}+Y_{i,87}-Y_{i,88}-Y_{i,89}+Y_{i,92}+Y_{i,93}-Y_{i,94}-Y_{i,95}+Y_{i,99}+Y_{i,100} \\
&\quad -Y_{i,101}-Y_{i,102}+Y_{i,106}-Y_{i,107}-Y_{i,108}+Y_{i,112}+2Y_{i,113}-Y_{i,114}-Y_{i,120}-Y_{i,126} \\
&\quad -Y_{i,127}+Y_{i,128}+Y_{i,130}-Y_{i,133}-Y_{i,135}+Y_{i,136}-Y_{i,138}+Y_{i,139}-Y_{i,141}+Y_{i,142}\\
&\quad-Y_{i,145}  +Y_{i,146}+Y_{i,149}-Y_{i,161}-Y_{i,165}+Y_{i,166}-Y_{i,170}-Y_ {171}+Y_{i,172}-Y_{i,175}\\
&\quad-Y_{i,176}+Y_{i,177}  +Y_{i,178}-Y_{i,181}-Y_{i,182}+Y_{i,183}+Y_{i,184}-Y_{i,187}+Y_{i,188}+Y_{i,189}\\
&\quad-Y_{i,193}-Y_{i,194}+Y_{i,195}  -Y_{i,199}-Y_{i,200}+Y_{i,201}+Y_{i,202}-Y_{i,206}-Y_{i,207}+Y_{i,208}\\
&\quad+Y_{i,209}-Y_{i,212}-Y_{i,213}+Y_{i,214} +Y_{i,215}-Y_{i,219}-Y_{i,220}+Y_{i,221}+Y_{i,222}-Y_{i,226}\\
&\quad+Y_{i,227}+Y_{i,228}-Y_{i,232}-2Y_{i,233}+Y_{i,234} +Y_{i,240} \big) 
\end{align*}
\begin{align*}
& \quad +Z_6+Z_7-Z_8-Z_{10}+Z_{13}+Z_{15}-Z_{16}+Z_{18}-Z_{19}+Z_{21}-Z_{22}+Z_{25}-Z_{26}-Z_{29} \\
&\quad +Z_{41}+Z_{45}-Z_{46}+Z_{50}+Z_{51}-Z_{52}+Z_{55}+Z_{56}-Z_{57}-Z_{58}+Z_{61}+Z_{62}-Z_{63}-Z_{64} \\
&\quad +Z_{67}-Z_{68}-Z_{69}+Z_{73}+Z_{74}-Z_{75}+Z_{79}+Z_{80}-Z_{81}-Z_{82}+Z_{86}+Z_{87}-Z_{88}-Z_{89} \\
&\quad +Z_{92}+Z_{93}-Z_{94}-Z_{95}+Z_{99}+Z_{100}-Z_{101}-Z_{102}+Z_{106}-Z_{107}-Z_{108}+Z_{112}+2Z_{113} \\
&\quad -Z_{114}-Z_{120}-Z_{126}-Z_{127}+Z_{128}+Z_{130}-Z_{133}-Z_{135}+Z_{136}-Z_{138}+Z_{139}-Z_{141}\\ 
&\quad +Z_{142} -Z_{145}+Z_{146}+Z_{149}-Z_{161}-Z_{165}+Z_{166}-Z_{170}-Z_{171}+Z_{172}-Z_{175}-Z_{176} \\
&\quad +Z_{177}+Z_{178}-Z_{181}-Z_{182}+Z_{183}+Z_{184}-Z_{187}+Z_{188}+Z_{189}-Z_{193}-Z_{194}+Z_{195} \\
&\quad -Z_{199}-Z_{200}+Z_{201}+Z_{202}-Z_{206}-Z_{207}+Z_{208}+Z_{209}-Z_{212}-Z_{213}+Z_{214}+Z_{215} \\
&\quad -Z_{219}-Z_{220}+Z_{221}+Z_{222}-Z_{226}+Z_{227}+Z_{228}-Z_{232}-2 Z_{233}+Z_{234}+Z_{240},
\end{align*}
\begin{align*}
\mathcal{C}^3= &\sum_{i=1}^n \big( Y_{i,5}-Y_{i,6}+Y_{i,10}+Y_{i,11}-Y_{i,12}-Y_{i,13}+Y_{i,19}+Y_{i,22}-Y_{i,23}+Y_{i,26}-Y_{i,27} \\
&\quad +Y_{i,29}-Y_{i,30}-Y_{i,33}+Y_{i,36}+Y_{i,40}-Y_{i,41}+Y_{i,43}+Y_{i,44}-Y_{i,45}+Y_{i,49}-Y_{i,50} \\
&\quad -Y_{i,51}+Y_{i,54}-Y_{i,55}-Y_{i,61}+Y_{i,64}+Y_{i,66}+Y_{i,69}-Y_{i,70}+Y_{i,72}-Y_{i,73}+Y_{i,75} \\
&\quad -Y_{i,76}+Y_{i,78}-Y_{i,79}+Y_{i,82}-Y_{i,83}+Y_{i,85}-Y_{i,86}+Y_{i,89}-Y_{i,90}-Y_{i,92}+Y_{i,95} \\
&\quad -Y_{i,96}+Y_{i,102}-Y_{i,103}+Y_{i,105}+Y_{i,106}+Y_{i,107}+Y_{i,108}-Y_{i,109}-Y_{i,112}-Y_{i,113} \\
&\quad +2Y_{i,114}-Y_{i,115}-Y_{i,116}-Y_{i,125}+Y_{i,126}-Y_{i,130}-Y_{i,131}+Y_{i,132}+Y_{i,133}-Y_{i,139} \\
&\quad -Y_{i,142}+Y_{i,143}-Y_{i,146}+Y_{i,147}-Y_{i,149}+Y_{i,150}+Y_{i,153}-Y_{i,156}-Y_{i,160}+Y_{i,161} \\
&\quad -Y_{i,163}-Y_{i,164}+Y_{i,165}-Y_{i,169}+Y_{i,170}+Y_{i,171}-Y_{i,174}+Y_{i,175}+Y_{i,181}-Y_{i,184} \\
&\quad -Y_{i,186}-Y_{i,189}+Y_{i,190}-Y_{i,192}+Y_{i,193}-Y_{i,195}+Y_{i,196}-Y_{i,198}+Y_{i,199}-Y_{i,202} \\
&\quad +Y_{i,203}-Y_{i,205}+Y_{i,206}-Y_{i,209}+Y_{i,210}+Y_{i,212}-Y_{i,215}+Y_{i,216}-Y_{i,222}+Y_{i,223} \\
&\quad -Y_{i,225}-Y_{i,226}-Y_{i,227}-Y_{i,228}+Y_{i,229}+Y_{i,232}+Y_{i,233}-2Y_{i,234}+Y_{i,235}+Y_{i,236}
\big) 
\end{align*}
\begin{align*}
& \quad +  Z_5-Z_6+Z_{10}+Z_{11}-Z_{12}-Z_{13}+Z_{19}+Z_{22}-Z_{23}+Z_{26}-Z_{27}+Z_{29}-Z_{30} \\
&\quad -Z_{33}+Z_{36}+Z_{40}-Z_{41}+Z_{43}+Z_{44}-Z_{45}+Z_{49}-Z_{50}-Z_{51}+Z_{54}-Z_{55}-Z_{61} \\
&\quad +Z_{64}+Z_{66}+Z_{69}-Z_{70}+Z_{72}-Z_{73}+Z_{75}-Z_{76}+Z_{78}-Z_{79}+Z_{82}-Z_{83}+Z_{85} \\
&\quad -Z_{86}+Z_{89}-Z_{90}-Z_{92}+Z_{95}-Z_{96}+Z_{102}-Z_{103}+Z_{105}+Z_{106}+Z_{107}+Z_{108} \\
&\quad -Z_{109}-Z_{112}-Z_{113}+2Z_{114}-Z_{115}-Z_{116}-Z_{125}+Z_{126}-Z_{130}-Z_{131}+Z_{132} \\
&\quad +Z_{133}-Z_{139}-Z_{142}+Z_{143}-Z_{146}+Z_{147}-Z_{149}+Z_{150}+Z_{153}-Z_{156}-Z_{160} \\
&\quad +Z_{161}-Z_{163}-Z_{164}+Z_{165}-Z_{169}+Z_{170}+Z_{171}-Z_{174}+Z_{175}+Z_{181}-Z_{184} \\
&\quad -Z_{186}-Z_{189}+Z_{190}-Z_{192}+Z_{193}-Z_{195}+Z_{196}-Z_{198}+Z_{199}-Z_{202}+Z_{203} \\
&\quad -Z_{205}+Z_{206}-Z_{209}+Z_{210}+Z_{212}-Z_{215}+Z_{216}-Z_{222}+Z_{223}-Z_{225}-Z_{226} \\
&\quad -Z_{227}-Z_{228}+Z_{229}+Z_{232}+Z_{233}-2 Z_{234}+Z_{235}+Z_{236},
\end{align*}
\begin{align*}
\mathcal{C}^4= &\sum_{i=1}^n \big( Y_{i,4}-Y_{i,5}+Y_{i,12}+Y_{i,13}-Y_{i,14}-Y_{i,15}+Y_{i,16}-Y_{i,19}+Y_{i,27}+Y_{i,30}-Y_{i,31} \\
&\quad +Y_{i,32}+Y_{i,33}-Y_{i,34}+Y_{i,35}-Y_{i,36}-Y_{i,37}+Y_{i,39}-Y_{i,40}-Y_{i,43}+Y_{i,51}+Y_{i,55} \\
&\quad -Y_{i,56}+Y_{i,58}+Y_{i,60}+Y_{i,61}-Y_{i,62}+Y_{i,63}-Y_{i,64}+Y_{i,65}-Y_{i,66}-Y_{i,67}+Y_{i,68} \\
&\quad -Y_{i,69}+Y_{i,71}-Y_{i,72}-Y_{i,75}-Y_{i,78}+Y_{i,90}+Y_{i,92}+Y_{i,96}-Y_{i,97}+Y_{i,98}-Y_{i,99} \\
&\quad +Y_{i,100}+Y_{i,101}+Y_{i,103}-Y_{i,104}-Y_{i,105}-Y_{106}-Y_{i,107}+Y_{i,108}+Y_{i,109}-Y_{i,110} \\
&\quad -Y_{i,114}+2Y_{i,116}-Y_{i,117}-Y_{i,124}+Y_{i,125}-Y_{i,132}-Y_{i,133}+Y_{i,134}+Y_{i,135}-Y_{i,136} \\
&\quad +Y_{i,139}-Y_{i,147}-Y_{i,150}+Y_{i,151}-Y_{i,152}-Y_{i,153}+Y_{i,154}-Y_{i,155}+Y_{i,156}+Y_{i,157} \\
&\quad -Y_{i,159}+Y_{i,160}+Y_{163}-Y_{i,171}-Y_{i,175}+Y_{i,176}-Y_{i,178}-Y_{i,180}-Y_{i,181}+Y_{i,182} \\
&\quad -Y_{i,183}+Y_{i,184}-Y_{i,185}+Y_{i,186}+Y_{i,187}-Y_{i,188}+Y_{i,189}-Y_{i,191}+Y_{i,192}+Y_{i,195} \\
&\quad +Y_{i,198}-Y_{i,210}-Y_{i,212}-Y_{i,216}+Y_{i,217}-Y_{i,218}+Y_{i,219}-Y_{i,220}-Y_{i,221}-Y_{i,223} \\
&\quad +Y_{i,224}+Y_{i,225}+Y_{i,226}+Y_{i,227}-Y_{228}-Y_{i,229}+Y_{i,230}+Y_{i,234}-2Y_{i,236}+Y_{i,237}
\big) 
\end{align*}
\begin{align*}
&\quad + Z_4-Z_5+Z_{12}+Z_{13}-Z_{14}-Z_{15}+Z_{16}-Z_{19}+Z_{27}+Z_{30}-Z_{31}+Z_{32}+Z_{33} \\
&\quad -Z_{34}+Z_{35}-Z_{36}-Z_{37}+Z_{39}-Z_{40}-Z_{43}+Z_{51}+Z_{55}-Z_{56}+Z_{58}+Z_{60} \\
&\quad +Z_{61}-Z_{62}+Z_{63}-Z_{64}+Z_{65}-Z_{66}-Z_{67}+Z_{68}-Z_{69}+Z_{71}-Z_{72}-Z_{75} \\
&\quad -Z_{78}+Z_{90}+Z_{92}+Z_{96}-Z_{97}+Z_{98}-Z_{99}+Z_{100}+Z_{101}+Z_{103}-Z_{104} \\
&\quad -Z_{105}-Z_{106}-Z_{107}+Z_{108}+Z_{109}-Z_{110}-Z_{114}+2Z_{116}-Z_{117}-Z_{124} \\
&\quad +Z_{125}-Z_{132}-Z_{133}+Z_{134}+Z_{135}-Z_{136}+Z_{139}-Z_{147}-Z_{150}+Z_{151} \\
&\quad -Z_{152}-Z_{153}+Z_{154}-Z_{155}+Z_{156}+Z_{157}-Z_{159}+Z_{160}+Z_{163}-Z_{171} \\
&\quad -Z_{175}+Z_{176}-Z_{178}-Z_{180}-Z_{181}+Z_{182}-Z_{183}+Z_{184}-Z_{185}+Z_{186} \\
&\quad +Z_{187}-Z_{188}+Z_{189}-Z_{191}+Z_{192}+Z_{195}+Z_{198}-Z_{210}-Z_{212}-Z_{216} \\
&\quad +Z_{217}-Z_{218}+Z_{219}-Z_{220}-Z_{221}-Z_{223}+Z_{224}+Z_{225}+Z_{226}+Z_{227} \\
&\quad -Z_{228}-Z_{229}+Z_{230}+Z_{234}-2 Z_{236}+Z_{237},
\end{align*}
\begin{align*}
\mathcal{C}^5= &\sum_{i=1}^n \big( Y_{i,3}-Y_{i,4}+Y_{i,14}+Y_{i,15}-Y_{i,17}-Y_{i,18}+Y_{i,19}-Y_{i,22}+Y_{i,23}-Y_{i,27}+Y_{i,28} \\
&\quad -Y_{i,32}+Y_{i,34}+Y_{i,37}-Y_{i,38}+Y_{i,40}-Y_{i,42}+Y_{i,43}-Y_{i,44}+Y_{i,45}-Y_{i,49}+Y_{i,50} \\
&\quad -Y_{i,51}+Y_{i,52}+Y_{i,53}-Y_{i,55}+Y_{i,57}-Y_{i,58}+Y_{i,59}-Y_{i,60}-Y_{i,63}-Y_{i,65}+Y_{i,67} \\
&\quad -Y_{i,74}+Y_{i,75}+Y_{i,78}-Y_{i,82}+Y_{i,83}-Y_{i,85}+Y_{i,86}-Y_{i,90}+Y_{i,91}-Y_{i,92}+Y_{i,93} \\
&\quad +Y_{i,94}-Y_{i,98}-Y_{i,100}-Y_{i,101}+Y_{i,102}+Y_{i,104}-Y_{i,108}+Y_{i,109}+Y_{i,110}-Y_{i,111} \\
&\quad -Y_{i,116}+2Y_{i,117}-Y_{i,118}-Y_{i,123}+Y_{i,124}-Y_{i,134}-Y_{i,135}+Y_{i,137}+Y_{i,138}-Y_{i,139} \\
&\quad +Y_{i,142}-Y_{i,143}+Y_{i,147}-Y_{i,148}+Y_{i,152}-Y_{i,154}-Y_{i,157}+Y_{i,158}-Y_{i,160}+Y_{i,162} \\
&\quad -Y_{i,163}+Y_{i,164}-Y_{165}+Y_{i,169}-Y_{i,170}+Y_{i,171}-Y_{i,172}-Y_{i,173}+Y_{i,175}-Y_{i,177} \\
&\quad +Y_{i,178}-Y_{i,179}+Y_{i,180}+Y_{i,183}+Y_{i,185}-Y_{i,187}+Y_{i,194}-Y_{i,195}-Y_{i,198}+Y_{i,202} \\
&\quad -Y_{i,203}+Y_{i,205}-Y_{i,206}+Y_{i,210}-Y_{i,211}+Y_{i,212}-Y_{i,213}-Y_{i,214}+Y_{i,218}+Y_{i,220} \\
&\quad +Y_{i,221}-Y_{i,222}-Y_{i,224}+Y_{i,228}-Y_{229}-Y_{i,230}+Y_{i,231}+Y_{i,236}-2Y_{i,237}+Y_{i,238}
\big) 
\end{align*}
\begin{align*}
&\quad+  Z_3-Z_4+Z_{14}+Z_{15}-Z_{17}-Z_{18}+Z_{19}-Z_{22}+Z_{23}-Z_{27}+Z_{28}-Z_{32}+Z_{34} \\
&\quad +Z_{37}-Z_{38}+Z_{40}-Z_{42}+Z_{43}-Z_{44}+Z_{45}-Z_{49}+Z_{50}-Z_{51}+Z_{52}+Z_{53} \\
&\quad -Z_{55} +Z_{57}-Z_{58}+Z_{59}-Z_{60}-Z_{63}-Z_{65}+Z_{67}-Z_{74}+Z_{75}+Z_{78}-Z_{82} \\
&\quad +Z_{83}-Z_{85}+Z_{86}-Z_{90}+Z_{91}-Z_{92}+Z_{93}+Z_{94}-Z_{98}-Z_{100}-Z_{101}+Z_{102} \\
&\quad +Z_{104}-Z_{108}+Z_{109}+Z_{110}-Z_{111}-Z_{116}+2Z_{117}-Z_{118}-Z_{123}+Z_{124}-Z_{134} \\
&\quad -Z_{135}+Z_{137}+Z_{138}-Z_{139}+Z_{142}-Z_{143}+Z_{147}-Z_{148}+Z_{152}-Z_{154}-Z_{157} \\
&\quad +Z_{158}-Z_{160}+Z_{162}-Z_{163}+Z_{164}-Z_{165}+Z_{169}-Z_{170}+Z_{171}-Z_{172}-Z_{173} \\
&\quad +Z_{175}-Z_{177}+Z_{178}-Z_{179}+Z_{180}+Z_{183}+Z_{185}-Z_{187}+Z_{194}-Z_{195}-Z_{198} \\
&\quad +Z_{202}-Z_{203}+Z_{205}-Z_{206}+Z_{210}-Z_{211}+Z_{212}-Z_{213}-Z_{214}+Z_{218}+Z_{220} \\
&\quad +Z_{221}-Z_{222}-Z_{224}+Z_{228}-Z_{229}-Z_{230}+Z_{231}+Z_{236}-2 Z_{237}+Z_{238},
\end{align*}
\begin{align*}
 \mathcal{C}^6= & \sum_{i=1}^n \big( Y_{i,2}-Y_{i,3}+Y_{i,17}+Y_{i,18}-Y_{i,20}-Y_{i,21}+Y_{i,22}-Y_{i,26}+Y_{i,27}-Y_{i,30}+Y_{i,31} \\
&\quad +Y_{i,32}-Y_{i,34}-Y_{i,35}+Y_{i,36}-Y_{i,40}+Y_{i,41}+Y_{i,42}-Y_{i,45}+Y_{i,46}+Y_{i,47}-Y_{i,48} \\
&\quad +Y_{i,49}-Y_{i,52}-Y_{i,53}-Y_{i,54}+Y_{i,55}-Y_{i,61}+Y_{i,62}+Y_{i,63}+Y_{i,65}-Y_{i,67}-Y_{i,68} \\
&\quad +Y_{i,69}-Y_{i,71}+Y_{i,72}-Y_{i,75}+Y_{i,76}-Y_{i,78}+Y_{i,79}-Y_{i,83}+Y_{i,84}-Y_{i,86}+Y_{i,87} \\
&\quad +Y_{i,88}-Y_{i,91}-Y_{i,93}-Y_{i,94}+Y_{i,95}-Y_{i,102}+Y_{i,103}-Y_{i,109}+Y_{i,110}+Y_{i,111} \\
&\quad -Y_{i,117}+2Y_{i,118}-Y_{i,119}-Y_{i,122}+Y_{i,123}-Y_{i,137}-Y_{i,138}+Y_{i,140}+Y_{i,141}-Y_{i,142} \\
&\quad +Y_{i,146}-Y_{i,147}+Y_{i,150}-Y_{i,151}-Y_{i,152}+Y_{i,154}+Y_{i,155}-Y_{i,156}+Y_{i,160}-Y_{i,161} \\
&\quad -Y_{i,162}+Y_{i,165}-Y_{166}-Y_{i,167}+Y_{i,168}-Y_{i,169}+Y_{i,172}+Y_{i,173}+Y_{i,174}-Y_{i,175} \\
&\quad +Y_{i,181}-Y_{i,182}-Y_{i,183}-Y_{i,185}+Y_{i,187}+Y_{i,188}-Y_{i,189}+Y_{i,191}-Y_{i,192}+Y_{i,195} \\
&\quad -Y_{i,196}+Y_{i,198}-Y_{i,199}+Y_{i,203}-Y_{i,204}+Y_{i,206}-Y_{i,207}-Y_{i,208}+Y_{i,211}+Y_{i,213} \\
&\quad +Y_{i,214}-Y_{i,215}+Y_{i,222}-Y_{i,223}+Y_{229}-Y_{i,230}-Y_{i,231}+Y_{i,237}-2Y_{i,238}+Y_{i,239}
\big)
\end{align*}
\begin{align*}
&\quad +  Z_2-Z_3+Z_{17}+Z_{18}-Z_{20}-Z_{21}+Z_{22}-Z_{26}+Z_{27}-Z_{30}+Z_{31}+Z_{32}-Z_{34} \\
&\quad -Z_{35}+Z_{36}-Z_{40}+Z_{41}+Z_{42}-Z_{45}+Z_{46}+Z_{47}-Z_{48}+Z_{49}-Z_{52}-Z_{53}-Z_{54} \\
&\quad +Z_{55}-Z_{61}+Z_{62}+Z_{63}+Z_{65}-Z_{67}-Z_{68}+Z_{69}-Z_{71}+Z_{72}-Z_{75}+Z_{76}-Z_{78} \\
&\quad +Z_{79}-Z_{83}+Z_{84}-Z_{86}+Z_{87}+Z_{88}-Z_{91}-Z_{93}-Z_{94}+Z_{95}-Z_{102}+Z_{103}-Z_{109} \\
&\quad +Z_{110}+Z_{111}-Z_{117}+2Z_{118}-Z_{119}-Z_{122}+Z_{123}-Z_{137}-Z_{138}+Z_{140}+Z_{141} \\
&\quad -Z_{142}+Z_{146}-Z_{147}+Z_{150}-Z_{151}-Z_{152}+Z_{154}+Z_{155}-Z_{156}+Z_{160}-Z_{161} \\
&\quad -Z_{162}+Z_{165}-Z_{166}-Z_{167}+Z_{168}-Z_{169}+Z_{172}+Z_{173}+Z_{174}-Z_{175}+Z_{181} \\
&\quad -Z_{182}-Z_{183}-Z_{185}+Z_{187}+Z_{188}-Z_{189}+Z_{191}-Z_{192}+Z_{195}-Z_{196}+Z_{198} \\
&\quad -Z_{199}+Z_{203}-Z_{204}+Z_{206}-Z_{207}-Z_{208}+Z_{211}+Z_{213}+Z_{214}-Z_{215}+Z_{222} \\
&\quad -Z_{223}+Z_{229}-Z_{230}-Z_{231}+Z_{237}-2 Z_{238}+Z_{239},
\end{align*}
\begin{align*}
\mathcal{C}^7= & \sum_{i=1}^n \big( Y_{i,1}-Y_{i,2}+Y_{i,20}+Y_{i,21}-Y_{i,24}-Y_{i,25}+Y_{i,26}-Y_{i,29}+Y_{i,30}-Y_{i,33}+Y_{i,34} \\
&\quad +Y_{i,35}-Y_{i,37}+Y_{i,38}-Y_{i,39}+Y_{i,40}-Y_{i,42}-Y_{i,43}+Y_{i,44}+Y_{i,45}-Y_{i,49}-Y_{i,50} \\
&\quad +Y_{i,51}+Y_{i,52}+Y_{i,53}-Y_{i,55}+Y_{i,56}-Y_{i,57}+Y_{i,58}-Y_{i,59}+Y_{i,60}-Y_{i,62}-Y_{i,63} \\
&\quad +Y_{i,64}-Y_{i,65}+Y_{i,66}-Y_{i,69}+Y_{i,70}-Y_{i,72}+Y_{i,73}-Y_{i,76}+Y_{i,77}-Y_{i,79}+Y_{i,80} \\
&\quad +Y_{i,81}-Y_{i,84}-Y_{i,87}-Y_{i,88}+Y_{i,89}-Y_{i,95}+Y_{i,96}-Y_{i,103}+Y_{i,104}-Y_{i,110}+Y_{i,111} \\
&\quad -Y_{i,118}+2Y_{i,119}-Y_{i,121}+Y_{i,122}-Y_{i,140}-Y_{i,141}+Y_{i,144}+Y_{i,145}-Y_{i,146}+Y_{i,149} \\
&\quad -Y_{i,150}+Y_{i,153}-Y_{i,154}-Y_{i,155}+Y_{i,157}-Y_{i,158}+Y_{i,159}-Y_{i,160}+Y_{i,162}+Y_{i,163} \\
&\quad -Y_{i,164}-Y_{i,165}+Y_{i,169}+Y_{i,170}-Y_{i,171}-Y_{i,172}-Y_{i,173}+Y_{i,175}-Y_{i,176}+Y_{i,177} \\
&\quad -Y_{i,178}+Y_{i,179}-Y_{i,180}+Y_{i,182}+Y_{i,183}-Y_{i,184}+Y_{i,185}-Y_{i,186}+Y_{i,189}-Y_{i,190} \\
&\quad +Y_{i,192}-Y_{i,193}+Y_{i,196}-Y_{i,197}+Y_{i,199}-Y_{i,200}-Y_{i,201}+Y_{i,204}+Y_{i,207}+Y_{i,208} \\
&\quad -Y_{i,209}+Y_{i,215}-Y_{i,216}+Y_{i,223}-Y_{i,224}+Y_{i,230}-Y_{i,231}+Y_{i,238}-2Y_{i,239}
\big)
\end{align*}
\begin{align*}
& \quad +Z_1-Z_2+Z_{20}+Z_{21}-Z_{24}-Z_{25}+Z_{26}-Z_{29}+Z_{30}-Z_{33}+Z_{34}+Z_{35}-Z_{37} \\
&\quad +Z_{38}-Z_{39}+Z_{40}-Z_{42}-Z_{43}+Z_{44}+Z_{45}-Z_{49}-Z_{50}+Z_{51}+Z_{52}+Z_{53} \\
&\quad -Z_{55}+Z_{56}-Z_{57}+Z_{58}-Z_{59}+Z_{60}-Z_{62}-Z_{63}+Z_{64}-Z_{65}+Z_{66}-Z_{69} \\
&\quad +Z_{70}-Z_{72}+Z_{73}-Z_{76}+Z_{77}-Z_{79}+Z_{80}+Z_{81}-Z_{84}-Z_{87}-Z_{88}+Z_{89} \\
&\quad -Z_{95}+Z_{96}-Z_{103}+Z_{104}-Z_{110}+Z_{111}-Z_{118}+2Z_{119}-Z_{121}+Z_{122}-Z_{140} \\
&\quad -Z_{141}+Z_{144}+Z_{145}-Z_{146}+Z_{149}-Z_{150}+Z_{153}-Z_{154}-Z_{155}+Z_{157}-Z_{158} \\
&\quad +Z_{159}-Z_{160}+Z_{162}+Z_{163}-Z_{164}-Z_{165}+Z_{169}+Z_{170}-Z_{171}-Z_{172}-Z_{173} \\
&\quad +Z_{175}-Z_{176}+Z_{177}-Z_{178}+Z_{179}-Z_{180}+Z_{182}+Z_{183}-Z_{184}+Z_{185}-Z_{186} \\
&\quad +Z_{189}-Z_{190}+Z_{192}-Z_{193}+Z_{196}-Z_{197}+Z_{199}-Z_{200}-Z_{201}+Z_{204}+Z_{207} \\
&\quad +Z_{208}-Z_{209}+Z_{215}-Z_{216}+Z_{223}-Z_{224}+Z_{230}-Z_{231}+Z_{238}-2 Z_{239},
\end{align*}
\begin{align*}
\mathcal{C}^8= &\sum_{i=1}^n \big( Y_{i,6}-Y_{i,7}+Y_{i,8}+Y_{i,9}-Y_{i,10}-Y_{i,11}+Y_{i,23}+Y_{i,27}-Y_{i,28}+Y_{i,30}+Y_{i,31} \\
&\quad -Y_{i,32}+Y_{i,33}+Y_{i,34}-Y_{i,35}-Y_{i,36}+Y_{i,37}+Y_{i,38}-Y_{i,39}-Y_{i,40}+Y_{i,42}-Y_{i,43} \\
&\quad -Y_{i,44}+Y_{i,48}-Y_{i,49}-Y_{i,54}+Y_{i,70}+Y_{i,73}+Y_{i,76}-Y_{i,77}+Y_{i,79}-Y_{i,80}+Y_{i,81} \\
&\quad +Y_{i,83}-Y_{i,84}+Y_{i,86}-Y_{i,87}+Y_{i,88}-Y_{i,89}+Y_{i,90}-Y_{i,91}+Y_{i,92}-Y_{i,93}+Y_{i,94} \\
&\quad -Y_{i,95}+Y_{i,97}-Y_{i,98}+Y_{i,99}-Y_{i,100}+Y_{i,101}-Y_{i,102}-Y_{i,105}-Y_{i,106}+Y_{i,107} \\
&\quad -Y_{i,108}-Y_{i,114}+2Y_{i,115}-Y_{i,126}+Y_{i,127}-Y_{i,128}-Y_{i,129}+Y_{i,130}+Y_{i,131}-Y_{i,143} \\
&\quad -Y_{i,147}+Y_{i,148}-Y_{i,150}-Y_{i,151}+Y_{i,152}-Y_{i,153}-Y_{i,154}+Y_{i,155}+Y_{i,156}-Y_{i,157} \\
&\quad -Y_{i,158}+Y_{i,159}+Y_{i,160}-Y_{i,162}+Y_{i,163}+Y_{i,164}-Y_{i,168}+Y_{i,169}+Y_{i,174}-Y_{i,190} \\
&\quad -Y_{i,193}-Y_{i,196}+Y_{i,197}-Y_{i,199}+Y_{i,200}-Y_{i,201}-Y_{i,203}+Y_{i,204}-Y_{i,206}+Y_{i,207} \\
&\quad -Y_{i,208}+Y_{i,209}-Y_{i,210}+Y_{i,211}-Y_{i,212}+Y_{i,213}-Y_{i,214}+Y_{i,215}-Y_{i,217}+Y_{i,218} \\
&\quad -Y_{i,219}+Y_{i,220}-Y_{i,221}+Y_{i,222}+Y_{i,225}+Y_{i,226}-Y_{i,227}+Y_{i,228}+Y_{i,234}-2Y_{i,235}
\big) 
\end{align*}
\begin{align*}
&\quad + Z_6-Z_7+Z_8+Z_9-Z_{10}-Z_{11}+Z_{23}+Z_{27}-Z_{28}+Z_{30}+Z_{31}-Z_{32}+Z_{33}+Z_{34} \\
&\quad -Z_{35}-Z_{36}+Z_{37}+Z_{38}-Z_{39}-Z_{40}+Z_{42}-Z_{43}-Z_{44}+Z_{48}-Z_{49}-Z_{54}+Z_{70} \\
&\quad +Z_{73}+Z_{76}-Z_{77}+Z_{79}-Z_{80}+Z_{81}+Z_{83}-Z_{84}+Z_{86}-Z_{87}+Z_{88}-Z_{89}+Z_{90} \\
&\quad -Z_{91}+Z_{92}-Z_{93}+Z_{94}-Z_{95}+Z_{97}-Z_{98}+Z_{99}-Z_{100}+Z_{101}-Z_{102}-Z_{105} \\
&\quad -Z_{106}+Z_{107}-Z_{108}-Z_{114}+2Z_{115}-Z_{126}+Z_{127}-Z_{128}-Z_{129}+Z_{130}+Z_{131} \\
&\quad -Z_{143}-Z_{147}+Z_{148}-Z_{150}-Z_{151}+Z_{152}-Z_{153}-Z_{154}+Z_{155}+Z_{156}-Z_{157} \\
&\quad -Z_{158}+Z_{159}+Z_{160}-Z_{162}+Z_{163}+Z_{164}-Z_{168}+Z_{169}+Z_{174}-Z_{190}-Z_{193} \\
&\quad -Z_{196}+Z_{197}-Z_{199}+Z_{200}-Z_{201}-Z_{203}+Z_{204}-Z_{206}+Z_{207}-Z_{208}+Z_{209} \\
&\quad -Z_{210}+Z_{211}-Z_{212}+Z_{213}-Z_{214}+Z_{215}-Z_{217}+Z_{218}-Z_{219}+Z_{220}-Z_{221} \\
&\quad +Z_{222}+Z_{225}+Z_{226}-Z_{227}+Z_{228}+Z_{234}-2 Z_{235}.
\end{align*}

\subsection{Coulomb ring relations}

For the group $E_8$, we obtain eight Coulomb ring relations,
using the same methods as before.  The results for these ring relations
are listed below.

The first ring relation is
\begin{align*}
\prod_{i=1}^n &  (\tilde{m}_i+2 \sigma _1-\sigma _2)^2 (-\tilde{m}_i-\sigma _1+2 \sigma _2-\sigma_3) (-\tilde{m}_i-\sigma _1-\sigma _2+\sigma _3) (-\tilde{m}_i-\sigma_1+\sigma _3-\sigma _4) \\
&\quad \cdot (-\tilde{m}_i-\sigma _1+\sigma _2-\sigma _3+\sigma_4) (-\tilde{m}_i-\sigma _1+\sigma _3-\sigma _5) (-\tilde{m}_i-\sigma_1+\sigma _4-\sigma _5)  \\
&\quad \cdot (-\tilde{m}_i-\sigma _1+\sigma _2-\sigma _3+\sigma_5) (-\tilde{m}_i-\sigma _1+\sigma _2-\sigma _4+\sigma _5)(-\tilde{m}_i-\sigma _1+\sigma _2-\sigma _6)  \\
&\quad \cdot (-\tilde{m}_i-\sigma _1+\sigma_4-\sigma _6) (-\tilde{m}_i-\sigma _1+\sigma _5-\sigma _6)(-\tilde{m}_i-\sigma _1+\sigma _3-\sigma _4+\sigma _5-\sigma _6)  \\
&\quad \cdot (-\tilde{m}_i-\sigma _1+\sigma _6) (-\tilde{m}_i-\sigma _1+\sigma _2-\sigma_4+\sigma _6) (-\tilde{m}_i-\sigma _1+\sigma _2-\sigma _5+\sigma _6)  \\
&\quad \cdot (-\tilde{m}_i-\sigma _1+\sigma _2-\sigma _3+\sigma _4-\sigma _5+\sigma _6)(-\tilde{m}_i-\sigma _1+\sigma _2-\sigma _7)   \\
&\quad \cdot  (-\tilde{m}_i-\sigma _1+\sigma _5-\sigma_7) (-\tilde{m}_i-\sigma _1+\sigma _2-\sigma _4+\sigma _5-\sigma _7)(-\tilde{m}_i-\sigma _1+\sigma _6-\sigma _7)  \\
&\quad \cdot (-\tilde{m}_i-\sigma _1+\sigma_3-\sigma _4+\sigma _6-\sigma _7) (-\tilde{m}_i-\sigma _1+\sigma _2-\sigma_5+\sigma _6-\sigma _7)  \\
&\quad \cdot (-\tilde{m}_i-\sigma _1+\sigma _7) (-\tilde{m}_i-\sigma_1+\sigma _3-\sigma _4+\sigma _7) (-\tilde{m}_i-\sigma _1+\sigma _2-\sigma_5+\sigma _7) \cdot \\
&\quad \cdot (-\tilde{m}_i-\sigma _1+\sigma _4-\sigma _5+\sigma _7)(-\tilde{m}_i-\sigma _1+\sigma _2-\sigma _6+\sigma _7)  \\
&\quad \cdot (-\tilde{m}_i-\sigma_1+\sigma _5-\sigma _6+\sigma _7) (-\tilde{m}_i-\sigma _1+\sigma _2-\sigma_4+\sigma _5-\sigma _6+\sigma _7) \cdot \\
&\quad \cdot (-\tilde{m}_i-\sigma _1+\sigma _3-\sigma _8) (-\tilde{m}_i-\sigma_1+\sigma _2+\sigma _3-\sigma _4-\sigma _8) (-\tilde{m}_i-\sigma _1+\sigma_4-\sigma _8)  \\
&\quad \cdot (-\tilde{m}_i-\sigma _1+\sigma _2+\sigma _4-\sigma _5-\sigma_8) (-\tilde{m}_i-\sigma _1+\sigma _3-\sigma _4+\sigma _5-\sigma _8)\\
&\quad \cdot (-\tilde{m}_i-\sigma _1+\sigma _3-\sigma _5+\sigma _6-\sigma _8)(-\tilde{m}_i-\sigma _1+\sigma _3-\sigma _7-\sigma _8) \\
&\quad \cdot (-\tilde{m}_i-\sigma _1+\sigma_2+\sigma _7-\sigma _8) (-\tilde{m}_i-\sigma _1+\sigma _3-\sigma _6+\sigma_7-\sigma _8) (-\tilde{m}_i-\sigma _1+\sigma _8)  \\
&\quad \cdot (-\tilde{m}_i-\sigma_1+\sigma _2-\sigma _3+\sigma _8) (-\tilde{m}_i-\sigma _1+\sigma _2-\sigma_4+\sigma _8) (-\tilde{m}_i-\sigma _1-\sigma _3+\sigma _4+\sigma _8)  \\
&\quad \cdot (-\tilde{m}_i-\sigma _1+\sigma _2-\sigma _3+\sigma _4-\sigma _5+\sigma _8)(-\tilde{m}_i-\sigma _1-\sigma _4+\sigma _5+\sigma _8) \cdot \\
&\quad \cdot (-\tilde{m}_i-\sigma_1-\sigma _5+\sigma _6+\sigma _8) (-\tilde{m}_i-\sigma _1-\sigma _7+\sigma_8)   \\
&\quad \cdot (-\tilde{m}_i-\sigma _1+\sigma _2-\sigma _3+\sigma _7+\sigma _8)(-\tilde{m}_i-\sigma _1-\sigma _6+\sigma _7+\sigma _8)   \\
&\quad \cdot  (-\tilde{m}_i-\sigma _1+\sigma_2-\sigma _3+\sigma _4-\sigma _7) (-\tilde{m}_i-\sigma _1+\sigma _4-\sigma _5+\sigma_6-\sigma _7)   \\
&\quad \cdot  (-\tilde{m}_i-\sigma_1+\sigma _2-\sigma _3+\sigma _4-\sigma _6+\sigma _7)  (-\tilde{m}_i-\sigma _1+\sigma _2-\sigma_8)    \\
&\quad \cdot  (-\tilde{m}_i-\sigma_1+\sigma _2-\sigma _3+\sigma _5-\sigma _6+\sigma _8)  (-\tilde{m}_i-\sigma_1+\sigma _2+\sigma _6-\sigma _7-\sigma _8)     \\
&\quad \cdot  (-\tilde{m}_i-\sigma _1+\sigma _2+\sigma _5-\sigma _6-\sigma _8) (-\tilde{m}_i-\sigma _1+\sigma _2-\sigma _3+\sigma _6-\sigma _7+\sigma_8)
\end{align*}
\begin{align*}
& = \prod_{i=1}^n (\tilde{m}_i-2 \sigma _1+\sigma _2)^2 (-\tilde{m}_i+\sigma _1+\sigma _2-\sigma_3) (-\tilde{m}_i+\sigma _1-2 \sigma _2+\sigma _3)  \\
&\quad \cdot (-\tilde{m}_i+\sigma_1-\sigma _2+\sigma _3-\sigma _4) (-\tilde{m}_i+\sigma _1-\sigma _3+\sigma_4) (-\tilde{m}_i+\sigma _1-\sigma _2+\sigma _3-\sigma _5)  \\
&\quad \cdot (-\tilde{m}_i+\sigma _1-\sigma _2+\sigma _4-\sigma _5) (-\tilde{m}_i+\sigma_1-\sigma _3+\sigma _5) (-\tilde{m}_i+\sigma _1-\sigma _4+\sigma _5) \\
&\quad \cdot (-\tilde{m}_i+\sigma _1-\sigma _6) (-\tilde{m}_i+\sigma _1-\sigma _2+\sigma_4-\sigma _6) (-\tilde{m}_i+\sigma _1-\sigma _2+\sigma _5-\sigma _6)  \\
&\quad \cdot (-\tilde{m}_i+\sigma _1-\sigma _2+\sigma _3-\sigma _4+\sigma _5-\sigma _6)(-\tilde{m}_i+\sigma _1-\sigma _2+\sigma _6) (-\tilde{m}_i+\sigma _1-\sigma_4+\sigma _6)  \\
&\quad \cdot (-\tilde{m}_i+\sigma _1-\sigma _5+\sigma _6)(-\tilde{m}_i+\sigma _1-\sigma _3+\sigma _4-\sigma _5+\sigma _6)(-\tilde{m}_i+\sigma _1-\sigma _7)  \\
&\quad \cdot (-\tilde{m}_i+\sigma _1-\sigma _3+\sigma_4-\sigma _7) (-\tilde{m}_i+\sigma _1-\sigma _2+\sigma _5-\sigma _7)(-\tilde{m}_i+\sigma _1-\sigma _4+\sigma _5-\sigma _7)  \\
&\quad \cdot (-\tilde{m}_i+\sigma_1-\sigma _2+\sigma _6-\sigma _7) (-\tilde{m}_i+\sigma _1-\sigma _2+\sigma_3-\sigma _4+\sigma _6-\sigma _7)  \\
&\quad \cdot (-\tilde{m}_i+\sigma _1-\sigma _2+\sigma _4-\sigma _5+\sigma_6-\sigma _7) (-\tilde{m}_i+\sigma _1-\sigma _2+\sigma _7) \\
&\quad \cdot (-\tilde{m}_i+\sigma _1-\sigma _5+\sigma _7) (-\tilde{m}_i+\sigma _1-\sigma_2+\sigma _4-\sigma _5+\sigma _7) (-\tilde{m}_i+\sigma _1-\sigma _6+\sigma_7)  \\
&\quad \cdot  (-\tilde{m}_i+\sigma _1-\sigma _3+\sigma _4-\sigma _6+\sigma _7)(-\tilde{m}_i+\sigma _1-\sigma _2+\sigma _5-\sigma _6+\sigma _7)\\
&\quad \cdot (-\tilde{m}_i+\sigma _1-\sigma _8) (-\tilde{m}_i+\sigma _1-\sigma _2+\sigma_3-\sigma _8)  (-\tilde{m}_i+\sigma _1-\sigma_2+\sigma _3-\sigma _6+\sigma _7-\sigma _8)  \\
&\quad \cdot (-\tilde{m}_i+\sigma _1-\sigma _2+\sigma _4-\sigma _8) (-\tilde{m}_i+\sigma_1+\sigma _4-\sigma _5-\sigma _8)  \\
&\quad \cdot (-\tilde{m}_i+\sigma _1+\sigma _5-\sigma_6-\sigma _8) (-\tilde{m}_i+\sigma _1-\sigma _2+\sigma _3-\sigma _5+\sigma_6-\sigma _8)   \\
&\quad \cdot (-\tilde{m}_i+\sigma _1+\sigma _6-\sigma _7-\sigma _8)(-\tilde{m}_i+\sigma _1+\sigma _7-\sigma _8)(-\tilde{m}_i+\sigma _1+\sigma _3-\sigma _4-\sigma _8)  \\
&\quad \cdot  (-\tilde{m}_i+\sigma _1-\sigma_2+\sigma _8) (-\tilde{m}_i+\sigma _1-\sigma _3+\sigma _8)(-\tilde{m}_i+\sigma _1-\sigma _4+\sigma _8)  \\
&\quad \cdot  (-\tilde{m}_i+\sigma _1-\sigma_2-\sigma _3+\sigma _4+\sigma _8) (-\tilde{m}_i+\sigma _1-\sigma _3+\sigma_4-\sigma _5+\sigma _8)  \\
&\quad \cdot (-\tilde{m}_i+\sigma _1-\sigma _3+\sigma _5-\sigma _6+\sigma_8) (-\tilde{m}_i+\sigma _1-\sigma _2-\sigma _5+\sigma _6+\sigma _8)   \\
&\quad \cdot (-\tilde{m}_i+\sigma_1-\sigma _3+\sigma _6-\sigma _7+\sigma _8) (-\tilde{m}_i+\sigma _1-\sigma_3+\sigma _7+\sigma _8)  \\
&\quad \cdot (-\tilde{m}_i+\sigma _1-\sigma _2-\sigma _7+\sigma _8)(-\tilde{m}_i+\sigma _1-\sigma _2-\sigma _6+\sigma_7+\sigma _8)  \\
&\quad \cdot (-\tilde{m}_i+\sigma _1-\sigma _2-\sigma _4+\sigma_5+\sigma _8)  (-\tilde{m}_i+\sigma _1-\sigma _2+\sigma _3-\sigma _7-\sigma_8)  \\
&\quad \cdot (-\tilde{m}_i+\sigma _1-\sigma _4+\sigma _5-\sigma _6+\sigma _7)  (-\tilde{m}_i+\sigma _1-\sigma _2+\sigma_3-\sigma _4+\sigma _5-\sigma _8) \\
&\quad \cdot (-\tilde{m}_i+\sigma _1-\sigma _2+\sigma _3-\sigma _4+\sigma _7)(-\tilde{m}_i+\sigma _1-\sigma _5+\sigma_6-\sigma _7) .
\end{align*}

The second ring relation is
\begin{align*}
\prod_{i=1}^n & (-\tilde{m}_i+2 \sigma _1-\sigma _2) (-\tilde{m}_i+\sigma _1-2 \sigma _2+\sigma_3)^2 (-\tilde{m}_i-\sigma _1-\sigma _2+\sigma _3) \\
&\quad \cdot (-\tilde{m}_i+\sigma_1-\sigma _2+\sigma _3-\sigma _4) (-\tilde{m}_i-\sigma _2+\sigma _4)(-\tilde{m}_i+\sigma _1-\sigma _2+\sigma _3-\sigma _5) \\
&\quad \cdot (-\tilde{m}_i+\sigma_1-\sigma _2+\sigma _4-\sigma _5) (-\tilde{m}_i-\sigma _2+\sigma _5)(-\tilde{m}_i-\sigma _2+\sigma _3-\sigma _4+\sigma _5) \\
&\quad \cdot (-\tilde{m}_i-\sigma_2+\sigma _3-\sigma _6) (-\tilde{m}_i+\sigma _1-\sigma _2+\sigma _4-\sigma_6) (-\tilde{m}_i+\sigma _1-\sigma _2+\sigma _5-\sigma _6) \\
&\quad \cdot (-\tilde{m}_i+\sigma _1-\sigma _2+\sigma _3-\sigma _4+\sigma _5-\sigma _6)(-\tilde{m}_i+\sigma _1-\sigma _2+\sigma _6) \\
&\quad \cdot (-\tilde{m}_i-\sigma _2+\sigma _3-\sigma _5+\sigma_6) (-\tilde{m}_i-\sigma _2+\sigma _4-\sigma _5+\sigma _6)(-\tilde{m}_i-\sigma _2+\sigma _3-\sigma _7) \\
&\quad \cdot (-\tilde{m}_i-\sigma _2+\sigma_4-\sigma _7) (-\tilde{m}_i+\sigma _1-\sigma _2+\sigma _5-\sigma _7)(-\tilde{m}_i-\sigma _2+\sigma _3-\sigma _4+\sigma _5-\sigma _7) \\
&\quad \cdot (-\tilde{m}_i+\sigma _1-\sigma _2+\sigma _6-\sigma _7) (-\tilde{m}_i+\sigma_1-\sigma _2+\sigma _3-\sigma _4+\sigma _6-\sigma _7)\\
&\quad \cdot (-\tilde{m}_i+\sigma _1-\sigma_2+\sigma _4-\sigma _5+\sigma _6-\sigma _7) (-\tilde{m}_i+\sigma _1-\sigma_2+\sigma _7)  \\
&\quad \cdot (-\tilde{m}_i-\sigma _2+\sigma _3-\sigma _5+\sigma _7)(-\tilde{m}_i+\sigma _1-\sigma _2+\sigma _4-\sigma _5+\sigma _7) \\
&\quad \cdot (-\tilde{m}_i-\sigma_2+\sigma _4-\sigma _6+\sigma _7) (-\tilde{m}_i+\sigma _1-\sigma _2+\sigma_5-\sigma _6+\sigma _7) \\
&\quad \cdot (-\tilde{m}_i-\sigma _2+\sigma _3-\sigma _8)(-\tilde{m}_i+\sigma _1-\sigma _2+\sigma _3-\sigma _8) (-\tilde{m}_i-\sigma _2+2\sigma _3-\sigma _4-\sigma _8) \\
&\quad \cdot (-\tilde{m}_i+\sigma _1-\sigma _2+\sigma_4-\sigma _8) (-\tilde{m}_i-\sigma _2+\sigma _3+\sigma _4-\sigma _5-\sigma_8)  \\
&\quad \cdot (-\tilde{m}_i-\sigma _2+\sigma _3+\sigma _5-\sigma _6-\sigma _8)(-\tilde{m}_i+\sigma _1-\sigma _2+\sigma _3-\sigma _5+\sigma _6-\sigma _8) \\
&\quad \cdot (-\tilde{m}_i+\sigma _1-\sigma _2+\sigma _3-\sigma _7-\sigma _8)(-\tilde{m}_i+\sigma _1-\sigma _2+\sigma _3-\sigma _4+\sigma _5-\sigma_8) \\
&\quad \cdot (-\tilde{m}_i-\sigma _2+\sigma _3-\sigma _4+\sigma_5-\sigma _6+\sigma _7)  (-\tilde{m}_i-\sigma_2+\sigma _3-\sigma _5+\sigma _6-\sigma _7) \\
&\quad \cdot (-\tilde{m}_i+\sigma _1-\sigma _2+\sigma _3-\sigma _4+\sigma_7)(-\tilde{m}_i-\sigma _2+\sigma_3-\sigma _4+\sigma _6)  \\
&\quad \cdot (-\tilde{m}_i-\sigma _2+\sigma _3+\sigma _6-\sigma _7-\sigma _8)(-\tilde{m}_i-\sigma _2+\sigma _3+\sigma _7-\sigma _8)  \\
&\quad \cdot (-\tilde{m}_i-\sigma_2+\sigma _8) (-\tilde{m}_i+\sigma _1-\sigma _2+\sigma _8)(-\tilde{m}_i-\sigma _2+\sigma _3-\sigma _4+\sigma _8) \\
&\quad \cdot (-\tilde{m}_i+\sigma_1-\sigma _2-\sigma _3+\sigma _4+\sigma _8) (-\tilde{m}_i-\sigma _2+\sigma_4-\sigma _5+\sigma _8)  \\
&\quad \cdot (-\tilde{m}_i-\sigma _2+\sigma _5-\sigma _6+\sigma _8)(-\tilde{m}_i+\sigma _1-\sigma _2-\sigma _5+\sigma _6+\sigma _8) \\
&\quad \cdot (-\tilde{m}_i-\sigma_2+\sigma _6-\sigma _7+\sigma _8) (-\tilde{m}_i-\sigma _2+\sigma _7+\sigma_8)(-\tilde{m}_i-\sigma _2+\sigma _3-\sigma _6+\sigma _7)  \\
&\quad \cdot (-\tilde{m}_i+\sigma_1-\sigma _2+\sigma _3-\sigma _6+\sigma _7-\sigma _8)(-\tilde{m}_i+\sigma _1-\sigma _2-\sigma _4+\sigma_5+\sigma _8) \\
&\quad \cdot (-\tilde{m}_i+\sigma _1-\sigma _2-\sigma _7+\sigma _8)(-\tilde{m}_i+\sigma _1-\sigma _2-\sigma _6+\sigma _7+\sigma _8) 
\\
\end{align*}
\begin{align*}
& =
\prod_{i=1}^n 
(-\tilde{m}_i-2 \sigma _1+\sigma _2) (-\tilde{m}_i+\sigma _1+\sigma _2-\sigma _3)(\tilde{m}_i+\sigma _1-2 \sigma _2+\sigma _3)^2 \\
&\quad \cdot (-\tilde{m}_i+\sigma _2-\sigma_4) (-\tilde{m}_i-\sigma _1+\sigma _2-\sigma _3+\sigma _4)(-\tilde{m}_i+\sigma _2-\sigma _5) \\
&\quad \cdot (-\tilde{m}_i+\sigma _2-\sigma _3+\sigma_4-\sigma _5) (-\tilde{m}_i-\sigma _1+\sigma _2-\sigma _3+\sigma _5)(-\tilde{m}_i-\sigma _1+\sigma _2-\sigma _4+\sigma _5) \\
&\quad \cdot (-\tilde{m}_i-\sigma_1+\sigma _2-\sigma _6) (-\tilde{m}_i+\sigma _2-\sigma _3+\sigma _4-\sigma_6) (-\tilde{m}_i+\sigma _2-\sigma _3+\sigma _5-\sigma _6) \\
&\quad \cdot (-\tilde{m}_i+\sigma _2-\sigma _4+\sigma _5-\sigma _6) (-\tilde{m}_i+\sigma_2-\sigma _3+\sigma _6) (-\tilde{m}_i-\sigma _1+\sigma _2-\sigma _4+\sigma_6) \\
&\quad \cdot (-\tilde{m}_i-\sigma _1+\sigma _2-\sigma _5+\sigma _6)(-\tilde{m}_i-\sigma _1+\sigma _2-\sigma _3+\sigma _4-\sigma _5+\sigma _6) \\
&\quad \cdot (-\tilde{m}_i-\sigma _1+\sigma_2-\sigma _3+\sigma _4-\sigma _7) (-\tilde{m}_i+\sigma _2-\sigma _3+\sigma_5-\sigma _7) \\
&\quad \cdot (-\tilde{m}_i+\sigma _2-\sigma _3+\sigma _6-\sigma _7)(-\tilde{m}_i+\sigma _2-\sigma _4+\sigma _6-\sigma _7) \\
&\quad \cdot (-\tilde{m}_i+\sigma _2-\sigma_3+\sigma _4-\sigma _5+\sigma _6-\sigma _7) (-\tilde{m}_i+\sigma _2-\sigma_3+\sigma _7)  \\
&\quad \cdot (-\tilde{m}_i-\sigma _1+\sigma _2-\sigma _5+\sigma _7) (-\tilde{m}_i+\sigma_2-\sigma _3+\sigma _4-\sigma _5+\sigma _7) \\
&\quad \cdot (-\tilde{m}_i-\sigma _1+\sigma _2-\sigma _3+\sigma_4-\sigma _6+\sigma _7) (-\tilde{m}_i+\sigma _2-\sigma _3+\sigma _5-\sigma_6+\sigma _7)  \\
&\quad \cdot (-\tilde{m}_i+\sigma _2-\sigma _8) (-\tilde{m}_i-\sigma_1+\sigma _2-\sigma _8) (-\tilde{m}_i-\sigma _1+\sigma _2+\sigma _3-\sigma_4-\sigma _8) \\
&\quad \cdot (-\tilde{m}_i+\sigma _2-\sigma _3+\sigma _4-\sigma _8)(-\tilde{m}_i-\sigma _1+\sigma _2+\sigma _4-\sigma _5-\sigma _8) \\
&\quad \cdot (-\tilde{m}_i-\sigma_1+\sigma _2+\sigma _5-\sigma _6-\sigma _8) (-\tilde{m}_i+\sigma _2-\sigma_5+\sigma _6-\sigma _8) (-\tilde{m}_i+\sigma _2-\sigma _7-\sigma _8) \\
&\quad \cdot (-\tilde{m}_i-\sigma _1+\sigma _2+\sigma _6-\sigma _7-\sigma _8) (-\tilde{m}_i+\sigma_2-\sigma _6+\sigma _7-\sigma _8) \\
&\quad \cdot (-\tilde{m}_i+\sigma _2-\sigma _3+\sigma_8) (-\tilde{m}_i-\sigma _1+\sigma _2-\sigma _3+\sigma _8)(-\tilde{m}_i-\sigma _1+\sigma _2-\sigma _4+\sigma _8) \\
&\quad \cdot (-\tilde{m}_i+\sigma _2-2\sigma _3+\sigma _4+\sigma _8) (-\tilde{m}_i-\sigma _1+\sigma _2-\sigma_3+\sigma _4-\sigma _5+\sigma _8) \\
&\quad \cdot (-\tilde{m}_i-\sigma _1+\sigma _2-\sigma _3+\sigma_5-\sigma _6+\sigma _8) (-\tilde{m}_i+\sigma _2-\sigma _3-\sigma _5+\sigma_6+\sigma _8)  \\
&\quad \cdot (-\tilde{m}_i-\sigma _1+\sigma _2-\sigma _3+\sigma _6-\sigma _7+\sigma _8)(-\tilde{m}_i-\sigma _1+\sigma _2-\sigma _3+\sigma _7+\sigma _8)\\
&\quad \cdot (-\tilde{m}_i+\sigma _2-\sigma _3-\sigma _6+\sigma _7+\sigma _8) (-\tilde{m}_i+\sigma _2-\sigma _3-\sigma _7+\sigma _8) \\
&\quad \cdot (-\tilde{m}_i+\sigma _2-\sigma _3-\sigma_4+\sigma _5+\sigma _8) (-\tilde{m}_i-\sigma _1+\sigma _2-\sigma _4+\sigma _5-\sigma_6+\sigma _7) \\
&\quad \cdot (-\tilde{m}_i-\sigma _1+\sigma _2-\sigma _4+\sigma _5-\sigma_7)  (-\tilde{m}_i-\sigma_1+\sigma _2-\sigma _5+\sigma _6-\sigma _7)  \\
&\quad \cdot (-\tilde{m}_i-\sigma _1+\sigma_2-\sigma _6+\sigma _7) (-\tilde{m}_i+\sigma _2-\sigma _4+\sigma _5-\sigma _8) \\
&\quad \cdot (-\tilde{m}_i-\sigma _1+\sigma _2-\sigma _7)(-\tilde{m}_i+\sigma _2-\sigma _4+\sigma _7)(-\tilde{m}_i-\sigma _1+\sigma _2+\sigma _7-\sigma _8) .
\end{align*}

The third ring relation is
\begin{align*}
\prod_{i=1}^n &  (-\tilde{m}_i+\sigma_1+\sigma_2-\sigma_3) (-\tilde{m}_i-\sigma_1+2 \sigma_2-\sigma_3) (-\tilde{m}_i-\sigma_3+\sigma_4)  \\
&\quad \cdot (-\tilde{m}_i+\sigma_1-\sigma_3+\sigma_4) (-\tilde{m}_i-\sigma_1+\sigma_2-\sigma_3+\sigma_4) (-\tilde{m}_i+\sigma_2-\sigma_3+\sigma_4-\sigma_5)  \\
&\quad \cdot (-\tilde{m}_i-\sigma_3+2 \sigma_4-\sigma_5) (-\tilde{m}_i+\sigma_1-\sigma_3+\sigma_5) (-\tilde{m}_i-\sigma_1+\sigma_2-\sigma_3+\sigma_5)   \\
&\quad \cdot (-\tilde{m}_i+\sigma_2-\sigma_3+\sigma_4-\sigma_6) (-\tilde{m}_i+\sigma_2-\sigma_3+\sigma_5-\sigma_6) (-\tilde{m}_i-\sigma_3+\sigma_4+\sigma_5-\sigma_6)   \\
&\quad \cdot (-\tilde{m}_i+\sigma_2-\sigma_3+\sigma_6) (-\tilde{m}_i+\sigma_1-\sigma_3+\sigma_4-\sigma_5+\sigma_6)\\
&\quad \cdot (-\tilde{m}_i+\sigma_1-\sigma_3+\sigma_4-\sigma_7) (-\tilde{m}_i-\sigma_1+\sigma_2-\sigma_3+\sigma_4-\sigma_7) (-\tilde{m}_i+\sigma_2-\sigma_3+\sigma_5-\sigma_7)  \\
&\quad \cdot (-\tilde{m}_i+\sigma_2-\sigma_3+\sigma_6-\sigma_7) (-\tilde{m}_i-\sigma_3+\sigma_4+\sigma_6-\sigma_7)   \\
&\quad \cdot (-\tilde{m}_i+\sigma_2-\sigma_3+\sigma_4-\sigma_5+\sigma_6-\sigma_7) (-\tilde{m}_i+\sigma_2-\sigma_3+\sigma_7)   \\
&\quad \cdot (-\tilde{m}_i-\sigma_3+\sigma_4+\sigma_7)(-\tilde{m}_i+\sigma_2-\sigma_3+\sigma_4-\sigma_5+\sigma_7)(-\tilde{m}_i+\sigma_1-\sigma_3+\sigma_4-\sigma_6+\sigma_7)  \\
&\quad \cdot (-\tilde{m}_i-\sigma_1+\sigma_2-\sigma_3+\sigma_4-\sigma_6+\sigma_7)(-\tilde{m}_i+\sigma_2-\sigma_3+\sigma_5-\sigma_6+\sigma_7)  \\
&\quad \cdot (-\tilde{m}_i+\sigma_1-\sigma_3+\sigma_8) (-\tilde{m}_i+\sigma_2-\sigma_3+\sigma_8)(-\tilde{m}_i-\sigma_1+\sigma_2-\sigma_3+\sigma_8)  \\
&\quad \cdot (-\tilde{m}_i+\sigma_2-2\sigma_3+\sigma_4+\sigma_8)^2 (-\tilde{m}_i-\sigma_1-\sigma_3+\sigma_4+\sigma_8)   \\
&\quad \cdot (-\tilde{m}_i+\sigma_1-\sigma_3+\sigma_4-\sigma_5+\sigma_8)(-\tilde{m}_i-\sigma_1+\sigma_2-\sigma_3+\sigma_4-\sigma_5+\sigma_8)   \\
&\quad \cdot (-\tilde{m}_i+\sigma_2-\sigma_3-\sigma_4+\sigma_5+\sigma_8) (-\tilde{m}_i-\sigma_3+\sigma_4-\sigma_6+\sigma_8)   \\
&\quad \cdot (-\tilde{m}_i-\sigma_1+\sigma_2-\sigma_3+\sigma_5-\sigma_6+\sigma_8) (-\tilde{m}_i-\sigma_3+\sigma_6+\sigma_8)   \\
&\quad \cdot (-\tilde{m}_i-\sigma_3+\sigma_4-\sigma_5+\sigma_6+\sigma_8) (-\tilde{m}_i+\sigma_2-\sigma_3-\sigma_7+\sigma_8) (-\tilde{m}_i-\sigma_3+\sigma_4-\sigma_7+\sigma_8)   \\
&\quad \cdot (-\tilde{m}_i-\sigma_3+\sigma_5-\sigma_7+\sigma_8) (-\tilde{m}_i+\sigma_1-\sigma_3+\sigma_6-\sigma_7+\sigma_8)   \\
&\quad \cdot (-\tilde{m}_i-\sigma_3+\sigma_4-\sigma_5+\sigma_6-\sigma_7+\sigma_8) (-\tilde{m}_i-\sigma_1+\sigma_2-\sigma_3+\sigma_7+\sigma_8)  \\
&\quad \cdot (-\tilde{m}_i+\sigma_1-\sigma_3+\sigma_7+\sigma_8)(-\tilde{m}_i-\sigma_3+\sigma_4-\sigma_5+\sigma_7+\sigma_8) \\
&\quad \cdot (-\tilde{m}_i-\sigma_3+\sigma_4-\sigma_6+\sigma_7+\sigma_8)(-\tilde{m}_i-\sigma_3+\sigma_5-\sigma_6+\sigma_7+\sigma_8)(-\tilde{m}_i-\sigma_3+2 \sigma_8)  \\
&\quad \cdot (-\tilde{m}_i-\sigma_1+\sigma_2-\sigma_3+\sigma_6-\sigma_7+\sigma_8)(-\tilde{m}_i-\sigma_1+\sigma_2-\sigma_3+\sigma_6-\sigma_7+\sigma_8)\\
&\quad \cdot (-\tilde{m}_i+\sigma_2-\sigma_3-\sigma_6+\sigma_7+\sigma_8)(-\tilde{m}_i+\sigma_2-\sigma_3-\sigma_5+\sigma_6+\sigma_8)\\
&\quad \cdot  (-\tilde{m}_i+\sigma_1-\sigma_3+\sigma_5-\sigma_6+\sigma_8)(-\tilde{m}_i+\sigma_1-\sigma_2-\sigma_3+\sigma_4+\sigma_8)  \\
&\quad \cdot (-\tilde{m}_i-\sigma_1+\sigma_2-\sigma_3+\sigma_4-\sigma_5+\sigma_6)(-\tilde{m}_i+\sigma_2-\sigma_3+\sigma_4-\sigma_8) \\
&\quad \cdot(-\tilde{m}_i-\sigma_3+\sigma_5+\sigma_8)(-\tilde{m}_i+\sigma_2-\sigma_3-\sigma_6+\sigma_7+\sigma_8)
\end{align*}
\begin{align*}
& =\prod_{i=1}^n  (-\tilde{m}_i+\sigma_1-2 \sigma_2+\sigma_3 )  (-\tilde{m}_i-\sigma_1-\sigma_2+\sigma_3 )  (-\tilde{m}_i+\sigma_3-\sigma_4 )  \\
&\quad \cdot (-\tilde{m}_i-\sigma_1+\sigma_3-\sigma_4 )  (-\tilde{m}_i+\sigma_1-\sigma_2+\sigma_3-\sigma_4 )  (-\tilde{m}_i-\sigma_1+\sigma_3-\sigma_5 )    \\
&\quad \cdot (-\tilde{m}_i+\sigma_1-\sigma_2+\sigma_3-\sigma_5 )  (-\tilde{m}_i+\sigma_3-2 \sigma_4+\sigma_5 )  (-\tilde{m}_i-\sigma_2+\sigma_3-\sigma_4+\sigma_5 )    \\
&\quad \cdot (-\tilde{m}_i-\sigma_2+\sigma_3-\sigma_6 )  (-\tilde{m}_i-\sigma_1+\sigma_3-\sigma_4+\sigma_5-\sigma_6 )    \\
&\quad \cdot (-\tilde{m}_i+\sigma_1-\sigma_2+\sigma_3-\sigma_4+\sigma_5-\sigma_6 )  (-\tilde{m}_i-\sigma_2+\sigma_3-\sigma_4+\sigma_6 )    \\
&\quad \cdot (-\tilde{m}_i-\sigma_2+\sigma_3-\sigma_5+\sigma_6 )  (-\tilde{m}_i+\sigma_3-\sigma_4-\sigma_5+\sigma_6 )  \\
&\quad \cdot (-\tilde{m}_i-\sigma_2+\sigma_3-\sigma_7 )  (-\tilde{m}_i+\sigma_3-\sigma_4-\sigma_7 )  (-\tilde{m}_i-\sigma_2+\sigma_3-\sigma_4+\sigma_5-\sigma_7 )  \\
&\quad \cdot (-\tilde{m}_i-\sigma_1+\sigma_3-\sigma_4+\sigma_6-\sigma_7 )  (-\tilde{m}_i+\sigma_1-\sigma_2+\sigma_3-\sigma_4+\sigma_6-\sigma_7 )    \\ 
&\quad \cdot (-\tilde{m}_i-\sigma_1+\sigma_3-\sigma_4+\sigma_7 )  (-\tilde{m}_i+\sigma_1-\sigma_2+\sigma_3-\sigma_4+\sigma_7 )    \\
&\quad \cdot (-\tilde{m}_i-\sigma_2+\sigma_3-\sigma_6+\sigma_7 )  (-\tilde{m}_i+\sigma_3-\sigma_4-\sigma_6+\sigma_7 )     \\
&\quad \cdot (-\tilde{m}_i+\sigma_3-2 \sigma_8 )  (-\tilde{m}_i-\sigma_1+\sigma_3-\sigma_8 )  (-\tilde{m}_i-\sigma_2+\sigma_3-\sigma_8 )    \\
&\quad \cdot (-\tilde{m}_i+\sigma_1-\sigma_2+\sigma_3-\sigma_8 )  (-\tilde{m}_i+\sigma_1+\sigma_3-\sigma_4-\sigma_8 )  (-\tilde{m}_i-\sigma_1+\sigma_2+\sigma_3-\sigma_4-\sigma_8 )    \\
&\quad \cdot (-\tilde{m}_i+\sigma_3-\sigma_5-\sigma_8 )  (-\tilde{m}_i-\sigma_2+\sigma_3+\sigma_4-\sigma_5-\sigma_8 )  (-\tilde{m}_i-\sigma_1+\sigma_3-\sigma_4+\sigma_5-\sigma_8 )    \\
&\quad \cdot (-\tilde{m}_i+\sigma_1-\sigma_2+\sigma_3-\sigma_4+\sigma_5-\sigma_8 )  (-\tilde{m}_i+\sigma_3-\sigma_6-\sigma_8 )     \\
&\quad \cdot (-\tilde{m}_i+\sigma_3-\sigma_4+\sigma_5-\sigma_6-\sigma_8 )  (-\tilde{m}_i+\sigma_3-\sigma_4+\sigma_6-\sigma_8 )    \\
&\quad \cdot (-\tilde{m}_i+\sigma_1-\sigma_2+\sigma_3-\sigma_5+\sigma_6-\sigma_8 )  (-\tilde{m}_i-\sigma_1+\sigma_3-\sigma_7-\sigma_8 )     \\
&\quad \cdot (-\tilde{m}_i+\sigma_3-\sigma_4+\sigma_5-\sigma_7-\sigma_8 )  (-\tilde{m}_i-\sigma_2+\sigma_3+\sigma_6-\sigma_7-\sigma_8 )    \\
&\quad \cdot (-\tilde{m}_i+\sigma_3-\sigma_5+\sigma_6-\sigma_7-\sigma_8 )  (-\tilde{m}_i-\sigma_2+\sigma_3+\sigma_7-\sigma_8 )  (-\tilde{m}_i+\sigma_3-\sigma_4+\sigma_7-\sigma_8 )   \\
&\quad \cdot (-\tilde{m}_i+\sigma_3-\sigma_5+\sigma_7-\sigma_8 )  (-\tilde{m}_i-\sigma_1+\sigma_3-\sigma_6+\sigma_7-\sigma_8 )    \\
&\quad \cdot (-\tilde{m}_i+\sigma_3-\sigma_4+\sigma_5-\sigma_6+\sigma_7-\sigma_8 )  (-\tilde{m}_i-\sigma_2+\sigma_3-\sigma_4+\sigma_8 )  \\
&\quad \cdot (-\tilde{m}_i+\sigma_1-\sigma_2+\sigma_3-\sigma_6+\sigma_7-\sigma_8 )(\tilde{m}_i+\sigma_2-2 \sigma_3+\sigma_4+\sigma_8 )^2  \\
&\quad \cdot (-\tilde{m}_i+\sigma_3-\sigma_4+\sigma_6-\sigma_7-\sigma_8 ) (-\tilde{m}_i+\sigma_1-\sigma_2+\sigma_3-\sigma_7-\sigma_8 )  \\
&\quad \cdot (-\tilde{m}_i-\sigma_2+\sigma_3+\sigma_5-\sigma_6-\sigma_8 ) (-\tilde{m}_i-\sigma_2+\sigma_3-\sigma_4+\sigma_5-\sigma_6+\sigma_7 ) \\
&\quad \cdot (-\tilde{m}_i-\sigma_2+\sigma_3-\sigma_5+\sigma_7 )  (-\tilde{m}_i-\sigma_2+\sigma_3-\sigma_5+\sigma_6-\sigma_7 ) \\
& \quad \cdot (-\tilde{m}_i-\sigma_1+\sigma_3-\sigma_5+\sigma_6-\sigma_8 )   .
\end{align*}

The fourth ring relation is 
\begin{align*}
\prod_{i=1}^n &
(-\tilde{m}_i+\sigma_2-\sigma_4)(-\tilde{m}_i+\sigma_3-\sigma_4)(-\tilde{m}_i-\sigma_1+\sigma_3-\sigma_4)  \\
&\quad \cdot (-\tilde{m}_i+\sigma_1-\sigma_2+\sigma_3-\sigma_4)(-\tilde{m}_i+\sigma_3-2\sigma_4+\sigma_5)^2  \\
&\quad \cdot (-\tilde{m}_i-\sigma_4+\sigma_5)(-\tilde{m}_i+\sigma_1-\sigma_4+\sigma_5)(-\tilde{m}_i-\sigma_1+\sigma_2-\sigma_4+\sigma_5) \\
&\quad \cdot (-\tilde{m}_i-\sigma_2+\sigma_3-\sigma_4+\sigma_5)(-\tilde{m}_i+\sigma_2-\sigma_4+\sigma_5-\sigma_6)  \\%
&\quad \cdot (-\tilde{m}_i-\sigma_1+\sigma_3-\sigma_4+\sigma_5-\sigma_6)(-\tilde{m}_i+\sigma_1-\sigma_2+\sigma_3-\sigma_4+\sigma_5-\sigma_6)  \\
&\quad \cdot (-\tilde{m}_i-\sigma_4+2\sigma_5-\sigma_6)(-\tilde{m}_i+\sigma_1-\sigma_4+\sigma_6)(-\tilde{m}_i-\sigma_1+\sigma_2-\sigma_4+\sigma_6)  \\
&\quad \cdot (-\tilde{m}_i-\sigma_2+\sigma_3-\sigma_4+\sigma_6)(-\tilde{m}_i+\sigma_3-\sigma_4-\sigma_5+\sigma_6)(-\tilde{m}_i+\sigma_3-\sigma_4-\sigma_7) \\
&\quad \cdot (-\tilde{m}_i+\sigma_1-\sigma_4+\sigma_5-\sigma_7)(-\tilde{m}_i-\sigma_1+\sigma_2-\sigma_4+\sigma_5-\sigma_7) \\
&\quad \cdot (-\tilde{m}_i-\sigma_2+\sigma_3-\sigma_4+\sigma_5-\sigma_7)(-\tilde{m}_i+\sigma_2-\sigma_4+\sigma_6-\sigma_7) \\
&\quad \cdot (-\tilde{m}_i-\sigma_1+\sigma_3-\sigma_4+\sigma_6-\sigma_7)(-\tilde{m}_i+\sigma_1-\sigma_2+\sigma_3-\sigma_4+\sigma_6-\sigma_7)  \\
&\quad \cdot (-\tilde{m}_i-\sigma_4+\sigma_5+\sigma_6-\sigma_7)(-\tilde{m}_i+\sigma_2-\sigma_4+\sigma_7)(-\tilde{m}_i-\sigma_1+\sigma_3-\sigma_4+\sigma_7) \\
&\quad \cdot (-\tilde{m}_i+\sigma_1-\sigma_2+\sigma_3-\sigma_4+\sigma_7)(-\tilde{m}_i-\sigma_4+\sigma_5+\sigma_7)(-\tilde{m}_i+\sigma_3-\sigma_4-\sigma_6+\sigma_7) \\
&\quad \cdot (-\tilde{m}_i+\sigma_1-\sigma_4+\sigma_5-\sigma_6+\sigma_7)(-\tilde{m}_i-\sigma_1+\sigma_2-\sigma_4+\sigma_5-\sigma_6+\sigma_7) \\
&\quad \cdot (-\tilde{m}_i-\sigma_2+\sigma_3-\sigma_4+\sigma_5-\sigma_6+\sigma_7)(-\tilde{m}_i+\sigma_1+\sigma_3-\sigma_4-\sigma_8) \\
&\quad \cdot (-\tilde{m}_i-\sigma_1+\sigma_2+\sigma_3-\sigma_4-\sigma_8)(-\tilde{m}_i-\sigma_2+2\sigma_3-\sigma_4-\sigma_8) \\
&\quad \cdot (-\tilde{m}_i+\sigma_2-\sigma_4+\sigma_5-\sigma_8)(-\tilde{m}_i-\sigma_1+\sigma_3-\sigma_4+\sigma_5-\sigma_8)  \\
&\quad \cdot (-\tilde{m}_i+\sigma_1-\sigma_2+\sigma_3-\sigma_4+\sigma_5-\sigma_8)(-\tilde{m}_i+\sigma_3-\sigma_4+\sigma_5-\sigma_6-\sigma_8) \\
&\quad \cdot (-\tilde{m}_i+\sigma_3-\sigma_4+\sigma_6-\sigma_8)(-\tilde{m}_i+\sigma_3-\sigma_4+\sigma_5-\sigma_7-\sigma_8) \\
&\quad \cdot (-\tilde{m}_i+\sigma_3-\sigma_4+\sigma_6-\sigma_7-\sigma_8)(-\tilde{m}_i+\sigma_3-\sigma_4+\sigma_7-\sigma_8) \\
&\quad \cdot (-\tilde{m}_i+\sigma_3-\sigma_4+\sigma_5-\sigma_6+\sigma_7-\sigma_8)(-\tilde{m}_i+\sigma_1-\sigma_4+\sigma_8)  \\
&\quad \cdot (-\tilde{m}_i-\sigma_1+\sigma_2-\sigma_4+\sigma_8)(-\tilde{m}_i-\sigma_2+\sigma_3-\sigma_4+\sigma_8)(-\tilde{m}_i-\sigma_1-\sigma_4+\sigma_5+\sigma_8) \\
&\quad \cdot (-\tilde{m}_i+\sigma_1-\sigma_2-\sigma_4+\sigma_5+\sigma_8)(-\tilde{m}_i+\sigma_2-\sigma_3-\sigma_4+\sigma_5+\sigma_8) \\
&\quad \cdot (-\tilde{m}_i-\sigma_4+\sigma_5-\sigma_6+\sigma_8)(-\tilde{m}_i-\sigma_4+\sigma_6+\sigma_8)(-\tilde{m}_i-\sigma_4+\sigma_5-\sigma_7+\sigma_8) \\ 
&\quad \cdot (-\tilde{m}_i-\sigma_4+\sigma_6-\sigma_7+\sigma_8)(-\tilde{m}_i-\sigma_4+\sigma_7+\sigma_8)(-\tilde{m}_i-\sigma_4+\sigma_5-\sigma_6+\sigma_7+\sigma_8) 
\end{align*}
\begin{align*}
& = \prod_{i=1}^n
(-\tilde{m}_i-\sigma_2+\sigma_4)(-\tilde{m}_i-\sigma_3+\sigma_4)(-\tilde{m}_i+\sigma_1-\sigma_3+\sigma_4) \\ 
&\quad \cdot (-\tilde{m}_i-\sigma_1+\sigma_2-\sigma_3+\sigma_4)(-\tilde{m}_i+\sigma_4-\sigma_5)(-\tilde{m}_i-\sigma_1+\sigma_4-\sigma_5)  \\ 
&\quad \cdot (-\tilde{m}_i+\sigma_1-\sigma_2+\sigma_4-\sigma_5)(-\tilde{m}_i+\sigma_2-\sigma_3+\sigma_4-\sigma_5)(\tilde{m}_i+\sigma_3-2\sigma_4+\sigma_5)^2  \\ 
&\quad \cdot (-\tilde{m}_i-\sigma_1+\sigma_4-\sigma_6)(-\tilde{m}_i+\sigma_1-\sigma_2+\sigma_4-\sigma_6)(-\tilde{m}_i+\sigma_2-\sigma_3+\sigma_4-\sigma_6)\\ 
&\quad \cdot (-\tilde{m}_i-\sigma_3+\sigma_4+\sigma_5-\sigma_6)(-\tilde{m}_i+\sigma_4-2\sigma_5+\sigma_6)(-\tilde{m}_i-\sigma_2+\sigma_4-\sigma_5+\sigma_6) \\ 
&\quad \cdot (-\tilde{m}_i+\sigma_1-\sigma_3+\sigma_4-\sigma_5+\sigma_6)(-\tilde{m}_i-\sigma_1+\sigma_2-\sigma_3+\sigma_4-\sigma_5+\sigma_6)  \\ 
&\quad \cdot (-\tilde{m}_i-\sigma_2+\sigma_4-\sigma_7)(-\tilde{m}_i+\sigma_1-\sigma_3+\sigma_4-\sigma_7)(-\tilde{m}_i-\sigma_1+\sigma_2-\sigma_3+\sigma_4-\sigma_7)  \\ 
&\quad \cdot (-\tilde{m}_i+\sigma_4-\sigma_5-\sigma_7)(-\tilde{m}_i-\sigma_3+\sigma_4+\sigma_6-\sigma_7)(-\tilde{m}_i-\sigma_1+\sigma_4-\sigma_5+\sigma_6-\sigma_7) \\ 
&\quad \cdot (-\tilde{m}_i+\sigma_1-\sigma_2+\sigma_4-\sigma_5+\sigma_6-\sigma_7)(-\tilde{m}_i+\sigma_2-\sigma_3+\sigma_4-\sigma_5+\sigma_6-\sigma_7) \\ 
&\quad \cdot (-\tilde{m}_i-\sigma_3+\sigma_4+\sigma_7)(-\tilde{m}_i-\sigma_1+\sigma_4-\sigma_5+\sigma_7)(-\tilde{m}_i+\sigma_1-\sigma_2+\sigma_4-\sigma_5+\sigma_7) \\ 
&\quad \cdot (-\tilde{m}_i+\sigma_2-\sigma_3+\sigma_4-\sigma_5+\sigma_7)(-\tilde{m}_i-\sigma_2+\sigma_4-\sigma_6+\sigma_7)  \\ 
&\quad \cdot (-\tilde{m}_i+\sigma_1-\sigma_3+\sigma_4-\sigma_6+\sigma_7)(-\tilde{m}_i-\sigma_1+\sigma_2-\sigma_3+\sigma_4-\sigma_6+\sigma_7)  \\ 
&\quad \cdot (-\tilde{m}_i+\sigma_4-\sigma_5-\sigma_6+\sigma_7)(-\tilde{m}_i-\sigma_1+\sigma_4-\sigma_8)(-\tilde{m}_i+\sigma_1-\sigma_2+\sigma_4-\sigma_8)  \\ 
&\quad \cdot (-\tilde{m}_i+\sigma_2-\sigma_3+\sigma_4-\sigma_8)(-\tilde{m}_i+\sigma_1+\sigma_4-\sigma_5-\sigma_8)  \\ 
&\quad \cdot (-\tilde{m}_i-\sigma_1+\sigma_2+\sigma_4-\sigma_5-\sigma_8)(-\tilde{m}_i-\sigma_2+\sigma_3+\sigma_4-\sigma_5-\sigma_8) \\ 
&\quad \cdot (-\tilde{m}_i+\sigma_4-\sigma_6-\sigma_8)(-\tilde{m}_i+\sigma_4-\sigma_5+\sigma_6-\sigma_8)(-\tilde{m}_i+\sigma_4-\sigma_7-\sigma_8)  \\ 
&\quad \cdot (-\tilde{m}_i+\sigma_4-\sigma_5+\sigma_6-\sigma_7-\sigma_8)(-\tilde{m}_i+\sigma_4-\sigma_5+\sigma_7-\sigma_8)  \\ 
&\quad \cdot (-\tilde{m}_i+\sigma_4-\sigma_6+\sigma_7-\sigma_8)(-\tilde{m}_i+\sigma_2-2\sigma_3+\sigma_4+\sigma_8)(-\tilde{m}_i-\sigma_1-\sigma_3+\sigma_4+\sigma_8)  \\ 
&\quad \cdot (-\tilde{m}_i+\sigma_1-\sigma_2-\sigma_3+\sigma_4+\sigma_8)(-\tilde{m}_i-\sigma_2+\sigma_4-\sigma_5+\sigma_8) \\ 
&\quad \cdot (-\tilde{m}_i+\sigma_1-\sigma_3+\sigma_4-\sigma_5+\sigma_8)(-\tilde{m}_i-\sigma_1+\sigma_2-\sigma_3+\sigma_4-\sigma_5+\sigma_8)  \\ 
&\quad \cdot (-\tilde{m}_i-\sigma_3+\sigma_4-\sigma_6+\sigma_8)(-\tilde{m}_i-\sigma_3+\sigma_4-\sigma_5+\sigma_6+\sigma_8)  \\ 
&\quad \cdot (-\tilde{m}_i-\sigma_3+\sigma_4-\sigma_7+\sigma_8)(-\tilde{m}_i-\sigma_3+\sigma_4-\sigma_5+\sigma_6-\sigma_7+\sigma_8)  \\ 
&\quad \cdot (-\tilde{m}_i-\sigma_3+\sigma_4-\sigma_5+\sigma_7+\sigma_8)(-\tilde{m}_i-\sigma_3+\sigma_4-\sigma_6+\sigma_7+\sigma_8) .
\end{align*}

The fifth ring relation is
\begin{align*}
\prod_{i=1}^n &
(-\tilde{m}_i+\sigma_2-\sigma_5)(-\tilde{m}_i-\sigma_1+\sigma_3-\sigma_5)(-\tilde{m}_i+\sigma_1-\sigma_2+\sigma_3-\sigma_5)  \\ 
&\quad \cdot (-\tilde{m}_i+\sigma_4-\sigma_5)(-\tilde{m}_i-\sigma_1+\sigma_4-\sigma_5)(-\tilde{m}_i+\sigma_1-\sigma_2+\sigma_4-\sigma_5)  \\ 
&\quad \cdot (-\tilde{m}_i+\sigma_2-\sigma_3+\sigma_4-\sigma_5)(-\tilde{m}_i-\sigma_3+2\sigma_4-\sigma_5)(-\tilde{m}_i+\sigma_4-2\sigma_5+\sigma_6)^2  \\ 
&\quad \cdot (-\tilde{m}_i-\sigma_5+\sigma_6)(-\tilde{m}_i+\sigma_1-\sigma_5+\sigma_6)(-\tilde{m}_i-\sigma_1+\sigma_2-\sigma_5+\sigma_6) \\ 
&\quad \cdot (-\tilde{m}_i-\sigma_2+\sigma_3-\sigma_5+\sigma_6)(-\tilde{m}_i+\sigma_3-\sigma_4-\sigma_5+\sigma_6) \\ 
&\quad \cdot (-\tilde{m}_i-\sigma_2+\sigma_4-\sigma_5+\sigma_6)(-\tilde{m}_i+\sigma_1-\sigma_3+\sigma_4-\sigma_5+\sigma_6)  \\ 
&\quad \cdot (-\tilde{m}_i-\sigma_1+\sigma_2-\sigma_3+\sigma_4-\sigma_5+\sigma_6)(-\tilde{m}_i+\sigma_4-\sigma_5-\sigma_7)  \\ 
&\quad \cdot (-\tilde{m}_i+\sigma_1-\sigma_5+\sigma_6-\sigma_7)(-\tilde{m}_i-\sigma_1+\sigma_2-\sigma_5+\sigma_6-\sigma_7)  \\ 
&\quad \cdot (-\tilde{m}_i-\sigma_2+\sigma_3-\sigma_5+\sigma_6-\sigma_7)(-\tilde{m}_i-\sigma_1+\sigma_4-\sigma_5+\sigma_6-\sigma_7)  \\ 
&\quad \cdot (-\tilde{m}_i+\sigma_1-\sigma_2+\sigma_4-\sigma_5+\sigma_6-\sigma_7)(-\tilde{m}_i+\sigma_2-\sigma_3+\sigma_4-\sigma_5+\sigma_6-\sigma_7)  \\ 
&\quad \cdot (-\tilde{m}_i-\sigma_5+2\sigma_6-\sigma_7)(-\tilde{m}_i+\sigma_1-\sigma_5+\sigma_7)(-\tilde{m}_i-\sigma_1+\sigma_2-\sigma_5+\sigma_7)  \\ 
&\quad \cdot (-\tilde{m}_i-\sigma_2+\sigma_3-\sigma_5+\sigma_7)(-\tilde{m}_i-\sigma_1+\sigma_4-\sigma_5+\sigma_7)  \\ 
&\quad \cdot (-\tilde{m}_i+\sigma_1-\sigma_2+\sigma_4-\sigma_5+\sigma_7)(-\tilde{m}_i+\sigma_2-\sigma_3+\sigma_4-\sigma_5+\sigma_7)  \\ 
&\quad \cdot (-\tilde{m}_i+\sigma_4-\sigma_5-\sigma_6+\sigma_7)(-\tilde{m}_i-\sigma_5+\sigma_6+\sigma_7)(-\tilde{m}_i+\sigma_3-\sigma_5-\sigma_8)  \\ 
&\quad \cdot (-\tilde{m}_i+\sigma_1+\sigma_4-\sigma_5-\sigma_8)(-\tilde{m}_i-\sigma_1+\sigma_2+\sigma_4-\sigma_5-\sigma_8)  \\ 
&\quad \cdot (-\tilde{m}_i-\sigma_2+\sigma_3+\sigma_4-\sigma_5-\sigma_8)(-\tilde{m}_i+\sigma_2-\sigma_5+\sigma_6-\sigma_8)  \\ 
&\quad \cdot (-\tilde{m}_i-\sigma_1+\sigma_3-\sigma_5+\sigma_6-\sigma_8)(-\tilde{m}_i+\sigma_1-\sigma_2+\sigma_3-\sigma_5+\sigma_6-\sigma_8) \\ 
&\quad \cdot (-\tilde{m}_i+\sigma_4-\sigma_5+\sigma_6-\sigma_8)(-\tilde{m}_i+\sigma_3-\sigma_5+\sigma_6-\sigma_7-\sigma_8)  \\ 
&\quad \cdot (-\tilde{m}_i+\sigma_4-\sigma_5+\sigma_6-\sigma_7-\sigma_8)(-\tilde{m}_i+\sigma_3-\sigma_5+\sigma_7-\sigma_8)  \\ 
&\quad \cdot (-\tilde{m}_i+\sigma_4-\sigma_5+\sigma_7-\sigma_8)(-\tilde{m}_i-\sigma_5+\sigma_8)(-\tilde{m}_i-\sigma_2+\sigma_4-\sigma_5+\sigma_8) \\ 
&\quad \cdot (-\tilde{m}_i+\sigma_1-\sigma_3+\sigma_4-\sigma_5+\sigma_8)(-\tilde{m}_i-\sigma_1+\sigma_2-\sigma_3+\sigma_4-\sigma_5+\sigma_8)  \\ 
&\quad \cdot (-\tilde{m}_i-\sigma_1-\sigma_5+\sigma_6+\sigma_8)(-\tilde{m}_i+\sigma_1-\sigma_2-\sigma_5+\sigma_6+\sigma_8) \\ 
&\quad \cdot (-\tilde{m}_i+\sigma_2-\sigma_3-\sigma_5+\sigma_6+\sigma_8)(-\tilde{m}_i-\sigma_3+\sigma_4-\sigma_5+\sigma_6+\sigma_8)  \\ 
&\quad \cdot (-\tilde{m}_i-\sigma_5+\sigma_6-\sigma_7+\sigma_8)(-\tilde{m}_i-\sigma_3+\sigma_4-\sigma_5+\sigma_6-\sigma_7+\sigma_8)  \\ 
&\quad \cdot (-\tilde{m}_i-\sigma_5+\sigma_7+\sigma_8)(-\tilde{m}_i-\sigma_3+\sigma_4-\sigma_5+\sigma_7+\sigma_8) \\
\end{align*}
\begin{align*}
=& \prod_{i=1}^n (-\tilde{m}_i-\sigma_2+\sigma_5)(-\tilde{m}_i+\sigma_1-\sigma_3+\sigma_5)(-\tilde{m}_i-\sigma_1+\sigma_2-\sigma_3+\sigma_5)  \\ 
&\quad \cdot (-\tilde{m}_i+\sigma_3-2\sigma_4+\sigma_5)(-\tilde{m}_i-\sigma_4+\sigma_5)(-\tilde{m}_i+\sigma_1-\sigma_4+\sigma_5) \\ 
&\quad \cdot (-\tilde{m}_i-\sigma_1+\sigma_2-\sigma_4+\sigma_5)(-\tilde{m}_i-\sigma_2+\sigma_3-\sigma_4+\sigma_5)(-\tilde{m}_i+\sigma_5-\sigma_6) \\ 
&\quad \cdot (-\tilde{m}_i-\sigma_1+\sigma_5-\sigma_6)(-\tilde{m}_i+\sigma_1-\sigma_2+\sigma_5-\sigma_6)(-\tilde{m}_i+\sigma_2-\sigma_3+\sigma_5-\sigma_6)  \\ 
&\quad \cdot (-\tilde{m}_i+\sigma_2-\sigma_4+\sigma_5-\sigma_6)(-\tilde{m}_i-\sigma_1+\sigma_3-\sigma_4+\sigma_5-\sigma_6)  \\ 
&\quad \cdot (-\tilde{m}_i+\sigma_1-\sigma_2+\sigma_3-\sigma_4+\sigma_5-\sigma_6)(-\tilde{m}_i-\sigma_3+\sigma_4+\sigma_5-\sigma_6)  \\ 
&\quad \cdot (\tilde{m}_i+\sigma_4-2\sigma_5+\sigma_6){}^2(-\tilde{m}_i-\sigma_1+\sigma_5-\sigma_7)(-\tilde{m}_i+\sigma_1-\sigma_2+\sigma_5-\sigma_7)  \\ 
&\quad \cdot (-\tilde{m}_i+\sigma_2-\sigma_3+\sigma_5-\sigma_7)(-\tilde{m}_i+\sigma_1-\sigma_4+\sigma_5-\sigma_7)  \\ 
&\quad \cdot (-\tilde{m}_i-\sigma_1+\sigma_2-\sigma_4+\sigma_5-\sigma_7)(-\tilde{m}_i-\sigma_2+\sigma_3-\sigma_4+\sigma_5-\sigma_7)  \\ 
&\quad \cdot (-\tilde{m}_i+\sigma_5-\sigma_6-\sigma_7)(-\tilde{m}_i-\sigma_4+\sigma_5+\sigma_6-\sigma_7)(-\tilde{m}_i-\sigma_4+\sigma_5+\sigma_7)  \\ 
&\quad \cdot (-\tilde{m}_i+\sigma_5-2\sigma_6+\sigma_7)(-\tilde{m}_i-\sigma_1+\sigma_5-\sigma_6+\sigma_7)(-\tilde{m}_i+\sigma_1-\sigma_2+\sigma_5-\sigma_6+\sigma_7)  \\ 
&\quad \cdot (-\tilde{m}_i+\sigma_2-\sigma_3+\sigma_5-\sigma_6+\sigma_7)(-\tilde{m}_i+\sigma_1-\sigma_4+\sigma_5-\sigma_6+\sigma_7) \\ 
&\quad \cdot (-\tilde{m}_i-\sigma_1+\sigma_2-\sigma_4+\sigma_5-\sigma_6+\sigma_7)(-\tilde{m}_i-\sigma_2+\sigma_3-\sigma_4+\sigma_5-\sigma_6+\sigma_7)  \\ 
&\quad \cdot (-\tilde{m}_i+\sigma_5-\sigma_8)(-\tilde{m}_i+\sigma_2-\sigma_4+\sigma_5-\sigma_8)(-\tilde{m}_i-\sigma_1+\sigma_3-\sigma_4+\sigma_5-\sigma_8) \\ 
&\quad \cdot (-\tilde{m}_i+\sigma_1-\sigma_2+\sigma_3-\sigma_4+\sigma_5-\sigma_8)(-\tilde{m}_i+\sigma_1+\sigma_5-\sigma_6-\sigma_8)  \\ 
&\quad \cdot (-\tilde{m}_i-\sigma_1+\sigma_2+\sigma_5-\sigma_6-\sigma_8)(-\tilde{m}_i-\sigma_2+\sigma_3+\sigma_5-\sigma_6-\sigma_8) \\ 
&\quad \cdot (-\tilde{m}_i+\sigma_3-\sigma_4+\sigma_5-\sigma_6-\sigma_8)(-\tilde{m}_i+\sigma_5-\sigma_7-\sigma_8)  \\ 
&\quad \cdot (-\tilde{m}_i+\sigma_3-\sigma_4+\sigma_5-\sigma_7-\sigma_8)(-\tilde{m}_i+\sigma_5-\sigma_6+\sigma_7-\sigma_8)  \\ 
&\quad \cdot (-\tilde{m}_i+\sigma_3-\sigma_4+\sigma_5-\sigma_6+\sigma_7-\sigma_8)(-\tilde{m}_i-\sigma_3+\sigma_5+\sigma_8)  \\ 
&\quad \cdot (-\tilde{m}_i-\sigma_1-\sigma_4+\sigma_5+\sigma_8)(-\tilde{m}_i+\sigma_1-\sigma_2-\sigma_4+\sigma_5+\sigma_8)  \\ 
&\quad \cdot (-\tilde{m}_i+\sigma_2-\sigma_3-\sigma_4+\sigma_5+\sigma_8)(-\tilde{m}_i-\sigma_2+\sigma_5-\sigma_6+\sigma_8)  \\ 
&\quad \cdot (-\tilde{m}_i+\sigma_1-\sigma_3+\sigma_5-\sigma_6+\sigma_8)(-\tilde{m}_i-\sigma_1+\sigma_2-\sigma_3+\sigma_5-\sigma_6+\sigma_8)  \\ 
&\quad \cdot (-\tilde{m}_i-\sigma_4+\sigma_5-\sigma_6+\sigma_8)(-\tilde{m}_i-\sigma_3+\sigma_5-\sigma_7+\sigma_8)(-\tilde{m}_i-\sigma_4+\sigma_5-\sigma_7+\sigma_8)  \\ 
&\quad \cdot (-\tilde{m}_i-\sigma_3+\sigma_5-\sigma_6+\sigma_7+\sigma_8)(-\tilde{m}_i-\sigma_4+\sigma_5-\sigma_6+\sigma_7+\sigma_8) .
\end{align*}

The sixth ring relation is
\begin{align*}
\prod_{i=1}^n & (-\tilde{m}_i+\sigma_1-\sigma_6)(-\tilde{m}_i-\sigma_1+\sigma_2-\sigma_6)(-\tilde{m}_i-\sigma_2+\sigma_3-\sigma_6) \\ 
&\quad \cdot (-\tilde{m}_i-\sigma_1+\sigma_4-\sigma_6)(-\tilde{m}_i+\sigma_1-\sigma_2+\sigma_4-\sigma_6)(-\tilde{m}_i+\sigma_2-\sigma_3+\sigma_4-\sigma_6)  \\ 
&\quad \cdot (-\tilde{m}_i+\sigma_5-\sigma_6)(-\tilde{m}_i-\sigma_1+\sigma_5-\sigma_6)(-\tilde{m}_i+\sigma_1-\sigma_2+\sigma_5-\sigma_6) \\ 
&\quad \cdot (-\tilde{m}_i+\sigma_2-\sigma_3+\sigma_5-\sigma_6)(-\tilde{m}_i+\sigma_2-\sigma_4+\sigma_5-\sigma_6) \\ 
&\quad \cdot (-\tilde{m}_i-\sigma_1+\sigma_3-\sigma_4+\sigma_5-\sigma_6)(-\tilde{m}_i+\sigma_1-\sigma_2+\sigma_3-\sigma_4+\sigma_5-\sigma_6)  \\ 
&\quad \cdot (-\tilde{m}_i-\sigma_3+\sigma_4+\sigma_5-\sigma_6)(-\tilde{m}_i-\sigma_4+2\sigma_5-\sigma_6)(-\tilde{m}_i+\sigma_5-\sigma_6-\sigma_7)  \\ 
&\quad \cdot (-\tilde{m}_i+\sigma_5-2\sigma_6+\sigma_7){}^2(-\tilde{m}_i-\sigma_6+\sigma_7)(-\tilde{m}_i+\sigma_1-\sigma_6+\sigma_7) \\ 
&\quad \cdot (-\tilde{m}_i-\sigma_1+\sigma_2-\sigma_6+\sigma_7)(-\tilde{m}_i-\sigma_2+\sigma_3-\sigma_6+\sigma_7)  \\ 
&\quad \cdot (-\tilde{m}_i+\sigma_3-\sigma_4-\sigma_6+\sigma_7)(-\tilde{m}_i-\sigma_2+\sigma_4-\sigma_6+\sigma_7)  \\ 
&\quad \cdot (-\tilde{m}_i+\sigma_1-\sigma_3+\sigma_4-\sigma_6+\sigma_7)(-\tilde{m}_i-\sigma_1+\sigma_2-\sigma_3+\sigma_4-\sigma_6+\sigma_7)  \\ 
&\quad \cdot (-\tilde{m}_i+\sigma_4-\sigma_5-\sigma_6+\sigma_7)(-\tilde{m}_i-\sigma_1+\sigma_5-\sigma_6+\sigma_7)  \\ 
&\quad \cdot (-\tilde{m}_i+\sigma_1-\sigma_2+\sigma_5-\sigma_6+\sigma_7)(-\tilde{m}_i+\sigma_2-\sigma_3+\sigma_5-\sigma_6+\sigma_7) \\ 
&\quad \cdot (-\tilde{m}_i+\sigma_1-\sigma_4+\sigma_5-\sigma_6+\sigma_7)(-\tilde{m}_i-\sigma_1+\sigma_2-\sigma_4+\sigma_5-\sigma_6+\sigma_7) \\ 
&\quad \cdot (-\tilde{m}_i-\sigma_2+\sigma_3-\sigma_4+\sigma_5-\sigma_6+\sigma_7)(-\tilde{m}_i-\sigma_6+2\sigma_7)(-\tilde{m}_i+\sigma_3-\sigma_6-\sigma_8) \\ 
&\quad \cdot (-\tilde{m}_i+\sigma_4-\sigma_6-\sigma_8)(-\tilde{m}_i+\sigma_1+\sigma_5-\sigma_6-\sigma_8) \\ 
&\quad \cdot (-\tilde{m}_i-\sigma_1+\sigma_2+\sigma_5-\sigma_6-\sigma_8)(-\tilde{m}_i-\sigma_2+\sigma_3+\sigma_5-\sigma_6-\sigma_8)  \\ 
&\quad \cdot (-\tilde{m}_i+\sigma_3-\sigma_4+\sigma_5-\sigma_6-\sigma_8)(-\tilde{m}_i+\sigma_2-\sigma_6+\sigma_7-\sigma_8) \\ 
&\quad \cdot (-\tilde{m}_i-\sigma_1+\sigma_3-\sigma_6+\sigma_7-\sigma_8)(-\tilde{m}_i+\sigma_1-\sigma_2+\sigma_3-\sigma_6+\sigma_7-\sigma_8)  \\ 
&\quad \cdot (-\tilde{m}_i+\sigma_4-\sigma_6+\sigma_7-\sigma_8)(-\tilde{m}_i+\sigma_5-\sigma_6+\sigma_7-\sigma_8) \\ 
&\quad \cdot (-\tilde{m}_i+\sigma_3-\sigma_4+\sigma_5-\sigma_6+\sigma_7-\sigma_8)(-\tilde{m}_i-\sigma_6+\sigma_8)  \\ 
&\quad \cdot (-\tilde{m}_i-\sigma_3+\sigma_4-\sigma_6+\sigma_8)(-\tilde{m}_i-\sigma_2+\sigma_5-\sigma_6+\sigma_8)  \\ 
&\quad \cdot (-\tilde{m}_i+\sigma_1-\sigma_3+\sigma_5-\sigma_6+\sigma_8)(-\tilde{m}_i-\sigma_1+\sigma_2-\sigma_3+\sigma_5-\sigma_6+\sigma_8)  \\ 
&\quad \cdot (-\tilde{m}_i-\sigma_4+\sigma_5-\sigma_6+\sigma_8)(-\tilde{m}_i-\sigma_1-\sigma_6+\sigma_7+\sigma_8)  \\ 
&\quad \cdot (-\tilde{m}_i+\sigma_1-\sigma_2-\sigma_6+\sigma_7+\sigma_8)(-\tilde{m}_i+\sigma_2-\sigma_3-\sigma_6+\sigma_7+\sigma_8) \\ 
&\quad \cdot (-\tilde{m}_i-\sigma_3+\sigma_4-\sigma_6+\sigma_7+\sigma_8)(-\tilde{m}_i-\sigma_3+\sigma_5-\sigma_6+\sigma_7+\sigma_8)  \\ 
&\quad \cdot (-\tilde{m}_i-\sigma_4+\sigma_5-\sigma_6+\sigma_7+\sigma_8)
\\
\end{align*}
\begin{align*}
& =\prod_{i=1}^n 
(-\tilde{m}_i-\sigma_1+\sigma_6)(-\tilde{m}_i+\sigma_1-\sigma_2+\sigma_6)(-\tilde{m}_i+\sigma_2-\sigma_3+\sigma_6)  \\ 
&\quad \cdot (-\tilde{m}_i+\sigma_1-\sigma_4+\sigma_6)(-\tilde{m}_i-\sigma_1+\sigma_2-\sigma_4+\sigma_6)(-\tilde{m}_i-\sigma_2+\sigma_3-\sigma_4+\sigma_6)  \\ 
&\quad \cdot (-\tilde{m}_i+\sigma_4-2\sigma_5+\sigma_6)(-\tilde{m}_i-\sigma_5+\sigma_6)(-\tilde{m}_i+\sigma_1-\sigma_5+\sigma_6)  \\ 
&\quad \cdot (-\tilde{m}_i-\sigma_1+\sigma_2-\sigma_5+\sigma_6)(-\tilde{m}_i-\sigma_2+\sigma_3-\sigma_5+\sigma_6)  \\ 
&\quad \cdot (-\tilde{m}_i+\sigma_3-\sigma_4-\sigma_5+\sigma_6)(-\tilde{m}_i-\sigma_2+\sigma_4-\sigma_5+\sigma_6)  \\ 
&\quad \cdot (-\tilde{m}_i+\sigma_1-\sigma_3+\sigma_4-\sigma_5+\sigma_6)(-\tilde{m}_i-\sigma_1+\sigma_2-\sigma_3+\sigma_4-\sigma_5+\sigma_6) \\ 
&\quad \cdot (-\tilde{m}_i+\sigma_6-2\sigma_7)(-\tilde{m}_i+\sigma_6-\sigma_7)(-\tilde{m}_i-\sigma_1+\sigma_6-\sigma_7)  \\ 
&\quad \cdot (-\tilde{m}_i+\sigma_1-\sigma_2+\sigma_6-\sigma_7)(-\tilde{m}_i+\sigma_2-\sigma_3+\sigma_6-\sigma_7) \\ 
&\quad \cdot (-\tilde{m}_i+\sigma_2-\sigma_4+\sigma_6-\sigma_7)(-\tilde{m}_i-\sigma_1+\sigma_3-\sigma_4+\sigma_6-\sigma_7)  \\ 
&\quad \cdot (-\tilde{m}_i+\sigma_1-\sigma_2+\sigma_3-\sigma_4+\sigma_6-\sigma_7)(-\tilde{m}_i-\sigma_3+\sigma_4+\sigma_6-\sigma_7)  \\ 
&\quad \cdot (-\tilde{m}_i+\sigma_1-\sigma_5+\sigma_6-\sigma_7)(-\tilde{m}_i-\sigma_1+\sigma_2-\sigma_5+\sigma_6-\sigma_7)  \\ 
&\quad \cdot (-\tilde{m}_i-\sigma_2+\sigma_3-\sigma_5+\sigma_6-\sigma_7)(-\tilde{m}_i-\sigma_1+\sigma_4-\sigma_5+\sigma_6-\sigma_7) \\ 
&\quad \cdot (-\tilde{m}_i+\sigma_1-\sigma_2+\sigma_4-\sigma_5+\sigma_6-\sigma_7)(-\tilde{m}_i+\sigma_2-\sigma_3+\sigma_4-\sigma_5+\sigma_6-\sigma_7)  \\ 
&\quad \cdot (-\tilde{m}_i-\sigma_4+\sigma_5+\sigma_6-\sigma_7)(\tilde{m}_i+\sigma_5-2\sigma_6+\sigma_7){}^2(-\tilde{m}_i-\sigma_5+\sigma_6+\sigma_7)  \\ 
&\quad \cdot (-\tilde{m}_i+\sigma_6-\sigma_8)(-\tilde{m}_i+\sigma_3-\sigma_4+\sigma_6-\sigma_8)(-\tilde{m}_i+\sigma_2-\sigma_5+\sigma_6-\sigma_8) \\ 
&\quad \cdot (-\tilde{m}_i-\sigma_1+\sigma_3-\sigma_5+\sigma_6-\sigma_8)(-\tilde{m}_i+\sigma_1-\sigma_2+\sigma_3-\sigma_5+\sigma_6-\sigma_8)  \\ 
&\quad \cdot (-\tilde{m}_i+\sigma_4-\sigma_5+\sigma_6-\sigma_8)(-\tilde{m}_i+\sigma_1+\sigma_6-\sigma_7-\sigma_8)  \\ 
&\quad \cdot (-\tilde{m}_i-\sigma_1+\sigma_2+\sigma_6-\sigma_7-\sigma_8)(-\tilde{m}_i-\sigma_2+\sigma_3+\sigma_6-\sigma_7-\sigma_8)  \\ 
&\quad \cdot (-\tilde{m}_i+\sigma_3-\sigma_4+\sigma_6-\sigma_7-\sigma_8)(-\tilde{m}_i+\sigma_3-\sigma_5+\sigma_6-\sigma_7-\sigma_8)  \\ 
&\quad \cdot (-\tilde{m}_i+\sigma_4-\sigma_5+\sigma_6-\sigma_7-\sigma_8)(-\tilde{m}_i-\sigma_3+\sigma_6+\sigma_8)(-\tilde{m}_i-\sigma_4+\sigma_6+\sigma_8)  \\ 
&\quad \cdot (-\tilde{m}_i-\sigma_1-\sigma_5+\sigma_6+\sigma_8)(-\tilde{m}_i+\sigma_1-\sigma_2-\sigma_5+\sigma_6+\sigma_8) \\ 
&\quad \cdot (-\tilde{m}_i+\sigma_2-\sigma_3-\sigma_5+\sigma_6+\sigma_8)(-\tilde{m}_i-\sigma_3+\sigma_4-\sigma_5+\sigma_6+\sigma_8)  \\ 
&\quad \cdot (-\tilde{m}_i-\sigma_2+\sigma_6-\sigma_7+\sigma_8)(-\tilde{m}_i+\sigma_1-\sigma_3+\sigma_6-\sigma_7+\sigma_8) \\ 
&\quad \cdot (-\tilde{m}_i-\sigma_1+\sigma_2-\sigma_3+\sigma_6-\sigma_7+\sigma_8)(-\tilde{m}_i-\sigma_4+\sigma_6-\sigma_7+\sigma_8)  \\ 
&\quad \cdot (-\tilde{m}_i-\sigma_5+\sigma_6-\sigma_7+\sigma_8)(-\tilde{m}_i-\sigma_3+\sigma_4-\sigma_5+\sigma_6-\sigma_7+\sigma_8) .
\end{align*}

The seventh ring relation is 
\begin{align*}
\prod_{i=1}^n & (-\tilde{m}_i-\sigma_7)(-\tilde{m}_i+\sigma_1-\sigma_7)(-\tilde{m}_i-\sigma_1+\sigma_2-\sigma_7)(-\tilde{m}_i-\sigma_2+\sigma_3-\sigma_7)  \\ 
&\quad \cdot (-\tilde{m}_i+\sigma_3-\sigma_4-\sigma_7)(-\tilde{m}_i-\sigma_2+\sigma_4-\sigma_7)(-\tilde{m}_i+\sigma_1-\sigma_3+\sigma_4-\sigma_7)  \\ 
&\quad \cdot (-\tilde{m}_i-\sigma_1+\sigma_2-\sigma_3+\sigma_4-\sigma_7)(-\tilde{m}_i+\sigma_4-\sigma_5-\sigma_7)(-\tilde{m}_i-\sigma_1+\sigma_5-\sigma_7) \\ 
&\quad \cdot (-\tilde{m}_i+\sigma_1-\sigma_2+\sigma_5-\sigma_7)(-\tilde{m}_i+\sigma_2-\sigma_3+\sigma_5-\sigma_7)  \\ 
&\quad \cdot (-\tilde{m}_i+\sigma_1-\sigma_4+\sigma_5-\sigma_7)(-\tilde{m}_i-\sigma_1+\sigma_2-\sigma_4+\sigma_5-\sigma_7)  \\ 
&\quad \cdot (-\tilde{m}_i-\sigma_2+\sigma_3-\sigma_4+\sigma_5-\sigma_7)(-\tilde{m}_i+\sigma_5-\sigma_6-\sigma_7)(-\tilde{m}_i+\sigma_6-\sigma_7) \\ 
&\quad \cdot (-\tilde{m}_i-\sigma_1+\sigma_6-\sigma_7)(-\tilde{m}_i+\sigma_1-\sigma_2+\sigma_6-\sigma_7)(-\tilde{m}_i+\sigma_2-\sigma_3+\sigma_6-\sigma_7)  \\ 
&\quad \cdot (-\tilde{m}_i+\sigma_2-\sigma_4+\sigma_6-\sigma_7)(-\tilde{m}_i-\sigma_1+\sigma_3-\sigma_4+\sigma_6-\sigma_7)  \\ 
&\quad \cdot (-\tilde{m}_i+\sigma_1-\sigma_2+\sigma_3-\sigma_4+\sigma_6-\sigma_7)(-\tilde{m}_i-\sigma_3+\sigma_4+\sigma_6-\sigma_7)  \\ 
&\quad \cdot (-\tilde{m}_i+\sigma_1-\sigma_5+\sigma_6-\sigma_7)(-\tilde{m}_i-\sigma_1+\sigma_2-\sigma_5+\sigma_6-\sigma_7)  \\ 
&\quad \cdot (-\tilde{m}_i-\sigma_2+\sigma_3-\sigma_5+\sigma_6-\sigma_7)(-\tilde{m}_i-\sigma_1+\sigma_4-\sigma_5+\sigma_6-\sigma_7) \\ 
&\quad \cdot (-\tilde{m}_i+\sigma_1-\sigma_2+\sigma_4-\sigma_5+\sigma_6-\sigma_7)(-\tilde{m}_i+\sigma_2-\sigma_3+\sigma_4-\sigma_5+\sigma_6-\sigma_7) \\ 
&\quad \cdot (-\tilde{m}_i-\sigma_4+\sigma_5+\sigma_6-\sigma_7)(-\tilde{m}_i-\sigma_5+2\sigma_6-\sigma_7)(\tilde{m}_i-\sigma_6+2\sigma_7)^2 \\ 
&\quad \cdot (-\tilde{m}_i+\sigma_2-\sigma_7-\sigma_8)(-\tilde{m}_i-\sigma_1+\sigma_3-\sigma_7-\sigma_8)(-\tilde{m}_i+\sigma_1-\sigma_2+\sigma_3-\sigma_7-\sigma_8) \\ 
&\quad \cdot (-\tilde{m}_i+\sigma_4-\sigma_7-\sigma_8)(-\tilde{m}_i+\sigma_5-\sigma_7-\sigma_8)(-\tilde{m}_i+\sigma_3-\sigma_4+\sigma_5-\sigma_7-\sigma_8)  \\ 
&\quad \cdot (-\tilde{m}_i+\sigma_1+\sigma_6-\sigma_7-\sigma_8)(-\tilde{m}_i-\sigma_1+\sigma_2+\sigma_6-\sigma_7-\sigma_8) \\ 
&\quad \cdot (-\tilde{m}_i-\sigma_2+\sigma_3+\sigma_6-\sigma_7-\sigma_8)(-\tilde{m}_i+\sigma_3-\sigma_4+\sigma_6-\sigma_7-\sigma_8)  \\ 
&\quad \cdot (-\tilde{m}_i+\sigma_3-\sigma_5+\sigma_6-\sigma_7-\sigma_8)(-\tilde{m}_i+\sigma_4-\sigma_5+\sigma_6-\sigma_7-\sigma_8) \\ 
&\quad \cdot (-\tilde{m}_i-\sigma_1-\sigma_7+\sigma_8)(-\tilde{m}_i+\sigma_1-\sigma_2-\sigma_7+\sigma_8)(-\tilde{m}_i+\sigma_2-\sigma_3-\sigma_7+\sigma_8) \\ 
&\quad \cdot (-\tilde{m}_i-\sigma_3+\sigma_4-\sigma_7+\sigma_8)(-\tilde{m}_i-\sigma_3+\sigma_5-\sigma_7+\sigma_8)  \\ 
&\quad \cdot (-\tilde{m}_i-\sigma_4+\sigma_5-\sigma_7+\sigma_8)(-\tilde{m}_i-\sigma_2+\sigma_6-\sigma_7+\sigma_8)  \\ 
&\quad \cdot (-\tilde{m}_i+\sigma_1-\sigma_3+\sigma_6-\sigma_7+\sigma_8)(-\tilde{m}_i-\sigma_1+\sigma_2-\sigma_3+\sigma_6-\sigma_7+\sigma_8)  \\ 
&\quad \cdot (-\tilde{m}_i-\sigma_4+\sigma_6-\sigma_7+\sigma_8)(-\tilde{m}_i-\sigma_5+\sigma_6-\sigma_7+\sigma_8)  \\ 
&\quad \cdot (-\tilde{m}_i-\sigma_3+\sigma_4-\sigma_5+\sigma_6-\sigma_7+\sigma_8) \\
\end{align*}
\begin{align*}
& =\prod_{i=1}^n 
(\tilde{m}_i+\sigma_6-2\sigma_7)^2(\sigma_7-\tilde{m}_i)(-\tilde{m}_i-\sigma_1+\sigma_7)(-\tilde{m}_i+\sigma_1-\sigma_2+\sigma_7)  \\ 
&\quad \cdot (-\tilde{m}_i+\sigma_2-\sigma_3+\sigma_7)(-\tilde{m}_i+\sigma_2-\sigma_4+\sigma_7)(-\tilde{m}_i-\sigma_1+\sigma_3-\sigma_4+\sigma_7)  \\ 
&\quad \cdot (-\tilde{m}_i+\sigma_1-\sigma_2+\sigma_3-\sigma_4+\sigma_7)(-\tilde{m}_i-\sigma_3+\sigma_4+\sigma_7)(-\tilde{m}_i+\sigma_1-\sigma_5+\sigma_7)  \\ 
&\quad \cdot (-\tilde{m}_i-\sigma_1+\sigma_2-\sigma_5+\sigma_7)(-\tilde{m}_i-\sigma_2+\sigma_3-\sigma_5+\sigma_7)  \\ 
&\quad \cdot (-\tilde{m}_i-\sigma_1+\sigma_4-\sigma_5+\sigma_7)(-\tilde{m}_i+\sigma_1-\sigma_2+\sigma_4-\sigma_5+\sigma_7)  \\ 
&\quad \cdot (-\tilde{m}_i+\sigma_2-\sigma_3+\sigma_4-\sigma_5+\sigma_7)(-\tilde{m}_i-\sigma_4+\sigma_5+\sigma_7)(-\tilde{m}_i+\sigma_5-2\sigma_6+\sigma_7)  \\ 
&\quad \cdot (-\tilde{m}_i-\sigma_6+\sigma_7)(-\tilde{m}_i+\sigma_1-\sigma_6+\sigma_7)(-\tilde{m}_i-\sigma_1+\sigma_2-\sigma_6+\sigma_7)  \\ 
&\quad \cdot (-\tilde{m}_i-\sigma_2+\sigma_3-\sigma_6+\sigma_7)(-\tilde{m}_i+\sigma_3-\sigma_4-\sigma_6+\sigma_7) \\ 
&\quad \cdot (-\tilde{m}_i-\sigma_2+\sigma_4-\sigma_6+\sigma_7)(-\tilde{m}_i+\sigma_1-\sigma_3+\sigma_4-\sigma_6+\sigma_7)  \\ 
&\quad \cdot (-\tilde{m}_i-\sigma_1+\sigma_2-\sigma_3+\sigma_4-\sigma_6+\sigma_7)(-\tilde{m}_i+\sigma_4-\sigma_5-\sigma_6+\sigma_7)  \\ 
&\quad \cdot (-\tilde{m}_i-\sigma_1+\sigma_5-\sigma_6+\sigma_7)(-\tilde{m}_i+\sigma_1-\sigma_2+\sigma_5-\sigma_6+\sigma_7) \\ 
&\quad \cdot (-\tilde{m}_i+\sigma_2-\sigma_3+\sigma_5-\sigma_6+\sigma_7)(-\tilde{m}_i+\sigma_1-\sigma_4+\sigma_5-\sigma_6+\sigma_7)  \\ 
&\quad \cdot (-\tilde{m}_i-\sigma_1+\sigma_2-\sigma_4+\sigma_5-\sigma_6+\sigma_7)(-\tilde{m}_i-\sigma_2+\sigma_3-\sigma_4+\sigma_5-\sigma_6+\sigma_7) \\ 
&\quad \cdot (-\tilde{m}_i-\sigma_5+\sigma_6+\sigma_7)(-\tilde{m}_i+\sigma_1+\sigma_7-\sigma_8)(-\tilde{m}_i-\sigma_1+\sigma_2+\sigma_7-\sigma_8)  \\ 
&\quad \cdot (-\tilde{m}_i-\sigma_2+\sigma_3+\sigma_7-\sigma_8)(-\tilde{m}_i+\sigma_3-\sigma_4+\sigma_7-\sigma_8)  \\ 
&\quad \cdot (-\tilde{m}_i+\sigma_3-\sigma_5+\sigma_7-\sigma_8)(-\tilde{m}_i+\sigma_4-\sigma_5+\sigma_7-\sigma_8)  \\ 
&\quad \cdot (-\tilde{m}_i+\sigma_2-\sigma_6+\sigma_7-\sigma_8)(-\tilde{m}_i-\sigma_1+\sigma_3-\sigma_6+\sigma_7-\sigma_8)  \\ 
&\quad \cdot (-\tilde{m}_i+\sigma_1-\sigma_2+\sigma_3-\sigma_6+\sigma_7-\sigma_8)(-\tilde{m}_i+\sigma_4-\sigma_6+\sigma_7-\sigma_8)  \\ 
&\quad \cdot (-\tilde{m}_i+\sigma_5-\sigma_6+\sigma_7-\sigma_8)(-\tilde{m}_i+\sigma_3-\sigma_4+\sigma_5-\sigma_6+\sigma_7-\sigma_8)  \\ 
&\quad \cdot (-\tilde{m}_i-\sigma_2+\sigma_7+\sigma_8)(-\tilde{m}_i+\sigma_1-\sigma_3+\sigma_7+\sigma_8)(-\tilde{m}_i-\sigma_1+\sigma_2-\sigma_3+\sigma_7+\sigma_8) \\ 
&\quad \cdot (-\tilde{m}_i-\sigma_4+\sigma_7+\sigma_8)(-\tilde{m}_i-\sigma_5+\sigma_7+\sigma_8)(-\tilde{m}_i-\sigma_3+\sigma_4-\sigma_5+\sigma_7+\sigma_8) \\ 
&\quad \cdot (-\tilde{m}_i-\sigma_1-\sigma_6+\sigma_7+\sigma_8)(-\tilde{m}_i+\sigma_1-\sigma_2-\sigma_6+\sigma_7+\sigma_8) \\ 
&\quad \cdot (-\tilde{m}_i+\sigma_2-\sigma_3-\sigma_6+\sigma_7+\sigma_8)(-\tilde{m}_i-\sigma_3+\sigma_4-\sigma_6+\sigma_7+\sigma_8)  \\ 
&\quad \cdot (-\tilde{m}_i-\sigma_3+\sigma_5-\sigma_6+\sigma_7+\sigma_8)(-\tilde{m}_i-\sigma_4+\sigma_5-\sigma_6+\sigma_7+\sigma_8) .
\end{align*}

The eighth ring relation is 
\begin{align*}
\prod_{i=1}^n &
(-\tilde{m}_i+\sigma_1-\sigma_8)(-\tilde{m}_i+\sigma_2-\sigma_8)(-\tilde{m}_i-\sigma_1+\sigma_2-\sigma_8) \\ 
&\quad \cdot (-\tilde{m}_i-\sigma_1+\sigma_3-\sigma_8)(-\tilde{m}_i-\sigma_2+\sigma_3-\sigma_8)(-\tilde{m}_i+\sigma_1-\sigma_2+\sigma_3-\sigma_8)  \\ 
&\quad \cdot (-\tilde{m}_i+\sigma_1+\sigma_3-\sigma_4-\sigma_8)(-\tilde{m}_i-\sigma_1+\sigma_2+\sigma_3-\sigma_4-\sigma_8)  \\ 
&\quad \cdot (-\tilde{m}_i-\sigma_2+2\sigma_3-\sigma_4-\sigma_8)(-\tilde{m}_i-\sigma_1+\sigma_4-\sigma_8)(-\tilde{m}_i+\sigma_1-\sigma_2+\sigma_4-\sigma_8)  \\ 
&\quad \cdot (-\tilde{m}_i+\sigma_2-\sigma_3+\sigma_4-\sigma_8)(-\tilde{m}_i+\sigma_3-\sigma_5-\sigma_8)(-\tilde{m}_i+\sigma_1+\sigma_4-\sigma_5-\sigma_8) \\ 
&\quad \cdot (-\tilde{m}_i-\sigma_1+\sigma_2+\sigma_4-\sigma_5-\sigma_8)(-\tilde{m}_i-\sigma_2+\sigma_3+\sigma_4-\sigma_5-\sigma_8)\\ 
&\quad \cdot (-\tilde{m}_i+\sigma_5-\sigma_8)(-\tilde{m}_i+\sigma_2-\sigma_4+\sigma_5-\sigma_8)(-\tilde{m}_i-\sigma_1+\sigma_3-\sigma_4+\sigma_5-\sigma_8)  \\ 
&\quad \cdot (-\tilde{m}_i+\sigma_1-\sigma_2+\sigma_3-\sigma_4+\sigma_5-\sigma_8)(-\tilde{m}_i+\sigma_3-\sigma_6-\sigma_8) \\ 
&\quad \cdot (-\tilde{m}_i+\sigma_4-\sigma_6-\sigma_8)(-\tilde{m}_i+\sigma_1+\sigma_5-\sigma_6-\sigma_8)  \\ 
&\quad \cdot (-\tilde{m}_i-\sigma_1+\sigma_2+\sigma_5-\sigma_6-\sigma_8)(-\tilde{m}_i-\sigma_2+\sigma_3+\sigma_5-\sigma_6-\sigma_8)  \\ 
&\quad \cdot (-\tilde{m}_i+\sigma_3-\sigma_4+\sigma_5-\sigma_6-\sigma_8)(-\tilde{m}_i+\sigma_6-\sigma_8)(-\tilde{m}_i+\sigma_3-\sigma_4+\sigma_6-\sigma_8)  \\ 
&\quad \cdot (-\tilde{m}_i+\sigma_2-\sigma_5+\sigma_6-\sigma_8)(-\tilde{m}_i-\sigma_1+\sigma_3-\sigma_5+\sigma_6-\sigma_8)  \\ 
&\quad \cdot (-\tilde{m}_i+\sigma_1-\sigma_2+\sigma_3-\sigma_5+\sigma_6-\sigma_8)(-\tilde{m}_i+\sigma_4-\sigma_5+\sigma_6-\sigma_8)  \\ 
&\quad \cdot (-\tilde{m}_i+\sigma_2-\sigma_7-\sigma_8)(-\tilde{m}_i-\sigma_1+\sigma_3-\sigma_7-\sigma_8)(-\tilde{m}_i+\sigma_1-\sigma_2+\sigma_3-\sigma_7-\sigma_8)  \\ 
&\quad \cdot (-\tilde{m}_i+\sigma_4-\sigma_7-\sigma_8)(-\tilde{m}_i+\sigma_5-\sigma_7-\sigma_8)(-\tilde{m}_i+\sigma_3-\sigma_4+\sigma_5-\sigma_7-\sigma_8) \\ 
&\quad \cdot (-\tilde{m}_i+\sigma_1+\sigma_6-\sigma_7-\sigma_8)(-\tilde{m}_i-\sigma_1+\sigma_2+\sigma_6-\sigma_7-\sigma_8)  \\ 
&\quad \cdot (-\tilde{m}_i-\sigma_2+\sigma_3+\sigma_6-\sigma_7-\sigma_8)(-\tilde{m}_i+\sigma_3-\sigma_4+\sigma_6-\sigma_7-\sigma_8) \\ 
&\quad \cdot (-\tilde{m}_i+\sigma_3-\sigma_5+\sigma_6-\sigma_7-\sigma_8)(-\tilde{m}_i+\sigma_4-\sigma_5+\sigma_6-\sigma_7-\sigma_8)  \\ 
&\quad \cdot (-\tilde{m}_i+\sigma_1+\sigma_7-\sigma_8)(-\tilde{m}_i-\sigma_1+\sigma_2+\sigma_7-\sigma_8)(-\tilde{m}_i-\sigma_2+\sigma_3+\sigma_7-\sigma_8)  \\ 
&\quad \cdot (-\tilde{m}_i+\sigma_3-\sigma_4+\sigma_7-\sigma_8)(-\tilde{m}_i+\sigma_3-\sigma_5+\sigma_7-\sigma_8)(-\tilde{m}_i+\sigma_4-\sigma_5+\sigma_7-\sigma_8)  \\ 
&\quad \cdot (-\tilde{m}_i+\sigma_2-\sigma_6+\sigma_7-\sigma_8)(-\tilde{m}_i-\sigma_1+\sigma_3-\sigma_6+\sigma_7-\sigma_8)  \\ 
&\quad \cdot (-\tilde{m}_i+\sigma_1-\sigma_2+\sigma_3-\sigma_6+\sigma_7-\sigma_8)(-\tilde{m}_i+\sigma_4-\sigma_6+\sigma_7-\sigma_8)  \\ 
&\quad \cdot (-\tilde{m}_i+\sigma_5-\sigma_6+\sigma_7-\sigma_8)(-\tilde{m}_i+\sigma_3-\sigma_4+\sigma_5-\sigma_6+\sigma_7-\sigma_8)(\tilde{m}_i-\sigma_3+2\sigma_8)^2
\\
\end{align*}
\begin{align*}
& = \prod_{i=1}^n
(\tilde{m}_i+\sigma_3-2\sigma_8)^2(-\tilde{m}_i-\sigma_1+\sigma_8)(-\tilde{m}_i-\sigma_2+\sigma_8)(-\tilde{m}_i+\sigma_1-\sigma_2+\sigma_8)  \\ 
&\quad \cdot (-\tilde{m}_i+\sigma_1-\sigma_3+\sigma_8)(-\tilde{m}_i+\sigma_2-\sigma_3+\sigma_8)(-\tilde{m}_i-\sigma_1+\sigma_2-\sigma_3+\sigma_8)  \\ 
&\quad \cdot (-\tilde{m}_i+\sigma_1-\sigma_4+\sigma_8)(-\tilde{m}_i-\sigma_1+\sigma_2-\sigma_4+\sigma_8)(-\tilde{m}_i-\sigma_2+\sigma_3-\sigma_4+\sigma_8)  \\ 
&\quad \cdot (-\tilde{m}_i+\sigma_2-2\sigma_3+\sigma_4+\sigma_8)(-\tilde{m}_i-\sigma_1-\sigma_3+\sigma_4+\sigma_8)  \\ 
&\quad \cdot (-\tilde{m}_i+\sigma_1-\sigma_2-\sigma_3+\sigma_4+\sigma_8)(-\tilde{m}_i-\sigma_5+\sigma_8)(-\tilde{m}_i-\sigma_2+\sigma_4-\sigma_5+\sigma_8)  \\ 
&\quad \cdot (-\tilde{m}_i+\sigma_1-\sigma_3+\sigma_4-\sigma_5+\sigma_8)(-\tilde{m}_i-\sigma_1+\sigma_2-\sigma_3+\sigma_4-\sigma_5+\sigma_8)  \\ 
&\quad \cdot (-\tilde{m}_i-\sigma_3+\sigma_5+\sigma_8)(-\tilde{m}_i-\sigma_1-\sigma_4+\sigma_5+\sigma_8)  \\ 
&\quad \cdot (-\tilde{m}_i+\sigma_1-\sigma_2-\sigma_4+\sigma_5+\sigma_8)(-\tilde{m}_i+\sigma_2-\sigma_3-\sigma_4+\sigma_5+\sigma_8)  \\ 
&\quad \cdot (-\tilde{m}_i-\sigma_6+\sigma_8)(-\tilde{m}_i-\sigma_3+\sigma_4-\sigma_6+\sigma_8)(-\tilde{m}_i-\sigma_2+\sigma_5-\sigma_6+\sigma_8)  \\ 
&\quad \cdot (-\tilde{m}_i+\sigma_1-\sigma_3+\sigma_5-\sigma_6+\sigma_8)(-\tilde{m}_i-\sigma_1+\sigma_2-\sigma_3+\sigma_5-\sigma_6+\sigma_8)  \\ 
&\quad \cdot (-\tilde{m}_i-\sigma_4+\sigma_5-\sigma_6+\sigma_8)(-\tilde{m}_i-\sigma_3+\sigma_6+\sigma_8)(-\tilde{m}_i-\sigma_4+\sigma_6+\sigma_8)  \\ 
&\quad \cdot (-\tilde{m}_i-\sigma_1-\sigma_5+\sigma_6+\sigma_8)(-\tilde{m}_i+\sigma_1-\sigma_2-\sigma_5+\sigma_6+\sigma_8)  \\ 
&\quad \cdot (-\tilde{m}_i+\sigma_2-\sigma_3-\sigma_5+\sigma_6+\sigma_8)(-\tilde{m}_i-\sigma_3+\sigma_4-\sigma_5+\sigma_6+\sigma_8)  \\ 
&\quad \cdot (-\tilde{m}_i-\sigma_1-\sigma_7+\sigma_8)(-\tilde{m}_i+\sigma_1-\sigma_2-\sigma_7+\sigma_8)(-\tilde{m}_i+\sigma_2-\sigma_3-\sigma_7+\sigma_8)  \\ 
&\quad \cdot (-\tilde{m}_i-\sigma_3+\sigma_4-\sigma_7+\sigma_8)(-\tilde{m}_i-\sigma_3+\sigma_5-\sigma_7+\sigma_8)  \\ 
&\quad \cdot (-\tilde{m}_i-\sigma_4+\sigma_5-\sigma_7+\sigma_8)(-\tilde{m}_i-\sigma_2+\sigma_6-\sigma_7+\sigma_8)  \\ 
&\quad \cdot (-\tilde{m}_i+\sigma_1-\sigma_3+\sigma_6-\sigma_7+\sigma_8)(-\tilde{m}_i-\sigma_1+\sigma_2-\sigma_3+\sigma_6-\sigma_7+\sigma_8)  \\ 
&\quad \cdot (-\tilde{m}_i-\sigma_4+\sigma_6-\sigma_7+\sigma_8)(-\tilde{m}_i-\sigma_5+\sigma_6-\sigma_7+\sigma_8)  \\ 
&\quad \cdot (-\tilde{m}_i-\sigma_3+\sigma_4-\sigma_5+\sigma_6-\sigma_7+\sigma_8)(-\tilde{m}_i-\sigma_2+\sigma_7+\sigma_8)  \\ 
&\quad \cdot (-\tilde{m}_i+\sigma_1-\sigma_3+\sigma_7+\sigma_8)(-\tilde{m}_i-\sigma_1+\sigma_2-\sigma_3+\sigma_7+\sigma_8)  \\ 
&\quad \cdot (-\tilde{m}_i-\sigma_4+\sigma_7+\sigma_8)(-\tilde{m}_i-\sigma_5+\sigma_7+\sigma_8)(-\tilde{m}_i-\sigma_3+\sigma_4-\sigma_5+\sigma_7+\sigma_8)  \\ 
&\quad \cdot (-\tilde{m}_i-\sigma_1-\sigma_6+\sigma_7+\sigma_8)(-\tilde{m}_i+\sigma_1-\sigma_2-\sigma_6+\sigma_7+\sigma_8)  \\ 
&\quad \cdot (-\tilde{m}_i+\sigma_2-\sigma_3-\sigma_6+\sigma_7+\sigma_8)(-\tilde{m}_i-\sigma_3+\sigma_4-\sigma_6+\sigma_7+\sigma_8)  \\ 
&\quad \cdot (-\tilde{m}_i-\sigma_3+\sigma_5-\sigma_6+\sigma_7+\sigma_8)(-\tilde{m}_i-\sigma_4+\sigma_5-\sigma_6+\sigma_7+\sigma_8).
\end{align*}

\subsection{Pure gauge theory}

In this part we will consider the mirror to the pure $E_8$ gauge
theory.  For brevity, we will not rewrite the superpotential here,
explicitly omitting 
$Y$ fields, but instead merely refer to the expression~(\ref{eq:e8:mirror-sup})
given earlier,
leaving the reader to omit $Y$ fields.

Now, let us consider the critical locus of the superpotential above.
For each root $\mu$, the fields $X_{\mu}$ and $X_{-\mu}$ appear paired with
opoosite signs coupling to each $\sigma$.  Therefore, one impliciation of
the derivatives
\begin{displaymath}
\frac{\partial W}{\partial X_{\mu} } \: = \: 0
\end{displaymath}
is that, on the critical locus,
\begin{equation}
X_{\mu} \: = \: - X_{- \mu}.
\end{equation}
(Furthermore, on the critical locus, each $X_{\mu}$ is determined by
$\sigma$s.)
Next, each derivative
\begin{displaymath}
\frac{\partial W}{\partial \sigma_a}
\end{displaymath}
is a product of ratios of the form
\begin{displaymath}
\frac{ X_{\mu} }{ X_{-\mu} } \: = \: -1.
\end{displaymath}
It is straightforward to check that in the superpotential above that
each $\sigma_a$ is multiplied by an even number of such ratios
({\it i.e.} the number of $Z$'s is a multiple of four).  Specifically,
for each $\sigma$, the sum of the absolute values of the coefficients of
the $Z$'s multiplying it is $116 = 4 \cdot 29$.
Thus, the constraint implied by the $\sigma$'s is automatically satisfied.

As a result, following the same analysis as in \cite{GuSharpe} and
previous sections, we see that 
the critical locus is nonempty, and in fact
is determined by eight $\sigma$s.  In other words, at the level of these
topological field theory computations, we have evidence that the pure
supersymmetric $E_8$ gauge theory in two dimensions flows in the IR
to a theory of eight free twisted chiral superfields.

\section{Conclusions}

In this paper we applied the recent nonabelian mirrors proposal \cite{GuSharpe}
to examples of two-dimensional A-twisted gauge theories with 
exceptional gauge groups $G_2, F_4, E_{6,7,8}$.
In each case, we explicitly compute the proposed mirror Landau-Ginzburg
orbifold and derived the Coulomb ring relations (the analogue of
quantum cohomology ring relations).
In the cases of the $G_2$ and $F_4$ 
gauge theories, we studied the action of the Weyl group on the critical
locus equations, which allowed us to perform consistency checks on the
results here, and in the case of $G_2$, performed a detailed analysis
of Weyl group orbits of the critical locis (vacua).

We also studied pure gauge theories with each gauge group, and provided
evidence (at the level of these topological-field-theory-type computations)
that each pure gauge theory (with simply-connected gauge group) flows in the
IR to a free theory of as many twisted chiral multiplets as the rank of
the gauge group.

\section{Acknowledgements}

We would like to thank  C.~Closset, I.~Melnikov, and A.~Sarshar 
for useful conversations. 
E.S. was partially supported by NSF grant PHY-1720321.

\end{document}